\documentclass[tight,classic,final,pageclassic,orcish,nobm,amsmath,amssymb]{iopart}

\usepackage{graphicx,graphicx,epsfig} 
\usepackage{iopams}  
\usepackage{bm}
\usepackage{color}
\usepackage[linktocpage=true]{hyperref}
\newcommand{\nn}{\nonumber}
\newcommand{\beq}{\begin{eqnarray}}
\newcommand{\eeq}{\end{eqnarray}}

\begin{document}

\topical[Symmetry Protected Topological Superfluid $^3$He-B]{Symmetry Protected Topological Superfluid $^3$He-B}

\author{Takeshi Mizushima}
\address{Department of Physics, Okayama University, Okayama 700-8530, Japan}
\ead{mizushima@mp.okayama-u.ac.jp}
\author{Yasumasa Tsutsumi}
\address{Condensed Matter Theory Laboratory, RIKEN, Wako, Saitama 351-0198, Japan}
\ead{y.tsutsumi@riken.jp}
\author{Masatoshi Sato}
\address{Department of Applied Physics, Nagoya University, 464-8603, Japan}
\ead{msato@nuap.nagoya-u.ac.jp}
\author{Kazushige Machida}
\address{Department of Physics, Okayama University, Okayama 700-8530, Japan}
\ead{machida@mp.okayama-u.ac.jp}

\begin{abstract}

Owing to the richness of symmetry and well-established knowledge on the bulk superfluidity, the superfluid $^3$He has offered a prototypical system to study intertwining of topology and symmetry. This article reviews recent progress in understanding the topological superfluidity of $^3$He in a multifaceted manner, including symmetry consideration, the Jackiw-Rebbi's index theorem, and the quasiclassical theory. Special focus is placed on the symmetry protected topological superfuidity of the $^3$He-B confined in a slab geometry. The $^3$He-B under a magnetic field is separated to two different sub-phases: The symmetry protected topological phase and non-topological phase. The former phase is characterized by the existence of symmetry protected Majorana fermions. The topological phase transition between them is triggered off by the spontaneous breaking of a hidden discrete symmetry. The critical field is quantitatively determined from the microscopic calculation that takes account of magnetic dipole interaction of $^3$He nucleus. It is also demonstrated that odd-frequency even-parity Cooper pair amplitudes are emergent in low-lying quasiparticles. The key ingredients, symmetry protected Majorana fermions and odd-frequency pairing, bring an important consequence that the coupling of the surface states to an applied field is prohibited by the hidden discrete symmetry, while the topological phase transition with the spontaneous symmetry breaking is accompanied by anomalous enhancement and anisotropic quantum criticality of surface spin susceptibility. We also illustrate common topological features between topological crystalline superconductors and symmetry protected topological superfluids, taking UPt$_3$ and Rashba superconductors as examples.

\end{abstract}



\tableofcontents 

\maketitle

\section{Introduction}

\begin{figure}[tb!]
\begin{center}
\includegraphics[width=60mm]{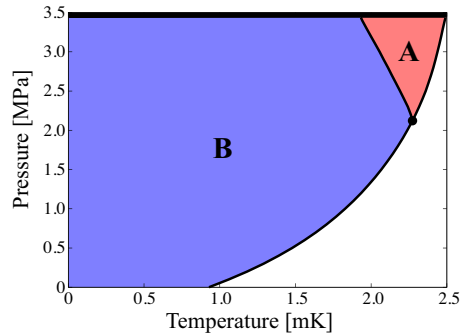}
\end{center}
\caption{Superfluid phase diagram of the bulk $^3$He in the plane of the temperature $T$ and pressure $P$.}
\label{fig:phase_bulk}
\end{figure}

Superfluid $^3$He has offered a prototypical system to study the intertwining of topology and symmetry, since the bulk properties has been well established as a spin-triplet $p$-wave superfluid. The $^3$He system that is composed of neutral fermions with nuclear spin $1/2$ remains liquid phase down to zero temperatures and possesses the typical properties of strongly correlated Fermi liquid, where the elementary excitations can be characterized by the concept of quasiparticles. The quantum liquid preserves the continuous rotational symmetry in spin and coordinate spaces independently. The huge symmetry group, $G$, held in the normal phase is responsible for the various types of spontaneous symmetry breaking that trigger off multiple superfluid phase transitions. As shown in Fig.~\ref{fig:phase_bulk}, two distinctive superfluid phases, called the A- and B-phases, are energetically competitive in the bulk $^3$He~\cite{leggettRMP,Wheatley,vollhardt}. The pairing symmetry of the B-phase has been established as the Balian-Werthamer state~\cite{bw} which is the spin-triplet $p$-wave pairing with the time-reversal symmetry, while the A-phase is the Anderson-Brinkman-Morel state~\cite{abm1,abm2,abm3} that spontaneously breaks the time-reversal symmetry. 

Superconductors and superfluids are generally composed of two key ingredients that are quasiparticles and Cooper pairs, where the former is the fermionic degrees of freedom of superconducting states and the latter is the order parameter associated with spontaneously breaking symmetry. It has been recognized that the topological property of these ingredients is intertwined with the remaining symmetry $H$ and spontaneously broken symmetry $R=G/H$. The broken symmetry $R$ that determines the order parameter degenerate space is responsible for the topological excitations of the {\it bosonic} ingredients that generate the intrinsic textural structure of the order parameter in the coordinate space. The possible types are classified by the homotopy group $\pi _{d}(R)$ that defines the group of homotopy classes of maps from a $d$-dimensional coordinate to the target space $R$, where the boundary of the $d$-dimensional cube is taken to be identical on $R$~\cite{volovik1976,volovik1977,mermin1978,mermin1979,nakaharabook,salomaaRMP,volovikbook1992,volovikbook}. The typical topological objects described by $\pi _{d}$ are vortices for $d=1$, monopoles and two-dimensional skyrmions for $d=2$, and skyrmions for $d=3$. In contrast to {\it conventional} superconductors and superfluid $^4$He, the huge degenerate space $R$ of the superfluid $^3$He ensures a considerably more complicated structure of topological excitations. This includes continuous vortices without a singularity~\cite{mermin1976,anderson1977}, nonaxisymmetric vortices~\cite{thuneberg1986,salomaa1986}, half-quantum vortices~\cite{volovik1976,cross}, skyrmions as Shanker monopoles~\cite{shanker,nakahara1987} and many other topological excitations~\cite{vollhardt,salomaaRMP,volovikbook1992,volovikbook}.

Apart from topological excitations associated with spontaneous broken symmetry, in this paper, we try to clarify the relation between the remaining symmetry $H$ and nontrivial topological structure of {\it fermionic} excitations. The nontrivial topology hiding in fermions was first uncovered in the quantum Hall state that occurs in two-dimensional electrons under high magnetic fields. This is a new state of matter that can not be explained by the concept of spontaneous symmetry breaking but can be characterized by the topological number. The topological number in this system is the first Chern number or the Thouless-Kohmoto-Nightingale-den Nijs (TKNN) number that measures the ``magnetic flux'' penetrating the magnetic Brillouin zone (BZ)~\cite{tknn,kohmoto}, 
\beq
\nu _{\rm Ch} = \frac{1}{2\pi} \int _{\rm BZ} d^2{\bm k}
\left[ 
\partial _{k_x}\mathcal{A}_y({\bm k})- \partial _{k_y}\mathcal{A}_x({\bm k}) \right].
\eeq
The vector potential is defined as $\mathcal{A}_{\mu} ({\bm k}) = i \sum _{E_n<E_{\rm F}}\langle u_n({\bm k})| \partial _{k_{\mu}}u_n({\bm k})\rangle $ with the wave function in occupied bands $|u_n({\bm k})\rangle$. The single-valuedness of the wave function up to ${\rm U}(1)$ phase requires the Chern number to be an integer value $\nu _{\rm Ch}\in \mathbb{Z}$. In accordance with the Kubo formula, the Hall conductivity $\sigma _{\rm H}$ is characterized by the Chern number as $\sigma _{\rm H} = \nu _{\rm Ch}e^2/h$, which explains the quantization of the Hall conductivity. The Chern number has the another physical meaning that the topological number is equal to the number of gapless edge channels~\cite{hatsugai1,hatsugai2}. This is what is known as the bulk-edge correspondence.

Although the original concept of the topological phase in the quantum Hall system is independent of the symmetry, it has recently been emphasized that the interplay between the topology and symmetry enriches the topological properties of the ordered state~\cite{volovikbook,tanakaJPSJ2012,qiRMP2011,andoJPSJ2013,grinevichJLTP1988,bernevig,fuPRB2007,qiPRB2008,schnyderPRB2008,roy2008,ryuNJP2010,qiPRL2009,satoPRB2009,satoPRBR2010,kitaev2009,volovik2009-1,volovik2009-2,guPRB2009,teoPRB2010,fuPRL2011,pollmannPRB2012,wenPRB2012,mizushimaPRL2012,chiuPRB2013,uenoPRL2013,mizushimaNJP2013,tsutsumiJPSJ2013,ezawa,morimoto,SatoPhysicaE2014,shiozaki2014,senthil2014,silaev2014}. The important step was the proposal of the topological table that categorizes topological insulators and superconductors to ten Altland-Zirnbauer (AZ) symmetry classes in terms of time-reversal, particle-hole, and chiral (sub-lattice) symmetries~\cite{schnyderPRB2008,ryuNJP2010}. 

The topological classification indicates that the superfluid $^3$He-B is categorized to the DIII class and is topologically nontrivial in three spatial dimensions~\cite{schnyderPRB2008,ryuNJP2010}. The nontrivial topology is ensured by the time-reversal symmetry as well as the particle-hole symmetry. The particle-hole symmetry also plays a crucial role in determining the nature of the topologically protected gapless quasiparticles. The remarkable consequence of the topological superconductivity and particle-hole symmetry is that the topologically protected gapless quasiparticles behave as Majorana fermions that are fermions equivalent to their own anti-fermions~\cite{read,semenoff,wilczek,wilczek2014,jackiw2014}. They are not coupled to density fluctuation, but possess the character of Ising spins which are detectable through anisotropic magnetic response~\cite{mizushimaPRL2012,chungPRL2009,nagatoJPSJ2009,shindouPRB2010,volovikJETP2010,mizushimaJLTP2011,silaevPRB2011,mizushimaPRB2012}. 

In accordance with the AZ classification, the topological superfluidity of $^3$He-B is protected by the time-reversal symmetry as well as the particle-hole symmetry. However, this does not necessarily mean that the topological superfluidity is fragile under a time-reversal breaking perturbation, {\it e.g.}, a magnetic field. Additional discrete symmetries arising from spin rotation and mirror reflection may support topologically nontrivial feature even in the presence of a time-reversal breaking perturbation. 

\begin{figure}[tb!]
\begin{center}
\includegraphics[width=60mm]{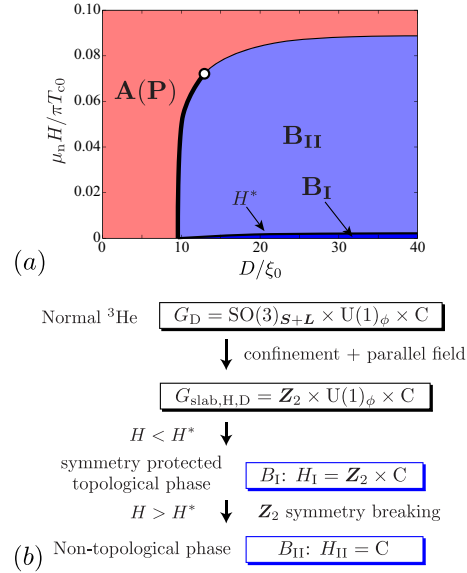}
\end{center}
\caption{(a) Phase diagram of the superfluid $^3$He confined in a slab geometry with a magnetic field parallel to the surface, where the temperature is set to be $T=0.4T_{\rm c0}$ and $D$ and $H$ are the thickness of the slab and the magnitude of an applied field, respectively. The thin (thick) curves is the first (second) order transition line. (b) Spontaneous symmetry breaking in ${\rm B}_{\rm I}$ and ${\rm B}_{\rm II}$. The former is the topological phase protected by the hidden ${\bm Z}_2$ symmetry, while the latter is topologically trivial. The details on symmetry consideration and microscopic calculation are discussed in Secs.~\ref{sec:symm_field} and \ref{sec:hidden} and Sec.~\ref{sec:numerical2}, respectively. }
\label{fig:phase_topo}
\end{figure}

It has recently been demonstrated that $^3$He-B confined in a slab survives as a topological phase in the presence of a weak magnetic field~\cite{mizushimaPRL2012}. Even in the presence of a magnetic field that explicitly breaks the time-reversal symmetry and continuous rotational symmetry in the spin space, the B-phase may hold the hidden ${\bm Z}_2$ symmetry that is the combined discrete symmetry of the time-inversion and joint $\pi$-rotation in spin and orbital spaces. The B-phase under a magnetic field is therefore classified to two phases, the ${\bm Z}_2$ symmetric phase, ${\rm B}_{\rm I}$, and ${\bm Z}_2$ symmetry breaking phase, ${\rm B}_{\rm II}$. The ${\rm B}_{\rm I}$ phase possesses topologically nontrivial superfluidity protected by the hidden ${\bm Z}_2$ symmetry, while the ${\rm B}_{\rm II}$ phase without the hidden ${\bm Z}_2$ symmetry is topologically trivial. The typical phase diagram of $^3$He-B confined in a slab geometry is displayed in Fig.~\ref{fig:phase_topo}. There exists the topological quantum critical point at a weak field, beyond which the hidden ${\bm Z}_2$ symmetry spontaneously breaks and it simultaneously triggers off the topological phase transition. The topological phase transition concomitant with spontaneous symmetry breaking can occur without closing the bulk energy gap, leading to the acquirement of the mass of surface Majorana fermions. It has also recently been predicted that the topological quantum critical point is accompanied by the emergent space-time supersymmetry that is the symmetry between fermion Green's function and spin-spin correlation function~\cite{grover}. 

This review attempts to clarify that the superfluid $^3$He confined in a restricted geometry provides a promising platform to study the interplay between the topology and additional discrete symmetries. In Sec.~\ref{sec:symmetry2}, we start with the summary of the symmetry that is preserved in $^3$He-B confined in a slab geometry. We here extract the hidden ${\bm Z}_2$ symmetry hiding in the huge remaining symmetry of $^3$He-B and categorize the B-phase under a magnetic field to two different phases: The ${\bm Z}_2$ symmetric phase, ${\rm B}_{\rm I}$, and ${\bm Z}_2$ symmetry breaking phase, ${\rm B}_{\rm II}$. The topological superfluidity of $^3$He-B with and without a magnetic field is reviewed in Sec.~\ref{sec:topo}. The generic consequence of the symmetry protected topological phase, ${\rm B}_{\rm I}$, is the emergence of the Majorana fermion that is bounded to the surface. In Sec.~\ref{sec:topo}, we clarify that the emergent Majorana fermion generally yields the Ising anisotropic response to the magnetic field and is robust against the density fluctuation.

\begin{figure}[tb!]
\begin{center}
\includegraphics[width=60mm]{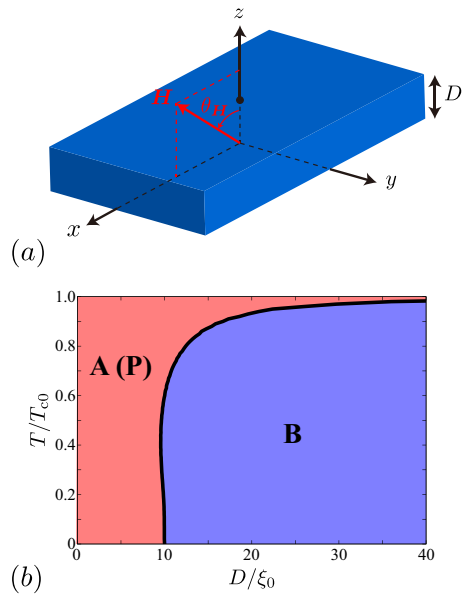}
\end{center}
\caption{(a) Schematic picture of the slab geometry which we consider, where $D$ is the thickness of the sample and ${\bm H}$ is the magnetic field. (b) Superfluid phase diagram of $^3$He confined in a slab geometry at zero pressures as a function of $D/\xi _0$, where $\xi _0$ is the coherence length of the bulk B-phase. The temperature $T$ is scaled by the superfluid transition temperature for the bulk B-phase $T_{\rm c0}$.}
\label{fig:phase_slab}
\end{figure}

Apart from the topological consideration, the low-lying quasiparticles emergent in the surface, interface, and vortices, have been recognized as a family of the Andreev bound state. In Sec.~\ref{sec:andreev}, we illustrate that the low-energy physics of unconventional superconductors and superfluids near a specular surface is in general mapped to a one-dimensional Dirac (or Majorana) equation with a spatially modulated mass term. Therefore, the existence of gapless quasiparticles is discussed in the basis of the Jackiw-Rebbi index theorem~\cite{jackiw}. Since the gapless states exist when the pair potential, namely the mass term, changes its sign~\cite{ohashiJPSJ1995,buchholtzPRB1981,haraPTP1986,hu}, they are expected to ubiquitously appear in various physics systems, such as superconducting junctions~\cite{kashiwayaRRP}, superconducting vortices~\cite{read,volovikJETP1989,kopninPRB1991,volovikJPCM1991,misirpashaev,gurarier,silaevJLTP2010,volovikJETP2011,tewariPRL2007}, unconventional superconductors~\cite{tankaPRL1995,matsumotoJPSJ1995,tanakaPRB1996-2,stone,asanoPRB2004}, and Fulde-Ferrell-Larkin-Ovchinnikov superconductors~\cite{machidaPRB1984,yoshii,mizushimaPRL2005-1,mizushimaPRL2005-2}. In addition, the same physics is shared with the solitons which emerge in polyacetylene~\cite{brazovskii,mertsching,horovitz,takayama,yamamoto,nakaharaPRB1981}, the incommensurate spin-density wave~\cite{machidaPRB1984-2,fujitaJPSJ1984}, and the stripe state in high-$T_{\rm c}$ cuprates~\cite{machida1989}.

In Sec.~\ref{sec:field}, the theory on topology and symmetry in the superfluid $^3$He-B is extended to the situation that takes account of the magnetic field and confinement in equal footing. The topology of the ${\rm B}_{\rm I}$ phase and its consequence are discussed in detail. We also apply the theory to superconducting states in which the rotational symmetry in spin space is generally absent. We will show that the mirror reflection symmetry originating from the crystalline symmetry may protect the topological nontriviality of time-reversal superconductors under a magnetic field. In Sec.~\ref{sec:field}, we will show that the prime candidates are the heavy-fermion superconductor UPt$_3$ and a quasi-one-dimensional Fermi gas with a synthetic gauge field~\cite{mizushimaNJP2013,tsutsumiJPSJ2013}. 

In a slab geometry which we consider here, confinement critically influences the self-consistent surface structure of the superfluid $^3$He~\cite{mizushimaPRB2012,buchholtzPRB1981,haraPTP1986,haraJLTP1988,vorontsovPRB2003,vorontsovPRL2007,nagai}. A strong pair breaking effect which is induced by confinement triggers off the quantum phase transition as displayed in Fig.~\ref{fig:phase_slab}(b). This indicates that the topological phase transition from the time-reversal invariant B-phase to time-reversal breaking A-phase occurs at the critical value of the ``thickness'' even at zero pressures. Several experiments have observed the confinement-induced A-B phase transition in a slab geometry with a thickness comparable to the superfluid coherence length~\cite{freemanPRL1988,freemanPRB1990,xu,kawae,miyawaki,kawasaki,levitin,bennettJLTP2010,levitin2013,levitinPRL2013}. We introduce in Sec.~\ref{sec:numerical} the quasiclassical theory for spin-triplet superconductors and superfluids~\cite{serene}. The quasiclassical theory offers a more tractable way to a quantitative study on microscopic structure and thermodynamics in a restricted geometry. 

The symmetry protected surface bound states have multifaceted properties, the Andreev bound state, odd-frequency pairing, and Majorana fermions~\cite{tanakaJPSJ2012,higashitaniPRB2012,tsutsumiJPSJ2012,dainoPRB2012,asanoPRB2013}. Using the quasiclassical theory, we will illustrate in Sec.~\ref{sec:numerical} the aspect of surface bound states as odd-frequency pairing. In Sec.~\ref{sec:exact2}, the relation between odd-frequency pairing and Majorana fermions is discussed in the basis of the exact solution of the quasiclassical transport equation. We also demonstrate that the odd-frequency pairing is responsible for the anomalous enhancement of surface spin susceptibility and surface spin current, where the former is consistent with that obtained from the concept of symmetry protected Majorana fermions in Secs.~\ref{sec:topo} and \ref{sec:field}.

The key parameter that characterizes the ${\rm B}_{\rm I}$ and ${\rm B}_{\rm II}$ phases is the $\hat{\bm n}$-vector. The $\hat{\bm n}$-vector is the order parameter of the B-phase associated with the broken spin-orbit symmetry, and the orientation favored in the equilibrium is determined by the competition of the magnetic Zeeman effect and magnetic dipole-dipole interaction, where the latter originates in the magnetic dipole moment of $^3$He nuclei. Carrying out full numerical calculation of the quasiclassical transport equation, we will present in Sec.~\ref{sec:numerical2} the complete phase diagram of $^3$He confined in a slab geometry in the presence of a magnetic field. The phase diagram has a critical field at which the topological phase transition concomitant with spontaneous symmetry breaking occurs. The possible ways to detect the topological superfluidity and Majorana fermions are discussed in Sec.~\ref{sec:detecting} and Sec.~\ref{sec:summary} is devoted to making a summary on the symmetry protected topological superfluidity of $^3$He in a restricted geometry.

We shall not discuss the topologically protected Majorana fermions that are bound to the topological defects, such as vortices and domain walls. The more comprehensive review will be given elsewhere.

Throughout this paper, we set $\hbar = k_{\rm B} = 1$ and the repeated Greek (Roman) indices imply the sum over $x, y, z$ (spins $\uparrow$ and $\downarrow$). The Pauli matrices in spin and particle-hole (Nambu) spaces are denoted by $\sigma _{\mu}$ and $\tau _{\mu}$, respectively.

\section{Symmetry and order parameters of $^3$He-B}
\label{sec:symmetry2}

The superfluid phase diagram of the bulk $^3$He is displayed in Fig.~\ref{fig:phase_bulk}, where the ground state phases are occupied by the so-called B-phase and A-phase. The B-phase known as the Balian-Werthamer (BW) state~\cite{bw} preserves the maximal symmetry group of the subset of $G$, while the A-phase which appears in the high pressure and high temperature region spontaneously breaks time-reversal symmetry. In the former case, an isotropic energy gap is opened on the Fermi surface, which gains the maximal condensation energy. In contrast, the latter phase known as the Anderson-Brinkman-Morel (ABM) state~\cite{abm1,abm2,abm3} has point nodes on the south and north poles of three-dimensional Fermi sphere and is stabilized by the spin-fluctuation feedback mechanism in the pairing interaction, which can be enhanced by increasing the pressure. Hence, these two states are thermodynamically distinguishable and competitive with each other. 

In this paper, we mainly concentrate our attention to the symmetry and topology of the B-phase. The topological superfluidity is ensured by the time-reversal symmetry which is held by the B-phase in the absence of time-reversal symmetry breaking perturbation, such as a magnetic field. Since the time-reversal symmetry is spontaneously broken in the A-phase, the topological properties are distinguishable from those in the B-phase. The topological phase in the A-phase is characterized by the mirror Chern number and non-abelian Majorana fermions are hosted by an integer and half quantum vortex~\cite{SatoPhysicaE2014}. 

In this section, we first summarize the symmetries which are preserved in the B-phase of the bulk $^3$He. We also clarify the symmetric properties of $^3$He confined in a slab geometry.  

\subsection{Basic properties of bulk superfluids and superconductors}

The total symmetry group $G$ relevant to liquid $^3$He in normal states is given by 
\beq
G = {\rm SO}(3)_{\bm L} \times {\rm SO}(3)_{\bm S} \times {\rm U}(1)_{\phi} \times {\rm T} \times {\rm C} \times {\rm P}.
\label{eq:full}
\eeq
The high symmetry group comprises the continuous symmetry groups, the group of three-dimensional rotations of coordinate ${\rm SO}(3)_{\bm L}$, the rotation group of spin spaces ${\rm SO}(3)_{\bm S}$, and the global phase transformation group ${\rm U}(1)_{\phi}$. The total group $G$ also comprises the following groups: The time-reversal ${\rm T}$, the particle-hole conversion ${\rm C}$, and the space parity ${\rm P}$. The superfluid phase transitions of $^3$He are accompanied by the spontaneous breaking of continuous symmetries and superfluid states are not invariant under the full symmetry group $G$ in Eq.~(\ref{eq:full}). The residual symmetry of superfluid phases is characterized by the subgroup $H\subset G$ and the coset group $R=G/H$ describes the degeneracy of the order parameters.

In a realistic situation of $^3$He, the dipole interaction originating from the magnetic moment of $^3$He nuclei provides the spin-orbit interaction, which reduces ${\rm SO}(3)_{\bm L} \times {\rm SO}(3)_{\bm S}$ to the joint rotation ${\rm SO}(3)_{{\bm L}+{\bm S}}$ in $G$. Since the interaction does not alter the symmetry held by $^3$He-B, we here neglect it. The effect of the nuclear-dipole interaction on topological superfluidity of $^3$He will be discussed in Sec.~\ref{sec:numerical2}. 

To clarify the nontrivial topology of the bulk superfluid $^3$He-B, we here consider a spatially uniform system, where the order parameter only depends on the relative coordinate. We then start with the second quantized Hamiltonian
\beq
\mathcal{H} &=& \frac{1}{V}\sum _{\bm k} c^{\dag}_{{\bm k},a} \varepsilon _{ab}({\bm k}) c_{{\bm k},b} \nn \\
&&+ \frac{1}{V^2}\sum _{{\bm k},{\bm k}^{\prime},{\bm q}} 
V^{c,d}_{a,b}({\bm k},{\bm k}^{\prime})
c^{\dag}_{{\bm k}+{\bm q}/2,a}c^{\dag}_{-{\bm k}+{\bm q}/2,b}
c_{-{\bm k}^{\prime}+{\bm q}/2,c} c_{{\bm k}^{\prime} + {\bm q}/2,d},
\label{eq:originalH}
\eeq
where $V$ is the volume of the system. This Hamiltonian describes spin-$1/2$ fermions interaction through the potential $V^{c,d}_{a,b}({\bm k},{\bm k}^{\prime})$, which preserves the symmetry group $G$ in Eq.~(\ref{eq:full}). In Eq.~(\ref{eq:originalH}), $c _{{\bm k},a}$ and $c^{\dag}_{{\bm k},a}$ denote the annihilation and creation operators of fermions with spin $a=\uparrow$ and $\downarrow$. 

Employing the standard procedure of the mean-field approximation, we introduce the pair potential in the bulk superfluids and superconductors as 
\beq
\Delta _{ab} ({\bm k}) = V^{c,d}_{a,b}({\bm k},{\bm k}^{\prime})\langle c_{{\bm k}^{\prime},c} c_{{\bm k}^{\prime},d}\rangle.
\label{eq:gap_bulk}
\eeq 
The mean-field Hamiltonian for superfluids and superconductors is expressed in terms of the BdG Hamiltonian density $\mathcal{H}({\bm k})$ as
\beq
\mathcal{H}_{\rm MF} = E_0 + \frac{1}{V} \sum _{\bm k} {\bm c}^{\dag}_{\bm k}\mathcal{H}({\bm k}) {\bm c}_{\bm k},
\eeq
where the fermionic field operator in the Nambu space is defined as ${\bm c}_{\bm k} \!\equiv\! (c_{{\bm k}\uparrow},c_{{\bm k}\downarrow},c^{\dag}_{-{\bm k}\uparrow},c^{\dag}_{-{\bm k}\downarrow})^{\rm T}$. The BdG Hamiltonian density is given by
\beq
\mathcal{H}({\bm k}) = \left(
\begin{array}{cc}
\varepsilon ({\bm k}) & \Delta ({\bm k}) \\
-\Delta^{\ast}(-{\bm k}) &- \varepsilon^{\rm T}(-{\bm k})
\end{array}
\right).
\label{eq:Hbdg}
\eeq
The single-particle Hamiltonian density $\varepsilon ({\bm k})$ may contain the diagonal elements of mean-field approximated self-energies and external potentials, which can be parameterized with the Pauli matrix $\sigma _{\mu}$ in the spin space as
\beq
\varepsilon ({\bm k}) = \varepsilon _0 ({\bm k}) + {\bm \sigma}\cdot{\bm \varepsilon}({\bm k}).
\label{eq:epsilon}
\eeq 
Owing to the emergence of the symmetry breaking term due to the order parameter $\Delta ({\bm k})$, the Hamiltonian for superfluid phases, $\mathcal{H}_{\rm MF}$, has the remaining symmetry $H \subset G$ and the order parameter manifold is characterized by $R=G/H$.

Following Refs.~\cite{salomaaRMP,volovik1990}, let us now summarize the action of the elements of $G$ on the creation and annihilation operators of $^3$He atoms, $c_{{\bm k}a}$ and $c^{\dag}_{{\bm k}a}$, and $\Delta ({\bm k})$ in Eq.~(\ref{eq:Hbdg}). The full symmetry group $G$ consists of the three continuous groups and three discrete groups. The continuous rotational groups in the coordinate space and spin space, $SO(3)_{\bm L}$ and $SO(3)_{\bm S}$, have the generators, the orbital angular momentum operator $\hat{\bm L}\!\equiv\! \hat{\bm r}\times \hat{\bm p}$ and the spin angular momentum operator $\hat{\bm S}$, which act on the field operators of $^3$He atoms as
\beq
\hat{\bm L} c_{{\bm k}a} = -i{\bm k}\times \partial _{\bm k} c_{{\bm k}a}, \hspace{3mm}
\hat{\bm S} c_{{\bm k}a} = \frac{1}{2}{\bm \sigma}_{ab}c_{{\bm k}b}.
\label{eq:LS}
\eeq
The element of $U(1)_{\phi}$ rotates the phase of the creation by $\phi$ and annihilation operators as
\beq
\hat{U}_{\phi} c_{{\bm k}a} = e^{-i\phi \hat{N}} c_{{\bm k}a} e^{i\phi \hat{N}} 
= e^{i\phi} c_{{\bm k}a},
\label{eq:gauge}
\eeq
where $\hat{N} = \sum _{\bm k,a} c^{\dag}_{{\bm k}a}c_{{\bm k}a}$ denotes the number operator of particles. The time-inversion operator $\hat{T}$ transforms a $^3$He atom with the momentum ${\bm k}$ and spin $\sigma$ to a atom with $-{\bm k}$ and spin rotated by $\pi$ by a unitary transformation, 
\beq
\hat{T} {c}_{{\bm k}a} = \Theta _{ab} {c}_{-{\bm k}b}, \hspace{3mm} 
\Theta = i\sigma _y.
\label{eq:time}
\eeq
This transformation exchanges the direction of the spin and momentum, $c_{{\bm k}\uparrow} \mapsto c_{-{\bm k}\downarrow}$ and $c_{{\bm k}\downarrow} \mapsto -c_{-{\bm k}\uparrow}$. The space inversion operator $\hat{P}$ rotates the momentum ${\bm k}$ by $\pi$ as 
\beq
\hat{P} c_{{\bm k}a} = c_{-{\bm k}a}. 
\label{eq:inv}
\eeq
The full symmetry group $G$ also contains the symmetry under the particle-hole conversion $C$, whose element $\hat{C}$ maps a quasiparticle with energy $E_{\bm k}$ to a quasiparticle with $-E_{-{\bm k}}$ as
\beq
\hat{C}c_{{\bm k}a} = c^{\dag}_{-{\bm k}a}.
\label{eq:ph}
\eeq
Note that the definition of the particle-hole conversion $\hat{C}$ is different from that in Ref.~\cite{fishman1985} which rotates the spin by $\pi$ in addition to the momentum.

The pair potential for bulk superfluids and superconductors is defined as Eq.~(\ref{eq:gap_bulk}) in the momentum space, where the interaction potential $V^{c,d}_{a,b}({\bm k},{\bm k}^{\prime})$ is supposed to be invariant under the full symmetry group $G$. For spin-triple superfluids and superconductors, we take only a spin-triplet $p$-wave (odd-parity) channel of the interaction potential, which obeys $V^{c,d}_{a,b} \!=\! V^{d,c}_{a,b} \!=\! V^{c,d}_{b,a}$ and $V^{c,d}_{a,b}({\bm k},{\bm k}^{\prime}) \!=\! -V^{c,d}_{a,b}(-{\bm k},{\bm k}^{\prime}) \!=\! -V^{c,d}_{a,b}({\bm k},-{\bm k}^{\prime})$. It is found from Eqs.~(\ref{eq:LS})-(\ref{eq:inv}) that each element of $G$ acts on $\Delta ({\bm k})$ as 
\beq
\hat{L}_{\mu}\Delta _{ab}({\bm k}) 
= -i \epsilon _{\mu\nu\eta} k_{\nu} \partial _{k_{\eta}} \Delta _{ab}({\bm k}),
\label{eq:L}
\eeq
\beq
\hat{S}_{\mu}\Delta _{ab}({\bm k}) = \frac{1}{2}(\sigma _{\mu})_{aa^{\prime}}\Delta _{a^{\prime}b}({\bm k}) 
+ \frac{1}{2}\Delta _{ab^{\prime}}({\bm k})(\sigma _{\mu})_{bb^{\prime}},
\eeq
\beq
\hat{U}_{\phi}\Delta _{ab}({\bm k}) = e^{2i\phi}\Delta _{ab}({\bm k}),
\eeq
\beq
\hat{T} \Delta _{ab}({\bm k}) = \Theta _{aa^{\prime}}\Delta^{\ast}_{a^{\prime}b^{\prime}}(-{\bm k})
\Theta _{bb^{\prime}}, 
\label{eq:T}
\eeq
\beq
\hat{P} \Delta _{ab}({\bm k}) = \Delta _{ab}(-{\bm k}),
\label{eq:parity}
\eeq
\beq
\hat{C} \Delta _{ab}({\bm k}) = \Delta^{\ast}_{ab}(-{\bm k}). 
\label{eq:phD}
\eeq

It is now convenient to parameterize the pair potential $\Delta ({\bm k})$ as
\beq
\Delta ({\bm k}) = \left(
\begin{array}{cc}
\Delta _{\uparrow\uparrow}({\bm k}) & \Delta _{\uparrow\downarrow}({\bm k}) \\
\Delta _{\downarrow\uparrow}({\bm k}) & \Delta _{\downarrow\downarrow} ({\bm k})
\end{array}
\right) = i\sigma _y \psi ({\bm k}) + i \sigma _{\mu}\sigma _y d_{\mu} ({\bm k}),
\label{eq:generalD}
\eeq
where $\psi ({\bm k}) =\psi (-{\bm k})$ and $d_{\mu}({\bm k}) = - d_{\mu}(-{\bm k})$ denote the spin-single and spin-triplet component of the pair potential, respectively. It is important to note that the pair potential is invariant under the time inversion in Eq.~(\ref{eq:T}), when $\psi({\bm k})$ and $d_{\mu}({\bm k})$ are real. The ${\bm d}({\bm k})$ vector is transformed by three-dimensional rotations in spin spaces as a vector, 
\beq
U({\bm n},\varphi) \Delta ({\bm k})U^{\rm T}({\bm n},\varphi) 
= i\sigma _{\mu}\sigma _yR_{\mu \nu}(\hat{\bm n},\varphi)d_{\nu}({\bm k}),
\label{eq:drot}
\eeq
while the spin-singlet part $\psi$ remains as a scalar. In Eq.~(\ref{eq:drot}), the ${\rm SU}(2)$ matrix, $U({\bm n},\varphi)=\cos(\varphi/2) - i \sigma _{\mu}\hat{n}_{\mu}\sin(\varphi/2)$, describes the rotation of a spin matrix about the $\hat{\bm n}$-axis by the angle $\varphi$ and $R_{\mu\nu}$ denotes the corresponding ${\rm SO}(3)$ matrix, $R_{\mu\nu}(\hat{\bm n},\varphi) = \cos\varphi\delta _{\mu,\nu} + (1-\cos\varphi)\hat{n}_{\mu}\hat{n}_{\nu} - \epsilon^{\mu\nu\eta}\hat{n}_{\eta}\sin\varphi$~\cite{jj}. These two matrices are related to each other through the identify $U\sigma _{\mu}U^{\dag} = \sigma _{\mu} R_{\mu\nu}$.

\subsection{Symmetry of the bulk $^3$He-B}
\label{sec:symmetry}

{\it Order parameter.}---
According to Eqs.~(\ref{eq:L})-(\ref{eq:parity}), the irreducible representations of $G$ are characterized by the values of orbital and spin moments $L$ and $S$, the space parity $P$, and the quantum number $N$ associated with the number of paired particles. We here concentrate our attention to superfluid $^3$He, that is spin-triplet $p$-wave pairing states having $L=S=1$, $P=-1$, $N=2$. 

The B-phase holds the maximal symmetry group of the subset of $G$, 
\beq
{ H}_{\rm B} = {\rm SO}(3)_{{\bm L}+{\bm S}} \times {\rm T} \times {\rm C} \times {\rm P}{\rm U}_{\pi/2}, 
\eeq
where ${\rm SO}(3)_{{\bm L}+{\bm S}}$ denotes the joint three-dimensional rotation in the coordinate and spin spaces. We have also introduced the combined discrete symmetry of the inversion ${\rm P}$ and the $\pi/2$-phase rotation ${\rm U}_{\pi/2}$. The broken symmetry $R$ in the B-phase is then given by 
\beq
{ R}=G/H_{\rm B} = {\rm SO}(3)_{{\bm L}-{\bm S}} \times {\rm U}(1)_{\phi}.
\label{eq:R}
\eeq
Hence, the B-phase is regarded as the spontaneous breaking phase of the spin-orbit symmetry~\cite{leggettRMP}. 
Since the generator of the joint rotational symmetry group ${\rm SO}(3)_{{\bm L}+{\bm S}}$ is the total angular momentum operator ${\bm J}$, the order parameter of the B-phase having $S=L=1$ is classified with the quantum number $J = 0$, $1$, and $2$. The simplest form of ${\bm d}({\bm k})$ is characterized with $J=0$, which was first proposed by Balian and Werthamer~\cite{bw} as
\beq
\Delta ({\bm k}) = \Delta _{0} \left[
(-\hat{k}_x+i\hat{k}_y) \left| \uparrow\uparrow \right\rangle
+ \hat{k}_z \left| \uparrow\downarrow + \downarrow \uparrow\right\rangle
+ (\hat{k}_x+i\hat{k}_y) \left| \downarrow\downarrow \right\rangle
\right].
\eeq
This form implies the ${\bm d}$-vector is parallel to the ${\bm k}$-vector. $d_{\mu}({\bm k}) = \Delta _0\hat{k}_{\mu}$. As shown in Eq.~(\ref{eq:R}), however, the degeneracy space is characterized in Eq.~(\ref{eq:R}) by the continuous rotations ${\rm SO}(3)_{{\bm L}-{\bm S}}$ and ${\rm U}(1)_{\phi}$. Therefore, the ${\bm d}({\bm k})$ vector in the B-phase can be deviated from the ${\bm k}$-vector as
\beq
d_{\mu}({\bm k}) = \Delta _0 e^{i\phi} R_{\mu\nu}(\hat{\bm n},\varphi) \hat{k}_{\nu},
\label{eq:dvecB}
\eeq
where the rotation matrix $R_{\mu\nu}$ is associated with the spin-orbit symmetry breaking ${\rm SO}(3)_{{\bm L}-{\bm S}}$ and $\phi$ arises from the breaking of the ${\rm U}(1)_{\phi}$ symmetry.

The equilibrium order parameter of the B-phase is eigenfunctions of the total twisted angular momentum $J_{\mu} =L_{\mu} + S_{\nu}R_{\nu\mu}(\hat{\bm n},\varphi)$~\cite{maki1974,fishman1986}. The diagonal representation of the pair potential, $\Delta _0({\bm k}) = i\Delta _0 {\bm k}\cdot{\bm \sigma}\sigma _y $, is obtained by using Eq.~(\ref{eq:drot}) from the general form of $\Delta ({\bm k})$ as $\Delta ({\bm k}) = U(\hat{\bm n},\varphi) \Delta _0 ({\bm k})U^{\rm T}(\hat{\bm n},\varphi)$. Let $\mathcal{U}$ be the $SU(2)$ matrix extended to the Nambu space, $\mathcal{U}(\hat{\bm n},\varphi) \!=\! {\rm diag}(U,U^{\ast})$. Using the $SU(2)$ matrix, one then finds
\beq
\mathcal{U}(\hat{\bm n},\varphi)\mathcal{H}_0({\bm k})\mathcal{U}^{\dag}(\hat{\bm n},\varphi) = \mathcal{H}({\bm k}), 
\label{eq:HB}
\eeq
where $\mathcal{H}_0({\bm k})$ is obtained from Eq.~(\ref{eq:Hbdg}) with $\Delta ({\bm k})\rightarrow \Delta _0({\bm k})$.

{\it Continuous symmetry.}---
Let us now consider the rotational symmetry of $\mathcal{H}({\bm k})$ associated with the joint rotation of spin and orbital spaces, ${\rm SO}(3)_{{\bm L}+{\bm S}}$. This continuously rotates the momentum and the ${\bm d}$-vector as $\hat{k}_{\mu} \mapsto \hat{k}^{\prime}_{\mu}=R^{(L)}_{\mu\nu}\hat{k}_{\nu}$ and $d_{\mu} \mapsto d^{\prime}_{\mu}=R^{(S)}_{\mu\nu}d_{\nu}$, where $R^{(S)}_{\mu\nu} =(RR^{(L)}R^{-1})_{\mu\nu}$ denotes the rotational matrix in the spin space. 
Then, the ${\rm SU}(2)$ representation of the ${\rm SO}(3)_{{\bm L}+{\bm S}}$ symmetry in the bulk B-phase is given as 
\beq
\mathcal{U}_{ S}\mathcal{H}({\bm k})\mathcal{U}^{\dag}_{ S} = \mathcal{H}(R^{(L)}{\bm k})
\label{eq:SU2}
\eeq
where we introduce $\mathcal{U}_{ S} = \mathcal{U}(\hat{\bm n},\varphi)\mathcal{U}_{ L} \mathcal{U}^{\dag}(\hat{\bm n},\varphi)$ and $\mathcal{U}_{ L}\equiv {\rm diag}(U_{ L},U^{\ast}_{ L})$ is the ${\rm SU}(2)$ representation of the rotation matrix associated with $R^{(L)}$. We will see below that the subset of the continuous rotation group provides a well-defined topological invariant that is protected even in the presence of confinement and time-reversal breaking perturbation.

{\it Discrete symmetries.}---
We now summarize the discrete symmetries that play crucial roles in determining the topological properties. Let us first consider the particle-hole conversion in Eq.~(\ref{eq:ph}) that exchange a particle with the momentum ${\bm k}$ to a hole with $-{\bm k}$. Since the particle-hole conversion acts on $\Delta ({\bm k})$ as that in Eq.~(\ref{eq:phD}), it is obvious that the BdG Hamiltonian $\mathcal{H}({\bm k})$ in Eq.~(\ref{eq:Hbdg}) holds the particle-hole symmetry,
\beq
\mathcal{C} \mathcal{H}({\bm k})\mathcal{C}^{-1} = -\mathcal{H}(-{\bm k}),
\hspace{3mm} \mathcal{C} = \tau _x K,
\label{eq:PHS}
\eeq
where $K$ is the complex conjugation operator. 

Let us suppose that the single-particle Hamiltonian density is invariant under the time-inversion, $\Theta\varepsilon^{\ast}({\bm k})\Theta^{\dag} = \varepsilon (-{\bm k})$, where $\Theta = i\sigma_y$ is the time-inversion operator defined in Eq.~(\ref{eq:time}). The general form of the pair potential in Eq.~(\ref{eq:generalD}) holds the time-reversal symmetry, $\Theta\Delta^{\ast}({\bm k})\Theta^{\rm T} = \Delta (-{\bm k})$, when $\psi({\bm k})$ and ${\bm d}({\bm k})$ are real. The ${\bm d}$-vector of the B-phase in Eq.~(\ref{eq:dvecB}) can be chosen to be real by rotating the ${\rm U}(1)$ phase. Therefore, the BdG Hamiltonian $\mathcal{H}({\bm k})$ in the B-phase holds the time-reversal symmetry 
\beq
\mathcal{T} \mathcal{H}({\bm k})\mathcal{T}^{-1} = \mathcal{H}(-{\bm k}),
\hspace{3mm} \mathcal{T} = \Theta K.
\label{eq:TRS}
\eeq
The discrete symmetries of the BdG Hamiltonian in Eqs.~(\ref{eq:PHS}) and (\ref{eq:TRS}) play a crucial role on determining the topological properties of the bulk B-phase as discussed in Sec.~\ref{sec:topo}. 

We also note that the inversion operator $\hat{P}$ acts on the B-phase
pair potential as $\hat{P}\Delta ({\bm k})=-\Delta (-{\bm k})$. Hence,
the inversion symmetry is realized up to the $U(1)_{\phi=\pi/2}$ gauge
symmetry as
\beq
\mathcal{P}\mathcal{H}({\bm k})\mathcal{P}^{\dag} = \mathcal{H}(-{\bm k}),
\hspace{3mm} \mathcal{P} = \tau _z,
\label{eq:inversion}
\eeq
In contrast to the time-reversal symmetry in Eq.~(\ref{eq:TRS}), this operation changes the sign of the momentum without changing the spin state. 

The discrete symmetries introduced above imposes the symmetric relation on quasiparticle states in the momentum space. The quasiparticle structure is obtained by diagonalizing the BdG Hamiltonian $\mathcal{H}({\bm k})$ in Eq.~(\ref{eq:Hbdg}) as
\beq
\mathcal{H}({\bm k}) |u_n({\bm k})\rangle = E_n({\bm k}) |u_n({\bm k})\rangle,
\label{eq:bdg}
\eeq
where the eigenfunction $|u_n({\bm k})\rangle$ must satisfy the normalization condition $\langle u_n ({\bm k})| u_n({\bm k})\rangle = 1$ and the subscript ``$n$'' denotes the band index. It is important to note that the additional discrete symmetries in Eqs.~(\ref{eq:PHS}) and (\ref{eq:TRS}) held by the BdG Hamiltonian $\mathcal{H}({\bm k})$ ensure the symmetric properties of the quasiparticle spectrum obtained from Eq.~(\ref{eq:bdg}). First, the particle-hole symmetry in Eq.~(\ref{eq:PHS}) ensures that there is a one-to-one correspondence between the positive and negative energy states, where the quasiparticle state with $E({\bm k})>0$ is associated with that with $-E({\bm k})$ through the relation,
\beq
|u_{-E}({\bm k})\rangle = \mathcal{C}| u_E(-{\bm k})\rangle.
\eeq
In addition, the quasiparticle states are two-fold degenerate as a consequence of the time-reversal symmetry characterized by the anti-unitary operator $\mathcal{T}^2=-1$ in Eq.~(\ref{eq:TRS}), which form a Kramers pair
\beq
|u_n({\bm k})\rangle = \mathcal{T}|u_n(-{\bm k})\rangle.
\eeq
This rotates the spin state of the quasiparticle together with changing the sign of the momentum. It is also found that the inversion symmetry in Eq.~(\ref{eq:inversion}) results in the relation between quasiparticle states with ${\bm k}$ and $-{\bm k}$,
\beq
|u_n({\bm k})\rangle = \mathcal{P}|u_n(-{\bm k})\rangle,
\eeq
which does not rotate the spin. The two-fold degenerate energy spectrum in the B-phase of superfluid $^3$He is obtained by diagonalizing $\mathcal{H}({\bm k})$ as
\beq
E({\bm k}) = \pm \sqrt{[\varepsilon ({\bm k})]^2 + \Delta^2_{\rm 0}},
\label{eq:EkB}
\eeq
which yields fully gapped excitation at the Fermi level.

We would like to note that the BdG Hamiltonian for the B-phase holds the additional discrete symmetry, which can be constructed by combining the time-reversal symmetry in Eq.~(\ref{eq:TRS}) and the particle-hole symmetry in Eq.~(\ref{eq:PHS}). Using the combination, we introduce the chiral operator $\Gamma$, which is anti-commutable with $\mathcal{H}({\bm k})$,  
\beq
\Gamma \mathcal{H}({\bm k}) \Gamma = - \mathcal{H}(\hat{\bm k}), 
\hspace{3mm} \Gamma = -i\mathcal{C} \mathcal{T},
\label{eq:chiral1}
\eeq
and $\Gamma = \Gamma^{\dag}$. In accordance with Ref.~\cite{satoPRB2011}, let $|v^{\pm}_n({\bm k})\rangle$ be an eigenstate of $\mathcal{H}^2({\bm k})$,
\beq
\mathcal{H}^2({\bm k})\left| v_n({\bm k})\right\rangle = E^2_n \left| v_n({\bm k})\right\rangle.
\eeq
Then, there exists a one-to-one correspondence between the eigenstates $|u_n({\bm k})\rangle$ and $|v_n({\bm k})\rangle$~\cite{satoPRB2011}. For a finite energy state $E^2_n\neq 0$, the eigenvector $|v_n({\bm k})\rangle$ is associated with the eigenstate of $\mathcal{H}({\bm k})$ as 
$| u_n({\bm k})\rangle = c(\mathcal{H}({\bm k})+E_n) | v_n({\bm k})\rangle$ for $(\mathcal{H}({\bm k})+E_n)|v_n({\bm k})\rangle\neq 0$
and as $| u_n({\bm k})\rangle = \Gamma |v_n({\bm k})\rangle$ for $(\mathcal{H}({\bm k})+E_n)|v_n({\bm k})\rangle=0$, where $c$ is the normalization constant. For the zero energy state, one finds $|u_0({\bm k})\rangle = |v_0({\bm k})\rangle$. 

The chiral symmetry ensures that the chiral operator $\Gamma$ is commutable with $\mathcal{H}^2({\bm k})$, $[\Gamma,\mathcal{H}^2({\bm k})]=0$. This indicates that $|v_n({\bm k})\rangle$ is the simultaneous eigenstate of $\Gamma$ and $\mathcal{H}^2({\bm k})$, 
\beq
\Gamma |v^{\pm}_n({\bm k})\rangle = \pm |v^{\pm}_n({\bm k})\rangle .
\eeq
It turns out that the eigenvector $|v^+_n({\bm k})\rangle$ is constructed from the counterpart $|v^-_n({\bm k})\rangle$ as 
\beq
|v^+_n({\bm k})\rangle = c^{\prime}\mathcal{H}({\bm k})|v^-_n({\bm k})\rangle, 
\label{eq:vn}
\eeq 
for $E^2_n \neq 0$, where $c^{\prime}$ is the normalization constant. Hence, the eigenstate with a finite energy $E^2_n \neq 0$ forms a chiral pair, where the quasiparticle with the chirality $\Gamma = +1$ is always paired with the quasiparticle with the opposite chirality $\Gamma = -1$. For zero energy states, however, the solution does not form a pair in general, since the right hand side of Eq.~(\ref{eq:vn}) vanishes. We will demonstrate in Sec.~\ref{sec:topo} that the difference between the number of zero energy state with $\Gamma = +1$ and that with $\Gamma = -1$ is associated with the topological winding number which is protected by the chiral symmetry.

\subsection{$^3$He-B in a slab geometry: Anisotropic pair breaking and a hidden discrete symmetry}
\label{sec:slab}

\begin{figure}[tb!]
\begin{center}
\includegraphics[width=85mm]{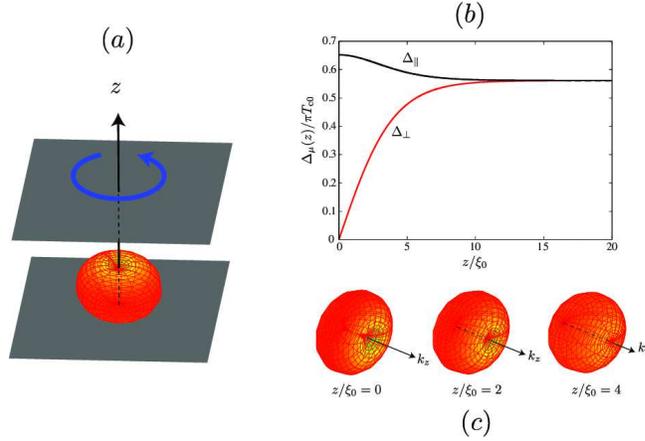}
\caption{(a) Schematic picture of the geometry considered here. The liquid $^3$He is confined to a slab geometry within $z\in[0,D]$, where the two specular surfaces are situated at $z = 0$ and $D$. 
(b) Spatial profile of the pair potentials in the superfluid $^3$He-B near the surface. The specular surface is set to be at $z=0$. (c) Local gap structure $\Delta (\hat{\bm k},z)$. All data are calculated in the framework of the quasiclassical theory. }
\label{fig:gap_slab}
\end{center}
\end{figure}

{\it Symmetries in a slab.}---
Let us consider the symmetric properties of $^3$He confined in a slab geometry, where $^3$He is sandwiched by two parallel surfaces, as depicted in Fig.~\ref{fig:phase_slab}(a). In such a geometry, confinement explicitly breaks the three-dimensional continuous rotational symmetry ${\rm SO}(3)$ in the coordinate space. Then, the symmetry group of the normal $^3$He in Eq.~(\ref{eq:full}) is reduced to 
\beq
G_{\rm slab} = {\rm SO}(2)_{L_z} \times {\rm SO}(3)_{\bm S} \times {\rm U}(1)_{\phi} \times {\rm T} \times {\rm C}. 
\label{eq:slab}
\eeq
As shown in Fig.~\ref{fig:gap_slab}(a), the B-phase confined in a slab geometry is invariant under the two dimensional rotation of the spin and orbital spaces about the surface normal axis ($\hat{\bm z}$-axis). Thus, the BdG Hamiltonian holds the ${\rm SO}(2)_{L_z+S_z}$ symmetry, 
\beq
H_{\rm slab} = {\rm SO}(2)_{L_z+S_z} \times {\rm T} \times {\rm C} \times {\rm PU}_{\pi/2}.
\eeq 
Note that the time-reversal symmetry, particle-hole symmetry, and inversion symmetry are still preserved in this situation. The ${\rm SO}(2)_{L_z+S_z}$ symmetric pair potential is parametrized as 
\beq
d_{\mu}({\bm k},z) = R_{\mu\nu}(\hat{\bm n},\varphi)d^{(0)}_{\nu\eta}(z)\hat{k}_{\eta},
\label{eq:dvec_slab}
\eeq
where 
\beq
d^{(0)}_{\mu\nu}(z) = \Delta _{\parallel}(z)\left( 
\delta _{\mu,\nu} - \hat{z}_{\mu}\hat{z}_{\nu}
\right) + \Delta _{\perp}(z)\hat{z}_{\mu}\hat{z}_{\nu}.
\label{eq:dvec_b2_initial}
\eeq
Hence, the quasiparticle excitation gap is distorted by confinement. When $\Delta _{\parallel}=\Delta _{\perp}$, the pair potential defined in Eq.~(\ref{eq:dvec_slab}) recovers the isotropic B-phase order parameter. 

In a slab geometry, the discrete symmetries in Eqs.~(\ref{eq:PHS}) and (\ref{eq:TRS}) are preserved, as well as the ${\rm SO}(2)_{{\bm L}+{\bm S}}$ rotational symmetry. The joint ${\rm SO}(2)$ symmetry indicates that the pair potential defined in Eq.~(\ref{eq:dvec_slab}) is invariant under the continuous rotation of the momentum and the ${\bm d}$-vector as $(k_x,k_y,k_z) \mapsto (k^{\prime}_x,k^{\prime}_y,k_z)$ and $(d_x,d_y,d_z) \mapsto (d^{\prime}_x,d^{\prime}_y,d^{\prime}_z)$, where $k^{\prime}_{\mu}= R^{(L)}_{\mu\nu}\hat{k}_{\nu}$ and $d^{\prime}_{\mu}= R^{(S)}_{\mu\nu}\hat{d}_{\nu}$ with the ${\rm SO}(2)$ rotation matrix about the surface normal axis, $R^{(L)}$ and $R^{(S)}\equiv R R^{(L)}R^{-1}$. Hence, the BdG Hamiltonian for the B-phase confined in a slab geometry holds the continuous symmetry
\beq
\mathcal{U}^{(2)}_{S}\mathcal{H}({\bm k})\mathcal{U}^{(2)\dag}_{S} = \mathcal{H}(k^{\prime}_x,k^{\prime}_y,k_z),
\label{eq:so2slab}
\eeq
where we inroduce $\mathcal{U}^{(2)}_{ S}\equiv \mathcal{U}(\hat{\bm n},\varphi)\mathcal{U}_{ L}\mathcal{U}^{\dag}(\hat{\bm n},\varphi)$ and $\mathcal{U}_{ L}$ corresponds to the ${\rm SU}(2)$ representation of the two dimensional rotation about the surface normal in the coordinate space. 

It is important to mention that the {\it hidden} ${\bm Z}_2$-symmetry is defined from the joint ${\rm SO}(2)$ symmetry, which enables us to introduce a one-dimensional winding number even in the presence of a time-reversal-breaking perturbation. Let $\mathcal{U}_z(\theta)$ be a unitary matrix describing the joint rotation of spin and orbital spaces by the angle $\theta$ about the surface normal axis (i.e., the $\hat{\bm z}$-axis), which is given as 
\beq
\mathcal{U}_z(\theta) = \mathcal{U}(\hat{\bm n},\varphi) \mathcal{U}_{ L}
(\hat{\bm z},\theta)\mathcal{U}^{\dag}(\hat{\bm n},\varphi). 
\eeq
Combining the $\pi$-rotation with the time-inversion operator $\mathcal{T}$, one can introduce the {\it hidden discrete} operator $\mathcal{T} \mathcal{U}_z(\pi)$ which transforms the ${\bm d}$ and $\hat{\bm k}$ vectors as ${\bm d}\mapsto(-d_x,-d_y,d_z)$ and $\hat{\bm k}\mapsto(-\hat{k}_x,-\hat{k}_y,\hat{k}_z)$. Therefore, the BdG Hamiltonian for $^3$He-B holds the hidden discrete symmetry, 
\beq
\mathcal{T} \mathcal{U}_z(\pi) \mathcal{H}(k_x,k_y,k_z) \mathcal{U}^{-1}_z(\pi)\mathcal{T}^{-1} 
= \mathcal{H}(k_x,k_y,-k_z). 
\label{eq:hidden}
\eeq
This discrete symmetry can be held even if each symmetry, $\mathcal{T}$ or $\mathcal{U}_z(\pi)$, is explicitly broken. As clarified in the subsequent subsection, the hidden ${\bm Z}_2$-symmetry is held by the B-phase even in the presence of a magnetic field for a particular order parameter manifold.

{\it Pair breaking effect and Andreev bound states.}---
We specifically consider the confined geometry depicted in Fig.~\ref{fig:phase_slab}(a), where the liquid $^3$He is sandwiched by two specular walls. As seen in Fig.~\ref{fig:phase_slab}(b), the confinement effect can drastically alter the superfluid phase diagram even at zero pressures, where the ABM state can be stable even in the absence of spin fluctuation feedback mechanism. The surface imposes a boundary condition on the quasiparticles and pair potentials. A quasiparticle incoming to the surface along the trajectory of ${\bm k}$ is specularly scattered by the wall to the quasiparticle state with $\underline{\bm k}= {\bm k} - 2\hat{\bm z}(\hat{\bm z}\cdot{\bm k})$. The specular scattering of quasiparticles on the surface imposes a strong constraint on the wavefunction of the Cooper pairs, 
\beq
\Delta ({\bm k},z) = \Delta (\underline{\bm k},z),
\eeq 
at the surface $z=z_{\rm surf}$. This implies that while the Cooper pairs with zero perpendicular momentum are insensitive to the surface condition, the pairs having a non-zero momentum perpendicular to the surface must vanish at the surface, namely, 
\beq
\frac{d}{dz}\Delta _{\parallel}(z)\bigg|_{z=z_{\rm surf}} = 0, \hspace{3mm}
\Delta _{\perp} (z_{\rm surf}) = 0.
\eeq
Hence, the surface induces anisotropic pair breaking mechanism that deviates $\Delta _{\parallel}$ and $\Delta _{\perp}$. 

The spatial profile of the pair potentials, $\Delta _{\parallel}(z)$ and $\Delta _{\perp}(z)$, is determined by self-consistently solving the equations for the pair potential and underlying quasiparticles. We display in Figs.~\ref{fig:gap_slab}(b) and \ref{fig:gap_slab}(c) the local gap structures of $\Delta _{\parallel}(z)$ and $\Delta _{\perp}(z)$, which is obtained from the numerical calculation of quasiclassical equations. The details will be discussed in Sec.~\ref{sec:exact}. It is clearly seen that the pair potential coupled to a momentum perpendicular to the surface vanishes at the surface, $\Delta _{\perp} = 0$. In contrast, the parallel component of the pair potential is rather enhanced at the surface, because of the maximal gain of the condensation energy. 

This nontrivial gap structure of $^3$He-B near the surface was first found by Buchholtz and Zwicknagl~\cite{buchholtzPRB1981}. The deviation of the gap structure on the surface is accompanied by the emergence of midgap bound states. Hara and Nagai~\cite{haraPTP1986} analyzed a $p$-wave polar state whose pair potential extends perpendicular to the slab wall. This model offers an exactly soluble model to study the confinement effect in unconventional superfluids and superconductors. Making use of an exact self-consistent solution, they demonstrated that the spatial profile of $\Delta$ is essentially the same as the order parameter in the continuum model of the polyacetylene with a soliton lattice and in Fulde-Ferrell-Larkin-Ovchinnikov superconductors~\cite{machidaPRB1984,brazovskii,mertsching,horovitz,nakaharaPRB1981}. The spatial profile also reflects the existence of low-lying quasiparticle states bound at the surface, that are called the surface Andreev bound states. As discussed in Sec.~\ref{sec:andreev}, the effective equation for describing the quasiparticles bound to the surface, that is the Andreev equation, is reduced to the one-dimensional Dirac equation with an effective mass associated with the pair potential~\cite{jackiw,ohashiJPSJ1995}. This effective theory gives a unified description for low-energy quasiparticle states and brings an important consequence that the zero-energy Andreev bound state always appears in the surface of unconventional superconductors and superfluids, when the  the pair potential which the quasiparticles feel changes its sign, namely, the mass domain wall. 


The phase diagram and the confinement effect on $^3$He confined in a slab geometry have quantitatively been discussed in Refs.~\cite{ambegaokar,brinkman,barton,smith,kjaldman,fujita,jacobsen,takagi,fetter,li,ullah,lin-liu,salomaa}
 with the Ginzburg-Landau theory and Refs.~\cite{mizushimaPRB2012,buchholtzPRB1981,haraPTP1986,haraJLTP1988,vorontsovPRB2003,vorontsovPRL2007,nagai,tsutsumiJPSJ2010,tsutsumiPRB2011} with the quasiclassical theory. The latter underlines the role of the surface Andreev bound states on thermodynamics. In addition, Vorontsov and Sauls~\cite{vorontsovPRL2007} proposed that there exists a new quantum phase in the vicinity of the A-B transition, that is, the crystalline phase that spontaneously breaks the translational symmetry. There have been numerous works on the effect of surface roughness that causes diffusive scattering of quasiparticles~\cite{zhangPRB1987,thuneberg1992,buchholtz1979,buchholtz1986,buchholtz1991,buchholtz1993,nagatoJLTP1996,nagatoJLTP1998,nagatoJLTP2007}. The surface roughness drastically changes the surface structure of order parameters and low-lying surface states in the superfluid $^3$He-B~\cite{vorontsovPRB2003,nagai,nagatoJLTP1998,nagatoJPSJ2011}. 
 
The pair-breaking effect and the enhancement of the surface density of states due to the existence of the surface Andreev bound states have been observed in several experiments~\cite{freemanPRL1988,freemanPRB1990,xu,kawae,miyawaki,kawasaki,levitin,bennettJLTP2010,levitin2013,levitinPRL2013,ahonenJLTP1976,osheroff,ishikawaJLTP1989,castelijns,ikegamiJPSJ2013}.
The contributions of surface Andreev bound states have also been detected through the deviation of the heat capacity from that in the bulk $^3$He-B~\cite{choiPRL2006} as well as the anomalous attenuation of transverse sound wave~\cite{davisPRL2008}. The surface acoustic impedance measurement provides another powerful tool to probe the surface structure~\cite{aokiPRL2005,saitoh,wada,murakawaPRL2009,murakawaJLTP2010,wasaiJLTP2010,murakawaJPSJ2011}. The spectrum of the surface bound states and its dependence on surface condition has been investigated by measuring surface acoustic impedance~\cite{murakawaPRL2009,murakawaJPSJ2011,okuda}. The further details will be described in Sec.~\ref{sec:detecting}.

\subsection{A hidden discrete symmetry: Bulk $^3$He-B under a magnetic field}
\label{sec:symm_field}

It is also important to mention the symmetry of the bulk B-phase in the presence of a spatially uniform magnetic field. Specifically, let us consider the case that the magnetic field is applied along the $\hat{\bm h}$-axis. The magnetic field explicitly breaks the time-reversal symmetry as well as the three-dimensional rotational symmetry in the spin space. The resultant symmetry group in the normal state is expressed as 
\beq
G_{\rm mag} = {\rm SO}(3)_{\bm L} \times {\rm SO}(2)^{(\hat{\bm h})}_{{\bm S}} \times {\rm U}(1)_{\phi} \times {\rm C}. 
\eeq
where ${\rm SO}(2)^{(\hat{\bm h})}_{{\bm S}}$ denotes the two-dimensional rotation about the $\hat{\bm h}$-axis in spin space. 

The pair potential for the bulk B-phase under a magnetic field is represented by the following form similar to Eqs.~(\ref{eq:dvec_slab}) and (\ref{eq:dvec_b2_initial}),
\beq
\Delta (\hat{\bm k}) = i\sigma _{\mu}\sigma _y R_{\mu\nu}(\hat{\bm n},\varphi) d^{(0)}_{\nu\eta}\hat{k}_{\eta},
\label{eq:BH}
\eeq 
where 
\beq
d^{(0)}_{\nu\eta} = \Delta _{\parallel}\left( 1 -\hat{h}_{\mu}\hat{h}_{\nu}\right) 
+ \Delta _{\perp}\hat{h}_{\mu}\hat{h}_{\nu}.
\label{eq:BH2}
\eeq
This pair potential still remains invariant under the continuous rotation, ${\rm SO}(2)^{(\hat{\bm h})}_{{\bm L}+{\bm S}}$, that is the joint rotation of the spin and orbital spaces about the $\hat{\bm h}$-axis. Now let us define the rotation matrix $R^{({\bm L})}(\hat{\bm h},\phi)$ that rotates the three-dimensional momentum vector $\hat{\bm k}$ about the $\hat{\bm h}$-axis by the angle $\phi$, as $\hat{k}_{\mu}\mapsto \hat{k}^{\prime}_{\mu} = R^{(L)}_{\mu\nu}\hat{k}_{\nu}$. In the same manner as Eq.~(\ref{eq:SU2}), the pair potential in Eq.~(\ref{eq:BH}) holds the ${\rm SO}(2)^{(\hat{\bm h})}_{{\bm L}+{\bm S}}$ symmetry,
\beq
U_{S}(\hat{\bm h},\phi)
\Delta ({\bm k})
U^{\rm T}_{S}(\hat{\bm h},\phi)
= \Delta \left(R^{(L)}(\hat{\bm h},\phi)\hat{\bm k}\right),
\eeq
where $U_S(\hat{\bm h},\phi)$ denotes the rotation matrix in the spin space corresponding to $R^{(L)}$ defined as 
\beq
U_S(\hat{\bm h},\phi) \equiv U(\hat{\bm n},\varphi) U_{L}(\hat{\bm h},\phi) U^{\dag}(\hat{\bm n},\varphi).
\eeq
The the ${\rm SU}(2)$ matrix, $U_S(\hat{\bm h},\phi)$, rotates the ${\bm d}$-vector as $d_{\mu}\mapsto d^{\prime}_{\mu} = (RR^{L}R^{-1})_{\mu\nu}d_{\nu}$.

It is worth mentioning that the squashed B-phase pair potential in Eq.~(\ref{eq:BH}) under a magnetic field holds the additional order-two discrete symmetry in addition to the particle-hole symmetry and inversion symmetry. The order-two symmetry is constructed by combining the $\pi$-rotation of spin and orbital spaces with the time-inversion, where the $\pi$-rotation is defined as a subgroup of ${\rm SO}(3)_{{\bm L}+{\bm S}}$. We call this symmetry as magnetic $\pi$-rotation symmetry or hidden ${\bm Z}_2$ symmetry. To clarify this, let us start by considering the $\pi$-rotation of $\Delta({\bm k})$ in the momentum space, where $(k_x,k_y,k_z)$ is rotated to $(-k_x,-k_y,k_z)$. The corresponding $\pi$-rotation matrix in the spin space is introduced as
\beq
U_z(\pi) \equiv U(\hat{\bm n},\varphi) U(\hat{\bm z},\pi) U^{\dag}(\hat{\bm n},\varphi),
\eeq
where $U(\hat{\bm z},\pi)$ denotes the ${\rm SU}(2)$ spin rotation matrix about the $\hat{\bm z}$-axis by the angle $\pi$. Without loss of generality, the rotation axis is taken to be the $\hat{\bm z}$-axis. Then, it turns out that the pair potential in Eq.~(\ref{eq:BH}) remains invariant under the $\pi$-rotation, 
\beq
U_z(\pi)\Delta ({\bm k})U^{\rm T}_z(\pi) = \Delta (-k_x,-k_y,k_z).
\eeq

In the presence of a magnetic field, however, the single-particle Hamiltonian ${ \varepsilon}({\bm k})$ in Eq.~(\ref{eq:epsilon}) contains the magnetic Zeeman term, which is recast into  
\beq
\varepsilon({\bm k})=\varepsilon _0 ({\bm k}) - \frac{\gamma \tilde{H}}{2}\hat{\bm h}\cdot{\bm \sigma},
\eeq
where $\gamma$ denotes the gyromagnetic ratio of $^3$He nuclei and $\tilde{H}$ is an effective magnetic field including the self-energy corrections. The $\pi$-rotation, then, changes the sign of the magnetic Zeeman term and is accompanied by the additional term as 
\beq
U_z(\pi) \left( \hat{\bm h}\cdot{\bm \sigma}\right)U^{\dag}_z(\pi) = -\hat{\bm h}\cdot{\bm \sigma} +2\hat{\ell}_z(\hat{\bm n},\varphi)\tilde{\sigma}_z,
\label{eq:piH}
\eeq
where $\tilde{\sigma}_z\equiv \sigma _{\nu}R_{\nu z}(\hat{\bm n},\varphi)$. The additional term induced by the $\pi$-rotation in Eq.~(\ref{eq:piH}) is characterized by the quantity $\hat{\ell}_z$ defined as  
\beq
\hat{\ell}_{\mu}(\hat{\bm n},\varphi) \equiv \hat{h}_{\nu}R_{\nu\mu}(\hat{\bm n},\varphi). 
\label{eq:ell}
\eeq
Similarly with the $\pi$-rotation, the time-inversion operator $\mathcal{T}=\Theta K$ with $\Theta = i\sigma _y$ does not alter the B-phase pair potential $\Delta({\bm k})$, but it changes the sign of the magnetic Zeeman term, $\Theta (\hat{\bm h}\cdot{\bm \sigma})^{\ast} \Theta^{\dag} = - \hat{\bm h}\cdot{\bm \sigma}$. Combining the $\pi$-rotation with the time-inversion, therefore, one can introduce the additional discrete symmetry held by the B-phase even in the presence of a magnetic field, 
\beq
\mathcal{T}\mathcal{U}(\pi) \mathcal{H}({\bm k})\mathcal{U}(\pi)^{-1}\mathcal{T}^{-1} = \mathcal{H}(k_x,k_y,-k_z),
\label{eq:Z2}
\eeq
when the quantity $\hat{\ell}_z$ vanishes, namely, $\hat{\ell}_z (\hat{\bm n},\varphi)=0$. We have here introduced $\mathcal{U}(\pi)\equiv {\rm diag}(U_z(\pi),U^{\ast}_z(\pi))$. For the case of a finite $\hat{\ell}_z$, the discrete symmetry is broken and the additional term associated with $\hat{\ell}_z$ remains, 
\beq
\mathcal{T}\mathcal{U}(\pi) \mathcal{H}({\bm k})\mathcal{U}(\pi)^{-1}\mathcal{T}^{-1} &=& \mathcal{H}(k_x,k_y,-k_z) \nn \\
&& - \gamma H \hat{\ell}_z (\hat{\bm n},\varphi) \left(
\begin{array}{cc} 
\tilde{\sigma}_z & 0 \\ 0 & - \tilde{\sigma}^{\ast}_z \end{array} \right).
\label{eq:Z2-2}
\eeq
We refere to the additional discrete symmetry in Eq.~(\ref{eq:Z2}) as the hidden ${\bm Z}_2$-symmetry, which plays a crucial role in extracting the nontrivial topology of $^3$He-B under a magnetic field.

To summarize, the B-phase in the presence of a magnetic field can be categorized to two different phases characterized by the hidden ${\bm Z}_2$-symmetry in Eq.~(\ref{eq:Z2}). The phase-I holds the symmetry 
\beq
H_{\rm mag, I} = {\rm SO}(2)^{(\hat{\bm h})}_{{\bm L}+{\bm S}} \times {\bm Z}_2 \times {\rm C} \times {\rm PU}_{\pi/2},
\eeq 
when $\hat{\ell}_z(\hat{\bm n},\varphi) = 0$ is satisfied. For a finite $\hat{\ell}_z$, however, the hidden ${\bm Z}_2$-symmetry is broken,
\beq
H_{\rm mag, II} = {\rm SO}(2)^{(\hat{\bm h})}_{{\bm L}+{\bm S}} \times {\rm C} \times {\rm PU}_{\pi/2}.
\eeq 
The condition that preserves the hidden ${\bm Z}_2$-symmetry,
\beq
\hat{\ell}_z(\hat{\bm n},\varphi) = 0,
\label{eq:ellz}
\eeq
is determined by the configuration of the order parameter of the B-phase $(\hat{\bm n},\varphi)$ and the direction of Zeeman magnetic field. Hence, altering $\hat{\ell}_z$ induces the phase transition from the ${\bm Z}_2$-symmetric phase-I to ${\bm Z}_2$-symmetry breaking phase-II. The quantity $\hat{\ell}_z$ mentioned above is transformed nontrivially as $\hat{\ell}_z \rightarrow -\hat{\ell}_z$ under the ${\bm Z}_2$ symmetry. Therefore, $\hat{\ell}_z$ is an order parameter of the hidden ${\bm Z}_2$ symmetry and it should be zero unless the discrete symmetry is spontaneously broken~\cite{mizushimaPRL2012}. 

It is well known that the $\hat{\bm n}$-vector is sensitive to the effects of a magnetic Zeeman field and surface boundary condition. This also depends on the presence of a spin-orbit interaction, such as the magnetic dipole interaction originating from the magnetic moment of $^3$He nuclei. As we will clarify in Sec.~\ref{sec:numerical2}, this implies that the phase transition associated with the spontaneous breaking the hidden ${\bm Z}_2$ symmetry can be induced by controlling an applied magnetic field and confinement geometry.

Note that the hidden ${\bm Z}_2$ symmetry in the B-phase of confined $^3$He remains even in the presence of superfluid flow~\cite{wuPRB2013}, where the phase bias that generates the flow field is parallel to the surface. The flow field explicitly breaks the ${\rm SO}(2)_{L_z+S_z}$ symmetry and the time-reversal symmetry, simultaneously. The BdG Hamiltonian, however, still remains invariant under the combined symmetry of the time-inversion and the joint $\pi$-rotation about the surface normal.

\section{Topological invariants and discrete symmetries}
\label{sec:topo}

We here make a brief review on the bulk topological properties and their physical consequences observed in the B-phase of superfluid $^3$He. The topological properties are intrinsic in the occupied states of eigenstates of $\mathcal{H}({\bm k})$, i.e., the quasiparticle states with $E_n({\bm k})<0$. Since the quasiparticle spectrum of $^3$He-B shown in Eq.~(\ref{eq:EkB}) has a finite excitation gap, the ground state is well separated from its excited states by the isotropic energy gap $\Delta _{0}$ and the occupied states are well-defined in low temperaturs. We will show that the particle-hole and time-reversal symmetries play a crucial role on determining the non-trivial topological properties in the bulk B-phase, which allows one to define the topological invariant~\cite{schnyderPRB2008,roy2008,ryuNJP2010,qiPRL2009,satoPRB2009,kitaev2009,volovik2009-1,volovik2009-2,silaevJETP2012}. Note that even for nodal superconductors and superfluids, the topological invariant is well defined in Ref.~\cite{satoPRL2010,sasakiPRL2011}.

It is also demonstrated in this section that the alternative topological invariant can be introduced as a one-dimensional winding number, by using the hidden ${\bm Z}_2$ symmetry introduced in Sec.~\ref{sec:slab} and Sec.~\ref{sec:symm_field}. Since the hidden ${\bm Z}_2$ symmetry is preserved in $^3$He-B even if the time-reversal symmetry is explicitly broken, it is applicable to the superfluid phase in the presence of a magnetic field. This will be discussed in Sec.~\ref{sec:field}.


\subsection{Topological invariant of the bulk $^3$He-B}
\label{sec:topo3d}

The BdG Hamiltonian in the B-phase holds two discrete symmetries that are the time-reversal symmetry in Eq.~(\ref{eq:TRS}) and the particle-hole symmetry in Eq.~(\ref{eq:PHS}). Combining these symmetries, as shown in Eq.~(\ref{eq:chiral1}), one can define the chiral operator $\Gamma$, which is anti-commutable with $\mathcal{H}({\bm k})$,  
\beq
\left\{ {\Gamma}, {\mathcal{H}}({\bm k}) \right\} = 0, 
\hspace{3mm} \Gamma = -i\mathcal{C} \mathcal{T}.
\label{eq:chiral}
\eeq
The discrete symmetries characterized by $\mathcal{T}^2=-1$,
$\mathcal{C}^2=+1$, and $\Gamma^2=+1$ categorize the B-phase to the DIII
class in the AZ table~\cite{schnyderPRB2008}.

In general, the topological properties of superfluids and superconductors are determined by the global structure of the Hilbert space spanned by the eigenvectors of the occupied band, $|u_n({\bm k})\rangle$ obtained from Eq.~(\ref{eq:bdg}). To capture the topological property of fully gapped superconductors and superfluids, it is convenient to introduce the so-called $Q$-matrix, 
\beq
Q({\bm k}) = \sum _{E_n>0} | u_n({\bm k})\rangle \langle u_n({\bm k}) | 
- \sum _{E_n<0} | u_n({\bm k})\rangle \langle u_n({\bm k}) | .
\eeq
Without loss of generality, we here suppose ${\rm dim}\mathcal{H}=2N$. Since the $Q$-matrix satisfies the conditions, $Q^2({\bm k})=+1$, $Q^{\dag}({\bm k})=Q({\bm k})$, and ${\rm tr}Q({\bm k})=0$, the eigenvalues are $+1$ and $-1$. The state $|u_n({\bm k})\rangle$ is a simultaneous eigenvector of the BdG Hamiltonian and $Q$-matrix, and the $Q$-matrix is obtained by continuously flattening the eigenvalues of $\mathcal{H}({\bm k})$ to $-1$ (occupied) and $+1$ (unoccupied). Therefore, the $Q$-matrix maps the Brillouin zone onto the target space spanned by the eigenvectors $|u_n({\bm k})\rangle$ of the BdG Hamiltonian. It turns out that the $Q$-matrix is anti-commutable with the chiral operator $\Gamma$, 
\beq
\left\{ {\Gamma}, Q({\bm k}) \right\} = 0,
\label{eq:chiralQ}
\eeq
where we use the symmetric relation $ |u_{-E}({\bm k})\rangle = \Gamma | u_{E}({\bm k})\rangle$ obtained from the particle-hole symmetry in Eq.~(\ref{eq:PHS}) and the time-reversal symmetry in Eq.~(\ref{eq:TRS}).

Let $U$ be a unitary matrix which diagonalize
$\Gamma$ as $U\Gamma U^{\dag} = {\rm diag}(+1_{N\times
N},-1_{N\times N})$. Then, it is obvious from Eq.~(\ref{eq:chiralQ})
that by using $U$, the $Q$-matrix becomes off-diagonal 
\beq
Q({\bm k}) = \left(
\begin{array}{cc}
0 & q({\bm k}) \\ q^{\dag}({\bm k}) & 0
\end{array}
\right),
\eeq
where $q({\bm k})$ is an element of the unitary group ${\rm U}(N)$ because
of $Q^2({\bm k})=+1$. 
Hence, the relevant homotopy group for the projector
$Q({\bm k})$ in three dimensions is given by $\pi _3 [{\rm U}(N)] = \mathbb{Z}$ . This homotopy group indicates that there
exists non-trivial mapping from the three-dimensional momentum space to
the Hilbert space. The topological invariant which characterizes the
classes of topologically distinct $q$-configurations in bulk
superconductors and superfluids is defined as the three-dimensional
winding number,  
\beq
w_{\rm 3d} = \int\frac{d{\bm k}}{24\pi^3}\epsilon _{\mu\nu\eta}{\rm tr}
\left[
(q^{\dagger}\partial _\mu q)
(q^{\dagger}\partial _\nu q)
(q^{\dagger}\partial _\eta q)
\right].
\label{eq:w3d}
\eeq
In terms of the $Q$-matrix, $w_{\rm 3d}$ is recast into
\begin{eqnarray}
w_{\rm 3d}=-\int\frac{d{\bm k}}{48\pi^3}\epsilon_{\mu\nu\eta}
{\rm tr}\left[\Gamma (Q\partial_\mu Q)(Q\partial_\nu Q)(Q\partial_\eta
	 Q)\right].
\label{eq:windingQ} 
\end{eqnarray}
For the DIII class with $\mathcal{T}^2=-1$ and $\mathcal{C}^2=+1$, the
winding number can be an arbitrary integer value. In general, however,
since the discrete symmetries prohibit certain types of maps from the
${\bm k}$ space to the target space, the value of $w_{3d}$ may be
restricted by the discrete symmetries~\cite{schnyderPRB2008}.

For single-band spin-triplet superfulids/superconductors with time-reversal invariance, there exists a simpler expression of $w_{\rm 3d}$.
From the BdG Hamiltonian 
\beq
{\cal H}({\bm k})=
\left(
\begin{array}{ccc}
\varepsilon_0({\bm k}) & i{\bm d}({\bm k})\cdot{\bm \sigma}\sigma_y \\
-i{\bm d}({\bm k})\sigma_y\cdot {\bm \sigma} &   -\varepsilon_0({\bm k})   \\  
\end{array}
\right),
\eeq
the $Q$-matrix is given by
\beq
Q({\bm k})
&=&\frac{1}{\sqrt{[\varepsilon_0({\bm k})]^2+[{\bm d}({\bm k})]^2}}
\left(
\begin{array}{ccc}
\varepsilon_0({\bm k}) & i{\bm d}({\bm k})\cdot{\bm \sigma}\sigma_y \\
-i{\bm d}({\bm k})\sigma_y\cdot {\bm \sigma} &   -\varepsilon_0({\bm k})   \\  
\end{array}
\right)
\nonumber\\
&=&\hat{\eta}_M({\bm k})\gamma_M,
\eeq
where $\hat{\eta}_M({\bm k})$ and $\gamma_M$ $(M=1,2,3,4)$ are given by
\begin{eqnarray}
\hat{\eta}_M({\bm k})=\frac{\left({\bm d}({\bm k}), \varepsilon_0({\bm k})\right)}{\sqrt{[\varepsilon_0({\bm k})]^2+[{\bm
 d}({\bm k})]^2}},
\end{eqnarray}
and
\begin{eqnarray}
\gamma_1=\left(
\begin{array}{cc}
0 & -\sigma_z\\
-\sigma_z & 0
\end{array}\right),
\quad
\gamma_2=\left(
\begin{array}{cc}
0 & i\\
-i & 0
\end{array}\right),
\nonumber\\
\gamma_3=\left(
\begin{array}{cc}
0 & \sigma_x\\
\sigma_x & 0
\end{array}\right),
\quad
\gamma_4=\left(
\begin{array}{cc}
1 & 0\\
0 & -1
\end{array}\right).
\end{eqnarray}
In the above representation of $\gamma$ matrices, the chiral operator
$\Gamma$ is written as
$\Gamma=\gamma_5\equiv\gamma_1\gamma_2\gamma_3\gamma_4$.
Substituting the $Q$-matrix for Eq.(\ref{eq:windingQ}), we obtain~\cite{satoPRB2009}
\begin{eqnarray}
w_{\rm 3d}=\int \frac{d{\bm k}}{12\pi^3}
\epsilon_{\mu\nu\eta}\epsilon_{MNPQ}
\hat{\eta}_M({\bm k})
\partial_\mu\hat{\eta}_N({\bm k})
\partial_\nu\hat{\eta}_P({\bm k})
\partial_\eta\hat{\eta}_Q({\bm k}). 
\end{eqnarray}
The four dimensional vector $\hat{\eta}_{\mu}({\bm k})$ satisfies
$\hat{\eta}_{\mu}\hat{\eta}_{\mu}=1$, so it defines a three dimensional
sphere with unit radius.
The above $w_{\rm 3d}$ counts the number of times the unit vector
$\hat{\eta}_\mu({\bm k})$ wraps the three dimensional sphere when one
sweeps the whole momentum space.
Using a topological nature of the integral, the above $w_{\rm 3d}$ can
be calculated as~\cite{satoPRB2009}
\begin{eqnarray}
w_{\rm 3d}=&-&\frac{1}{2}\sum_{{\bm k}_0} 
{\rm sgn}[\varepsilon_0({\bm k}_0)] 
{\rm sgn}[{\rm det}\{\partial_\mu d_\nu({\bm k}_0)\}]
\nonumber\\
&+&\frac{1}{2}
{\rm sgn}[\varepsilon_0(\infty)] 
{\rm sgn}[{\rm det}\{\partial_id_j(\infty)\}],
\end{eqnarray}
where the summation is taken for all ${\bm k}_0$ satisfying ${\bm d}({\bm
k}_0)=0$ and ${\rm det}\{\partial_\mu d_\nu({\bm k}_0)\}$ denotes the
determinant of the $3\times 3$ matrix
$\partial_\mu d_\nu({\bm k}_0)$. 
The second term in the right hand side
is the contribution from ${\bm k}_0\rightarrow \infty$.
From the above formula, the winding number of the B-phase is evaluated
as $w_{\rm 3d}=1$.

The physical consequence of the three-dimensional winding number is the emergence of the two-dimensional Majorana cone on the surface, whose effective Hamiltonian is described as~\cite{qiPRL2009,volovikJETP2010}
\beq
\mathcal{H}_{\rm surf} = \sum _{{\bm k}_{\parallel}} \psi^{\rm T}_{-{\bm k}_{\parallel}} c \left( {\bm k}_{\parallel} \times \tilde{\bm \sigma} \right)\cdot\hat{\bm s}\psi _{{\bm k}_{\parallel}},
\label{eq:Hsurf}
\eeq
where we set $\tilde{\sigma}_{\mu} = R_{\mu \nu}(\hat{\bm n},\varphi)\sigma _{\nu}$. The momentum parallel to the surface is ${\bm k}_{\parallel}$ and the unit vector normal to the surface is denoted by $\hat{\bm s}$. This effective surface Hamiltonian yields the gapless relativistic spectrum with the velocity $c = \Delta _0 /k_{\rm F}$, which is protected by the time-reversal symmetry. In addition, the field operator $\psi$ in the Nambu space is self-conjugate, that is, the Majorana fermion. The additional perturbation term, 
\beq
\mathcal{H}_{\rm mass} = M \sum _{{\bm k}_{\parallel}}  \psi^{\rm T}_{-{\bm k}_{\parallel}} \tilde{\bm \sigma}\cdot\hat{\bm s}\psi _{{\bm k}_{\parallel}},
\eeq
opens the mass gap on the surface Majorana cone, where the mass $M$ changes its sign under the time-inversion $\mathcal{T}$. The resultant surface spectrum is 
\beq
E_{\rm surf}({\bm k}_{\parallel}) = \sqrt{c^2 k^2_{\parallel} + M^2}. 
\eeq
The manifestation of three-dimensional topological superconductors is the coupling to the gravitational field through the gravitational instant term~\cite{wangPRB2011,ryuPRB2012}. Due to the nontrivial topological property, the gapped Majorana cone is responsible for the quantization of thermal Hall conductivity (see also Sec.~\ref{sec:thermal}). It has also been predicted that the coupling of the Majorana fermion to the gravitational field gives rise to cross correlated responses~\cite{nomuraPRL2012}.

The B-phase of the bulk superfluid $^3$He serves a concrete example of
three-dimensional topological superfluids and superconductors, where the
nontrivial topology is protected by the particle-hole symmetry and
time-reversal symmetry~\cite{schnyderPRB2008}. After a pioneering work
by Grinevich and Volovik~\cite{grinevichJLTP1988},
the topological superfluidity of the B-phase of the bulk $^3$He was discussed by
Schnyder {\it et al}.~\cite{schnyderPRB2008}, Roy~\cite{roy2008}, Qi {\it et al}.~\cite{qiPRL2009} and Sato~\cite{satoPRB2009}. We will demonstrate in Sec.~\ref{sec:field} that confined $^3$He-B under a magnetic field undergoes a topological phase transition concomitant with spontaneous ${\bm Z}_2$ symmetry breaking at which the surface Majorana fermion acquires the finite mass proportional to the Zeeman energy. In this situation, the mass term is parameterized by the quantity $\hat{\ell}_z(\hat{\bm n},\varphi)$ defined in Eq.~(\ref{eq:ell}) that determines the hidden ${\bm Z}_2$ symmetry. 

Another candidate of three-dimensional topological superfluids and superconductors 
is the carrier-doped topological insulator, that is,
Cu$_x$Bi$_2$Se$_3$. The recent point-contact experiment in
Ref.~\cite{sasakiPRL2011} observed a pronounced zero-bias conductance
peak supporting a topological odd-parity superconductivity of
Cu$_x$Bi$_2$Se$_3$. While the zero-bias conductance peak has been
observed in other experiments~\cite{kirzhner1,kirzhner2,kirzhner3},
conflicting experimental results have been reported in
Refs.~\cite{peng2013,levy} where the zero-bias conductance peak is
absent. Among four possible pairing symmetries proposed by Fu and Berg~\cite{fuPRL2010},
the $A_{1g}$ state that is the even-parity pairing is topologically
trivial, while the $A_{1u}$, $A_{2u}$, and $E_u$ states are topological
odd-parity pairing. The topological nontriviality for the $A_{1u}$ state
is characterized by $w_{\rm 3d}$ introduced in Eq.~(\ref{eq:w3d}) and
the topology of $A_{2u}$ and $E_{u}$ states is characterized by the
parity of $w_{\rm 3d}$~\cite{satoPRBR2010,sasakiPRL2011}. On the (111) surface which is naturally cleaved, the $A_{1u}$ and $E_u$ states are accompanied by topologically protected gapless states, when the Fermi surface is a sphere enclosing the $\Gamma$ point. The gapless states are responsible for zero-bias conductance peak~\cite{yamakagePRB2012,haoPRB2011,hsiehPRL2012}. On the other hand, if the Fermi surface is cylindrical, the gappless states may disappear~\cite{satoPRB2009,satoPRBR2010}. The recent conflicting experimental results are consistently understandable in the context of the topological odd-parity $A_{1u}$ or $E_u$ superconductivity with the Fermi surface evolution~\cite{mizushima2014}.



\subsection{Hidden symmetry and one-dimensional winding number} 
\label{sec:hidden}

The nontrivial topology of the superfluid $^3$He-B is ensured by the
time-reversal symmetry in Eq.~(\ref{eq:TRS}), being characterized by the
three-dimensional winding number. The B-phase of superfluid $^3$He holds
the three-dimensional rotational symmetry ${\rm SO}(3)_{{\bm L}+{\bm
S}}$ in addition to the time-reversal and particle-hole symmetries. The
microscopic Hamiltonian for the B-phase, therefore, holds the additional
discrete symmetry introduced in Eq.~(\ref{eq:hidden}). This hidden ${\bm
Z}_2$ symmetry is very robust and useful for determining the topological
properties of $^3$He-B under a magnetic field, because it can be
preserved even if each symmetry, $\mathcal{T}$ or $\mathcal{U}_z(\pi)$,
is explicitly broken, as was mentioned in Sec.~\ref{sec:symm_field}. As we will clarify below, the combination of the particle-hole symmetry and the hidden ${\bm Z}_2$ symmetry leads to the chiral symmetry, which ensures the nontrivial topological invariant and its intriguing physical consequences, such as Majorana Ising spins.

Combining the hidden discrete symmetry in Eq.~(\ref{eq:hidden}) with the particle-hole symmetry in Eq.~(\ref{eq:PHS}), one can define the hidden chiral symmetry on the one-dimensional line of the ${\bm k}$-space as
\beq
\left\{ {\Gamma}_1, {\mathcal{H}}(0,0,k_z) \right\} = 0, 
\hspace{3mm} \Gamma_1 = \mathcal{C}\mathcal{T}\mathcal{U}(\pi).
\label{eq:chiral1d}
\eeq
In the same manner as the procedure in Sec.~\ref{sec:topo3d}, the one-dimensional winding number is defined by using the $Q$-matrix as
\beq
w_{\rm 1d} = \frac{1}{2\pi} {\rm Im}\int^{\infty}_{-\infty} dk_z \left[
\partial _{k_z}\ln\det q({\bm k})\right]_{{\bm k}_{\parallel} = {\bm 0}},
\label{eq:w1d}
\eeq
where we set ${\bm k}_{\parallel} \equiv (k_x,k_y)$. 

It is now worth mentioning that $w_{\rm 1d}$ can be expressed in terms of the BdG Hamiltonian $\mathcal{H}({\bm k})$. For this purpose, let us first notice that the BdG Hamiltonian can becomes off-diagonal in the basis that the chiral operator in Eq.~(\ref{eq:chiral1d}) is diagonalized as $\Gamma = {\rm diag}(+1, -1)$,
\beq
\mathcal{H}({\bm k}_{\perp}) = \left(
\begin{array}{cc}
0 & h({\bm k}_{\perp}) \\ h^{\dag}({\bm k}_{\perp}) & 0
\end{array}
\right),
\eeq
where $h({\bm k}_{\perp})$ is given as $h({\bm k}_{\perp}) \!=\! -
\epsilon ({\bm k}_{\perp})\sigma_y - \Delta ({\bm k}_{\perp})$, and  we have introduced ${\bm k}_{\perp} \equiv (0,0,k_z)$. Using $h({\bm k}_{\perp})$, the eigenvector of $\mathcal{H}({\bm k}_{\perp})$ is given as
\beq
|u_n({\bm k}_{\perp}) \rangle = \frac{1}{\sqrt{2}}\left(
\begin{array}{c}
\varphi _n ({\bm k}_{\perp}) \\ 
h^{\dag}({\bm k}_{\perp})\varphi _n ({\bm k}_{\perp}) / |E_n({\bm k}_{\perp})|
\end{array}
\right),
\eeq
where $\varphi _n$ satisfies $h({\bm k}_{\perp})h^{\dag}({\bm k}_{\perp})\varphi _n = E^2({\bm k}_{\perp})\varphi _n$ and $\langle \varphi _n | \varphi _m \rangle = \delta _{n,m}$. The matrix $q({\bm k}_{\perp})$ is then associated with the BdG Hamiltonian $h({\bm k}_{\perp})$ through the relation
\beq
\left. q({\bm k})\right|_{{\bm k}_{\parallel}={\bm 0}} = \sum _n \frac{\varphi _n({\bm k}_{\perp}) \varphi^{\dag}_n({\bm k}_{\perp})}{|E_n({\bm k}_{\perp})|}
h({\bm k}_{\perp}).
\eeq
Therefore, the relation, ${\rm Im}\ln\det [\varphi _n\varphi^{\dag}_n/|E_n|] = 0$ yields ${\rm Im}[\ln\det q] = {\rm Im}[\ln\det h]$. This allows us to introduce the alternative expression of the one-dimensional winding number, $w_{1d} = \frac{1}{2\pi} {\rm Im}\int dk_z [\partial _{k_z}\ln\det h({\bm k})]_{{\bm k}_{\parallel} = {\bm 0}}$. Then, the one-dimensional winding number is given in terms of the BdG Hamiltonian as
\beq
w_{\rm 1d} = - \frac{1}{4\pi i}\int^{\infty}_{-\infty} dk_z {\rm tr}\left[
\Gamma _1 \mathcal{H}^{-1}({\bm k})\partial _{k_z} \mathcal{H}({\bm k})
\right]_{{\bm k}_{\parallel} = {\bm 0}}.
\label{eq:topo1d}
\eeq
The winding number is evaluated as~\cite{satoPRB2011}
\beq
w_{\rm 1d} = \frac{1}{2} \sum_{k_z} {}^{\prime}
{\rm sgn}\left[
\varepsilon _0 \psi - {\bm \varepsilon}\cdot{\bm d}
\right]
{\rm sgn}
\left[
\partial _{k_z}\left( 
\varepsilon^2_0 - 
{\bm \varepsilon}^2
\right)\right],
\eeq
where the summation $\sum^{\prime}_{k_z}$ is taken for $k_z$ satisfying $\varepsilon^2_0({\bm k})-{\bm \varepsilon}^2({\bm k})=0$ with ${\bm k}_{\parallel}={\bm 0}$. This winding number in the case of $^3$He-B is estimated as $w_{\rm 1d} = 2$ for $\mu > 0$ and $w_{\rm 1d} =0$ for $\mu < 0$, unless the hidden ${\bm Z}_2$ symmetry is broken.

This winding number was first introduced in a different context by Wen and Zee~\cite{wen} in order to clarify the topological stability of bulk nodes of an electron hopping Hamiltonian in a magnetic field, $\mathcal{H}=\sum _{ij} t_{ij}c^{\dag}_i c_j u_{ij}$. It can be proven that the winding number $w_{\rm 1d}$ is a topological invariant, implying that $w_{\rm 1d}$ is invariant under any continuous deformation of $\mathcal{H}$ without breaking the hidden chiral symmetry. Using the winding number, Sato and Fujimoto uncovered the nontrivial topological properties of noncentrosymmetric superconductors under a spatially uniform magnetic field~\cite{satoPRB2009-2}. The bulk-edge correspondence was proven in Ref.~\cite{satoPRB2011} which indicates that the nontrivial value of $w_{\rm 1d}$ corresponds to the number of zero energy quasiparticle states bound at the edge of the system. We will clarify in Sec.~\ref{sec:majorana} that the bulk-edge correspondence and chiral symmetry brings physical consequence that the magnetic response of zero energy states is highly anisotropic, called the Majorana Ising spins.

To give an intuitive understanding on $w_{\rm 1d}$, it is convenient to introduce the following two-dimensional unit vector ${\bm m}({\bm k}) = (m_1({\bm k}),m_2({\bm k}))$, 
\beq
m_1 ({\bm k}) = \frac{2[\varepsilon _0 ({\bm k})\psi ({\bm k})- \varepsilon _{\mu}({\bm k})d_{\mu}({\bm k})]}{N({\bm k})}, \\
m_2 ({\bm k}) = \frac{{\bm \varepsilon}^2({\bm k})-{\bm d}({\bm k})^2-\varepsilon^2({\bm k})+\psi^2_0({\bm k})}{N({\bm k})},
\eeq
where $N({\bm k})$ is a normalization constant determined by $|{\bm m}({\bm k})| \!=\! 1$. The single-particle Hamiltonian density $\varepsilon ({\bm k})$ is parametrized in terms of the scalar and vector parts in Eq.~(\ref{eq:epsilon}). By using this two-dimensional unit vector ${\bm m}({\bm k})$, the winding number in Eq.~(\ref{eq:topo1d}) is recast into the form,
\beq
w_{\rm 1d} = \frac{1}{2\pi} \int dk_z \epsilon^{ij} m_i ({\bm k})\partial _{k_z} m_j ({\bm k}) .
\eeq
This describes the map of he one-dimensional line of the ${\bm k}$-space onto the target space $S^1$ characterized by the two-dimensional unit vector ${\bm m}({\bm k}) = (m_1({\bm k}),m_2({\bm k}))$. The winding number $w_{\rm 1d}$, therefore, counts how many times the map warps the target space $S^1$ and $w_{\rm 1d} \in \mathbb{Z}$~\cite{tanakaJPSJ2012,satoPRB2011}.


\subsection{Majorana fermion and Ising spin}
\label{sec:majorana}

Now let us show that the topological invariant protected by the hidden
discrete symmetry brings an important consequence, that is, the Ising
spin character of Majorana fermions. Below we consider the low energy limit where only the zero energy quasiparticle states at ${\bm k}_{\parallel}\!=\!{\bm 0}$ contribute.

{\it Majorana field.}---
We now construct the quantized field ${\bm \Psi} = (\psi _{\uparrow},\psi _{\downarrow}, \psi^{\dag}_{\uparrow},\psi^{\dag}_{\downarrow})^{\rm T}$ in terms of energy eigenstates of the BdG Hamiltonian. The mean-field approximated Hamiltonian is rewritten in the coordinate space as
\beq
\mathcal{H} = \int d{\bm r}_1\int d{\bm r}_2 {\bm \Psi}^{\dag}({\bm r}_1)\mathcal{H}({\bm r}_1,{\bm r}_2){\bm \Psi}({\bm r}_2). 
\eeq
Since the four-component quantized field ${\bm \Psi}$ in the Nambu space satisfies the self-charge-conjugate constraint 
\beq
{\bm \Psi}({\bm r}) = \mathcal{C}{\bm \Psi}({\bm r}),
\label{eq:selfcharge}
\eeq
the BdG Hamiltonian holds the particle-hole symmetry as
$\mathcal{C}\mathcal
{H}({\bm r}_1,{\bm r}_2)\mathcal{C}^{-1}=-\mathcal{H}({\bm r}_1,{\bm
r}_2)$, as shown in Eq.~(\ref{eq:PHS}).
By choosing a phase factor, the energy eigenstates of the BdG Hamiltonian that is determined by the BdG equation in the coordinate space,
\beq
\int d{\bm r}_2 \mathcal{H}({\bm r}_1,{\bm r}_2){\bm \varphi}_{E}({\bm r}_2) = E{\bm \varphi}_E({\bm r}_1), 
\eeq
may satisfy the particle-hole symmetry
\beq
{\bm \varphi}_E({\bm r}) = \mathcal{C} {\bm \varphi}_{-E}({\bm r})
\label{eq:PHS2}. 
\eeq
The eigenfunctions obey the orthogonality and completeness relations 
\beq
\int d{\bm r} {\bm \varphi}^{\dag}_E({\bm r}) {\bm \varphi}_{E^{\prime}}({\bm r}) = \delta _{E,E^{\prime}}, \\
\sum _E {\bm \varphi}_E({\bm r}_1){\bm \varphi}^{\dag}_E({\bm r}_2) = \delta ({\bm r}_1-{\bm r}_2). 
\eeq
The quantized field also satisfies the anti-commutation relations, 
$\{ \psi _{\sigma}({\bm r}), \psi^{\dag}_{\sigma^{\prime}}({\bm r}^{\prime})\} 
=\delta _{\sigma,\sigma^{\prime}}\delta({\bm r}-{\bm r}^{\prime})$ 
and 
$\{  \psi _{\sigma}({\bm r}), \psi _{\sigma^{\prime}}({\bm
r}^{\prime})\} = \{ \psi^{\dag}_{\sigma}({\bm r}),
\psi^{\dag}_{\sigma^{\prime}}({\bm r}^{\prime}) \}= 0$. The further
details on the mean-field approximation in the coordinate space will be described in Sec.~\ref{sec:NG}.

Now we expand the quantized field ${\bm \Psi}$ by the eigenfunctions ${\bm \varphi}_E({\bm r})$,
\beq
{\bm \Psi}({\bm r}) = \sum_{E}{\bm \varphi}_E({\bm r})\eta_E.
\eeq
In the case of topological superconductors, there exist gapless quasiparticles that are bound to the surface. The fermion ${\bm \Psi}({\bm r})$ constructed from such gapless quasiparticles obeys the Dirac equation with the self-conjugate constraint in Eq.~(\ref{eq:selfcharge}), and thus it  is referred to as Majorana fermion~\cite{semenoff,wilczek,wilczek2014,jackiw2014,chamon}. 
Owing to the particle-hole symmetry in Eq.~(\ref{eq:PHS2}), the branch with positive energies is identical to that with negative energies. Hence, supposing that there exist $n$ zero-energy states ${\bm \varphi}^{(a)}_{0}({\bm r})$ ($a=1,\cdots,n$), we can rewrite the quantized field ${\bm \Psi}$ as 
\beq
{\bm \Psi}({\bm r}) = \sum^n_{a=1} {\bm \varphi}^{(a)}_{0}({\bm r}) \gamma^{(a)}
+ \sum _{E>0} \left[ 
{\bm \varphi}_E({\bm r}) \eta _E + \mathcal{C}{\bm \varphi}_E({\bm r}) \eta^{\dag}_E
\right],
\label{eq:PsiM}
\eeq
where we have used $\gamma^{(a)}$, instead of $\eta_{E=0}$, in order to distinguish these zero modes.
The self-conjugate constraint in Eq.~(\ref{eq:selfcharge}) imposes the following relations, 
\beq
\eta_E^{\dagger}=\eta_{-E},
\quad
\gamma^{(a)} = \gamma^{(a)\dag},
\eeq
the latter of which implies that the zero modes are composed of equivalent contributions from particle and hole excitations.
Since the zero modes satisfy the self-conjugate constraint by themselves, 
they are called as Majorana zero mode.
From the completeness relation of eigenfunctions, the quasiparticle operator $\eta_{E>0}$ satisfies the anti-commutation relations, $\{ \eta _E, \eta^{\dag}_{E^{\prime}}\} = \delta _{E,E^{\prime}}$ and $\{\eta _E, \eta _{E^{\prime}} \}= \{\eta^{\dag}_E, \eta ^{\dag}_{E^{\prime}} \}=0$. 
%
The operator of the Majorana zero modes, however,  does not obey the standard anti-commutation relation but has the following relation
\beq
\gamma^{(a)2} = 1,\hspace{3mm} \{ \gamma^{(a)}, \eta _{E} \} = \{ \gamma^{(a)}, \eta^{\dag}_E \} = 0.
\label{eq:anti}
\eeq 
As we discuss below, this unusual relation gives rise to a singnificant feature inherent to Majorana field. 

In the case that there exists two Majorana zero modes ($n=2$), the minimal representation of the algebra in Eq.~(\ref{eq:anti}) is two-dimensional.
The two dimensional representation is built up by introducing two degenerate vacua $|\pm \rangle$ that obey $\eta _{E>0} |\pm \rangle = 0$ and by defining new fermion operator $c$ and $c^{\dag}$ as
\beq
c = \frac{1}{\sqrt{2}} (\gamma^{(1)}+i\gamma^{(2)}), \hspace{3mm}
c^{\dag} = \frac{1}{\sqrt{2}} (\gamma^{(1)}-i\gamma^{(2)}).
\eeq 
It is obvious that this {\it complex fermion} operator obeys the standard anti-commutation relations, $\{c,c^{\dag}\}=1$ and $\{c,c\}=\{c^{\dag},c^{\dag}\}=0$. Since the complex fermion has the zero energy, the two degenerate vacua $|\pm \rangle$ are defined as the eigenstate of the fermion parity. We here suppose $| + \rangle$ ($| - \rangle$) to be the vacuum with the fermion parity even (odd), 
\beq
c|-\rangle = c^{\dag}|+\rangle = 0.
\eeq
Two Hilbert spaces are now spanned by using the vacuum state $|\pm \rangle$ and excited states that are constructed as $\eta^{\dag}_{E}\eta ^{\dag}_{E^{\prime}}\eta ^{\dag}_{E^{\prime\prime}} \cdots |\pm \rangle $. The complex zero mode operators, $c$ and $c^{\dag}$, connect two Hilbert spaces as
\beq
c | + \rangle = | - \rangle, \hspace{3mm}
c^{\dag} | - \rangle = | + \rangle .
\eeq

Note that the introduction of the complex fermion is indispensable to preservation of the fermion parity~\cite{semenoff,chamon}
. Indeed, even without introduction of $c$, $\gamma^{(a)}$ itself can be diagonalized as $\gamma^{(a)} |a \pm \rangle = \pm  | a \pm \rangle$ with the following 
$|a\pm\rangle$,
\beq
\left|1 \pm \right\rangle = \frac{1}{\sqrt{2}}(\left| + \right\rangle \pm \left| - \right\rangle), \hspace{3mm}
\left|2 \pm \right\rangle = \frac{1}{\sqrt{2}}(e^{i\frac{\pi}{4}}\left| + \right\rangle 
\pm e^{-i\frac{\pi}{4}}\left| - \right\rangle),
\label{eq:majovector}
\eeq
but the eigenstates are superpositions of two vacua with opposite fermion parity, so they cannot be physical states preserving the fermion parity. 

Another important consequence from Eq.~(\ref{eq:majovector}) is that the eigenstates of the single Majorana zero mode is intrinsically entangled. When two Majorana zero modes $\gamma^{(1)}$ and $\gamma^{(2)}$ are spatially separated and well isolated from other quasiparticle states with higher energies, they yield the ``teleportation'' and non-local correlation~\cite{semenoff,semenoff2006}. Moreover, when the Majorana zero modes bound to quantum vortices, these host vortices obey the non-Abelian statistics~\cite{ivanovPRL2001}. The representations of the braiding operation of vortices are obtained as a discrete set of the unitary group which manipulates the occupation of complex fermions~\cite{ivanovPRL2001,ohmi}. Since the zero modes are topologically protected against quantum decoherence, they offer a promising platform to realize topological quantum computation~\cite{ivanovPRL2001,ohmi,nayakRMP2008,kitaev,freedman,sarmaPRL2005,tewariPRL2007-1,zhangPRL2007}. 

An effective realization of non-local correlation and non-Abelian statistics requires Majorana fermions without internal degree of freedom. Possible realization of such Majorana fermions in effectively spinless superconductors includes axion strings~\cite{satoPL2003}, fermionic cold atoms with a $p$-wave Feshbach resonance~\cite{read,gurarieAP2007,mizushimaPRL2008}, proximity induced superconductivity on the surface of a topological insulator~\cite{fuPRL2008}, an $s$-wave superconductor with the Rashba spin-orbit interaction and the Zeeman field~\cite{satoPRL2009,satoPRB2010,sauPRL2010}, and one-dimensional nanowire systems~\cite{lutchynPRL2010,oregPRL2010,aliceaNP2011}.
In addition, the low-energy physics of half-quantized vortices in spinful chiral $p$-wave superconductors is describable as a spinless chiral superconductor~\cite{ivanovPRL2001}.

Recently, however, it has been clarified that additional discrete symmetries, such as mirror reflection symmetries, may help to retain the properties of non-local correlation and non-Abelian statistics even in spinful topological superconductors with spinful Majorana fermions~\cite{uenoPRL2013,SatoPhysicaE2014}. The additional discrete symmetry add strong constraint on Majorana zero modes and the zero modes are separated to subsectors that are eigenstates of the additional discrete symmetry. Note that statistics of vortices having multiple Majorana fermions has also been discussed in Refs.~\cite{yasui,hirono,eto}.

In the following, we will clarify the another important consequence of spinful Majorana zero modes, that is the Majorana Ising spin.  It turns out that this Ising character is a generic consequence obtained from an extra chiral symmetry like Eq.~(\ref{eq:chiral1d})~\cite{shiozaki2014}.  
This is expected to be realized in a variety of materials, such as non-centrosymmetric superconductors, the heavy fermion superconductor UPt$_3$, and the superconducting topological insulator Cu$_x$Bi$_2$Se$_3$ as well as $^3$He-B confined to a slab. 




{\it Index theorem.}---
We note that the commutation relation, $[\Gamma _1,\mathcal{H}^2(0,0,k_z)]=0$, is obtained from the chiral symmetry in Eq.~(\ref{eq:chiral1d}). As discussed in Sec.~\ref{sec:symmetry}, all the eigenstates with a finite energy $E\neq 0$ have chiral partners, where a state with the chirality $\Gamma _1 = +1$ is always paired with another state with $\Gamma _1 = -1$. 

According to Ref.~\cite{satoPRB2011}, however, the zero energy states do not form chiral pairs in general. They are eigenstates of $\Gamma_1$, and the number of the zero energy states with $\Gamma_1=+1$, $n_+$, is not same as the number of the $\Gamma_1=-1$ zero energy states, $n_-$. From the generalized index theorem in Ref.~\cite{satoPRB2011}, the non-zero value of $w_{1d}$ in Eq.~(\ref{eq:w1d}) is equal to the difference between the number of zero energy states in each chiral sector,
\beq 
w_{\rm 1d} = n_- - n_+.
\eeq 
Hence, 
at least $|w_{\rm 1d}|$ zero energy states exist under any disturbances preserving the chiral symmetry. 

For superfluid $^3$He-B, one finds 
\beq
w_{\rm 1d} = n_- = 2, \hspace{3mm} n_+ =0 .
\label{eq:winding3HeB}
\eeq  
There appear doubly degenerated zero energy states at ${\bm k}_{\parallel} \!=\! {\bm 0}$ bound to the surface of $^3$He-B.
The doubly degenerated zero energy states have no chiral partner, and the wavefunctions, ${\bm \varphi}^{(a)}_{{\bm
k}_{\parallel}\!=\!{\bm 0}}({\bm r})$ ($a=1, 2$), satisfy 
\beq
\Gamma _1{\bm \varphi}^{(a)}_{{\bm k}_{\parallel}={\bm 0}}({\bm r})
= -{\bm \varphi}^{(a)}_{{\bm k}_{\parallel}={\bm 0}} ({\bm r}),
\label{eq:zes1}
\eeq
with $\Gamma _1 = \mathcal{C}\mathcal{T}\mathcal{U}_z(\pi)=\mathcal{U}(\hat{\bm n},\varphi)\sigma _x\tau _y\mathcal{U}^{\dag}(\hat{\bm n},\varphi)$.
One can also impose the relation
\beq
{\bm \varphi}^{(a)}_{{\bm k}_{\parallel}={\bm 0}}({\bm r})
= {\cal C}{\bm \varphi}^{(a)}_{{\bm k}_{\parallel}={\bm 0}} ({\bm r}),
\label{eq:phse0}
\eeq
by the particle-hole symmetry.
From these two relations in Eqs.~(\ref{eq:zes1}) and (\ref{eq:phse0}), 
${\bm \varphi}^{(a)}_{{\bm k}_{\parallel} \!=\! {\bm 0}}$ has a generic form as 
\beq
{\bm \varphi}^{(a)}_{{\bm k}_{\parallel} = {\bm 0}} ({\bm r}) = 
\left( 
\begin{array}{c}
\chi^{(a)}({\bm r}) \\
\chi^{(a)\ast}({\bm r})
\end{array}
\right), 
\label{eq:general}
\eeq 
where the two-component spinor is defined with an arbitrary function $\xi^{(a)}({\bm r})$ as
\beq
\chi^{(a)}({\bm r})= U(\hat{\bm n},\varphi)
\left( 
\begin{array}{c}
\xi^{(a)}({\bm r}) \\
i\xi^{(a)\ast}({\bm r})
\end{array}
\right).
\eeq

{\it Majorana Ising spin.}---
Having clarified in Eq.~(\ref{eq:general}) that the particle-hole symmetry and chiral symmetry imposes constraint on the wavefunction of Majorana zero modes, 
we now show that the doubly degenerate zero modes behave as an Ising spin.
Since we concentrate our attention to the low-energy limit, we here ignore the contributions from non-zero energy modes. 
As shown in Eq.~(\ref{eq:PsiM}), the quantized field ${\bm \Psi} $
is then expanded in terms of topologically protected zero energy states as
${\bm \Psi}({\bm r}) \!=\! \sum_{a=1,2}{\bm \varphi}^{(a)}_{{\bm k}_{\parallel} \!=\! {\bm
0}}({\bm r})\gamma^{(a)}$ with a couple of real operators $\gamma^{(1)}$ and $\gamma^{(2)}$. 
From the general form of ${\bm \varphi}^{(a)}_{{\bm k}_{\parallel} \!=\! {\bm 0}}$ shown in Eq.~(\ref{eq:general}), one obtains the following condition, 
\beq
\left(
\begin{array}{c}
\psi _{\uparrow}({\bm r}) \\
\psi _{\downarrow}({\bm r})
\end{array}
\right) = i \sigma _{\mu}R_{\mu z}(\hat{\bm n},\varphi)
\left(
\begin{array}{c}
\psi^{\dag}_{\downarrow}({\bm r}) \\
-\psi^{\dag}_{\uparrow}({\bm r}) 
\end{array}
\right),
\label{eq:Mcondition}
\eeq
which is a general consequence of symmetry protected topological phases hosting the additional chiral symmetry in Eq.~(\ref{eq:chiral1d}) and the nontrivial winding number in Eq.~(\ref{eq:winding3HeB}). The rotation matrix $R_{\mu z}(\hat{\bm n},\varphi)$ is associated with the broken symmetry of the B-phase, ${\rm SO}(3)_{{\bm L}-{\bm S}}$. For $\hat{\bm n}=\hat{\bm z}$ and $\varphi = 0$, the relation reduces to $\psi _{\uparrow}({\bm r}) = i\psi^{\dag}_{\downarrow}({\bm r})$.

Now following Refs.~\cite{mizushimaPRL2012, chungPRL2009, stone}, one can show that the Majorana Ising condition in Eq.~(\ref{eq:Mcondition}) yields
the Ising character of the topologically protected zero energy states.
We now introduce the local density operator and the spin density operators in the Nambu space as 
\beq
\rho ({\bm r}) \equiv \frac{1}{2}\left[{\psi}^{\dagger}_a({\bm r}){\psi}_a
-{\psi}_a({\bm r}){\psi}^{\dagger}_a({\bm r})\right], 
\label{eq:densityOP} \\
{\bm S}({\bm r}) \equiv
\frac{1}{4}\left[
{\psi}^{\dagger}_a({\bm r}){\bm \sigma}_{ab}{\psi}_b({\bm r})
-{\psi}_a({\bm r}){\bm \sigma}^{\rm T}_{ab}{\psi}^{\dagger}_b({\bm r})\right].
\label{eq:spinOP}
\eeq
Substituting the Majorana condition of Eq.~(\ref{eq:Mcondition}) into Eqs.~(\ref{eq:densityOP}), one finds that the surface zero energy states do not contribute to the local density operator,
\beq
\rho^{({\rm surf})} ({\bm r}) = 0.
\label{eq:rho0}
\eeq 
This indicates that the Majorana fermions protected by the hidden chiral symmetry can not be coupled to the local density fluctuation and thus are very robust against non-magnetic impurities. Similarly, the local spin operator is constructed from the surface Majorana fermion in Eq.~(\ref{eq:Mcondition}) as 
\beq
S^{({\rm surf})}_{\mu} = R_{\mu z}(\hat{\bm n},\varphi) S^{\rm M}_z,
\label{eq:MIS}
\eeq
where $S^{\rm M}_z$ is the logical spin operator in the case of $\hat{\bm n}=\hat{\bm z}$ and $\varphi = 0$. Equation (\ref{eq:MIS}) implies that only the $S_z$ component is nonzero while the other components are identically zero when $\hat{\bm n}=\hat{\bm z}$ and $\varphi = 0$. 

These results indicate that the topologically protected zero energy states are not coupled to local density fluctuation, and the local spin density yields Ising-like anisotropy. Here note that we only use a general property of the hidden chiral symmetry and particle-hole symmetry, thus the Ising character is a direct consequence of our symmetry protected topological phase. In Secs.~\ref{sec:index2} and \ref{sec:odd}, we will demonstrate that the Majorana Ising character is retained not only by the zero energy states but also by the whole dispersion of the surface Majorana cone with $E_{\rm surf}({\bm k}_{\parallel}) $, when the Andreev approximation is valid, that is, $\Delta _{0}\ll E_{\rm F}$.

The anisotropic magnetic response of surface Majorana fermions was first predicted by Stone and Roy~\cite{stone} in the context of chiral $p$-wave superconductors. 
Sato and Fujimoto~\cite{satoPRB2009-2} demonstrated that the anisotropic magnetic responce also emerges in topological phases of noncentrosymmetric superconductors.
They also revealed that the anisotropic behavior is characterized by the one-dimensional winding number introduced in Eq.~(\ref{eq:topo1d}).
Majorana Ising spin in $^3$He-B was demonstrated by Chung and Zhang~\cite{chungPRL2009}, by directly solving the Bogoliubov-de Gennes equation with the Andreev approximation. They proposed an electron spin relaxation experiment to detect the Majorana nature of the surface bound states, whose details will be discussed in Sec.~\ref{sec:detecting}. 
To establish an experimental way to probe the Ising spin, Shindou {\it et al.}~\cite{shindouPRB2010} examined the coupling of a spin-$1/2$ magnetic impurity to Majorana fermions bound at the edge of two-dimensional topological superconductors. Due to quantum dissipation from the Majorana Ising spin, the quantum impurity spin yields strongly anisotropic and singular magnetic response. Hence, the electron spin resonance may serve as a local probe for Majorana Ising spins.




Before closing this section, we would like to mention that additional discrete symmetries other than the hidden ${\bm Z}_2$ symmetry may play a crucial role on the nontrivial topological properties of time-reversal invariant superconductors and superfluids. In the case of $^3$He-B, as shown in Eq.~(\ref{eq:chiral1d}), the chiral operator $\Gamma _1$ is constructed from the hidden ${\bm Z}_2$ symmetry $\mathcal{T}\mathcal{U}_z(\pi)$ with the particle-hole symmetry. For time-reversal invariant superconductors, however, the hidden ${\bm Z}_2$ symmetry may arise from other combinations such as $\mathcal{T}\mathcal{M}$ where $\mathcal{M}$ denotes the mirror reflection operator. The role of mirror symmetry has been clarified in quasi-one-dimensional fermionic gases with a synthetic gauge field~\cite{mizushimaNJP2013} and the $E_{1u}$ scenario of the B-phase of the heavy-fermion superconductor UPt$_3$~\cite{tsutsumiJPSJ2013}. The details will be discussed in Sec.~\ref{sec:mirror}. Recently, numerous works have extended the topological classification of bulk wave functions to include the point group symmetry and magnetic symmetry in topological insulators~\cite{slager,jadaun,paananen,cxliu2013,alexandradinata,mong,cxliu2013v2} and in topological superconductors~\cite{uenoPRL2013,SatoPhysicaE2014,teo,zhangPRL2013,kotetes,liu2013,hughes2013,benalcazar,fang2013,hsieh2014}. The topological classification enlarged from the Altland-Zirnbauer tenfold ways has been proposed in Refs.~\cite{chiuPRB2013,morimoto,shiozaki2014}. All these works have unveiled that such additional discrete symmetry enriches nontrivial topological properties and support the emergence of Majorana fermions in topological superconductors.

\section{Jackiw-Rebbi index theorem and surface Andreev bound states}
\label{sec:andreev}

We have shown in the previous section that the topological nontriviality and existence of gapless quasiparticles bound to a surface and interface are closely linked to underlying discrete symmetries. In this section, we will describe the existence of gapless quasiparticle states in the surface of unconventional superconductors, by explicitly solving the Bogoliubov-de Gennes equation with the Andreev approximation. This will give an intuitive understanding for the existence of gapless bound states. 

First, starting from the second quantized Hamiltonian for attractively interacting spin-$\frac{1}{2}$ fermions, we derive the Bogoliubov-de Gennes equation and gap equation which gives a closed set for describing self-consistent quasiparticle structures in spatially inhomogeneous superconductors and superfluids. In Sec.~\ref{sec:jackiw}, we demonstrate that the resultant Andreev equation for an unconventional superconductor with a specular surface can be mapped onto the one-dimensional Dirac equation with a mass domain wall. It is well known that the Dirac equation can be accompanied by the exact zero-energy state as a consequence of the index theorem~\cite{jackiw}. The index theorem indicates that the existence of the zero energy state localized at the domain wall is determined by the sign change of the mass domains. We extend the index theorem to the case of chiral superconductors with time-reversal symmetry breaking in Sec.~\ref{sec:index} and to the time-reversal invariant superfluid $^3$He-B in Sec.~\ref{sec:index2}, respectively. We reproduce the Majorana condition in Eq.~(\ref{eq:Mcondition}) and Majorana Ising spin shown in Eq.~(\ref{eq:MIS}).

\subsection{Nambu-Gor'kov formalism}
\label{sec:NG}

We start with the Hamiltonian for spin-$\frac{1}{2}$ fermions interacting through an attractive interaction $V^{c,d}_{a,b}({\bm r}_1,{\bm r}_2)$
\beq
\mathcal{H} &=& 
\int d{\bm r}\psi^{\dag} _{a}({\bm r})\varepsilon _{ab}(-i{\bm \nabla})\psi _{b}({\bm r}) \nn \\
&&+ \frac{1}{2}\int d{\bm r}_1\int d{\bm r}_2V^{c,d}_{a,b}({\bm r}_1,{\bm r}_2)
\psi^{\dag} _{a}({\bm r}_1)\psi^{\dag} _{b}({\bm r}_2)\psi _{c}({\bm r}_2)\psi _{d}({\bm r}_1).
\eeq
where $\varepsilon _{ab}(-i{\bm \nabla})$ is a single-particle Hamiltonian density in the coordinate space which in general contains external potentials. The fermion field operator with spin $a \!=\! \uparrow, \downarrow$, $\psi _{a}({\bm r})$, must obey the anti-commutation relations,
$\{ \psi _a ({\bm r}_1), \psi^{\dag}_b ({\bm r}_2) \} = \delta _{ab}\delta ({\bm r}_{12})$ and 
$\{ \psi _a ({\bm r}_1), \psi _b ({\bm r}_2) \} = 
\{ \psi^{\dag}_a ({\bm r}_1), \psi^{\dag}_b ({\bm r}_2) \} = 0$, where ${\bm r}_{12}={\bm r}_1-{\bm r}_2$ is the relative coordinate.
The interaction potential is parameterized as $V^{c,d}_{a,b}({\bm r}_1,{\bm r}_2)\!=\!V_{1}({\bm r}_{12})\delta _{a,d}\delta _{b,c} \!+\! V_2({\bm r}_{12}){\bm \sigma}_{ad}\cdot{\bm \sigma}_{bc}$, which is invariant under the spin space rotation ${\rm SO}(3)_{\bm S}$. It is now convenient to introduce the projection operators onto the spin-singlet and -triplet pairing states, $\mathcal{P}_{\rm s}$ and $\mathcal{P}_{\rm t}$. Using the projection operators, the interaction potential is recast into 
\beq
V^{c,d}_{a,b}({\bm r}_1,{\bm r}_2)=V_{\rm s}({\bm r}_{12})\mathcal{P}_{\rm s} + V_{\rm t}({\bm r}_{12})\mathcal{P}_{\rm t},
\eeq
where $V_{\rm s} \equiv V_1-\frac{3}{4}V_2$ and $V_{\rm t} \equiv V_1 + \frac{1}{4} V_2$ are the interaction potentials in spin-singlet and -triplet channels.

All the information on superfluid phases in equilibrium at a finite temperature $T$ are contained by the Matsubara Green's functions defined as
\beq
\underline{G}(x_1,x_2) &=& 
\left(
\begin{array}{cc}
\mathcal{G}(x_1,x_2) & \mathcal{F}(x_1,x_2) \\ \bar{\mathcal{F}}(x_1,x_2) & \bar{\mathcal{G}}(x_1,x_2)
\end{array}
\right) \nn \\
&=& - \langle\langle T_{\tau} [{\bm \Psi}(x_1)\bar{\bm \Psi}(x_2)] \rangle\rangle .
\label{eq:G}
\eeq 
We here set $x_j \!\equiv\! (\tau_j, {\bm r}_j)$ and $\langle\langle\cdots\rangle\rangle
\!\equiv\! {\rm Tr}[e^{(\Omega-\mathcal{H})/T}\cdots]$ with the
thermodynamic potential $\Omega$. We have introduced the field
operator in Nambu space as
${\bm \Psi} = (\psi _{\uparrow},\psi _{\downarrow},\psi^{\dag}_{\uparrow},\psi^{\dag}_{\downarrow})^{\rm T}$, where $\bar{\bm \Psi} = {\bm \Psi}^{\dag}({\bm r},-\tau)$. 
The Matsubara Green's functions are governed by the Nambu-Gor'kov equation derived from the Kadanoff-Baym conserving approximation~\cite{serene,baym1,baym2} as
\beq
\int dx_3
\left[
\underline{G}^{-1}_0(x_1,x_3)
-\underline{\Sigma}(x_1,x_3)
\right] \underline{G}(x_3,x_2) = \delta (x_1-x_2).
\label{eq:NG}
\eeq
The $4\times 4$ matrix form of the bare Green's function $\underline{G}^{-1}_0$ in Eq.~(\ref{eq:NG}) is given as
$\underline{G}^{-1}_0(x_1,x_2) \equiv \delta (x_{12})[ - \partial _{\tau_2} -
\underline{\varepsilon}(-i{\bm \nabla}_2)]$, where $\underline{\varepsilon} \equiv {\rm diag}(\varepsilon,-\varepsilon^{\ast})$.
The $\Phi$-functional generates the perturbation expansion for the skelton self-energy diagrams, 
\beq
\underline{\Sigma} [\underline{G}] = 2 \frac{\delta \Phi[\underline{G}]}{\delta \underline{G}^{\rm T}}. 
\eeq
The Luttinger-Ward thermodynamic potential $\Omega[\underline{G}]$ is obtained as a functional of $\underline{G}$ as~\cite{luttinger} 
\beq
\Omega [\underline{G}] = -\frac{1}{2}
{\rm Sp}\left[ \ln(-\underline{G}^{-1}_0+\underline{\Sigma})+ \underline{\Sigma}~ \underline{G} \right] + \Phi[\underline{G}],
\label{eq:LW}
\eeq
where ${\rm Sp} \cdots \equiv \int dx_1\int dx_2 {\rm Tr} \cdots$. 
Below, we expand $\underline{G}$ by the Matsubara frequency $\omega _n = (2n+1)\pi T$ as $\underline{G}(x_1,x_2) = T\sum _n \underline{G}({\bm r}_1,{\bm r}_2;\omega _n)e^{-i\omega _n \tau _{12}}$.

Now we show that the Gor'kov equation is reduced to the
Bogoliubov-de Gennes (BdG) equation,
\begin{eqnarray}
\int d{\bm r}_2 \underline{\mathcal{H}}({\bm r}_1,{\bm r}_2)
{\bm \varphi}_{i}({\bm r}_2)
= E_{i} {\bm \varphi}_{i}({\bm r}_1),
\label{eq:bdg3}
\end{eqnarray}
where $\underline{\mathcal{H}}$ is a $4\!\times\!4$ matrix in Nambu space, $\underline{\mathcal{H}}({\bm r}_1,{\bm r}_2) = \delta ({\bm r}_{12})\underline{\varepsilon}(-i{\bm \nabla}_2) + \Sigma ({\bm r}_1,{\bm r}_2)$. 
Here we have omitted the diagonal part of the self-energy matrix.
The four-component eigenvector ${\bm \varphi}_{i}({\bm r})$ fulfills the orthonormal condition,
$\int {\bm \varphi}^{\dag}_{i}({\bm r}){\bm \varphi}_{j}({\bm r}) d{\bm
r} = \delta _{i,j}$.  
We first note that 
the BdG Hamiltonian in the coordinate space is particle-hole symmetric,
$
\mathcal{C}\underline{\mathcal{H}}({\bm r}_1,{\bm r}_2)\mathcal{C}^{-1} = 
-\underline{\mathcal{H}}({\bm r}_1,{\bm r}_2)
$, as shown in Eq.~(\ref{eq:PHS}).
The particle-hole symmetry of the BdG Hamiltonian ensures that the
positive energy solution ${\bm \varphi}_{E}({\bm r})$ is
associated with the negative energy solution ${\bm \varphi}_{-E}({\bm
r}) = \mathcal{C}{\bm \varphi}_{E}({\bm r})$.  
Therefore, the following $4\times 4$ unitary matrix
$\underline{u}_{i} \equiv [{\bm \varphi}^{(1)}_{i},
{\bm \varphi}^{(2)}_{i}, 
\mathcal{C}{\bm\varphi}^{(1)}_{i}, 
\mathcal{C}{\bm \varphi}^{(2)}_{i}] 
$ diagonalizes the BdG Hamiltonian as
$ \int d{\bm r}_1 \int d{\bm r}_2\underline{u}^{\dag}_{i}({\bm r}_1) 
\underline{\mathcal{H}}({\bm r}_1,{\bm r}_2)
\underline{u}_{i}({\bm r}_2) = \underline{E}_{i}
$, 
with
$\underline{E}_{i} \equiv {\rm diag}( 
E^{(1)}_{i}, E^{(2)}_{i}, -E^{(1)}_{i}, -E^{(2)}_{i})$. 
The unitary matrix $\underline{u}_{n}({\bm r})$ satisfies the orthonormal
and completeness conditions,  
$\int \underline{u}^{\dag}_{i}({\bm r}) \underline{u}_{j}({\bm r})
d{\bm r}
= \delta _{i,j}$ and  
$\sum _{i} \underline{u}_{i}({\bm r}_1) \underline{u}^{\dag}_{i}({\bm r}_2)
= \delta ({\bm r}_{12})$.

By using the unitary matrix $\underline{u}_{i}$, the solution of the Gor'kov equation
(\ref{eq:NG}) is obtained as, 
\beq
\underline{G}({\bm r}_1,{\bm r}_2; \omega _n) = \sum _{i} \underline{u}_{i}({\bm r}_1) 
\left( i\omega _n \underline{\tau}_0 - \underline{E}_{i} 
\right)^{-1} \underline{u}^{\dag}_{i}({\bm r}_2),
\label{eq:G2}
\eeq
which can be recast into
\beq
\underline{G}({\bm r}_1,{\bm r}_2; \omega _n)
= \sum _{E_i>0} \left[ 
\frac{{\bm \varphi}_{i}({\bm r}_1){\bm \varphi}^{\dag}_{i}({\bm r}_2)}{i\omega _n - E_{i}}
+ 
\frac{\mathcal{C}{\bm \varphi}_{i}({\bm r}_1){\bm \varphi}^{\dag}_{i}({\bm r}_2)\mathcal{C}^{-1}}
{i\omega _n + E_{i}}
\right].
\label{eq:Gfinal}
\eeq
The pair potentials $\Delta _{ab}$ are defined by the anomalous part of the Green's functions as
\begin{eqnarray}
\Delta _{ab} ({\bm r}_1,{\bm r}_2)  
=
\lim _{\eta \rightarrow 0}
T \sum _n \mathcal{V}^{c,d}_{a,b}({\bm r}_{12})
\mathcal{F}_{cd}({\bm r}_2,{\bm r}_1;\omega _n)
e^{-i\omega _n \eta}.
\label{eq:gap}
\end{eqnarray}
From Eq.~(\ref{eq:Gfinal}),
the sum over the Matsubara frequency results in the
Fermi distribution function $f(x) \!\equiv\! 1/(e^{x/T}+1)$. 
The BdG equation Eq.(\ref{eq:bdg3}) and the gap equation
(\ref{eq:gap}) offers a self-consistent framework for describing superfluid phases in equilibrium. 

\subsection{Jackiw-Rebbi index theorem}
\label{sec:jackiw}

\begin{figure}[tb!]
\begin{center}
\includegraphics[width=70mm]{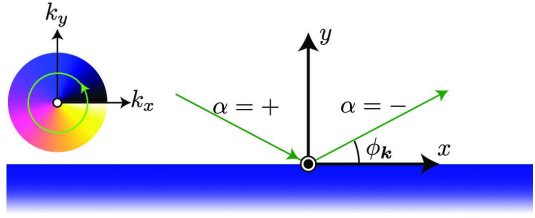}
\end{center}
\caption{Geometry of specular reflections at the surface of chiral $\ell$-wave superconductors. We also depict the phase profile of chiral $p$-wave pairing in the momentum space.}
\label{fig:specular}
\end{figure}

Let us consider a simple form of the single-particle Hamiltonian density with an effective mass $M^{\ast}$, which is given as $\varepsilon(-i{\bm \nabla}) = - \frac{1}{2M^{\ast}} \nabla^2 - E_{\rm F}$. 
Following the Andreev approximation~\cite{andreev}, we decompose the quasiparticle wavefunction ${\bm \varphi}_{i}({\bm r})$ to the slowly varying part $\tilde{\varphi}$ and the rapid oscillation part with the Fermi wave length $k^{-1}_{\rm F}$, 
\beq
{\bm \varphi}_{i}({\bm r}) 
= \sum _{\alpha = \pm }C_{\alpha} \tilde{\bm \varphi}_{\alpha}({\bm r})e^{i{k}_{\alpha}\cdot{\bm r}}, 
\label{eq:andreevwf}
\eeq
where ${\bm k}_{\alpha} \!=\! k_{\rm F}(\cos\phi _{\bm k}\sin\theta _{\bm k},\alpha\sin\phi _{\bm k}\sin\theta _{\bm k},\cos \theta _{\bm k})$ denotes the momentum of incoming ($\alpha \!=\! +$) and outgoing ($\alpha \!=\! -$) quasiparticles (Fig.~\ref{fig:specular}). The normalization condition is imposed on $\tilde{\bm \varphi}_{\alpha}({\bm r})$ as $\sum _{\alpha}\int d{\bm r} \tilde{\bm \varphi}^{\dag}_{\alpha}({\bm r}) \tilde{\bm \varphi}_{\alpha}({\bm r}) = 1$. The rigid boundary condition at ${\bm r} = {\bm R}$, ${\bm \varphi}({\bm R})=0$, leads to $C_+=-C_-$ and the continuity condition $\tilde{\varphi}_{+}({\bm R}) = \tilde{\varphi}_-({\bm R})$, where ${\bm R}=(R_x,0,R_z)$ is the coordinate on the surface. This Andreev approximation holds within the weak-coupling regime, $k_{\rm F}\xi = 2E_{\rm F}/\Delta _0 \gg 1$.
Substituting Eq.~(\ref{eq:andreevwf}) into the BdG equation (\ref{eq:bdg3}), one obtains the Andreev equation for $\tilde{\varphi}_{\alpha}({\bm r})$ as
\beq
\left[
-i {\bm v}_{{\rm F}}(\hat{\bm k}_{\alpha}) \cdot {\bm \nabla} \underline{\tau}_z  + \underline{v} + \underline{\Delta} (\hat{\bm k}_{\alpha},{\bm r})
\right]\tilde{\bm \varphi}_{\alpha}({\bm r}) = E \tilde{\bm \varphi}_{\alpha}({\bm r}), 
\label{eq:a}
\eeq
with $\hat{\bm k}_{\alpha}\equiv {\bm k}_{\alpha}/|{\bm k}_{\alpha}|$. 
The Fourier transformed form of the pair potential with respect to the relative coordinate ${\bm r}_{12}\equiv {\bm r}_1-{\bm r}_2$ is given as
${\Delta} ({\bm k},{\bm r}) = \int d{\bm r}_{12}e^{-i{\bm k}\cdot{\bm r}_{12}}\Delta ({\bm r}_1,{\bm r}_2)$, where the momentum dependence of the pair potential is dominated by the Fermi momentum, that is, ${\bm k}\approx{\bm k}_{\rm F} \equiv k_{\rm F}\hat{\bm k}$ and 
$\underline{\Delta} ({\bm k},{\bm r}) \approx \underline{\Delta} (\hat{\bm k},{\bm r})$. Here, we have introduced the $4\times 4$ matrix form of the pair potential, 
\beq
\underline{\Delta} (\hat{\bm k},{\bm r}) = 
\left(
\begin{array}{cc}
\hat{0} & {\Delta}(\hat{\bm k},{\bm r}) \\ -{\Delta}^{\ast}(-\hat{\bm k},{\bm r}) & \hat{0}
\end{array}
\right).
\eeq

Before going into the detailed analysis of the Andreev equation for unconventional superconductors, let us first mention a generic consequence arising from the topological structure of $\Delta (\hat{\bm k},{\bm r})$. To clarify this, we reduce Eq.~(\ref{eq:a}) to a minimal situation, that is, the spatially one-dimensional system along the $\hat{\bm y}$-axis with a spin-singlet $s$-wave pairing $\Delta (\hat{\bm k},{\bm r}) \!=\! \Delta (y)i\sigma _y$. Then, the Andreev equation (\ref{eq:a}) is recast into 
\beq
\left[
-i v_{\rm F}\partial _y \underline{\tau}_x  + \Delta (y)\underline{\tau}_z
\right]\tilde{\bm \varphi}(y) = E \tilde{\bm \varphi}(y), 
\label{eq:dirac}
\eeq
where we assume $\Delta (y) \!\in\! \mathbb{R}$ and carry out the unitary transformation $U\tau _xU^{\dag}\!=\!\tau _z$ and $U\tau _z U^{\dag} \!=\! \tau _x$. For simplicity, we omit the spin degrees of freedom. Hence, the one-dimensional Andreev equation is equivalent to the one-dimensional Dirac equation with a spatially inhomogeneous mass $\Delta (y)$. This serves as an effective model to study the low-energy quasiparticle structure of unconventional superconductors and superfluids with a specular surface, interface, and a quantized vortex.

Let us first consider fermionic excitations in a single domain wall of an $s$-wave superconductor, which offers an exactly solvable model of the BdG equation~\cite{takayama}. We first note that the Dirac equation (\ref{eq:dirac}) can be recast into $(-iv_{\rm F}\partial _y \pm i \Delta)f_{\pm} = E f_{\pm}$ with $(f_+,f_-)^{\rm T} = T\tilde{\varphi}$ by making the unitary transformation $T=(\tau _y + \tau _z)/\sqrt{2}$. Then, by multiplying it by $(-iv_{\rm F}\partial _y \mp i \Delta)$, the equation for the wavefunction $f_{\pm}$ is reduced to the Schr\"{o}dinger-type equation, 
\beq
\left[ v^2_{\rm F}\frac{d^2}{dy^2} + E^2 - \Delta^2(y) \pm v_{\rm F} \frac{d\Delta (y)}{dy} \right] f_{\pm} (y) = E f_{\pm}(y).
\label{eq:schro}
\eeq
The pair potential describing the domain wall structure is given by 
\beq
\Delta (y) = \Delta _0 \tanh \left(\frac{y}{\xi }\right),
\label{eq:kink}
\eeq
where we set $\xi = a\xi _0$ with the superconducting coherence length $\xi _0 = v_{\rm F} /\Delta _0$ and $a$ is the parameter self-consistently determined with the gap equation. Then, Eq.~(\ref{eq:schro}) with Eq.~(\ref{eq:kink}) is solvable with the help of the hypergeometric function. The bound state energy with $|E_n|< \Delta _0$ is then obtained as
\beq
E_n = \left\{
\begin{array}{ll}
0 & \mbox{for $n=0$ and $\forall{a}$} \\
\\
\displaystyle{\pm \Delta _0 \sqrt{\frac{n}{a}\left( 2-\frac{n}{a}\right)}} & \mbox{for $n=1,2,\cdots < a$}
\end{array}
\right. .
\eeq
The wavefunction of the zero energy state is given by 
\beq
f_+(y) =\frac{N}{[\cosh(y/\xi)]^a}, \hspace{3mm} f_-(y) = 0 ,
\eeq
where $N$ is the normalization constant. Therefore, there always exists a single zero energy state that is bound to a domain wall within the length scale of $\xi _0$. 

Jackiw and Rebbi \cite{jackiw} clarified the topologically nontrivial structure of the one-dimensional Dirac equation (\ref{eq:dirac}), which offers a generalization of the argument based on the kink solution with Eq.~(\ref{eq:kink}). The consequence is that at least, one zero energy solution exists when the mass term $\Delta (y)$ changes its sign. To clarify this, we concentrate our attention to a zero energy solution with $E=0$. Then, the eigenfunction of the zero energy state is obtained by integrating Eq.~(\ref{eq:dirac}) as
\beq
\tilde{\bm \varphi}(y) = N \exp\left( 
- \frac{1}{v_{\rm F}}\int^{y}_0 \Delta (y^{\prime})dy^{\prime}
\right)\left( 
\begin{array}{c}
1 \\ i
\end{array}
\right),
\eeq
where $N$ is a normalization constant. 
By assuming that the mass term approaches a uniform value in the limit of $y\rightarrow \pm \infty$, it is obvious that one of the zero energy solutions is normalizable only when the mass term changes its sign at $y\rightarrow \pm \infty$ as
\beq
{\rm sgn}\frac{\displaystyle{\lim _{y\rightarrow +\infty}\Delta(y)}}
{\displaystyle{\lim _{y\rightarrow -\infty}\Delta (y)}} = -1. 
\eeq
Hence, the existence of the zero energy solutions in the BdG equation is determined only by the sign change of the pair potential at $y\rightarrow\pm\infty$ and is independent of the detailed structure of the interface.

Equation (\ref{eq:dirac}) self-consistently coupled to the gap equation gives an effective theory for describing low-lying electronic states in the one-dimensional Peierls system~\cite{brazovskii,mertsching,horovitz,takayama,yamamoto,nakaharaPRB1981}, spin density waves~\cite{machidaPRB1984-2}, the spin-Peierls system~\cite{fujitaJPSJ1984}, the stripes in high-$T_{\rm c}$ cuprates~\cite{machida1989}, superconducting junction systems~\cite{kashiwayaRRP}, and Fulde-Ferrell-Larkin-Ovchinnikov states~\cite{machidaPRB1984,yoshii}. Indeed, the kink solution with Eq.~(\ref{eq:kink}) offers a prototype to determine the boundary between a spatially uniform ordered state and spatially modulated ordered state, such as the phase transition from the BCS state to Fulde-Ferrell-Larkin-Ovchinnikov state in a Pauli-limited superconductor. Using the hypergeometric function, Nakahara~\cite{nakahara} derived the dispersion and wave functions of quasiparticles that are bound to the chiral domain wall of the superfluid $^3$He-A film, where each domain has different chirality. The self-consistent pair potential is well described with the the kink potential in Eq.~(\ref{eq:kink}). In addition, note that low-lying quasiparticles bound at a quantized vortex of superconductors also share a common physics, because the low-energy theory can be mapped onto the one-dimensional Dirac or Majorana equation~\cite{tewariPRL2007}.

The counterpart of spatially inhomogeneous fermionic condensates in the quantum field theory corresponds to the Gross-Neveu model~\cite{gross} and Nambu-Jona-Lasinio model~\cite{nambuPR1961} which effectively describes dynamical chiral symmetry breaking~\cite{thies}. Recently, the general solutions in Eq.~(\ref{eq:dirac}) have been proposed by using a technique of the Ablowitz-Kaup-Newell-Segur hierarchy well-known in integrable systems~\cite{akns}. These include a single-kink state~\cite{hu,bar-sagi,shei}, multiple kinks (kink-anti-kink and kink-polaron states)~\cite{campbell,okuno,feinberg2003,feinberg2004}, and complex kinks and their crystalline states~\cite{basar1,basar2,basar3,takahashi} as well as a single kink solution in Eq.~(\ref{eq:kink}).




\subsection{Index theorem for chiral $\ell$-wave pairing states}
\label{sec:index}

Let us now apply the index theorem to the surface Andreev bound state in a chiral $\ell$-wave pairing state,
\beq
\Delta (\hat{\bm k},{\bm r}) = D(\theta _{\bm k})\left(\hat{k}_{x} + i{\rm sgn}(\ell)\hat{k}_y\right)^{|\ell|},
\eeq
where $D(\theta _{\bm k}) $ is supposed to be an arbitrary form of the pair potential depending on $\theta _{\bm k}$ and $\theta _{\bm k}$ is defined in Eq.~(\ref{eq:andreevwf}). Specifically let us set the specular wall at $y \!=\! 0$ and suppose the $x$- and $z$-axis to be parallel to the wall. For simplicity, we also do not take account of spin degrees of freedom, that is, a $S_z$-preserving chiral $\ell$-wave pairing state. Then, the Andreev equation (\ref{eq:a}) reduces to 
\beq
\left[ - i\alpha v_{\rm F}\sin\!\phi _{\bm k}\sin\!\theta _{\bm k}\partial _y {\sigma}_z 
+ \tilde{D}(\theta _{\bm k}){\sigma}_x e^{-i\alpha \ell \phi _{\bm k}\sigma _z} \right]
\tilde{\bm \varphi}_{\alpha} (y) = E \tilde{\bm \varphi}_{\alpha}(y),
\label{eq:ay}
\eeq
where $\tilde{D}(\theta _{\bm k}) \equiv D(\theta _{\bm k},y) (\sin\theta _{\bm k})^{|\ell|}$. 

The chiral pairing spontaneously breaks the time-reversal symmetry. The topologically nontrivial properties are characterized by the first Chern number, which is nonzero for $|k_z|< k_{\rm F}$ in chiral $\ell$-wave superconductors. This ensures the existence of the flat band zero energy states, forming the topologically protected Fermi arc extending in the $k_z$ direction.. 

Following the procedure in Refs.~\cite{stone,mizushimaPRB2012}, we introduce the following coordinate $\tilde{y}_{\pm}$, corresponding the distance along the classical trajectory:
$\tilde{y}_{\alpha } \equiv \alpha \frac{y}{v_{\rm F}\sin\phi _{\bm k}\sin\theta _{\bm k}} 
= \frac{y}{v_{\rm F}\sin (\alpha \phi _{\bm k})\sin\theta _{\bm k}}$. 
Now, let us introduce the new axis 
$\tilde{y}_+ \mapsto \rho > 0$ and 
$-\tilde{y}_- \mapsto \rho < 0$. 
Using this new axis, the Andreev equation (\ref{eq:ay}) reduces to the one-dimensional Dirac (or Majorana) equation with a mass domain wall
\beq
\left[
- i \partial _{\rho} {\sigma}_z + \tilde{D}(\theta _{\bm k},\rho) {\sigma}_xe^{-i{\sigma}_z\vartheta (\rho)}
\right]{\bm \psi}(\rho) = E {\bm \psi}(\rho),
\label{eq:diracl}
\eeq
where 
${\bm \psi}(\rho) =
\tilde{\bm \varphi}_{+}(\rho)$ for $\rho \!>\! 0$ and 
${\bm \psi}(\rho) =\tilde{\bm \varphi}_{-}(\rho)$ for $\rho \!<\! 0$.
The phase $\vartheta (\rho)$ is given as
$\vartheta (\rho) =
\phi _{\rm R} \equiv \ell \phi _{\bm k}$ for $\rho \!>\! 0$ and 
$\vartheta (\rho) =\phi _{\rm L} \equiv -\ell \phi _{\bm k}$ for $\rho \!<\! 0$, where $\phi _{\rm L}-\phi _{\rm R} \!\in\! [0,2\pi]$ is required.

According to the Jackiw-Rebbi's index theorem~\cite{jackiw}, the one-dimensional Dirac equation has a zero energy eigenstate when the mass domain wall has phase jump of $(2n-1)\pi$, where $n \!\in \! \mathbb{Z}$. 
Since the Andreev equation has the phase jump $2\ell\phi _{\bm k}$ at the surface ($\rho =0$), the Jackiw-Rebbi's index theorem ensures that the zero energy solution appears at 
$\phi _{{\bm k},n}  = \left( n - \frac{1}{2} \right) \frac{\pi}{|\ell|} \in [0,\pi]$.
As a result, the chiral $\ell$-wave pairing state has $|\ell|$ gapless points at 
\beq
\hat{k}_x = \sin\!\theta _{\bm k}\cos\left[ \left( n - \frac{1}{2}\right) \frac{\pi}{ |\ell| }\right] ,
\eeq
where $n \!\in\! \mathbb{N}$ satisfies $ n = 1, 2, \cdots, |\ell| $. This is a generic consequence of the Andreev equation for chiral $\ell$-wave superconductors. 

\begin{figure}[tb!]
\begin{center}
\includegraphics[width=80mm]{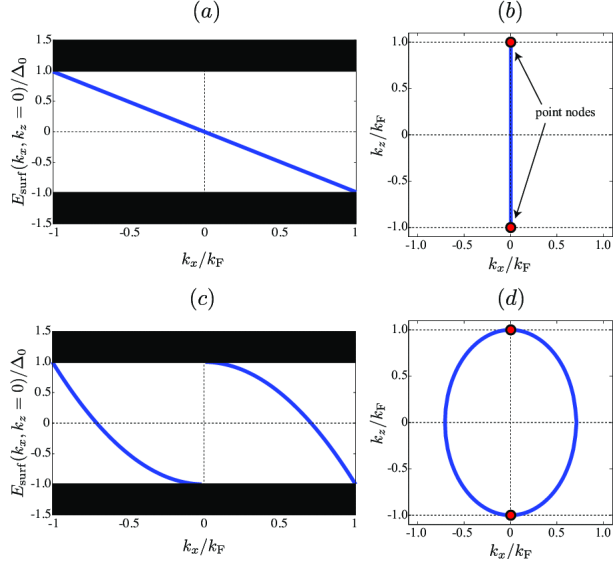}
\end{center}
\caption{Dispersions of the surface Andreev bound states at $k_z=0$, $E_{\rm surf}(k_x,k_z=0)$, for $\ell = 1$ (a) and $\ell = 2$ (c). Fermi arcs for $\ell=1$ (b) and $\ell = 2$ (d) which are defined as $(k_x,k_z)$ satisfying $E_{\rm surf}(k_x,k_z)=0$. The filled circles in (b) and (c) depict the point nodes. }
\label{fig:chirald}
\end{figure}

For spatially uniform pair potential $D(\hat{\bm k},y) = D(\hat{\bm k})$, Eq.~(\ref{eq:diracl}) gives
$E_{\rm surf}(k_x,k_y) = \tilde{D}(\theta)\cos[(\phi _{\rm L}-\phi _{\rm R})/2]$, which determines the dispersion of the Andreev bound state~\cite{mizushimaPRB2012,stone}.
For example, in the case of the $\ell\!=\! 2$ pairing state, that is, the chiral $d$-wave pairing state,
one finds that $\phi _{\rm R} = 2\phi _{\bm k}$ and 
$\phi _{\rm L} = 
2\pi - 2\phi _{\bm k}$ for $\phi _{\bm k}\in [0,\pi/2]$ and 
$\phi _{\rm L} = 4\pi - 2\phi _{\bm k}$ for $\phi _{\bm k}\in [\pi/2,\pi]$, respectively.
Thus, the dispersion of the surface Andreev bound state in a chiral $d$-wave superconductor is
\beq
E_{\rm surf}(k_x,k_y) = - {\rm sgn}(k_x) \frac{|D(\theta _{\bm k})|}{k^2_{\rm F}}
\left( 2 k^2_x - k^2_{\rm F}\sin^2\theta _{\bm k}
\right).
\eeq
The dispersion of the surface Andreev bound states in chiral $p$-wave ($\ell = 1$) and $d$-wave ($\ell=2$) superconductors, $E_{\rm surf}(k_x,k_z=0)$, is plotted in Fig.~\ref{fig:chirald}(a) and (c), respectively. Because of time-reversal symmetry breaking, the dispersion can be asymmetric in $k_x$, i.e., $E_{\rm surf}(k_x) = -E_{\rm surf}(-k_x)$. Note that as shown in Fig.~\ref{fig:chirald}(b) and (d), the Fermi arc appears in the momentum space $(k_x,k_z)$ parallel to the surface, where the zero energy flat band is terminated at the projection of two point nodes at $k_x = 0$ and $k_z = \pm k_{\rm F}$. Since the negative energy part of the branch is occupied in the ground state, the surface Andreev bound states carry the spontaneous mass current in equilibrium~\cite{stone,tsutsumiJPSJ2012,mizushimaPRL2008,furusaki,mizushimaPRA2010-1}.

\subsection{Surface Majorana fermions and Ising spins in superfluid $^3$He-B}
\label{sec:index2}

We now extend the results in Sec.~\ref{sec:index} to the surface Andreev bound states in the B-phase of superfluid $^3$He confined in a slab geometry. In accordance with the topological argument in Sec.~\ref{sec:topo}, the time-reversal symmetry ensures the existence of zero energy states that are bound to the surface of $^3$He-B. By explicitly solving the BdG equation with the Andreev approximation, we demonstrate that there exist gapless bound states. We also reproduce that the resultant gapless states shows the Majorana Ising nature. 

Let us now solve the Andreev equation (\ref{eq:a}) for the B-phase with a specular surface in the absence of a Zeeman field. As discussed in Sec.~\ref{sec:slab}, confinement explicitly breaks the three-dimensional rotation symmetry. The resultant pair potential in this restricted geometry is expressed in Eq.~(\ref{eq:dvec_slab}) with Eq.~(\ref{eq:dvec_b2_initial}), which is ${\rm SO}(2)_{L_z+S_z}$ symmetric. It is convenient to introduce the unitary matrix defined as
$S_{\phi_{\bm k}} \!\equiv\! (\sigma _x+\sigma _z)e^{i\vartheta\sigma _z}/\sqrt{2}$ with $\vartheta = \frac{\phi _{\bm k}}{2} - \frac{\pi}{4}$. Combining this with the ${\rm SU}(2)$ spin rotation $U(\hat{\bm n},\varphi)$, one finds that the B-phase pair potential can be transformed into the diagonal representation,
\begin{eqnarray}
{\Delta}(\hat{\bm k},{\bm r}) &=& U(\hat{\bm n},\varphi) {\Delta}_0(\hat{\bm k},{\bm r}) U^{\rm T}(\hat{\bm n},\varphi) \nn \\
&=& U(\hat{\bm n},\varphi) S_{\phi_{\bm k}}
\left( 
\begin{array}{cc}
\Delta _+ (\hat{\bm k},{\bm r}) & 0 \\
0 & \Delta _- (\hat{\bm k},{\bm r})
\end{array}
\right)
S^{\rm T}_{\phi_{\bm k}}U^{\rm T}(\hat{\bm n},\varphi).
\label{eq:unitaryB} 
\end{eqnarray}
The resultant pair potentials, $\Delta _{\pm}$, behave as chiral $p$-wave pairing in each sector, 
\beq
\Delta _{\pm} (\hat{\bm k},{\bm r})
= \pm \Delta _{\perp}({\bm r})\cos\theta _{\bm k} 
+ i \Delta _{\parallel}({\bm r})\sin\theta _{\bm k},
\label{eq:deltapm}
\eeq
where we set ${\bm k}_{{\rm F}} \!=\! k_{\rm F}(\cos\!\phi _{\bm k}\sin\!\theta _{\bm k},\sin\!\phi _{\bm k}\sin\!\theta _{\bm k},\cos _{\bm k}\!\theta _{\bm k})$. This reveals that the unitary transformation maps the B-phase order parameter onto two subsectors with different chirality. Hence, the Andreev equation (\ref{eq:a}) in the absence of a magnetic field reduces to that for spinless chiral $p$-wave superconductors~\cite{mizushimaPRB2012,stone}.

Without loss of generality, we set the specular surface to be normal to the $\hat{\bm z}$-axis. We also suppose the spatially uniform isotropic energy gap $\Delta _{\parallel}=\Delta _{\perp} \equiv \Delta _0$. In accordance with the consequence of the index theorem in Sec.~\ref{sec:index}, then, the bound state solution with $|E({\bm k}_{\parallel})| \!\le\! \Delta _0$ has the energy dispersion linear in the momentum ${\bm k}_{\parallel} \!=\! (k_x,k_y)$ as
\beq
E_0({\bm k}_{\parallel}) = \pm \frac{\Delta _0}{k_{\rm F}} |{\bm k}_{\parallel}|.
\label{eq:E0a}
\eeq 
This expression is independent of the orientation of $\hat{\bm n}$ and the angle $\varphi$. The corresponding wavefunctions for the quasiparticles bound at $z\!=\! 0$ are given by
\beq
{\bm \varphi}^{(+)}_{0,{\bm k}_{\parallel}} ({\bm r}) = N_{\bm k}
e^{i{\bm k}_{\parallel}\cdot{\bm r}_{\parallel}}f(k_{\perp},z)\mathcal{U}(\hat{\bm n},\varphi)
\left(
{\bm \Phi}_+
- e^{i\phi _{\bm k}}{\bm \Phi}_-
\right), 
\label{eq:varphi1}
\eeq
where $N_{\bm k}$ is the normalization constant and $\mathcal{U} \!\equiv\! {\rm diag}(U,U^{\ast})$. The wavefunction ${\bm \varphi}^{(+)}_{0,{\bm k}_{\parallel}}$ corresponds to the positive energy solution of $E_0({\bm k}_{\parallel})$ and ${\bm \varphi}^{(-)}_{0,{\bm k}_{\parallel}}$ is the negative branch. The particle-hole symmetry in Eq.~(\ref{eq:PHS}) ensures the one-to-one correspondence between the two branches of the energy eigenstates through ${\bm \varphi}^{(-)}_{0,{\bm k}_{\parallel}}({\bm r}) \!=\! \mathcal{C}{\bm \varphi}^{(+)}_{0,-{\bm k}_{\parallel}}({\bm r})$. In Eq.~(\ref{eq:varphi1}), we also set $f(k_{\perp},z)\!=\! \sin\left( k_{\perp}z\right)e^{-z/\xi}$ with $k^2_{\perp} \!\equiv\! k^2_{\rm F}-k^2_{\parallel}$. The spinors,  
${\bm \Phi}_+ \!\equiv\! (1,0,0,-i)^{\rm T}$ and ${\bm \Phi}_- \!\equiv\! (0,i,1,0)^{\rm T}$, are the eigenvectors of the spin operator $S_{z}\equiv \frac{1}{2}{\rm diag}(\sigma _z, - \sigma^{\rm T}_z)$ in the Nambu space, 
\beq
S_z {\bm \Phi}_{\pm} = \pm \frac{1}{2} {\bm \Phi}_{\pm}. 
\label{eq:sz}
\eeq

The quantized field ${\bm \Psi}\!=\! (\psi _{\uparrow}, \psi _{\downarrow},\psi^{\dag}_{\uparrow}, \psi^{\dag}_{\downarrow})^{\rm T}$ in spin-triplet superfluids can be expanded in terms of the positive energy states of the surface Andreev bound states with $E_{\rm surf}({\bm k}_{\parallel}) \!\ge\!0$ and ${\bm \varphi}_{{\bm k}_{\parallel}}({\bm r})$ in addition to continuum states. For low temperature regimes $T\!\ll\!\Delta _0$, the field operator can be constructed from the contributions of only the surface Andreev bound states as 
\beq
{\bm \Psi}({\bm r}) 
= \sum _{{\bm k}_{\parallel}} 
\left[ {\bm \varphi}^{(+)}_{0,{\bm k}_{\parallel}}({\bm r})\eta _{{\bm k}_{\parallel}}
+\mathcal{C}{\bm \varphi}^{(+)}_{0,{\bm k}_{\parallel}}({\bm r})\eta^{\dag}_{{\bm k}_{\parallel}}
\right] + (E>\Delta _0), 
\eeq
where $\eta _{{\bm k}_{\parallel}}$ and $\eta^{\dag}_{{\bm k}_{\parallel}}$ denote the Bogoliubov quasiparticle operators obeying the anti-commutation relations, $\{ \eta _{\bm k}, \eta^{\dag}_{{\bm k}^{\prime}}\}=\delta _{{\bm k},{\bm k}^{\prime}}$ and $\{\eta _{\bm k},\eta _{{\bm k}^{\prime}}\}\!=\! \{\eta^{\dag}_{\bm k},\eta^{\dag}_{{\bm k}^{\prime}}\}\!=\!0$. Substituting Eq.~(\ref{eq:varphi1}) into the expansion form of ${\bm \Psi}$, the quantized field operator contributed from the surface Andreev bound states is recast into 
${\bm \Psi}({\bm r})\approx \sum _{{\bm k}_{\parallel}} 
{\bm \varphi}^{(+)}_{0,{\bm k}_{\parallel}}({\bm r}) [
\eta _{{\bm k}_{\parallel}} - e^{-i\phi _{\bm k}} e^{-2i{\bm k}_{\parallel}\cdot{\bm r}_{\parallel}}
\eta^{\dag}_{{\bm k}_{\parallel}}
] $. 
This reproduce the Majorana Ising condition in Eq.~(\ref{eq:Mcondition}) 
\beq
\left(
\begin{array}{c}
\psi _{\uparrow}({\bm r}) \\
\psi _{\downarrow}({\bm r})
\end{array}
\right) = i \sigma _{\mu}R_{\mu z}(\hat{\bm n},\varphi)
\left(
\begin{array}{c}
\psi^{\dag}_{\downarrow}({\bm r}) \\
-\psi^{\dag}_{\uparrow}({\bm r}) 
\end{array}
\right).
\label{eq:Mcondition2}
\eeq
Hence, the Majorana fields constructed from the surface Andreev bound states reproduce the Ising spin property and the surface bound states are not coupled to the local density operators, as discussed in Eqs.~(\ref{eq:rho0}) and (\ref{eq:MIS}). Using Eq.~(\ref{eq:Mcondition2}), the dynamical spin susceptibility is obtained as 
\beq
\chi _{\mu\nu}({\bm r}_1,{\bm r}_2; \omega) 
= \chi^{\rm M}_{zz}({\bm r}_1,{\bm r}_2;\omega) R_{\mu z}(\hat{\bm n},\varphi) R_{\nu z} (\hat{\bm n},\varphi). 
\label{eq:chiM}
\eeq
Therefore, it originates from Majorana Ising spins ${S}^{\rm M}_{z}({\bm r})$ and $\chi^{({\rm M})}_{zz}({\bm r}_1,{\bm r}_2;\omega)\!\equiv\! \langle S^{\rm M}_{z}({\bm r}_1)S^{\rm M}_{z}({\bm r}_2)\rangle _{\omega}$ with the ${\rm SO}(3)$ rotation $R_{\mu\nu}(\hat{\bm n},\varphi)$. The property of $\chi^{({\rm M})}_{zz}({\bm r}_1,{\bm r}_2;\omega)$ was discussed in Refs.~\cite{chungPRL2009,mizushimaJLTP2011} and the further details will be discussed in Sec.~\ref{sec:detecting}.

We emphasize that, in the absence of a magnetic field, the whole branch of the surface Andreev bound states can retain the Ising spin character, contrary to the argument based on the chiral symmetry. The Majorana nature of the whole branch is a consequence of the Andreev approximation, and the approximated Majorana nature of the whole branch may enable to realize the macroscopic Ising-like spin correlation. 

Whereas the gapless spectrum of the surface Andreev bound states is protected by the nontrivial topological invariant defined in the bulk region of the B-phase~\cite{schnyderPRB2008,qiPRL2009,mizushimaPRL2012}, if two specular surfaces at $z \!=\! 0$ and $D$ get close to each other, the interference between surface Majorana cones distorts the surface cone spectrum in Eq.~(\ref{eq:E0a}). Then, the hybridization of two surface states exponentially splits the zero energy state at $|{\bm k}_{\parallel} |\!=\! 0$ with quantum oscillation on the scale of $k^{-1}_{\rm F}$ as $\delta E({\bm k}_{\parallel} \!=\! {\bm 0}) \!\sim\! e^{-D/\xi}\sin(k_{\rm F}D)$~\cite{mizushimaPRA2010-2,kawakamiJPSJ2011,cheng1,cheng2}. The numerical calculation based on the quasiclassical theory confirmed the appearance of a mini-gap generated by the hybridization of two surface states in $^3$He-B confined in a slab geometry, where $^3$He is confined in two directions~\cite{tsutsumiPRB2011}. For $^3$He confined in one direction, i.e., in parallel plates, however, no finite excitation gap is generated by the hybridization even for strong confinement~\cite{wuPRB2013}. 

In addition to the splitting due to the quasiparticle tunneling, the finite size of the system with the thickness $D \!=\! \mathcal{O}(\xi)$ gives rise to the pair breaking effect, which may also cause a change of the gapless spectrum. The distortion of the gapless Majorana cone due to the quasiparticle tunneling and pair breaking effect may break the Majorana Ising nature of the surface bound states.


\section{Symmetry protected topological phase and topological quantum critical point in $^3$He-B under a magnetic field}
\label{sec:field}

In the previous section, having explicitly solved the Andreev equation, we have shown that the B-phase of superfluid $^3$He is accompanied by gapless quasiparticle states localized at the surface within the superfluid coherence length $\xi _0$. The surface bound states behave as a Majorana fermion that is a particle equivalent to its own anti-particle. 
As shown in Eqs.~(\ref{eq:rho0}) and (\ref{eq:MIS}), the surface Majorana fermion yields the
Ising spin, which indicates that the surface states are coupled to only
a particular direction of applied magnetic fields. It has been clarified in
Sec.~\ref{sec:topo} that the Majorana Ising character is a consequence
of a hidden ${\bm Z}_2$ symmetry. 
 
We here consider the topological superfluidity of $^3$He-B confined in a slab geometry under a magnetic field. See Fig.~\ref{fig:phase_slab}. As shown in Eqs.~(\ref{eq:Z2}) and (\ref{eq:Z2-2}), the hidden ${\bm Z}_2$ symmetry may survive in a slab even if the time-reversal symmetry is explicitly broken. Hence, we extend our argument based on the topology in Sec.~\ref{sec:topo} and the Andreev equation in Sec.~\ref{sec:andreev} in the case with a finite magnetic field.

\subsection{The hidden discrete symmetry and topological invariant}
\label{sec:hidden}

\begin{figure}[!t]
\begin{center}
\includegraphics[width=120mm]{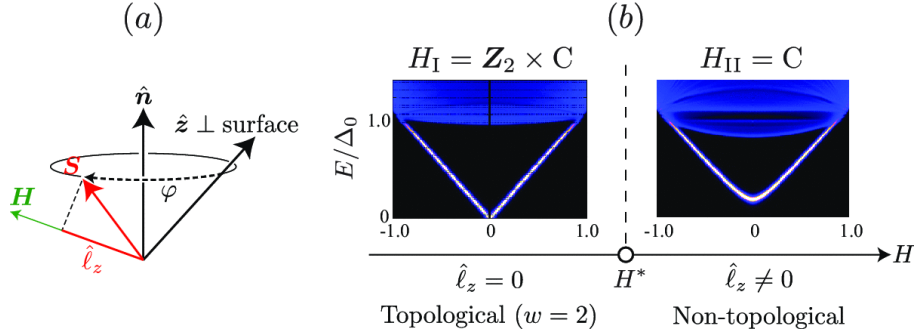}
\end{center}
\caption{(a) Relation between $\hat{\ell}_z$ and the orientation of ${\bm d}(0,0,k_z)$ for an arbitrary $(\hat{\bm n},\varphi)$ at the surface. (b) Schematic phase diagram of $^3$He-B under a parallel magnetic field, where $H$ involves the topological phase transition with spontaneous symmetry breaking. In (b), the panels show the momentum-resolved surface density of states, $\mathcal{N}(\hat{\bm k},z; E)$ for $\hat{\ell}_z=0$ and $\hat{\ell} \neq 0$. The definition of $\mathcal{N}(\hat{\bm k},z; E)$ is given in Eq.~(\ref{eq:dosk}).}
\label{fig:tpt}
\end{figure}

Now let us consider $^3$He-B under a finite magnetic field ${\bm H}$. We
also suppose that the liquid $^3$He is confined in a slab geometry,
where the $\hat{\bm n}$-vector and $\varphi$ are
spatially uniform. We first would like to mention that
Eq.~(\ref{eq:MIS}) gives an intuitive interpretation of the quantity
$\hat{\ell}_z$ which is directly linked to the breaking of the hidden
${\bm Z}_2$ symmetry.
The orientation of an applied magnetic field is denoted by
$\hat{h}_{\nu} \!=\! H_{\nu}/H$. Then, it turns out that the
$\hat{\ell}_z(\hat{\bm n}, \varphi)$ describes the projection of the
Majorana Ising spin ${\bm S}({\bm r})$ in Eq.~(\ref{eq:MIS}) onto the
orientation of the applied magnetic field ${\bm H}$ as
\beq
\hat{\ell}_z(\hat{\bm n},\varphi) = \frac{\hat{\bm h}\cdot{\bm S}({\bm r})}{\left| {\bm S}({\bm r}) \right|}.
\label{eq:lz}
\eeq
Figure~\ref{fig:tpt}(a) depicts the schematic picture for ${\bm S}$, ${\bm H}$, and $\hat{\ell}_z$. For $\hat{\ell}_z \!=\! 0$, the Majorana Ising spin ${\bm S}$ is perpendicular to the applied magnetic field, which implies that the SABS does not contribute to the magnetic response. However, the surface bound states may be responsible to ${\bm H}$ when $\hat{\ell}_z\!\neq\! 0$.  

To explicitly solve the Andreev equation in the presence of a magnetic field, let us suppose the wavefunction to be 
${\bm \varphi}_{{\bm k}_{\parallel}}({\bm r}) = a_+{\bm \varphi}^{(+)}_{0,{\bm k}_{\parallel}}({\bm r}) + a_-{\bm \varphi}^{(-)}_{0,{\bm k}_{\parallel}}({\bm r})$, where ${\bm \varphi}^{(\pm)}_{0,{\bm k}_{\parallel}}$ are the wave functions of the positive and negative energy states and their corresponding energies are given by $\pm |E_0({\bm k}_{\parallel})|$ in Eq.~(\ref{eq:E0a}).
The normalization condition for ${\bm \varphi}_{{\bm k}_{\parallel}}({\bm r})$ requires $|a_+|^2 + |a_-|^2\!=\! 1$. 
The coefficients $a_{\pm}$ and energy $E({\bm k}_{\parallel})$ are determined by solving the eigenvalue equation
\beq
\left(
\begin{array}{cc}
|E_0| & e^{-i\phi _{\bm k}} \gamma _z \\
e^{i\phi _{\bm k}} \gamma _z & -|E_0|
\end{array}
\right)\left( 
\begin{array}{c} a_+ \\ a_- \end{array}
\right) = E \left( 
\begin{array}{c} a_+ \\ a_- \end{array}
\right),
\label{eq:eigen}
\eeq
where $\gamma _z \equiv \mu _{\rm n}H_{\mu}R_{\mu z}(\hat{\bm n},\varphi)$ denotes the gap of the surface cone. The wave function of the negative energy branch is obtained from Eq.~(\ref{eq:varphi1}) with the particle-hole symmetry, ${\bm \varphi}^{(-)}_{0,{\bm k}_{\parallel}} = \mathcal{C}{\bm \varphi}^{(+)}_{0,-{\bm k}_{\parallel}}$, as 
$
{\bm \varphi}^{(-)}_{0,{\bm k}_{\parallel}} ({\bm r}) = N_{\bm k}
e^{i{\bm k}_{\parallel}\cdot{\bm r}_{\parallel}}f(k_{\perp},z)\mathcal{U}(\hat{\bm n},\varphi)
\left(
{\bm \Phi}_-
+ e^{i\phi _{\bm k}}{\bm \Phi}_+
\right)$. The off-diagonal term in Eq.~(\ref{eq:eigen}), $e^{i\phi _{\bm k}} \gamma _z$, results from Eq.~(\ref{eq:sz}) that the spinors $\Phi _{\pm}$ are the eigenstates of $S _z$, implying that $\Phi^{\dag}_{\pm}S_{\mu}\Phi _{\pm} = \pm \delta _{\mu, z}$. 
The dispersion of the surface bound state is given by diagonalizing the matrix as 
\beq
E({\bm k}_{\parallel}) = \pm \sqrt{ \left| E_0({\bm k}_{\parallel})\right|^2 + \left|\mu _{\rm n}H\hat{\ell}_z(\hat{\bm n}, \varphi)\right|^2}.
\label{eq:sabs}
\eeq

The resulting dispersion in Eq.~(\ref{eq:sabs}) implies that the energy gap of the surface state depends on the $\hat{\bm \ell}$-vector as
$\min |E ({\bm k}_{\parallel})| = \mu _{\rm n}H  |\hat{\ell}_z(\hat{\bm n},\varphi)|$.
For $\hat{\ell}_z = 0$, as we showed in Eq.~(\ref{eq:sabs}), the surface bound state remains gapless even in the presence of a magnetic field, which is responsible for the Ising nature of the Majorana cone. 
In particular, the case of $\hat{\bm n}\!=\! \hat{\bm z}$ always yields $\hat{\ell}_z = 0$, which leads to consistent results to the previous works~\cite{volovik2009-1,chungPRL2009,nagatoJPSJ2009,shindouPRB2010}. 
For $\hat{\ell}_z \!\neq\! 0$, however, an arbitrary orientation of the magnetic field opens a finite energy gap and the Majorana Ising nature of the surface bound states disappears~\cite{volovikJETP2010}. From topological point of view, the former behavior might seem to be a puzzle: Because the magnetic field breaks the time-reversal invariance, topological protection as a time-reversal invariant topological superfluid does not work any more. Nevertheless, as discussed in Sec.~\ref{sec:symm_field}, the B-phase still retains the hidden ${\bm Z}_2$ symmetry which protects the topologically nontrivial phase in the presence of a magnetic field.

Confinement, magnetic Zeeman field, and dipole interaction explicitly break the continuous rotational symmetry in both spin and coordinate spaces. The group symmetry subject to the confined $^3$He under a magnetic field is composed of the hidden ${\bm Z}_2$ symmetry, ${\rm U}(1)$ gauge symmetry, and the particle-hole symmetry, 
\beq
G_{\rm slab, H, D} = {\bm Z}_2 \times {\rm U}(1)_{\phi} \times {\rm C}.
\eeq
There are two possible subgroups of $G_{\rm slab, H, D}$ relevant to the superfluid $^3$He-B, $H_{\rm I}\in G_{\rm slab, H, D}$ and $H_{\rm II} \subset G_{\rm slab, H, D}$, which indicates that there exists two different phases in a confined $^3$He-B under a magnetic field, depending on $\hat{\ell}_z$. The ${\bm Z}_2$ symmetric phase that holds ${H}_{I}$ is realized when $\hat{\ell}_z = 0$,
\beq
{H}_{\rm I} = {\bm Z}_2\times {\rm C}.
\eeq
Once $\hat{\ell}_z$ becomes nonzero, however, the hidden ${\bm Z}_2$ symmetry is broken spontaneously and the remaining symmetry is 
\beq
H_{\rm II}= {\rm C}.
\eeq
Similarly with Eq.~(\ref{eq:Z2-2}), the ${\bm Z}_2$ symmetry of the microscopic Hamiltonian is expressed as
\beq
\mathcal{T}\mathcal{U}(\pi) \mathcal{H}({\bm k})\left[\mathcal{T}\mathcal{U}(\pi)\right]^{-1} 
= \mathcal{H}(\underline{\bm k}) 
- \gamma H \hat{\ell}_z (\hat{\bm n},\varphi) \left(
\begin{array}{cc} 
\tilde{\sigma}_z & 0 \\ 0 & - \tilde{\sigma}^{\ast}_z \end{array} \right).
\eeq
%
%
Remarkably, one can introduce a topological invariant if the discrete symmetry is not spontaneously broken, that is, the case of $\hat{\ell}_z = 0$. Combining it with the particle-hole symmetry of the BdG Hamiltonian, one obtains the relation
$
\underline{\Gamma}_1 \underline{\cal H}(k_x,k_y,k_z)\underline{\Gamma}^{-1}_1\!=\!-\underline{\cal H}(-k_x,-k_y, k_z)
$
with $\underline{\Gamma}_1 \!=\! \mathcal{C}\mathcal{T}\mathcal{U}_z(\pi)$. 
Hence, the so-called chiral symmetry is preserved in the confined $^3$He-B under a magnetic field as
\beq
\{\underline{\Gamma}_1, \underline{\cal H}(0,0,k_z)\} = 0.
\label{chiral}
\eeq
Following Refs.~\cite{satoPRB2009,satoPRB2011}, one can introduce the one-dimensional winding number as
\beq
w = -\frac{1}{4\pi i}\int_{-\infty}^{\infty}dk_z\left. 
{\rm tr}[\underline{\Gamma}_1\underline{\cal H}^{-1}({\bm k})\partial_{k_z}\underline{\cal H}({\bm k})]
\right|_{{\bm k}_{\parallel}\!=\!{\bm 0}},
\label{eq:winding}
\eeq
which are evaluated as $w \!=\! 2$ for $\gamma H \! <\! E_{\rm F}$ ($\Delta_{\perp}>0$). 
The bulk-edge correspondence shown in Sec.~\ref{sec:majorana} implies that in the case of $\hat{\ell}_z=0$, the surface bound state remains gapless as $E({\bm k}_{\parallel}) \!=\! 0$ at ${\bm k}_{\parallel}\!=\! {\bm 0}$ even in the presence of the magnetic field. Hence, the chiral symmetry and the bulk-edge correspondence still bring the physical consequence that the gapless bound states yields the Ising anisotropic magnetic response. 

Since the topological invariant $w$ requires the hidden ${\bm Z}_2$
symmetry to be preserved, the winding number is ill-defined in the ${\bm
Z}_2$ symmetry breaking phase.
Therefore, the ``order parameter'' $\hat{\ell}_z$ of the hidden ${\bm Z}_2$
symmetry also characterizes a topological phase transition at which
$w$ becomes ill-defined. 
In other words, there may exist a quantum critical point in $^3$He
confined in a slab at which the topological phase transition occurs
together with spontaneous symmetry breaking. 
At the quantum critical point $H^{\ast}$, the topologically nontrivial properties of $^3$He-B can be changed by
the spontaneous symmetry breaking without closing the bulk excitation
gap. 


The order parameter $\hat{\ell}_z$ for the hidden ${\bm Z}_2$ symmetry must be determined by minimizing the thermodynamic potential in the equilibrium. In Sec.~\ref{sec:numerical2}, we will clarify that the magnetic dipole-dipole interaction arising from the magnetic moment of $^3$He nuclei is indispensable for the thermodynamic stability of the symmetry protected topological phase and the quantum critical point $H^{\ast}$ is quantitatively obtained from microscopic calculation. As a result, a magnetic field can drive the topological phase transition concomitant with the spontaneous breaking of the hidden ${\bm Z}_2$ symmetry, as shown in Fig.~\ref{fig:tpt}(b).

\subsection{Topology of surface Majorana fermions}
\label{sec:topoMajo}

Once the order parameter $\hat{\ell}_z$ becomes finite, the topological invariant is ill-defined in the B-phase, implying that surface Majorana fermions acquire an effective mass associated with the Zeeman energy. The gapped Majorana fermions, however, have their own nontrivial topology. To see this, we first introduce the effective Hamiltonian of the surface states introduced in Eq.~(\ref{eq:Hsurf}), which can be extended to contain the mass term associated with the magnetic Zeeman energy as
\beq
\mathcal{H}_{\rm surf} = \sum _{{\bm k}_{\parallel}} {\psi}^{\rm T}_{\rm M}(-{\bm k}_{\parallel}) \left[
c \left( {\bm k}_{\parallel} \times {\bm \sigma} \right)\cdot\hat{\bm z}
+M(\hat{\bm n},\varphi)\sigma _z
\right] {\psi}_{\rm M}({\bm k}_{\parallel}),
\label{eq:Hsurf2}
\eeq
where $c = \Delta _0/k_{\rm F}$. The Majorana field ${\psi}_{\rm M}$ is associated with the original quantized field for surface states, $\psi$, as  
\beq
\psi({\bm k}) \equiv U(\hat{\bm n},\varphi){\psi}_{\rm M}({\bm k}), 
\eeq
which obeys $\{ \psi _{a}, \psi _b \} = \delta _{ab}$. The effective mass of the surface Majorana fermion, $M$, is determined by the single parameter $\hat{\ell}_z$ as 
\beq
M(\hat{\bm n},\varphi) = \frac{\gamma H}{2}\hat{\ell}_z(\hat{\bm n},\varphi).
\label{eq:Mmass}
\eeq 
The equation of motion for the surface Majorana fermions ${\psi}_{\rm M}(x)$ is governed by the $2+1$-dimensional Majorana equation,
\beq
\left(-i \gamma^{\mu}\partial _{\mu} + M \right){\psi}_{\rm M} (x) = 0,
\label{eq:Majoranaeq}
\eeq
where we replace $(k_x,k_y)$ to $(-i\partial _x, -i\partial _y)$ and set $x\equiv ({\bm r},t)$. Without loss of generality, we set $M/c\rightarrow M$. The $\gamma$-matrix is introduced as $(\gamma^0,\gamma^1,\gamma^2) = (\sigma_z, i\sigma _x, i\sigma _y)$, which satisfies $\{\gamma^{\mu},\gamma^{\nu}\}=2g^{\mu\nu}$ with the metric $g^{\mu\nu} = g_{\mu\nu} = {\rm diag}(+1,-1,-1)$ ($\mu, \nu = 0, 1, 2$). The effective action for the $2+1$-dimensional Majorana equation (\ref{eq:Majoranaeq}) is given as
\beq
\mathcal{S}_{\rm surf} = \int dx^3 \bar{\psi}_{\rm M}(x)\left( -i \gamma^{\mu}\partial _{\mu} + M \right) {\psi}_{\rm M}(x).
\label{eq:action}
\eeq
As shown in Eq.~(\ref{eq:Mmass}), the effective mass $M$ is parameterized with $\hat{\ell}_z$ that is the order parameter associated with the hidden ${\bm Z}_2$ symmetry breaking. At the quantum critical point $H^{\ast}$, the quantum fluctuation of the order parameter $\hat{ell}_z$ may be enhanced. Grover {\it et al.}~\cite{grover} proposed the $2+1$-dimensional effective action that describes the Majorana fermion coupled to the Ising field. They found the emergence of supersymmetry (SUSY) at the quantum critical point $H^{\ast}$. In contrast to previous works~\cite{friedan,fendley,huijse,yu,bauer} that the emergence of SUSY requires the fine-tuning of two or more parameters, the terms that break SUSY become irrelevant at the critical point and SUSY emerges without enforcing the conditions microscopically.


For $\hat{\ell}_z \neq 0$, the Majorana cone in Eq.~(\ref{eq:Hsurf2}) has a finite energy gap. In the the effective Hamiltonian of such a quasi-two-dimensional system, the topological invariant can be introduced as~\cite{volovikbook}
\beq
N_{\rm 2} = \frac{1}{4\pi^2}\int d{\bm k}_{\parallel} \int d\omega {\rm tr}\left[G \partial _{k_x} G^{-1} G \partial _{k_y} G^{-1}
G\partial _{\omega} G^{-1}\right].
\label{eq:topologicalN}
\eeq
Here, the Green's function of the surface Majorana fermion is defined as $G^{-1}\equiv i\omega - \mathcal{H}_{\rm surf}({\bm k}_{\parallel})$, where $\mathcal{H}_{\rm surf}({\bm k}_{\parallel})\!\equiv\! c ( {\bm k}_{\parallel} \times {\bm \sigma} )\cdot\hat{\bm z} +M\sigma _z$. 

To evaluate $N_2$, we note  that the eigenstates of the effective Hamiltonian in Eq.~(\ref{eq:Hsurf2}) can be expressed as spinor coordinates of a sphere $S^2$, when the time-reversal symmetry is absent, corresponding to the case of $\hat{\ell}_z \neq 0$. The $S^2$ sphere of the Hilbert space can be parameterized using the three-dimensional unit vector $\hat{\bm m}({\bm k}_{\parallel})$ as 
\beq
\mathcal{H}_{\rm surf}({\bm k}_{\parallel}) = |E({\bm k}_{\parallel})|{\bm \sigma}\cdot\hat{\bm m}({\bm k}_{\parallel}).
\eeq
The unit vector $\hat{\bm m}({\bm k}_{\parallel})$ points to $\hat{\bm m} = (0,0,{\rm sgn}(\hat{\ell}_z))$ for ${\bm k}_{\parallel} = {\bm 0}$, while $\hat{\bm m} = (\cos\phi _{\bm m},\sin \phi _{\bm m},0)$ for $|{\bm k}_{\parallel}| \rightarrow \infty$. Hence, the eigenfunction of $\mathcal{H}_{\rm surf}({\bm k}_{\parallel})$ gives a mapping from the two-dimensional momentum space to the upper or lower half of $S^2$. The unit vector covers the upper (lower) half of the Bloch sphere for $\hat{\ell}_z > 0$ (for $\hat{\ell}_z<0$), which implies that the effective Hamiltonian intrinsically has the half-skyrmion or meron texture in the two-dimensional momentum space. The topologically nontrivial texture in the momentum space is characterized by the two-dimensional winding number
\beq
w_{\rm 2d} = \frac{1}{4\pi}\int dk_x\int dk_y \hat{\bm m}\cdot
\left( 
\frac{\partial \hat{\bm m}}{\partial k_x} \times \frac{\partial \hat{\bm m}}{\partial k_y}
\right) .
\eeq
It turns out that this winding number is equivalent to the topological invariant introduced in Eq.~(\ref{eq:topologicalN}) and thus, the topological invariant is estimated for the massive Majorana fermion bound at the surface of $^3$He-B as
\beq
N_{2} = w_{\rm 2d} = \frac{{\rm sgn}(\hat{\ell}_z)}{2}.
\eeq
This topological invariant $N_{2}$ was originally introduced in $2+1$ dimensional relativistic field theory which is responsible for intrinsic quantum Hall conductivity~\cite{ishikawa1,ishikawa2,matsuyama}. The topological invariant is also applied to superfluid $^3$He-A film~\cite{volovik1988,volovik1989,goryo}.

\subsection{Topological crystalline superconductors: Role of mirror symmetries}
\label{sec:mirror}


Up to now, we have used $\pi$-rotation symmetry to define the ${\bm Z}_2$ symmetry. In this section, we discuss roles of another material dependent symmetry, namely mirror reflection symmetry in topological phases.

\begin{figure}[t!]
\begin{center}
\includegraphics[width=70mm]{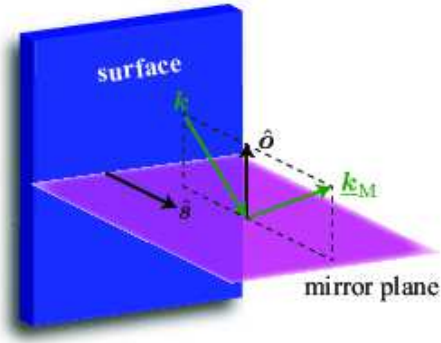}
\end{center}
\caption{Configuration of the specular surface and mirror reflection plane. The unit vectors, $\hat{\bm o}$ and $\hat{\bm s}$, are normal to the mirror plane and surface, respectively.}
\label{fig:mirror}
\end{figure}

We first summarize the relation between the mirror reflection symmetry
 $M$ and topological invariant in time-reversal invariant
 superconductors. Let us suppose that the Hamiltonian of the normal
 electrons, $\varepsilon ({\bm k})$, preserves a mirror reflection
 symmetry, $M\varepsilon({\bm k})M^{\dag}=\varepsilon(\underline{\bm
 k}_{\rm M})$. We also temporarily neglect a magnetic Zeeman field, that
 is, ${\bm H}={\bm 0}$. Let $\hat{\bm o}$ be a unit vector that is normal to
 the mirror plane. The mirror reflection operator $M$ changes the
 momentum ${\bm k}$ and spin ${\bm \sigma}$ to $\underline{\bm k}_{\rm
 M}\equiv{\bm k}-2\hat{\bm o}({\bm k}\cdot\hat{\bm o})$ and $-{\bm
 \sigma}+2\hat{\bm o}({\bm \sigma}\cdot\hat{\bm o})$ on the mirror
 plane (see Fig.~\ref{fig:mirror}). Without loss of generality, the mirror reflection operator is
 defined as $M=i ({\bm \sigma}\cdot\hat{\bm o})$. We also consider that
 the specular surface is normal to the mirror plane.

The superconducting state retains the mirror symmetry if the gap function $\Delta({\bm k})$ is even or odd under the mirror reflection, $M\Delta({\bm k})M^{\rm T} = \pm \Delta (\underline{\bm k}_{\rm M})$. In this situation, the BdG Hamiltonian $\mathcal{H}_0({\bm k})$ preserves the mirror reflection symmetry 
\beq
\mathcal{M}^{\pm}\mathcal{H}_0({\bm k})\mathcal{M}^{{\pm}\dag} = \mathcal{H}(\underline{\bm k}_{\rm M}),
\label{eq:mirror}
\eeq
where $\mathcal{M}$ is the mirror reflection operator extended to the Nambu space as
\beq
\mathcal{M}^{\pm} = \left(
\begin{array}{cc}
M & 0 \\ 0 & \pm M^{\ast}
\end{array}
\right).
\eeq

Let us now apply a magnetic field, where we replace $\mathcal{H}_0({\bm k})$ to the Hamiltonian $\mathcal{H}({\bm k})$ including the Zeeman term. Under this situation, we can define two different topological numbers by using mirror reflection. Once the Zeeman fields are applied, the mirror symmetry with respect to the plane parallel to the Zeeman filed is lost and the time-reversal symmetry is explicitly broken, but a combination of them can be still preserved if $\hat{\bm h}\cdot\hat{\bm o} = 0$. This is because the combination of the mirror reflection and the time-reversal rotates the magnetic field ${\bm H}$ to ${\bm H} - 2\hat{\bm o}({\bm H}\cdot\hat{\bm o})$. Consequently, the Hamiltonian $\mathcal{H}({\bm k})$ with $\hat{\bm h}\cdot\hat{\bm o} = 0$ holds the following ${\bm Z}_2$ symmetry,
\beq
\mathcal{T}\mathcal{M}^{\pm}\mathcal{H}({\bm k})
\left[\mathcal{T}\mathcal{M}^{\pm}\right]^{-1} = \mathcal{H}(-{\bm k}_{\rm M}),
\label{eq:z2}
\eeq
where $\mathcal{T} \!=\! i\sigma_y K$ is the time-reversal operator
with the complex conjugate operator $K$. Combining the ${\bm Z}_2$ symmetry with the particle-hole symmetry, 
$\mathcal{C}\mathcal{H}_{\rm eff}({\bm k})\mathcal{C}^{-1} = -
\mathcal{H}^{\ast}_{\rm eff}(-{\bm k})$, 
we define the chiral symmetry operator, 
$\Gamma _1 \!=\! \mathcal{C}\mathcal{T}\mathcal{M}^{\pm}$. Then, it turns out that $\Gamma _1$ is anti-commutable with the effective Hamiltonian
\beq
\left\{ \Gamma _1, \mathcal{H} ({\bm k})\right\} = 0,
\eeq
when ${\bm k}\cdot\hat{\bm o} = 0$. Then, the one-dimensional winding
number is defined as Eq.~(\ref{eq:winding})~\cite{satoPRB2009,mizushimaPRL2012,satoPRB2011}.

In the case of $\hat{\bm h}\parallel \hat{\bm o}$, the mirror reflection symmetry is preserved, while the time-reversal symmetry is explicitly broken. In this situation, the
mirror Chern number ensures the existence of a zero energy mode~\cite{uenoPRL2013}. 
When the mirror operator is commutable with the particle-hole
operator $\mathcal{C}$, the zero mode behaves as a non-Abelian Majorana fermion. 
The mirror Chern number is also applicable to 
the zero mode that is bound to a quantum vortex of 
superconductors and superfluids~\cite{tsutsumiJPSJ2013,SatoPhysicaE2014}.

\subsection{Promising examples of topological crystalline superconductors}

{\it Spin-orbit coupled quasi-one-dimensional superfluids.}--- 
As a promising example, we clarify the mirror-symmetry-protected topological superfluidity of a fermionic gas with a synthetic gauge field, where the gas is confined to a quasi-one-dimensional potential. As we will see below, the effective Hamiltonian is equivalent to that of semiconductor-superconductor nanowire with multichannels and thus, our theory with one-dimensional winding number protected by a mirror symmetry is also applicable to semiconductor-superconductor nanowires.

We here start with the Hamiltonian for spin-orbit coupled two-component fermionic atoms with an $s$-wave attractive interaction, $g$,
\beq
\mathcal{H} = \int d{\bm r} \left[ {\bm \Psi}^{\dag}({\bm r})
\varepsilon ({\bm r}){\bm \Psi}({\bm r})
+ g\psi^{\dag}_{\uparrow}({\bm r})\psi^{\dag}_{\downarrow}({\bm r})
\psi _{\downarrow}({\bm r})\psi _{\uparrow}({\bm r})\right],
\label{eq:H}
\eeq
where ${\bm \Psi} \!\equiv\! [\psi _{\uparrow}, \psi_{\downarrow}]^{\rm T}$ denotes the fermionic field operators with up- and down-spins. The single-particle Hamiltonian density is defined as
$\varepsilon ({\bm r}) 
\!=\! -\frac{1}{2m}{\bm \nabla}^2  - \mu _{\rm cp} + V_{\rm pot}({\bm r}) 
- H_{\mu}\sigma _{\mu} + \mathcal{S}({\bm r})$ with a confinement potential $V_{\rm pot}$ and $\sigma _{\mu}$ being the Pauli matrices in spin space. The
Zeeman field ${\bm H}$ is naturally induced by
implementing the spin-orbit coupling through two-photon Raman
process~\cite{nist}. We here consider a two-dimensional Rashba-type spin-orbit coupling~\cite{satoPRL2009}
\beq
\mathcal{S}({\bm r}) = i \kappa _x \sigma _y \partial _x 
- i \kappa _y \sigma _x \partial _y.
\eeq
Recently, a spin-orbit coupling with equal Rashba and Dresselhaus strengths can be synthetically induced by applying Raman lasers to atomic gases with hyperfine spin degrees of freedom~\cite{nist}. This scheme has also been implemented using fermionic $^6$Li~\cite{mit} and $^{40}$K atoms~\cite{china}. Although the spin–orbit coupling realized in experiments is one-dimensional, schemes for creating two- and three-dimensional analogue to Rashba spin-orbit coupling have theoretically been proposed~\cite{juzeliunas,campbell,anderson,anderson2,xu,hui,hui2,galitskiNature}.

Let us suppose that fermions are confined by a
two-dimensional optical lattice in the $y$-$z$ plane in addition to a
shallow harmonic potential along the $x$-direction, as shown in
Fig.~\ref{fig:pot}. The system under this confinement potential is
regarded as a two-dimensional array of $N_y \! \times \! N_z$ one-dimensional
tubes. Such a quasi-one-dimensional geometry has been reported in Ref.~\cite{rice},
where fermionic atoms are confined in a two-dimensional optical lattice potential. 
The number of tubes is typically about 
$N_y \!\times\! N_z \! \sim \mathcal{O}(10\times 10)$, so the system
should be treated as a finite system. 
We also note that confinement can vary an effective scattering length of fermions, that is, a confinement-induced Feshbach resonance~\cite{olshanii,liuPRA2007}. The properties of interatomic interaction can be drastically changed by confinement. 
The confined geometry is expected to isolate Majorana zero modes from the higher energy
quasiparticle states. 


\begin{figure}
\centering
\includegraphics[width=80mm]{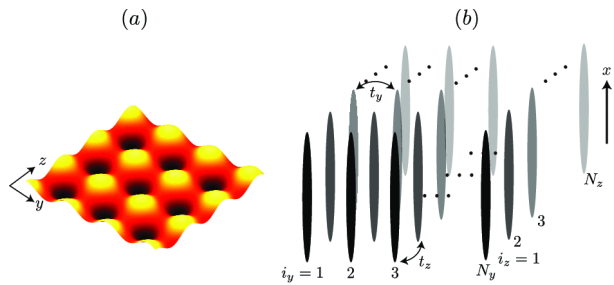}
\caption{(a) Two-dimensional optical lattice potential and (b) schematic picture of the calculated system. 
Figures adapted from Ref.~\cite{mizushimaNJP2013}.}
\label{fig:pot}
\end{figure}

To clarify the topological property of the effective Hamiltonian, we
here ignore the shallow trap potential along $x$-direction. 
Within the mean-field approximation, the Hamiltonian in Eq.~(\ref{eq:H})
can be diagonalized in terms of the quasiparticle states. 
As mentioned above, the system is regarded as
a two-dimensional array of $N_y \times N_z$ one-dimensional tubes. 
Employing mean-field approximation and tight-binding approximation in Eq.~(\ref{eq:H}), therefore, one obtains the BdG equation as an effectively one-dimensional equation~\cite{mizushimaNJP2013}, 
\beq
\mathcal{H}^{\rm eff}_{{\bm i}, {\bm j}}(k) {\bm \varphi}_{n,{\bm j}}(k) 
= E_n {\bm \varphi}_{n,{\bm i}}(k) ,
\eeq
where $k$ denotes the momentum along the $x$-axis. 
The BdG Hamiltonian density is given as
\beq
\mathcal{H}^{\rm eff}_{{\bm i}, {\bm j}}(k) 
&=& \left[ 
\varepsilon^{(0)}_{{\bm i},{\bm j}}(k) 
- \left\{ H_x\sigma _x +H_z \sigma _z +\kappa _x k \sigma
_y \right\}
\delta _{{\bm i},{\bm j}}\right]\tau _z 
- \Delta _{{\bm i},{\bm j}}\sigma _y \tau _y  \nn \\
&& + \left[  H_y\sigma_y \delta _{{\bm i},{\bm j}} + i \tilde{\kappa}_y\sigma _x 
\left\{\delta _{{\bm i},{\bm j}+ \hat{\bm e}_y}
-\delta _{{\bm i},{\bm j}- \hat{\bm e}_y}\right\} \right] \tau_0.
\label{eq:Heffk}
\eeq
The single-particle Hamiltonian density $\varepsilon^{(0)}_{{\bm i},{\bm
j}}(k) $ is
\beq
\varepsilon^{(0)}_{{\bm i},{\bm j}}(k) =& \left( 
\frac{k^2}{2m} - \mu _{\rm cp}
\right)\delta _{{\bm i},{\bm j}}   \nn \\
&- t_y \left( \delta _{{\bm i},{\bm j}+\hat{\bm e}_y} + \delta _{{\bm i},{\bm j}-\hat{\bm e}_y} \right) 
- t_z \left( \delta _{{\bm i},{\bm j}+\hat{\bm e}_z} + \delta _{{\bm i},{\bm j}-\hat{\bm e}_z} \right).
\eeq
Here, we set ${\bm i} = i_y\hat{\bm e}_y + i_z \hat{\bm e}_z$ and ${\bm j} = j_y\hat{\bm e}_y + j_z \hat{\bm e}_z$ with $\hat{\bm e}_y = (1,0)$ and $\hat{\bm e}_z = (0,1)$. The hopping energies between intertubes are denoted by $t_y$ and $t_z$.
We define an $s$-wave pair potential $\underline{\Delta}$ that is a
$N_yN_z\times N_yN_z$ matrix for ${\bm i}$ and ${\bm j}$, where
$\Delta _{{\bm i},{\bm j}} = \delta _{{\bm i},{\bm j}}\Delta _{\bm i}$
is assumed to be real without the loss of generality. 


Let us now consider the topological invariant in this quasi-one-dimensional Fermi gas under the time-reversal symmetry breaking potential. 
If one temporary neglects the Zeeman fields ${\bm H}$, 
our system is
invariant under the mirror reflection with respect to the $zx$-plane, as
well as the time-reversal, as shown in
Eq.~(\ref{eq:mirror}). 
Once the Zeeman fields are applied, the mirror symmetry is lost, but a
combination of the mirror reflection and the time-reversal is still
preserved if $H_y=0$.
Consequently, the Hamiltonian $\mathcal{H}_{\rm eff}(k)$ with $H_y=0$ holds the
following ${\bm Z}_2$ symmetry in Eq.~(\ref{eq:z2}), 
\beq
\mathcal{T}\mathcal{M}_{zx}\mathcal{H}_{\rm eff}(k)
\mathcal{M}_{zx}^{\dagger}\mathcal{T}^{-1} = \mathcal{H}^{\ast}_{\rm eff}(-k),
\label{eq:z22}
\eeq
The mirror reflection operator is given by $\mathcal{M}_{zx}=i\sigma_yU$, 
where $U$ is the operator flipping the $y$-component of ${\bm
i}=(i_y, i_z)$, 
\begin{eqnarray}
U_{{\bm i}, {\bm j}}=\delta_{i_y, N_y+1-j_y} 
\,\delta_{i_z,j_z}.
\end{eqnarray}

Combining the ${\bm Z}_2$ symmetry with the particle-hole symmetry in Eq.~(\ref{eq:PHS}), 
$\mathcal{C}\mathcal{H}_{\rm eff}(k)\mathcal{C}^{-1} = - \mathcal{H}^{\ast}_{\rm eff}(-k)$, 
we define the chiral symmetry operator, 
$\Gamma _1\!=\! \mathcal{T}\mathcal{M}_{zx}\mathcal{C} \!=\! \tau _xU$. 
Then, it turns out that $\Gamma_1$ is anti-commutable with the effective Hamiltonian
\beq
\left\{ \Gamma _1, \mathcal{H}_{\rm eff} (k)\right\} = 0.
\eeq
This implies that the BdG Hamiltonian $\mathcal{H}_{\rm eff}(k)$ holds
the chiral symmetry.
Then, the one-dimensional winding number is defined as~\cite{satoPRB2009,mizushimaPRL2012,mizushimaNJP2013,satoPRB2011}
\beq
w = -\frac{1}{4\pi i} \int^{\infty}_{-\infty} dk 
{\rm tr}\left[\Gamma _1{\cal H}^{-1}_{\rm eff}(k) \partial_k {\cal H}_{\rm
eff}(k)\right],
\label{eq:w}
\eeq
which takes an integer. We note that a similar one-dimensional winding number was considered for two-dimensional 
and pure one-dimensional Rashba superconductors,  
where $\Gamma _1$ in Eq.(\ref{eq:w}) is replaced by
$\tau_x$~\cite{satoPRB2009,TS2012}.
The above expression (\ref{eq:w}) is a generalization of these cases into
multi-tube or multi-band systems. 
Like the winding number $w$ in Sec.~\ref{sec:hidden}, the mirror winding number is also responsible for the Ising character of Majorana zero modes. The direction of the Ising spin is, however, different from the previous one: When $w\neq 0$, the Ising direction is parallel to the $\pi$-rotation axis, but $w_m\neq 0$, it is perpendicular to the mirror plane~\cite{shiozaki2014}.


We now evaluate the topological number $w$ and clarify the robustness of the zero-energy states 
against the intertube tunneling. 
As we mentioned above, even in the presence of intertube tunneling, the
winding number $w$ is well-defined for a whole system of tubes. 
Since one can turn off the intertube tunneling without the bulk gap
closing, the value of $w$ can be evaluated by setting
$t_y=t_z=\tilde{\kappa}_y=0$ in Eq.(\ref{eq:w}). Then one obtains 
\begin{eqnarray}
w={\rm tr}U\frac{1}{2\pi i}\int_{-\infty}^{\infty}dk \partial_k \ln
\left[\det{\cal Q}(k)\right],
\end{eqnarray}
where ${\cal Q}(k) = \varepsilon^{(0)}(k) - (H_x\sigma _x + H_z\sigma _z + \kappa k \sigma _y) + i\Delta \sigma _y$.
Noting ${\rm tr}U=0$ for even $N_y$'s and ${\rm tr}U=N_z$ for odd $N_y$'s,
one can valuate $w$ as
\begin{eqnarray}
|w|=\left\{ 
\begin{array}{ll}
0, & \mbox{for even $N_y$'s} \\
\\
N_z, & \mbox{for odd $N_y$'s}
\end{array}
\right. ,
\label{eq:w0}
\end{eqnarray}
when $\sqrt{H_x^2+H_z^2}>H_{\rm c}$. For a spatially uniform $\Delta$, the critical field $H_{\rm c}$ is given as $H_{\rm c}\!=\! \sqrt{\mu^2 + \Delta^2}$~\cite{satoPRL2009,satoPRB2010}. The winding number is trivial when $\sqrt{H_x^2+H_z^2}<H_{\rm c}$. Hence, the topological phase transition occurs at the critical field $H_c$ and Majorana zero modes survive for odd $N_y$'s in the topological regime $\sqrt{H_x^2+H_z^2}>H_{\rm c}$.

For a realistic situation of ultracold atomic gases, however, a more careful consideration is needed. Since fermionic atoms are confined along the $x$-axis by a trap potential, a spatially inhomogeneous superfluid is realized naturally. In addition, the inhomogeneous pair potential $\Delta (x)$, which is self-consistently determined by the gap equation and the BdG equation, depends on the Zeeman fields significantly. In contrast to semiconductor–superconductor junction systems, these two characteristics cannot be neglected. This means that within the local density approximation, the critical field $H_{\rm c}$ and one-dimensional winding number $w$ should be replaced by $H_{\rm c}(x) = \sqrt{\mu^2(x) + \Delta^2(x)}$ and $w(x)$, where $\mu (x)$ is the local chemical potential including the confinement potential along the $x$-axis. The inhomogeneity and self-consistency of $\Delta (x)$ and $\mu (x)$ play a critical role on the topological property of atomic gases~\cite{mizushimaNJP2013,liuPRA2012,liuPRA2012v2,weiPRA2012}.

The intertube tunneling effect can be understood as follows.
As was shown in Eq.~(\ref{eq:w0}), 
when one neglects the intertube couplings $t_y$, $t_z$ and $\tilde{\kappa}_y$, 
each tube supports Majorana
zero modes localized at its end points.
Now let us denote them
as $\gamma_{i_y}$ ($i_y=1, \cdots, N_y$), and consider how the
intertube couplings affect on them. 
When $t_y$ and $\tilde{\kappa}_y$ are turned on, 
the zero modes on neighboring tubes are coupled by the intertube tunneling
and its tight-binding Hamiltonian is given as
\begin{eqnarray}
{\cal
 H}=it\gamma_1\gamma_2+it\gamma_2\gamma_3+\cdots+it\gamma_{N_y-1}\gamma_{N_y}
 \equiv {\bm \gamma}^{\rm T}H{\bm \gamma}/2,
\label{eq:effective}
\end{eqnarray} 
where ${\bm \gamma}^{\rm T} = (\gamma_1, \gamma_2, \cdots, \gamma_{N_y-1}, \gamma_{N_y})$
is the operators of Majorana zero modes and $t$ denotes the induced tunneling coupling. 
Note that $t$ is real since $\gamma_{i_y}$ is a Majorana zero mode
satisfying $\gamma_{i_y}=\gamma_{i_y}^{\dagger}$.
Diagonalizing the $N_y\times N_y$ matrix ${H}$, one can examine the
effects of the intertube tunneling. It can be easily shown that ${H}$ has a single zero
eigenvalue for odd $N_y$'s, while it does not have for even
$N_y$'s.
This result naturally explain why Majorana zero modes survive only for
odd $N_y$'s.
One also finds that the zero eigenstate of ${H}$ has the following form
\begin{eqnarray}
(1,0, -1)^t, \quad \mbox{for $N_y=3$} 
\nonumber\\
(1,0, -1,0, 1)^t, \quad \mbox{for $N_y=5$} 
\nonumber\\
(1,0, -1,0, 1, 0, -1)^t, \quad \mbox{for $N_y=7$}.
\label{eq:majoranaWF}
\end{eqnarray}

To confirm the topological argument described above, we numerically solve the BdG equation for the $N_y\times N_z$ bundle of one-dimensional tubes,
\beq
\mathcal{H}^{\rm eff}_{{\bm i},{\bm j}}(x){\bm \varphi}_{E,{\bm j}}(x) = E {\bm \varphi}_{n,{\bm i}}(x),
\eeq
where $\mathcal{H}^{\rm eff}_{{\bm i},{\bm j}}(x)$ is obtained from the effective Hamiltonian (\ref{eq:Heffk}) by replacing $k\rightarrow - i \partial _x$. We also introduce the shallow trap potential along the $x$-axis, $V(x) = \frac{1}{2}m\omega^2x^2$ with the trap frequency $\omega$, which generates the inhomogeneity of the local pair potential $\Delta _{\bm i}(x)$. The detail on the inhomogeneous effect is discussed in Ref.~\cite{mizushimaNJP2013}. For numerical calculation, we apply the magnetic field along the $\hat{\bm z}$-axis, ${\bm H}=(0,0,H)$, which does not break the ${\bm Z}_2$ symmetry (\ref{eq:z22}). The other parameters are the same as those in Ref.~\cite{mizushimaNJP2013}. For a pure one-dimensional system with $t_y=t_z =\tilde{\kappa}_y = 0$, we find that in the set of current parameters the topological phase transition occurs at the critical field $H_c \approx 0.3E_{\rm F}$. For $H>H_{\rm c}$, the zero energy states appear at the end points of one-dimensional tubes. 

\begin{figure}
\centering
\includegraphics[width=65mm]{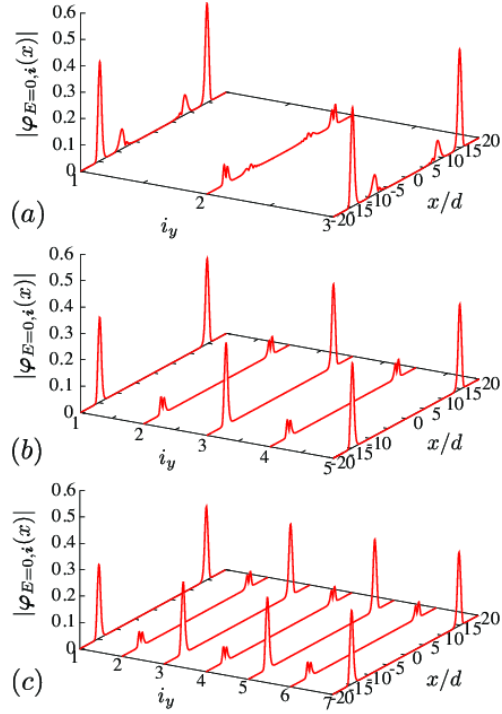}
\caption{Amplitudes of wavefunctions $|{\bm \varphi} _{E=0},{\bm i}(x)|$ for the lowest energy (zero energy) state in the case of (a) $N_y=3$, (b) $N_y=5$, and (c) $N_y=7$, respectively. The magnetic field is set to be $H=0.36E_{\rm F}$ which corresponds to the topological phase. In all the data, the hopping $t_y$ and the strength of the spin-orbit interaction $\tilde{\kappa}_y$ is set to be $t_y =0.01 E_{\rm F}$ and $\tilde{\kappa}_y/\kappa _x = 0.5$. The length is scaled by the harmonic oscillator length $d$.
Figures adopted from Ref.~\cite{mizushimaNJP2013}.} 
\label{fig:ldos}
\end{figure}

In Fig.~\ref{fig:ldos}, we plot the spatial profiles of wavefunctions of the zero energy states, $|{\bm \varphi} _{{E=0},{\bm i}}(x)|$, for $N_y=3$, $5$, and $7$. The zero energy quasiparticles are bound to the end points that are correspond to the Thomas-Fermi radius $x=R_{\rm TF}\approx \pm 15 d$. The satellite peaks around $x\approx \pm 8d$ originates in the inhomogeneity effect~\cite{mizushimaNJP2013}. As discussed in Eq.~(\ref{eq:majoranaWF}), because of the intertube tunneling of Majorana zero modes, the zero-energy wavefunctions have large amplitudes on tubes only at odd numbers of $i_y$, while the amplitudes at even $i_y$'s vanish. These behaviors of low-lying quasiparticles associated with the hidden ${\bm Z}_2$ symmetry protected topology might be detectable through momentum-resolved radio-frequency spectroscopy~\cite{stewart,peng}.

As a one-dimensional class D system, the Rashba superfluid also has a one-dimensional ${\mathbb Z}_2$ topological number. 
The one-dimensional ${\mathbb Z}_2$ number is defined
as~\cite{tanakaJPSJ2012} 
\begin{eqnarray}
\nu=\frac{1}{\pi}\int_{-\pi}^{\pi}dk A(k)+\mbox{mod. 2},
\label{eq:z2top}
\end{eqnarray}
with $A(k)$ being the geometrical phase, 
\begin{eqnarray}
A(k)=i\sum_{E_n(k)<0}\sum _{\bm i}
\langle {\bm \varphi}_{n,{\bm i}}(k)|\partial_k {\bm \varphi}_{n,{\bm i}}(k)\rangle.
\end{eqnarray} 
where $|{\bm \varphi}_{n,{\bm i}}(k)\rangle$ is the Bloch wave function of an negative energy state of the BdG Hamiltonian $\mathcal{H}(k)$, and ${\bm i}$ denotes the multichannels of fermions.  When $\nu$ is odd (even), the system is topologically non-trivial (trivial).

As a consequence of the bulk-edge correspondence, these two one-dimensional topological numbers ensure the existence of zero energy states appearing in the end points of one-dimensional segments. Here we note that the parities of these two topological numbers coincide with each other,
\begin{eqnarray}
(-1)^{\nu}=(-1)^w, 
\end{eqnarray}
which implies that $w$ can be nonzero even when $\nu$ is trivial, but the opposite is not true. Therefore, the actual number of the zero energy states is determined by $w$ unless the ${\bm Z}_2$ symmetry (\ref{eq:z2}) is broken macroscopically. 
Once the ${\bm Z}_2$ symmetry is broken, however, the one-dimensional ${\mathbb Z}_2$ number $\nu$ in Eq.(\ref{eq:z2top}) determines the topological stability of the Majorana zero modes.

We finally would like to mention that the results obtained here are not straightforwardly applicable to Fermi gases under a spin-orbit coupling with equal Rashba and Dresselhaus strengths~\cite{nist,mit,china}. This is because an additional symmetry specific to the equal Rashba and Dresselhaus spin-orbit coupling ensures the existence of zero energy states, regardless of even-odd parity of $N_y$.

{\it A superconducting nanowire with multichannels.}---
The one-dimensional winding number introduced in Eq.~(\ref{eq:w}) is directly applicable to semiconductor-superconductor nanowire with multichannels: It is seen from Eq.~(\ref{eq:Heffk}) that the array of one-dimensional tubes
with a spin-orbit interaction is analogous to a
semiconductor-superconductor nanowire with $N$-th electron
bands~\cite{potter,lutchynPRL2011,stanescuPRB2011,kellsPRB2012,Tewari}, where $N  \!\equiv\! N_y \!\times\!N_z$. If we consider the nanowire extending in the $x$-direction on top of an $s$-wave superconductor in the $xy$-plane, as shown in Fig.~\ref{fig:pot}, the system is naturally supposed to be invariant under the mirror reflection, $y\rightarrow -y$, to the $xz$-plane. This mirror symmetry could be broken under Zeeman fields, but the ${\bm Z}_2$ symmetry (\ref{eq:z22}) remains if the Zeeman fields are applied in the $x$- or $z$-direction. Then, the topological number $w$ in Eq.~(\ref{eq:w}) is defined in the same manner. From arguments similar to the above, one finds that $w$ is non-zero if the Zeeman field $H$ satisfies $|H|>H_{\rm c}$ and the number of channels in the $y$-direction of the nanowire is odd. Indeed, under this condition, $|w|$ is equal to the number of channels in the $z$-direction of the nanowire. When an additional transverse field $H_y$ is applied, however, it externally breaks the ${\bm Z}_2$ symmetry and the resultant topological property is characterized by the one-dimensional ${\bm Z}_2$ number in Eq.~(\ref{eq:z2top}). 

\begin{figure}
\centering
\includegraphics[width=70mm]{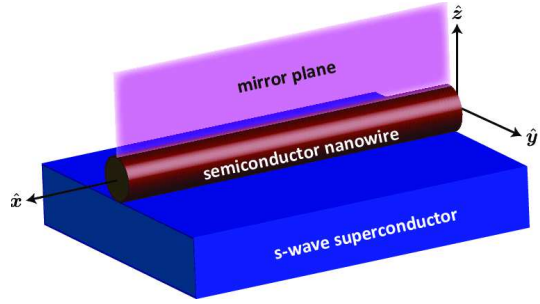}
\caption{The experimental setup consisting of a semiconductor nanowire on top of an $s$-wave superconductor.}
\label{fig:pot}
\end{figure}

Note that, in contrast to the one-dimensional ${\bm Z}_2$ number in Eq.~(\ref{eq:z2top}), $w$ can be non-zero even when the total number of channels in the nanowire is even, since it is given by the sum of the channels in the $z$- and $y$-direction. As clarified in Sec.~\ref{sec:majorana}, the local density operator of the Majorana zero modes vanishes~\cite{mizushimaPRL2012}. This implies that the coupling between the Majorana zero modes and non-magnetic local disorder potential also vanishes, and thus the Majorana zero modes are stable against weak non-magnetic disorders.

{\it Topological crystalline superconductivity in UPt$_3$.}---
The another candidate for topological crystalline superconductivity is the $E_{1u}$ scenario of the heavy fermion superconductor UPt$_3$~\cite{tsutsumiJPSJ2012-2,tsutsumiJPSJ2013}. In low fields, UPt$_3$ undergoes the double superconducting transitions from normal- to A-phases at $T_{\rm c1} \approx 550$mK and from A- to B-phases at $T_{\rm c2} \approx 500$mK. In spite of numerous works over three decades after the discovery of superconductivity in UPt$_3$, the puzzles on the pairing mechanism and gap function have not been fully solved yet~\cite{reviewupt3}.

Recent experiments on angle-resolved thermal conductivity have observed the two-fold rotational symmetry in the $a$-$b$ plane, which convincingly suggests a spin-triplet $f$-wave function in the $E_{1u}$ representation~\cite{ymachida}. Based on the measurement, the pairing symmetry in the B-phase is described by two components of the ${\bm d}$-vectors. The $E_{1u}$ scenario also explains the rotation of the ${\bm d}$-vectors in the Knight shift measurement for ${\bm H}\parallel\hat{\bm c}$. As ${\bm H}\parallel \hat{\bm c}$ increases, the ${\bm d}$-vectors rotate from ${\bm d}_{\rm I}$ to ${\bm d}_{\rm II}$ at the critical field $H_{\rm rot}\sim 2$kG. The ${\bm d}$-vectors for the B-phase are obtained as
\beq
&{\bm d}_{\rm I}({\bm k}) =  (\Delta _1\hat{k}_a \hat{\bm b} + \Delta _2\hat{k}_b \hat{\bm c})(5\hat{k}^2_c-1), 
\label{eq:upt31} \\
&{\bm d}_{\rm II}({\bm k}) = (\Delta _1\hat{k}_a \hat{\bm b} + \Delta _3\hat{k}_b \hat{\bm a})(5\hat{k}^2_c-1),
\label{eq:upt32}
\eeq
where $\hat{\bm a}$, $\hat{\bm b}$, and ${\bm c}$ are the unit vectors
in hexagonal crystal and we set $\hat{k}_a\equiv \hat{\bm
k}\cdot\hat{\bm a}$. Note that most bulk thermodynamic quantities are
understandable with the another candidate based on the $E_{2u}$
representation, as well as the $E_{1u}$ scenario. The pairing symmetry
in the $E_{2u}$ representation is given as ${\bm d}^{\prime}({\bm k}) =
\Delta _0 (\hat{k}_a+i\hat{k}_b)^2\hat{k}_c\hat{\bm c}$. While the $E_{1u}$ state preserves the time-reversal symmetry, the $E_{2u}$ does not. The multiple phase diagram in UPt$_3$ is understandable with the multicomponent ${\bm d}$-vectors for the $E_{1u}$ representation and with the orbital degrees of freedom for the $E_{2u}$ scenario. 

We here clarify the topological aspect of the $E_{1u}$ scenario. We
start with the BdG Hamiltonian for bulk spin-triplet superconductors,
\beq
\widehat{\mathcal{H}}({\bm k}) = 
\left( 
\begin{array}{cc}
\hat{\epsilon} ({\bm k}) & \hat{\Delta} ({\bm k}) \\ \hat{\Delta}^{\dag}({\bm k})
& - \hat{\epsilon}^{\rm T}(-{\bm k})
\end{array}
\right),
\eeq
where $\hat{\epsilon}({\bm k})$ is the Hamiltonian in the normal state of UPt$_3$ which holds the D$_{\rm 6h}$ hexagonal symmetry. 
We here consider the low-lying quasiparticles that are bound to the surface normal to the $a$-axis. 
Then, we find that the mirror symmetry with respect to the $ca$-plane plays a key role on the protection of surface Majorana fermions in the B-phase of UPt$_3$. We also notice that the other mirror symmetry, that is the mirror reflection $M_{ab}$ with respect to the
$ab$-plane, also protects Majorana zero modes in a vortex along the $c$-axis. Below we concentrate our attention to the surface Majorana fermions protected by the mirror reflection symmetry in the $ca$-plane.

\begin{figure}
\centering
\includegraphics[width=120mm]{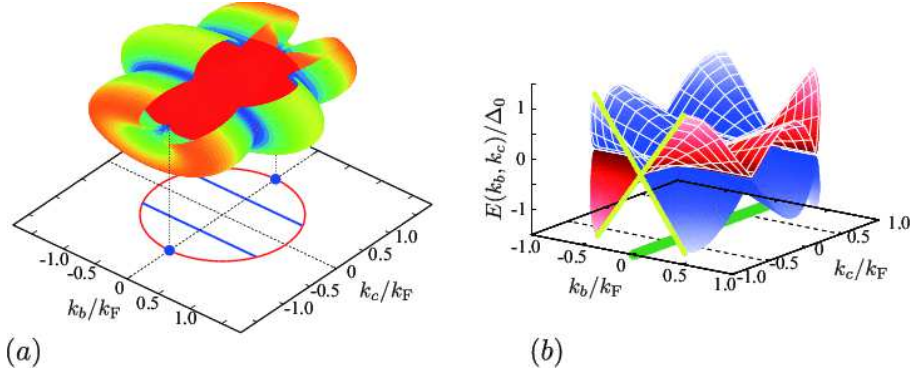}
\caption{A stereographic view of the gap function (a) and the dispersion (b) of surface bound states, the Majorana valley, in the $E_{1u}$ state of UPt$_3$-B. In (b), the blue (red) color corresponds to the $\rightarrow$-spin ($\leftarrow$-spin) sector. The broken lines denote the positions of line nodes and the point nodes are at $\hat{k}_b = 0$ and $\hat{k}_c = \pm 1$. }
\label{fig:edge}
\end{figure}

We start to clarify that the gap function in the B-phase of UPt$_3$, Eqs.~(\ref{eq:upt31}) and (\ref{eq:upt32}), is invariant
under the mirror reflection ${M}_{ca}=i\sigma _b$, that is, ${M}_{ca}\Delta({\bm k})M^{\dag}_{ca} = \Delta (k_a,-k_b,k_c)$.
The BdG Hamiltonian $\mathcal{H}({\bm k})$ then satisfies
${\mathcal{M}}^+_{ca}\mathcal{H}({\bm k})\mathcal{M}_{ca}^{+\dagger}={\mathcal{H}}(k_a,-k_b,k_c)$
with ${\mathcal{M}}^+_{ca}\equiv{\rm diag}(M_{ca},M^{\ast}_{ca})$, and thus, combining with the time-reversal symmetry $\mathcal{T}$
and the particle hole symmetry $\mathcal{C}$, we have ``mirror  chiral
symmetry''
\beq
\{\Gamma_1, 
\mathcal{H}(k_a,k_b=0,k_c)\}=0,
\eeq 
with $\Gamma _1=\mathcal{T}\mathcal{C} {\mathcal{M}}^+_{ca}$ at $k_b=0$ or $\pi$~\cite{mizushimaNJP2013}. 
The mirror chiral symmetry enables us to define the one-dimensional
winding number (\ref{eq:w}) for $k_b=0$ and $\pi$ as
\beq
w(k_c)=-\frac{1}{4\pi i}\int_{-\pi}^{\pi}dk_a{\rm
tr}[\Gamma _1 {\mathcal{H}}^{-1}({\bm k})\partial_{k_a}{\mathcal{H}}({\bm k})].
\eeq 
This is evaluated as 
\beq
|w(k_c)| =\left\{
\begin{array}{ll}
2 & \mbox{for $k_b=0$ and $|k_c| < k_{\rm F}$} \\
\\
0 & \mbox{for other $k_b$ and $k_c$}
\end{array}
\right. 
\eeq
Thus, the system is topologically non-trivial and the bulk-edge
correspondence ensures the existence of the Majorana valley in
Fig.~\ref{fig:edge} with a flat dispersion connecting the point nodes as
$E=0$ at $k_b=0$ and $|k_c|<k _{\rm F}$. 
Owing to the mirror chiral symmetry, the Majorana valley shows the
Majorana Ising anisotropy that the surface bound states are gapped only
by a magnetic field along the $b$-axis~\cite{mizushimaPRL2012}.
A magnetic field in the $ca$-plane or the ${\bm d}$-vector
rotation in the high field phase in the B-phase does not obscure the
topological protection since the combination of the mirror reflection
${\mathcal {M}}_{ca}$ and the time-reversal is not broken while each
of them is not.
Here note that while the Majorana valley has a close similarity to the
topological Fermi arcs in $^3$He-A phase \cite{silaev2014,silaevPRB2012,volovik2013}, their topological origins are
totally different: 
The time-reversal breaking is essential for the topological Fermi arcs
in $^3$He-A phase,
but the time-reversal 
breaking is not necessary for the Majorana valley.

\section{Majorana Ising spins and odd-frequency pairs in the quasiclassical theory}
\label{sec:numerical}

As mentioned in Sec.~\ref{sec:andreev}, the equilibrium properties of superfluids and superconductors are determined by the Matsubara Green's function $\underline{G}$. The topological nontriviality emerges in superconductors and superfluids as gapless quasiparticles that are bound to the surface, and spatial inhomogeneity is indispensable for understanding the microscopic structure of topologically protected gapless states. In general, however, solving the BdG euqation (\ref{eq:bdg3}) for spatially inhomogeneous systems is arduous. We here introduce the quasiclassical theory which is the natural extension of the Landau's Fermi liquid theory to superfluid phases. This theory offers a tractable way for studying the microscopic structure of superconductors and superfluids and is applicable to a wide range within the weak coupling regime $k_{\rm F}\xi \ll 1$~\cite{serene}. 

In this section, we reproduce in the context of the quasiclassical theory that the Majorana Ising spins emerge as a consequence of surface Andreev bound states. It is demonstrated that the Majorana Ising spin is attributed to the strong constraint of odd-frequency pairing that is imposed by discrete symmetries. 

\subsection{Quasiclassical theory}

The central object of the quasiclassical theory is the propagator that contains all the informations on both quasiparticles and superfluidity. The quasiclassical propagator $\underline{g} \!\equiv\! \underline{g}(\hat{\bm k},{\bm r};\omega _n)$ is obtained from the Green's function $\underline{G}$ introduced in Sec.~\ref{sec:NG}, by integrating $\underline{G}$ over a shell $v_{\rm F}|k-k_{\rm F}| < E_{\rm c} \ll E_{\rm F}$~\cite{serene}
\beq
\underline{g}(\hat{\bm k},{\bm r};\omega _n) = \frac{1}{a} \int^{+E_{\rm c}}_{-E_{\rm c}} d\xi _{\bm k}
\underline{\tau}_z\underline{G}({\bm k},{\bm r};\omega _n),
\eeq
where $\underline{G}({\bm k},{\bm r};\omega _n) 
= \int d{\bm r}_{12} e^{-i{\bm k}\cdot{\bm r}_{12}} \underline{G}({\bm r}_1,{\bm r}_2;\omega _n) $. The normalization constant $a$ corresponds to the weight of the quasiparticle pole in the spectral function. 

The quasiclassical propagator $\underline{g}\equiv\underline{g}(\hat{\bm k},{\bm r};\omega _n)$ is governed by the transport-like equation~\cite{serene,eilenberger,larkin1,eliashberg,larkin2,larkin3}. Following the procedure in Ref.~\cite{serene}, one obtains the quasiclassical transport equation from the Nambu-Gor'kov equation (\ref{eq:NG}) as
\beq
\left[i\omega _n \underline{\tau}_z - \underline{v}(\hat{\bm k},{\bm r})
- \underline{\Delta}(\hat{\bm k},{\bm r}), 
\underline{g}
\right] =- i {\bm v}_{\rm F} \!\cdot{\bm \nabla}
\underline{g}.
\label{eq:eilen}
\eeq
The Fermi velocity is defined as ${\bm v}_{\rm F}(\hat{\bm k}) \!=\! \partial \varepsilon _0({\bm k})/\partial {\bm k}|_{{\bm k}=k_{\rm F}\hat{\bm k}}$. The self-energy term in Eq.~(\ref{eq:NG}) is replaced to $\underline{\tau}_z\underline{\Sigma} ({\bm k},{\bm r}) \approx\underline{\tau}_z\underline{\Sigma} (k_{\rm F}\hat{\bm k},{\bm r}) = [\underline{\nu} (\hat{\bm k},{\bm r}) + \underline{\Delta}(\hat{\bm k},{\bm r})]/a$. The term $ \underline{v}$ in Eq.~(\ref{eq:eilen}) consists of an external potential $\underline{v}_{\rm ext}$ and quasiclassical self-energy $\underline{\nu}$ associated with Fermi liquid corrections, as $\underline{v}(\hat{\bm k},{\bm r}) = \underline{v}_{\rm ext}({\bm r}) + \underline{\nu}(\hat{\bm k},{\bm r})$, where 
\beq
\underline{\nu} = 
\left(
\begin{array}{cc}
\nu _{0} + {\sigma}_{\mu} \nu _{\mu} &   \\ 
 & \bar{\nu}_{0} + {\sigma}^{\rm T}_{\mu} \bar{\nu}_{\mu}
\end{array}
\right).
\eeq
The off-diagonal component of the quasiclassical self-energies is given as
\beq
\underline{\Delta}(\hat{\bm k},{\bm r}) 
= \left(
\begin{array}{cc}
& \Delta(\hat{\bm k},{\bm r}) \\
\Delta^{\dag}(-\hat{\bm k},{\bm r}) & 
\end{array}
\right). 
\eeq

The quasiclassical transport equation (\ref{eq:eilen}) is a first-order ordinary differential equation along a trajectory in the direction of ${\bm v}_{\rm F}(\hat{\bm k})$. For spatially uniform system with $\underline{\Delta} (\hat{\bm k},{\bm r})= \underline{\Delta} (\hat{\bm k})$ and $\underline{v}(\hat{\bm k},{\bm r}) = 0$, the solution of Eq.~(\ref{eq:eilen}) is given by 
\beq
\underline{g}(\hat{\bm k},\omega _n) = -\pi\frac{i\omega _n \underline{\tau}_z - \underline{\Delta} (\hat{\bm k})}{\sqrt{\omega^2_n + \frac{1}{2}{\rm Tr}[\Delta (\hat{\bm k})\Delta^{\dag}(\hat{\bm k})]}},
\label{eq:gbulk}
\eeq
where we suppose unitary states obeying ${\bm d}\times {\bm d}^{\ast}=0$. 

The solution of the quasiclassical transport equation (\ref{eq:eilen}) is not uniquely determined per se, because $a+bg$ satisfies the same equation as $g$ ($a$ and $b$ are arbitrary constants). To obtain a unique solution for $g$, Eq.~(\ref{eq:eilen}) must be supplemented by the normalization condition on the quasiclassical propagator as~\cite{eilenberger,larkin3,shelankov2}
\beq
\left[ \underline{g} (\hat{\bm k},{\bm r};\omega _n)\right]^2 = -\pi^2.
\label{eq:norm}
\eeq 
It is obvious that since $g^2$ is the solution of the quasiclassical transport equation (\ref{eq:eilen}), it can be parametrized as $g^2=a+bg$. In accordance with direct calculation of Eq.~(\ref{eq:eilen}) for spatially uniform systems, the arbitrary constants $a$ and $b$ are found to be $a = -\pi^2$ and $b=0$. The general solutions for nonuniform systems should be determined without any contradiction to uniform solutions in Eq.~(\ref{eq:gbulk}). The normalization condition (\ref{eq:norm}) was proven by Shelankov~\cite{shelankov2} in the more direct manner. 

The quasiclassical propagator $\underline{g}$ that is a $4\times 4$ matrix in particle-hole and spin spaces is parameterized with spin Pauli matrices $\sigma _{\mu}$ as
\beq
\underline{g} = \left(
\begin{array}{cc}
g_{0} + {\sigma}_{\mu} g_{\mu} & i\sigma _y f_0 + i {\sigma}_{\mu} {\sigma}_y f_{\mu}  \\ 
i\sigma _y \bar{f}_0 +i \sigma _y {\sigma}_{\mu}\bar{f}_{\mu}  & \bar{g}_{0} + {\sigma}^{\rm T}_{\mu} \bar{g}_{\mu}
\end{array}
\right).
\label{eq:g}
\eeq
Here, $\sigma^{\rm T}_{\mu}$ denotes the transpose of the Pauli matrices $\sigma _{\mu}$. The off-diagonal propagators are composed of spin-singlet and triplet Cooper pair amplitudes, $f_0$ and $f_{\mu}$. 
The quasiclassical propagators must satisfy the following relations arising from the Fermi statistics in Eq.~(\ref{eq:G}),
\beq
\left[ \underline{g}(\hat{\bm k},{\bm r};\omega_n)\right]^{\dag} 
= \underline{\tau}_z\underline{g}(\hat{\bm k},{\bm r};-\omega_n) \underline{\tau}_z
\label{eq:sym1} , \\
\left[ \underline{g}(\hat{\bm k},{\bm r};\omega_n)\right]^{\rm T} 
= \underline{\tau}_y\underline{g}(-\hat{\bm k},{\bm r};-\omega_n) \underline{\tau}_y.
\label{eq:sym2}
\eeq
From the normalization condition in Eq.~(\ref{eq:norm}), one obtains $gf = - f\bar{g}$ and $\bar{g}\bar{f} = - \bar{f}g$, leading to the relation between $\bar{g}_0$ and $g_0$,
\beq
\bar{g}_0(\hat{\bm k},{\bm r};\omega _n) = -g_0(\hat{\bm k},{\bm r};\omega _n).
\label{eq:trs2}
\eeq 

The quasiclassical self-energies are associated with the quasiclassical propagator. The diagonal components are determined by the propagators $g_0$ and ${\bm g}$ as 
\beq
\nu _0 (\hat{\bm k},{\bm r}) = \sum _{\ell} A^{({\rm s})}_{\ell}\left\langle
P_{\ell}(\hat{\bm k}\cdot\hat{\bm k}^{\prime})g_0(\hat{\bm k}^{\prime},{\bm r};\omega _n)
\right\rangle _{\hat{\bm k}^{\prime},n}, \label{eq:nu0} \\
{\bm \nu} (\hat{\bm k},{\bm r}) = \sum _{\ell} A^{({\rm a})}_{\ell}\left\langle
P_{\ell}(\hat{\bm k}\cdot\hat{\bm k}^{\prime}){\bm g}(\hat{\bm k}^{\prime},{\bm r};\omega _n)
\right\rangle _{\hat{\bm k}^{\prime},n}, \label{eq:nu}
\eeq
where $P_{\ell}(x)$ is the Legendre polynomials with $\ell = 0, 1, 2, \cdots$. The coefficients $A^{({\rm s})}_{\ell}$ and $A^{({\rm a})}_{\ell}$ are symmetric and antisymmetric quasiparticle scattering amplitudes, respectively, which are parametrized with the Landau's Fermi-liquid parameters $F^{({\rm s},{\rm a})}_{\ell}$ as 
$A^{({\rm s, a})}_{\ell} = F^{({\rm s,a})}_{\ell}/[1+F^{({\rm s,a})}_{\ell}/(2\ell+1)]$. The terms with the coefficients $F^{({\rm s})}_{\ell = 0}$ and $F^{({\rm s})}_{\ell = 1}$ are self-energy potentials originating from the local particle density and superfluid mass flow in the equilibrium, respectively. The self-energy potentials with $F^{({\rm a})}_{\ell = 0}$ and $F^{({\rm a})}_{\ell = 1}$ arise from the local magnetization density and spin current density. Among the various parameters $F^{\rm s,a}_{\ell}$, $F^{\rm s}_{\ell =1}$ and $F^{\rm a}_{\ell = 0}$ give Fermi liquid corrections to the effective mass and spin susceptibility. Throughout this paper, we use the following abbreviation for the Matsubara sum and the average over the Fermi surface, 
$\langle\cdots\rangle _{\hat{\bm k},n} = \frac{T}{\mathcal{N}_{\rm F}}\sum _{n}
\int \frac{d\hat{\bm k}}{(2\pi)^3 |{\bm v}_{\rm F}(\hat{\bm k})|}\cdots$, 
where $\mathcal{N}_{\rm F} \!=\! \int \frac{d\hat{\bm k}}{(2\pi)^3|{\bm v}_{\rm F}(\hat{\bm k})|}$ is the total density of states at the Fermi surface in the normal state. For three dimensional Fermi sphere, one finds ${\bm v}_{\rm F}(\hat{\bm k}) \!=\! v_{\rm F}\hat{\bm k}$ and $\mathcal{N}_{\rm F} \!=\! \frac{1}{2\pi^2 v_{\rm F}}$, leading to $\langle\cdots\rangle _{\hat{\bm k},n} = T\sum _{n} \int \frac{d{\bm k}}{4\pi}\cdots $. 

The complete set of the self-consistent quasiclassical theory is composed of the transport equation (\ref{eq:eilen}), the normalization condition in Eq.~(\ref{eq:norm}), and the Fermi liquid corrections in Eqs.~(\ref{eq:nu0}) and (\ref{eq:nu}) in addition to the gap equation
\beq
\Delta _{ab} (\hat{\bm k},{\bm r}) = \left\langle V^{cd}_{ab}(\hat{\bm k},\hat{\bm k}^{\prime})
\left[i\sigma _{\mu}\sigma _y f_{\mu}(\hat{\bm k}^{\prime},{\bm r};\omega _n)\right]_{cd} \right\rangle _{\hat{\bm k}^{\prime},n}.
\label{eq:gap_qct}
\eeq
This is obtained from the gap equation (\ref{eq:gap}).

Even for spatially uniform $\underline{\Delta}$ and $\underline{\nu}$, the quasiclassical transport equation (\ref{eq:eilen}) generally has two solutions in addition to Eq.~(\ref{eq:gbulk}) that exponentially grow and decay along trajectories in the direction of $\hat{\bm k}$. These exploding solutions are not normalizable in spatially uniform superconductors and superfluids, which make Eq.~(\ref{eq:eilen}) numerically unstable. Thuneberg {\it et al}.~\cite{thuneberg} proposed a numerically accessible method using the explosion trick, where the physical solution is constructed from the commutation relation of two exploding solutions. In Sec.~\ref{sec:exact}, however, we give an overview of an alternative scheme for solving the transport equation (\ref{eq:eilen}). The scheme based on the projection operator found by Shelankov~\cite{shelankov1,shelankov2} was established by Eschrig {\it et al}.~\cite{eschrigPRB1999,eschrigPRB2000}. The projection operator maps the transport equation (\ref{eq:eilen}) having exploding solutions onto the Riccati-type differential equation which does not have non-normalizable solutions and is numerically stable. We also notice that the Riccati-type differential equation can be derived directly from the BdG equation (\ref{eq:bdg}) within the Andreev approximation~\cite{nagatoJLTP1993}. 



In this paper, we concentrate our attention to a slab geometry with perfectly specular surfaces. The surface specularity is experimentally controllable by coating the container with $^4$He layers~\cite{okuda}, and the surface roughness may change the surface structure of the superfluid $^3$He-B. Boundary conditions that describe quasiparticle scattering from an atomically rough surface were developed in several manners. This includes the scattering of quasiparticles from a thin layer of atomic-size impurities~\cite{zhangPRB1987}, a distribution of randomly oriented mirror on the surface~\cite{thuneberg1992}, and randomly rippled wall~\cite{buchholtz1979,buchholtz1986,buchholtz1991,buchholtz1993}. Nagato {\it et al.} implemented boundary conditions that describe a partially diffusive surface, by using a random $S$-matrix~\cite{nagatoJLTP1996,nagatoJLTP1998,nagatoJLTP2007}.

\subsection{Odd-frequency pairing and magnetization}
\label{eq:odd2}

All the informations on Cooper pair correlation are included in the anomalous propagator $f=i\sigma _y f_0 + i{\bm \sigma}\cdot{\bm f}\sigma _y$. It is important to mention that in accordance with the Fermi-Dirac statistics, a wave function of Cooper pairs must change its sign after a permutation of two paired fermions. Then, as summarized in Table~\ref{table3}, the symmetry of Cooper pairing in a single-band centrosymmetric superconductor is naturally categorized to the four-fold way. Two of them are conventional spin-singlet even-parity pairing and spin-triplet odd-parity pairing, which do not change the sign of Cooper pair wave function by the exchange of times of paired fermions. These are referred to as even-frequency spin-singlet even-parity (ESE) and spin-triplet odd-parity (ETO) pairings. There still remain two possibilities of Cooper pair symmetries, odd-frequency spin-singlet odd-parity (OSO) and spin-triplet even-parity (OTE) pairs. Conclusive evidence of odd-frequency pairing in bulk materials has not been observed experimentally yet since the first prediction by Berezinskii~\cite{berezinskiiJETP1974}. Nevertheless, OSO and OTE pair amplitudes can emerge ubiquitously in spatially non-uniform systems accompanied by Andreev bound states and anomalous proximity effect. In particular, anomalous charge and spin transport, electromagnetic responses, proximity effects via Andreev bound states have also been clarified in the light of odd-frequency Cooper pairing~\cite{tanakaPRB1996,barashPRL1996,tanakaPRB1997,higashitaniJPSJ1997,WalterPRL1998,kashiwayaPRB1999,bergeretPRL2001,tanakaPRB2005,tanakaPRL2007,tanakaPRL2007v2,linderPRL2009,linderPRB2010,yokoyamaPRL2011,asanoPRL2011}.

{\it Odd-frequency pairing.}---
In general, the Cooper pair amplitudes are separated to even-frequency and odd-frequency components, $
f_{\mu} \!=\! f^{{\rm EF}}_{\mu} + f^{{\rm OF}}_{\mu}$ and $f_{0} \!=\! f^{{\rm EF}}_{0} + f^{{\rm OF}}_{0}$, where even- and odd-frequency pair amplitudes are defined as ($j \!=\! 0, x, y, z$)
\beq
f^{\rm EF}_j(\hat{\bm k},{\bm r};\omega _n) = \frac{1}{2} \left[
f_{j}(\hat{\bm k},{\bm r};\omega _n) + f_j(\hat{\bm k},{\bm r};-\omega _n)
\right], 
\label{eq:even}
\eeq
\beq
f^{\rm OF}_j(\hat{\bm k},{\bm r};\omega _n) = \frac{1}{2} \left[
f_j(\hat{\bm k},{\bm r};\omega _n) - f_j(\hat{\bm k},{\bm r};-\omega _n)
\right].
\label{eq:odd}
\eeq
To categorize the possible types of Cooper pair amplitudes in terms of the basic discrete symmetries preserved by superfluid/superconducting states, we here introduce three operators, $\hat{P}_{\omega}$, $\hat{P}_{\sigma}$, and $\hat{P}_k$, that act on the quasiclassical propagators as $\hat{P}_{\omega} f(\hat{\bm k},{\bm r};\omega _n)
= f(\hat{\bm k},{\bm r};-\omega _n)$, $\hat{P}_{\sigma}f_{ab}(\hat{\bm k},{\bm r};\omega _n) = f_{ba}(\hat{\bm k},{\bm r};\omega _n)$, and $\hat{P}_{k}f(\hat{\bm k},{\bm r};\omega _n) = (-\hat{\bm k},{\bm r};\omega _n)$. 

In Table~\ref{table3}, we summarize the four possible classes of Cooper pair amplitudes in bulk superconductors and superfluids, and the additional Cooper pairs induced by a symmetry breaking field~\cite{tanakaJPSJ2012,dainoPRB2012,tanakaPRL2007,tanakaPRB2007,yokoyamaPRB2008,yokoyamaJPSJ2010}. In the case of spin-triplet superconductors and superfluids, ETO components $f^{{\rm EF}}_{\mu}$ exist in the bulk, where $f^{{\rm EF}}_{\mu}({\bm k})=-f^{{\rm EF}}_{\mu}(-{\bm k})$. A time-reversal breaking perturbation, such as a magnetic Zeeman field, can induce the mixing of spin-singlet Cooper pair amplitudes. Since the induced spin-singlet pairing must have a odd parity unless the translational symmetry is broken, the Cooper pairs in bulk ETO superconductors and superfluids are OSO pairing in addition to ETO pairing. A translational symmetry breaking field, such as a surface boundary condition and vortices, induces OTE components $f^{\rm OF}_{\mu}$ in bulk ETO superconductors and superfluids. We also note that all four pairings can emerge when both time-reversal and translational symmetries are broken.

It has recently been demonstrated that in a bulk ETO superconductor and superfluid, OTE Cooper pair amplitudes, $f^{\rm OF}_{\mu}(-\hat{\bm k},{\bm r};\omega _n)=-f^{\rm OF}_{\mu}(\hat{\bm k},{\bm r};\omega _n)$, are equivalent to the low-energy density of states originating from Andreev bound states that are bound to the surface or vortices~\cite{higashitaniPRB2012,dainoPRB2012,asanoPRB2013,tsutsumiJPSJ2012}. In particular, at the zero energy limit, $f^{\rm OF}_{\mu}$ is equivalent to the Majorana zero modes~\cite{dainoPRB2012,asanoPRB2013}.

\begin{table}
\centering
\begin{tabular}{c|ccc|cc}
\hline\hline
& \multicolumn{3}{c}{Parity} 
& \multicolumn{2}{c}{Broken symmetry} \\
$\Delta$ & frequency & spin & parity & time-inversion & translational \\
\hline
ESE & $+$ & $-$ & $+$ & OTE & OSO \\
ETO & $+$ & $+$ & $-$ & OSO & OTE \\
OSO & $-$ & $-$ & $-$ & ETO & ESE \\
OTE & $-$ & $+$ & $+$ & ESE & ETO \\
\hline\hline
\end{tabular}
\caption{Classification of possible Cooper pairing in bulk superconductors: ESE, ETO, OSO, and OTE pairs. The fifth and sixth columns show the Cooper pair amplitudes emergent in systems with the breaking of time-reversal symmetry and translational symmetry, respectively.  
}
\label{table3}
\end{table}


{\it Discrete symmetries.}---
The discrete symmetries that are preserved by the BdG Hamiltonian can be extended to the quasiclassical formalism, which add constraint on the quasiclassical propagator.
First, the particle-hole symmetry in Eq.~(\ref{eq:PHS}) is recast into 
\beq
\underline{\mathcal{C}}~\underline{g}(\hat{\bm k},{\bm r};\omega _n) \underline{\mathcal{C}}^{-1}
= \underline{g}(-\hat{\bm k},{\bm r};\omega _n).
\eeq
This symmetry can be obtained from the basic relations of the quasiclassical propagator in Eqs.~(\ref{eq:sym1}) and (\ref{eq:sym2}). For time-reversal invariant superconductors and superfluids that yeild $\Theta \Delta ({\bm k}) \Theta^{\rm T} = \Delta (-{\bm k})$, the quasiclassical propagator holds the time-reversal symmetry as 
\beq
\underline{\mathcal{T}}~
\underline{g}(\hat{\bm k},{\bm r};\omega _n) \underline{\mathcal{T}}^{-1}
= \underline{g}(-\hat{\bm k},{\bm r};-\omega _n) .
\label{eq:trs1}
\eeq
We here suppose that the single-particle Hamiltonian $\varepsilon({\bm k})$ is invariant under the time-inversion. In the context of the quasiclassical formalism, this implies that $\underline{v}$ does not contains a magnetic Zeeman energy.

In addition, the BdG Hamiltonian may hold the simultaneous $\pi$-rotational symmetry in spin and orbital spaces,
\beq
\mathcal{U}(\pi)\mathcal{H}({\bm k})\mathcal{U}^{\dag}(\pi)
= \mathcal{H}(-\underline{\bm k}). 
\eeq
As we have seen in Eq.~(\ref{eq:so2slab}), this can be realized in the superfluid $^3$He-B confined in a slab geometry, where a magnetic field is absent. Then, the quasiclassical propagator also holds the $\pi$-rotational symmetry,
\beq
\underline{\mathcal{U}}(\pi)\underline{g}(\hat{\bm k},{\bm r};\omega _n) \underline{\mathcal{U}}^{\dag}(\pi)
= \underline{g}(R\hat{\bm k},{\bm r};\omega _n).
\label{eq:pi2}
\eeq

In superconducting materials, the rotational symmetry in the spin space may be absent, because a crystal field usually lowers the symmetry. Such a material, however, may preserve additional discrete symmetry that arises from the crystalline symmetry. In Sec.~\ref{sec:mirror}, we have seen that the mirror reflection symmetry may protect the nontrivial topological property, even when the $\pi$-rotational symmetry in the spin space is absent. The possible candidate of topological crystalline superconductors is the $E_{1u}$ scenario of the heavy-fermion superconductor UPt$_3$~\cite{tsutsumiJPSJ2013,tsutsumiJPSJ2012-2}. In the context of the quasiclassical theory, the mirror symmetry for the propagator is described as 
\beq
\underline{\mathcal{M}}^{\pm}_{\mu\nu}\underline{g}(\hat{\bm k},{\bm r};\omega _n) 
\underline{\mathcal{M}}^{\pm\dag}_{\mu\nu}
= \underline{g}(\underline{\hat{\bm k}}_{\rm M},{\bm r};\omega _n).
\eeq
We will show below that these additional discrete symmetries add a strong constraint to Cooper pair amplitudes emergent in superconductors and superfluids with broken translational symmetry. The strong constraint gives rise to the Ising anisotropy of surface spin susceptibility. 

{\it Magnetization.}---
Let us now derive the generic form of the magnetization density $M_{\mu}({\bm r})$ for superfluids and superconductors under a magnetic Zeeman field ${\bm H}\!=\!H\hat{\bm h}$. The potential term $\underline{v}(\hat{\bm k},{\bm r})$ in the quasiclassical equation (\ref{eq:eilen}) is composed of a magnetic Zeeman field and quasiclassical self-energies $\underline{\nu}$,
\beq
\underline{v}(\hat{\bm k},{\bm r}) = - \frac{1}{1+F^{\rm a}_0}\mu _{\rm n} H_{\mu}\left(
\begin{array}{cc}
\sigma _{\mu} & \\ &  \sigma^{\rm T}_{\mu} 
\end{array}
\right) + \underline{\nu}(\hat{\bm k},{\bm r}), 
\label{eq:v2}
\eeq
where $F^{\rm a}_0$ is the Fermi liquid parameter associated with the enhancement of spin susceptibility and $\mu _{\rm n}$ is the magnetic moment of $^3$He nuclei. In the quasiclassical formalism, the magnetization density is given by~\cite{mizushimaPRL2012,mizushimaPRB2012,serene}
\beq
M_{\mu}(z) = M_{\rm N}\left[
\hat{h}_{\mu} + \frac{1}{\mu _{\rm n} H}\langle g_{\mu}(\hat{\bm k},z;\omega_n)\rangle _{\hat{\bm k},n}
\right].
\label{eq:M}
\eeq
This is also applicable to the surface region of superconductors in the type-II limit where the surface region within the coherence length $\xi$ is much thinner than the penetration depth of the external field. For superconductors, $\mu _{\rm n}$ in Eq.~(\ref{eq:M}) is replaced to the Bohr magneton $\mu _{\rm B}$. The magnetization in normal $^3$He is
$M_{\rm N} \!=\! \chi _{\rm N}H \!=\! \frac{2\mu^2_{\rm n}}{1+F^{\rm a}_0}N_{\rm F}H$. 

The quasiclassical propagator must satisfy a constraint given in Eq.~(\ref{eq:norm}) which requires the propagators to hold the relation, 
$g_{\mu} \!=\! ( f_0 \bar{f}_{\mu} + \bar{f}_0 f_{\mu}
+ i\epsilon _{\mu\nu\eta}f_{\nu}\bar{f}_{\eta})/2g_0$. This relates the spin component of quasiclassical propagators to spin-singlet and -triplet Cooper pair amplitudes. 
Using the relation and the symmetries in Eqs.~(\ref{eq:sym1}), (\ref{eq:sym2}), and (\ref{eq:trs2}), the magnetization density in Eq.~(\ref{eq:M}) reduces to
\beq
\frac{M_{\mu}({\bm r})}{M_{\rm N}} = \hat{h}_{\mu} + \frac{1}{\mu _{\rm n}H}
\left\langle \frac{f_0 \bar{f}_{\mu} + \bar{f}_0 f_{\mu}}{2g_0} \right\rangle _{\hat{\bm k},n}.
\label{eq:m2}
\eeq
This indicates that only the mixing term of spin-singlet and triplet Cooper pair amplitudes contributes to the spin susceptibilities. This expression is a quite generic form for $M_{\mu}({\bm r})$ in superfluids and also applicable to the surface region of type-II superconductors. This was first derived in Ref.~\cite{higashitaniPRL2013} for the aerogel-superfluid $^3$He-B system.

We now clarify the relation between OTE Cooper pairs and spin susceptibility in spin-triplet superfluids and superconductors. It is supposed that the system holds the time-reversal symmetry at zero field. We here deal with a magnetic field perturbatively in parameter, $\mu _{\rm n}H/\Delta \!\ll\! 1$. Then, we formally expand $g_0$, $f_0$, and $f_{\mu}$ in powers of $\mu _{\rm n}H/\Delta$: $g_0 \!=\! g^{(0)}_0 + g^{(1)}_0 +\cdots$, $f_0 \!=\! f^{(1)}_0 + \cdots$, and $f_{\mu} \!=\! f^{(0)}_{\mu} + f^{(1)}_{\mu} + \cdots$. At zero field, time-reversal invariant superfluids and superconductors hold Eq.~(\ref{eq:trs1}).
Combining the symmetric property in Eq.~(\ref{eq:trs1}) with Eqs.~(\ref{eq:sym1}), (\ref{eq:sym2}), and (\ref{eq:trs2}), one finds 
\beq
g^{(0)}_0(\hat{\bm k},z; \omega _n) = - g^{(0)}_0(\hat{\bm k},z; -\omega _n).
\label{eq:g0} 
\eeq
Substituting this in Eq.~(\ref{eq:m2}) and using the symmetry in Eq.~(\ref{eq:g0}), one finds that the spin susceptibility $\chi \equiv \hat{h}_{\mu}\chi _{\mu\nu}\hat{h}_{\nu}$ is composed of the contributions of odd- and even-parity Cooper pair amplitudes~\cite{mizushima2014-2}, 
\beq
\frac{\chi(z)}{\chi _{\rm N}} \equiv 
\frac{\hat{h}_{\mu}\chi _{\mu\nu}(z)\hat{h}_{\nu}}{\chi _{\rm N}}
= 1 + \frac{\chi^{\rm OP}(z)}{\chi _{\rm N}}
+ \frac{\chi^{\rm EP}(z)}{\chi _{\rm N}},
\label{eq:chi2}
\eeq
where we set $M_{\mu} = \chi _{\mu \nu}H_{\nu}$. The odd-parity contribution $\chi^{\rm OP}(z)$ is given by the mixing term of the OSO pair amplitude $f^{\rm OF}_0$ and the ETO pair ${\bm f}^{\rm EF}$,
\beq
\frac{\chi^{\rm OP}(z)}{\chi _{\rm N}} \equiv
\frac{1}{\mu _{\rm n}H}{\rm Re}\left\langle 
\frac{f^{{\rm OF}(1)}_0\hat{h}_{\mu}f^{{\rm EF}(0)\ast}_{\mu}}{g^{(0)}_0}
\right\rangle _{\hat{\bm k},n}. \label{eq:chiOP} 
\eeq
The even-parity contribution $\chi^{\rm EP}(z)$ is given by the mixing term of the ESE pair amplitude $f^{\rm EF}_0$ and the OTE pair ${\bm f}^{\rm OF}$,
\beq
\frac{\chi^{\rm EP}(z)}{\chi _{\rm N}} \equiv
-\frac{1}{\mu _{\rm n}H}{\rm Re}\left\langle 
\frac{f^{{\rm EF}(1)}_0\hat{h}_{\mu}f^{{\rm OF}(0)\ast}_{\mu}}{g^{(0)}_0}
\right\rangle _{\hat{\bm k},n}. \label{eq:chiEP}
\eeq

Equation (\ref{eq:chi2}) indicates that the spin susceptibility in time-reversal invariant superconductors and superfluids is composed of the contributions from odd-parity Cooper pair amplitudes, $\chi^{\rm OP}$, and even-parity pairing, $\chi^{\rm EP}$.   Only the spin-triplet pairings $f^{(0)}_{\mu}$ at zero fields can be directly coupled to the applied field. 

The ETO pairings $f^{{\rm EF}(0)}_{\mu}$ exist in the bulk of spin-triplet superfluids and superconductors, which are responsible for the ${\bm d}$-vector. The situation $\hat{\bm h}\cdot{\bm f}^{\rm EF}$ corresponds to ${\hat{\bm h}}\cdot {\bm d}$. The behavior of $\chi^{\rm OP}$ is then understandable with the rotation of the ${\bm d}$-vector, where $\chi^{\rm OP}=0$ for $\hat{\bm h}\perp{\bm d}$ and $\chi^{\rm OP} \le 0$ for ${\bm d}\cdot\hat{\bm h}\neq 0$. In contrast, the OTE Cooper pairs $f^{{\rm OF}(0)}_{\mu}$ are absent in the bulk and induced by the breaking of translational symmetry at surfaces, interfaces, or vortices. As a result, the total spin susceptibility at surfaces is determined by the OTE pairing $f^{{\rm OF}(0)}_{\mu}$ directly coupled to the applied field in addition to the ordinary contribution from the relative orientation of the ${\bm d}$-vectors to $\hat{\bm h}$. 


\subsection{Constraint on emergent Cooper pairs and Ising spin susceptibility}
\label{sec:odd}

{\it Constraint on emergent Cooper pairs.}---
Let us now consider the spin susceptibility in the superfluid $^3$He-B confined in a slab geometry. The specular boundary condition is imposed on the quasiclassical propagators at the surfaces $z_0 = 0$ and $D$, 
\beq
\underline{g}(\hat{\bm k},z = z_0; \omega _n) = \underline{g}(\underline{\hat{\bm k}},z = z_0; \omega _n),
\label{eq:bc}
\eeq
where $\underline{\hat{\bm k}}\equiv \hat{\bm k} - 2 \hat{\bm z} (\hat{\bm z}\cdot\hat{\bm k})$ denotes the momentum specularly scattered by the surface. As shown in Eqs.~(\ref{eq:chiOP}) and (\ref{eq:chiEP}), the spin susceptibility in time-reversal invariant superfluids is determined by the Cooper pair amplitudes in the absence of a magnetic field, $f^{(0)}_{\mu}$. In particular, the OTE pairing is a key ingredient for understanding the anomalous behavior of the surface spin susceptibility. 

We here clarify that the discrete symmetries impose strong constraint on the Cooper pair amplitudes $f^{(0)}_{\mu}$ emergent at surfaces. As discussed in Sec.~\ref{sec:slab}, the superfluid $^3$He-B holds the discrete symmetry that ariases from the continuous ${\rm SO}(2)_{L_z+S_z}$ rotation symmetry in the slab geometry without a magnetic field. The $\pi$-rotational symmetry $U(\pi)$ is defined as the subgroup of ${\rm U}(1)^{(\hat{\bm z})}_{{\bm S}+{\bm L}}$, $U(\pi)\equiv U(\phi \!=\! \pi)$. This imposes the discrete symmetry on the quasiclassical propagator as shown in Eq.~(\ref{eq:pi2}), 
$\underline{\mathcal{U}}(\pi)
\underline{g}^{(0)}(\hat{\bm k},z;\omega _n) \underline{\mathcal{U}}^{\dag}(\pi)
= \underline{g}^{(0)}(-\underline{\hat{\bm k}},z;\omega _n)$. Combining this with the boundary condition in Eq.~(\ref{eq:bc}) and the relation in Eq.~(\ref{eq:sym2}), one obtains the relation between $\underline{g}(\omega _n)$ and $\underline{g}(-\omega _n)$ at the surface $z=z_0$ as
\beq
\underline{g}^{(0)}(\hat{\bm k},z_0;-\omega _n) = 
\underline{\mathcal{U}}(\pi)\underline{\tau}_y
\left[\underline{g}^{(0)}(\hat{\bm k},z_0;\omega _n)\right]^{\rm T} \underline{\tau}_y
\underline{\mathcal{U}}^{\dag} (\pi).
\label{eq:relation_g0}
\eeq

It is convenient to introduce $\underline{\tilde{g}}^{(0)}$ obtained by the unitary transformation of the original quasiclassical propagator as
\beq
\underline{\tilde{g}}^{(0)}(\hat{\bm k},z;\omega _n) = 
\underline{\mathcal{U}}^{\dag}(\hat{\bm n},\varphi) 
\underline{g}(\hat{\bm k},z;\omega _n) \underline{\mathcal{U}}(\hat{\bm n},\varphi). 
\eeq
The propagator $\underline{\tilde{g}}^{(0)}$ obeys the quasiclassical equation (\ref{eq:eilen}) with the definition $\underline{\tilde{\Delta}} \equiv U^{\dag}(\hat{\bm n},\varphi)\Delta (\hat{\bm k},z)U(\hat{\bm n},\varphi)^{\ast} = i\sigma _{\mu}\sigma _y d_{\mu \nu}(z)\hat{k}_{\nu}$.
This is equivalent to Eq.~(\ref{eq:eilen}) in the case of $\hat{\bm n} = \hat{\bm z}$ and $\varphi =0$. Then, Eq.~(\ref{eq:relation_g0}) imposes the constraint on the pair amplitudes $\tilde{f}_{\mu}$ at the surfaces as
\beq
\tilde{f}^{{\rm OF}(0)}_{\parallel}(\theta _{\bm k},z_0;\omega _n) = \tilde{f}^{{\rm EF}(0)}_{z}(\theta _{\bm k},z_0;\omega _n) = 0.
\label{eq:tildef}
\eeq
The factorization of the parallel components, $(\tilde{f}^{(0)}_x,\tilde{f}^{(0)}_y) = \tilde{f}^{(0)}_{\parallel}(\cos\phi _{\bm k},\sin\phi _{\bm k})$, results from the ${\rm SO}(2)_{L_z+S_z}$ symmetry that is preserved in the slab. It turns out from Eq.~(\ref{eq:tildef}) that in the configuration of $\hat{\bm n}\!\parallel\! \hat{\bm z}$, only $\tilde{f}_{z}$ has odd-frequency Cooper pairs at the surfaces. By using this notation, the pair amplitudes $f^{(0)}_{\mu}\equiv f^{(0)}_{\mu}(\hat{\bm k},z_0;\omega _n)$ at the surfaces of the superfluid $^3$He-B are expressed as 
\beq
f^{{\rm EF}(0)}_{\mu} = \left( R_{\mu x}(\hat{\bm n},\varphi)\cos\phi _{\bm k} 
+ R_{\mu y}(\hat{\bm n},\varphi)\sin\phi _{\bm k}\right)
\tilde{f}^{{\rm EF}(0)}_{\parallel}, 
\label{eq:fEF} 
\eeq
\beq
f^{{\rm OF}(0)}_{\mu} = R_{\mu z}(\hat{\bm n},\varphi) \tilde{f}^{{\rm OF}(0)}_z .
\label{eq:fOF}
\eeq

Hence, the discrete symmetry in Eq.~(\ref{eq:pi2}) arising from the ${\rm SO}(2)_{L_z+S_z}$ symmetry imposes strong constraint on the possible symmetry of Cooper pair amplitudes $f^{(0)}_{\mu}$ in the superfluid $^3$He-B. The relative orientation of the ETO pairing $f^{\rm EF}_{\mu}$ to the applied field is parameterized by $\hat{\ell}_x(\hat{\bm n},\varphi)$ and $\hat{\ell}_y(\hat{\bm n},\varphi)$. The OTE pairing ${\bm f}^{{\rm OF}(0)}$ which is responsible for $\chi^{\rm EP}$ is forced by the discrete symmetry to point to the surface normal direction ${\bm f}^{{\rm OF}(0)} \parallel \hat{\bm z}$. The magnetic response of the OTE pairing is characterized by $\hat{\ell}_z(\hat{\bm n},\varphi)$ that is the topological order associated with the hidden ${\bm Z}_2$ symmetry.

{\it Ising spin susceptibility}.---
Substituting the expression of pairing amplitudes into Eqs.~(\ref{eq:chiOP}) and (\ref{eq:chiEP}), the spin susceptibility in Eq.~(\ref{eq:chi2}) is recast into the following forms:
\beq
\chi _{\rm surf} 
= \chi _{\rm N} + \sqrt{1-\hat{\ell}^2_{z}}\tilde{\chi}^{\rm OP}_{\rm surf}
+ \hat{\ell}_z \tilde{\chi}^{\rm EP}_{\rm surf}.
\label{eq:chi_final}
\eeq
The contributions from odd-parity and even-parity pair amplitudes are given as
\beq
\frac{\tilde{\chi}^{\rm OP}_{\rm surf}}{\chi _{\rm N}} =
-\frac{1}{\mu _{\rm n}H}{\rm Re}\left\langle 
\frac{\cos(\phi _{\bm \ell}-\phi _{\bm k})f^{{\rm OF}(1)}_0\tilde{f}^{{\rm EF}(0)\ast}_{\parallel}}{g^{(0)}_0}
\right\rangle _{\hat{\bm k},n} ,
\label{eq:chiOP2} \\
\frac{\tilde{\chi}^{\rm EP}_{\rm surf}}{\chi _{\rm N}} =
\frac{1}{\mu _{\rm n}H}{\rm Re}\left\langle 
\frac{f^{{\rm EF}(1)}_0\tilde{f}^{{\rm OF}(0)\ast}_{z}}{g^{(0)}_0}
\right\rangle _{\hat{\bm k},n}.
\label{eq:chiEP2} 
\eeq
where $\hat{\ell}_{\mu}(\hat{\bm n},\varphi)$ is defined in Eq.~(\ref{eq:ell})~\cite{mizushimaPRL2012,volovikJETP2010}. We have also introduced $\hat{\ell}_{\parallel} \equiv \sqrt{\hat{\ell}^2_x + \hat{\ell}^2_y}$ and the azimuthal angle $\phi _{\bm \ell} = \tan^{-1}(\hat{\ell}_y/\hat{\ell}_x)$. 

The OTE Cooper pair, $\tilde{f}^{{\rm OF}(0)}_{z}$, is equivalent to the surface density of states in the low-energy regime,~\cite{higashitaniPRB2012,tsutsumiJPSJ2012}
\beq
\mathcal{N}(\hat{\bm k},z;E)
\approx \frac{1}{\pi}
\left|{\rm Re}{\bm f}^{{\rm OF}(0)}(\hat{\bm k},z;\omega _n \rightarrow -iE+0_+)\right|,
\label{eq:equality}
\eeq
where the momentum resolved local density of states is defined as 
\beq
\mathcal{N}(\hat{\bm k},z;E) = -\frac{\mathcal{N}_{\rm F}}{\pi}{\rm Im}g^{(0)}_0(\hat{\bm k},z;\omega _n \rightarrow -iE+0_+).
\label{eq:dosk}
\eeq 
Equation (\ref{eq:equality}) implies that the OTE Cooper pair amplitudes always appear on the surface of $^3$He-B when the surface Andreev bound state exists. 
The coupling of the OTE pairing with an applied magnetic field at the surface is parameterized by $\hat{\ell}_z(\hat{\bm n},\varphi)$. For $^3$He-B in a slab geometry, the $\hat{\bm n}$-texture and the angle $\varphi$ are determined by the applied magnetic field, the dipole-dipole interaction arising from the magnetic moment of nuclei, and surface boundary condition. In the limit of a weak magnetic field, the dipole-dipole interaction and surface boundary condition favors the configuration of $\hat{\bm n} = \hat{\bm z}$~\cite{mizushimaPRL2012,vollhardt}. Hence, one finds $\hat{\ell}_z (\hat{\bm n}= \hat{\bm z},\varphi) = \cos\theta _{\bm H}$ for a magnetic field ${\bm H}\cdot \hat{\bm z} = H \cos\theta _{\bm H}$. This configuration of the $\hat{\bm n}$-texture gives rise to the Ising anisotropy of the spin susceptibility, 
\beq
\chi _{\rm surf} 
= \chi _{\rm N} + \tilde{\chi}^{\rm OP}_{\rm surf}\sin\theta _{\bm H}
+ \tilde{\chi}^{\rm EP}_{\rm surf}\cos\theta _{\bm H}.
\eeq
This indicates that for a magnetic field parallel to the surface ($\theta _{\bm H}=\pi /2$), although the OTE pairings exists at the surfaces, it does not couple to the applied field. The resultant spin susceptibility is contributed from only the ETO pairing, which stays about the same as that in the bulk. The OTE pairing contributes to the surface spin susceptibility when the applied field is tilted from the surface normal direction or $\hat{\ell}_z$ is nonzero. Therefore, the Ising magnetic anisotropy of surface bound states is describable with the context of odd-frequency Cooper pair amplitudes.

Equation (\ref{eq:chi_final}) shows that only the OTE pairs contribute to the surface spin susceptibility when $\hat{\ell}_z=0$, while $\chi$ for $\hat{\ell}_z=1$ is composed of only the ETO Cooper pairs,
\beq
\chi = \left\{
\begin{array}{ll}
\displaystyle{\chi _{\rm N}+ \tilde{\chi}^{\rm OP}} & \mbox{for $\hat{\ell}_z = 0$} \\
\\
\displaystyle{\chi _{\rm N}+ \tilde{\chi}^{\rm EP}} & \mbox{for $\hat{\ell}_z = 1$} 
\end{array}
\right. .
\eeq
In the case of bulk superfluid $^3$He-B, since the OTE pairing is absent, the spin susceptibility is given as $\chi = \chi _{\rm N}+\chi^{\rm OP}$, where $\chi^{\rm OP}<0$ suppreses the spin susceptibility. In contrast, the spin susceptibility contributed from the OTE pairs, $\chi^{\rm EP}$, is expected to increase the spin susceptibility, which comes up to $\chi > \chi _{\rm N}$~\cite{higashitaniPRB2014}. As we will discuss below, there is the critical magnetic field beyond which $\hat{\ell}_z$ becomes nonzero and the OTE pair contribute to the surface spin susceptibility.

{\it Surface spin susceptibility in the Ginzburg-Landau regime.}---
To capture the essential part of the relation between the surface spin susceptibility and emergent Cooper pairs, we explicitly solve the quasiclassical Eilenberger equation (\ref{eq:eilen}) within the Ginzburg-Landau approximation. The full analytic solution of the Eilenberger equation (\ref{eq:eilen}) at zero fields is described in Sec.~\ref{sec:exact}. 

In the Ginzburg-Landau regime near $T_{\rm c0}$, we may replace the diagonal component of the quasiclassical operator $g$ to the normal-state propagator $g_{\rm N} = -i\pi {\rm sgn}(\omega _n)$. For simplicity, the pair potential is assumed to be spatially uniform. In addition, we formally expand the anomalous propagator $f$ and the ${\bm d}$-vector ${\bm d}$ in powers of the applied field: $f = f^{(0)}+f^{(1)}+\cdots$ and ${\bm d} = {\bm d}^{(0)} + {\bm d}^{(1)}+\cdots$. We first solve the equation with $H=0$ and then the finite field corrections are obtained, order by order of $(\mu _n H/\Delta _0)$. In the zero field, it is obvious that the spin-singlet pair amplitudes are absent, that is, $f^{(0)}_{0}=0$. We also note that the spatially uniform pair potential is distorted by order $(\mu _{\rm n}H/\Delta _0)^2$ and we neglect ${\bm d}^{(1)}$. The pair potential at zero fields preserves the ${\rm SO}(2)_{L_z+S_z}$ symmetry, which is given in Eq.~(\ref{eq:dvec_slab}). The Cooper pair amplitudes at zero field are obtained by solving the equation for $\tilde{f}^{(0)}_{\parallel}$~\cite{higashitaniJPSJ2014}, 
\beq
v_{\rm F}\hat{k}_z\partial _z \tilde{f}^{(0)}_{\parallel}
= - 2\omega _n \tilde{f}^{(0)}_{\parallel} - 2\pi {\rm sgn}(\omega _n)\Delta _{\parallel}, 
\eeq
and for $\tilde{f}^{(0)}_{z}$, 
\beq
v_{\rm F}\hat{k}_z\partial _z \tilde{f}^{(0)}_{\perp}
= - 2\omega _n \tilde{f}^{(0)}_{\perp} - 2\pi {\rm sgn}(\omega _n)\Delta _{\perp}.
\eeq
Using the specular boundary conditions at $z=0$ and $z=D$, one obtains the ETO pair amplitudes at zero field as
\beq
\tilde{f}^{{\rm EF}(0)}_{\parallel}(\theta _{\bm k},z;\omega _n) = -\pi \frac{\Delta _{\parallel}}{|\omega _n|} \sin\theta _{\bm k}, \\
\tilde{f}^{{\rm EF}(0)}_{\perp}(\theta _{\bm k},z;\omega _n) = -\pi \frac{\Delta _{\perp}\cos\theta _{\bm k}}{|\omega _n|}
\left[
1 - \frac{\cosh[(z-D/2)/\lambda]}{\cosh(D/2\lambda)} 
\right],
\eeq
where we have introduced $\lambda = v_{\rm F}|\cos\theta _{\bm k}|/2|\omega _n|$. The OTE component emerges in the surface region as 
\beq
\tilde{f}^{{\rm OF}(0)}_{\perp}(\theta _{\bm k},z;\omega _n) = -\pi \frac{\Delta _{\perp}|\cos\theta _{\bm k}|}{\omega _n}
\frac{\sinh[(z-D/2)/\lambda]}{\cosh(D/2\lambda)} ,
\eeq
and $f^{{\rm OF}(0)}_{\parallel} = 0$. 

The ESE and OSO pair amplitudes are induced by the linear Zeeman corrections. The field-induced spin-singlet pair amplitudes are governed by the following equation that are obtained from Eq.~(\ref{eq:eilen}),
\beq
iv_{\rm F}\hat{k}_z \partial _z f^{(1)}_0 = -2i\omega _n f^{(1)}_0 - \tilde{\omega}_{\rm L}\hat{\ell}_z \tilde{f}^{(0)}_{\perp}.
\eeq
The magnetic Zeeman term is parameterized by the topological order $\hat{\ell}_z$ and the effective Lamor frequency $\tilde{\omega}_{\rm L}$ is defined as 
$\tilde{\omega}_{\rm L} = \frac{2\mu _{\rm n}H}{1+F^{\rm a}_0}$. Solving the equation shown above, one finds that the ESE Cooper pair amplitude is induced at the surface $z=0$ by the magnetic Zeeman field as
\beq
f^{{\rm EF}(1)}_0(\hat{\bm k},0;\omega _n) = i\frac{\pi}{2} \hat{\ell}_z\frac{\tilde{\omega}_{\rm L}\Delta _{\perp}|\hat{k}_z|}{|\omega _n|^2}
\left[
\tanh\!\left( \frac{D}{2\lambda}\right) + \frac{D}{2\lambda}{\rm sech}^2\!\left( \frac{D}{2\lambda}\right)
\right],
\eeq
while the OSO pair amplitude does not appear at the surface, $f^{{\rm OF}(1)}_0(\hat{\bm k},0;\omega _n) = 0$. It is also found that the intensity of the ESE Cooper pair amplitude in the central region of the system ($z\approx D/2$) exponentially decreases with increasing $D/\lambda$. Therefore, the ESE pair amplitude that are induced by the linear Zeeman corrections is localized in the surface region. 

For $D\gg \lambda$, the OTE and ESE pair amplitudes emergent at the surface are simplified as 
\beq
\tilde{f}^{{\rm OF}(0)}_{\perp}(\theta _{\bm k},0;\omega _n) = \pi \frac{\Delta _{\perp}|\cos\theta _{\bm k}|}{\omega _n},
\eeq
and 
\beq
f^{{\rm EF}(1)}_0(\hat{\bm k},0;\omega _n) 
= i\frac{\pi}{2} \hat{\ell}_z|\hat{k}_z|\frac{\tilde{\omega}_{\rm L}\Delta _{\perp}}{|\omega _n|^2}.
\eeq
Substituting these expressions of OTE and ETO pair amplitudes into Eq.~(\ref{eq:chiEP2}), one obtains the first order correction to the even-parity Cooper pair contribution as 
\beq
\chi^{(1)\rm EP}_{\rm surf} = \frac{7\zeta(3)}{12(1+F^{\rm a}_0)}\left( \frac{\Delta _{\perp}}{\pi T}\right)^2   > 0,
\eeq
where $\zeta(3)$ is the Riemann zeta function. 
This clearly shows that the even-parity Cooper pairs carry the paramagnetic response $\chi^{(1)\rm EP}_{\rm surf} > 0$. Note that the odd-parity Cooper pair contribution $\chi^{(1)\rm OP}_{\rm surf}$ is absent in the Ginzburg-Landau regime. To this end, the surface spin susceptibility in the superfluid $^3$He-B is anomalously enhanced by the coupling of emergent OTE Cooper pairs to the field-induced ESE pair as 
\beq
\chi _{\rm surf} = \chi _{\rm N} + \hat{\ell}^2_z(\hat{\bm n},\varphi)\chi^{(1)\rm EP}_{\rm surf} > \chi _{\rm N}.
\eeq
This implies that although the OTE pair amplitudes always exist in the surface of ETO superconductors and superfluids and yield paramagnetic response, they do not necessarily couple to the applied magnetic field. The topological order $\hat{\ell}_z$ that is associated with the spontaneous breaking of the hidden ${\bm Z}_2$ symmetry determines the contribution of odd-parity Cooper pairs to the surface spin susceptibility.

\section{Surface Andreev bound states in the quasiclassical theory}
\label{sec:exact2}

The quasiclassical transport equation (\ref{eq:eilen}) offers a more tractable way for studying microscopic structures of spatially inhomogeneous superfluids and superconductors, while it always contains an unphysical solution that is not normalizable~\cite{thuneberg}. 
Following Refs.~\cite{eschrigPRB1999} and \cite{eschrigPRB2000}, in this section, we first introduce projection operators that map an arbitrary normal state quasiparticle onto the Nambu space in which the particle-like and hole-like subspaces are coherently coupled through the superconducting pair potential. This projector transfers the quasiclassical transport equation to the so-called Riccati-type equation. The equation contains the nonlinear term that makes it numerically stable. Therefore, the Riccati parameterization of the quasiclassical transport equation has offered a tractable way to study spatially inhomogeneous superconducting and superfluid states~\cite{eschrigPRB1999,nagatoJLTP1993,schopohlPRB1995}. The Riccati equation is also exactly solvable, which gives a self-consistent solution of the quasiclassical transport equation and gap equation. Using the solution, we examine the detailed properties of time-reversal invariant topological superfluids and superconductors, such as Majorana fermions, odd-frequency pairing, and spin current density.

\subsection{Projection operators and Riccati equations}
\label{eq:projection}

Before starting to introduce the projection operators, it is worth mentioning the retarded propagator for spatially uniform superconductors and superfluids, which is obtained from Eq.~(\ref{eq:gbulk}) with the analytic continuation $i\omega _n \rightarrow E+i0_+$ as 
\beq
\underline{g}^{\rm R}_{\rm bulk}(\hat{\bm k},E) = \alpha (\hat{\bm k},E) \underline{\tau}_z 
+ \beta (\hat{\bm k},E)\underline{\Delta}(\hat{\bm k}),
\eeq
where 
\beq
\alpha (\hat{\bm k},E)=-E\beta(\hat{\bm k},E) \nn \\
= -  \frac{\pi E}{\sqrt{|{\bm \Delta}|^2-E^2}}\Theta (|{\bm \Delta}|^2 - E^2)
-i \frac{\pi |E|}{\sqrt{E^2-|{\bm \Delta}|^2}}\Theta (E^2-|{\bm \Delta}|^2),
\eeq
where we introduce $|{\bm \Delta}|^2\!\equiv\! \frac{1}{2}{\rm Tr}[\Delta(\hat{\bm k})\Delta^{\dag}(\hat{\bm k})] \!=\! |\psi _0(\hat{\bm k})|^2 + |{\bm d}(\hat{\bm k})|^2$. The second term in $\alpha$ is associated with the density of states. 

The projection operators $\underline{\mathcal{P}}_{\pm}$ are defined with the quasiclassical propagator $\underline{g}$ as
\beq
\underline{\mathcal{P}}_{\pm}(\hat{\bm k},{\bm r};\omega _n) = \frac{1}{2}\left[
1 \pm \frac{i}{\pi}\underline{g}(\hat{\bm k},{\bm r};\omega _n)
\right].
\eeq
This projector was first introduced by Shelankov~\cite{shelankov2} to directly derive the normalization condition (\ref{eq:norm}) on the quasiclassical propagator. It is obvious that the operators satisfy the following conditions: 
\beq
\underline{\mathcal{P}}_+ + \underline{\mathcal{P}}_- =1, \hspace{3mm}
\underline{\mathcal{P}}_+\underline{\mathcal{P}}_- \!=\! 0, \hspace{3mm}
\underline{\mathcal{P}}^2_{\pm} = \underline{\mathcal{P}}_{\pm}.
\label{eq:P}
\eeq
Using the retarded and advanced propagators $\underline{g}^{\rm R}(E) = \underline{g}(\omega _n\!\rightarrow\!-iE+0_+)$ and $\underline{g}^{\rm A}(E) = \underline{g}(\omega _n\!\rightarrow\! -iE-0_+)$, the operators $\underline{\mathcal{P}}^{\rm R,A}_{\pm}$ are introduced as
\beq
\underline{\mathcal{P}}^{\rm R,A}_{\pm}(\hat{\bm k},{\bm r};E) = \frac{1}{2}\left[
1 \pm \frac{i}{\pi}\underline{g}^{\rm R,A}(\hat{\bm k},{\bm r};E)
\right].
\label{eq:PR}
\eeq
The operators must satisfy the orthonormal conditions in Eqs.~(\ref{eq:P}).

It turns out that the operators $\underline{\mathcal{P}}^{\rm R,A}_{\pm}$ project an arbitrary Nambu spinor onto the particle-like and hole-like subspaces of the superconducting state {\it locally}. To see this, let us start with the propagator in the normal state, $\underline{g}_{\rm N}(\hat{\bm k},{\bm r};\omega _n) = -i\pi{\rm sgn}(\omega _n)\underline{\tau}_z$, which is independent of both $\hat{\bm k}$ and ${\bm r}$. Then, the projection operator in the normal state for $E>0$ is given as
\beq
\underline{\mathcal{P}}^{\rm R}_+(E>0) = \left( 
\begin{array}{cc}
1_{2\times 2} & 0 \\ 0 & 0
\end{array}
\right), \hspace{3mm}
\underline{\mathcal{P}}^{\rm R}_-(E>0) = \left( 
\begin{array}{cc}
0 & 0 \\ 0 & 1_{2\times 2}
\end{array}
\right).
\eeq
This indicates that the operator $\mathcal{P}^{\rm R}_+(E>0)$ projects an arbitrary Nambu spinor onto the particle branch. Likewise, $\mathcal{P}^{\rm R}_-(E>0)$ is the projector onto the hole branch. 

The projectors for superconducting states are derived by substituting the solutions in Eq.~(\ref{eq:gbulk}) into Eq.~(\ref{eq:PR}). Then, the projection operator for the positive energy state $E>|{\bm \Delta}|$ is recast into 
\beq
\underline{\mathcal{P}}^{\rm R}_{\pm} (E) = \frac{1}{\xi (\hat{\bm k},E)}
\left( 
\begin{array}{cc}
\xi (\hat{\bm k},E) \pm |E| & \mp \Delta (\hat{\bm k}) \\ 
\mp \Delta^{\dag}(-\hat{\bm k}) & \xi (\hat{\bm k},E)\mp |E|
\end{array}
\right),
\eeq
where $\xi (\hat{\bm k},E) \!\equiv\! \sqrt{E^2-|{\bm \Delta}|^2} $. For $E\gg \Delta _0$
\beq
{\varphi} _{+}(\hat{\bm k},E) 
\equiv \underline{\mathcal{P}}^{\rm R}_+(E) \left( \begin{array}{c} 1 \\ 0 \\ 0 \\ 0 \end{array}\right)
\approx \left( \begin{array}{c} 1 \\ 0 \\ 0 \\ 0 \end{array}\right)
+ \frac{\Delta _0}{2E} \left( \begin{array}{c} 0 \\ 0 \\ -\hat{k}_x-i\hat{k}_y \\ \hat{k}_z \end{array}\right), 
\label{eq:psi1} \\
{\varphi} _{-}(\hat{\bm k},E) 
\equiv \underline{\mathcal{P}}^{\rm R}_-(E) \left( \begin{array}{c} 0 \\ 0 \\ 1 \\ 0 \end{array}\right)
\approx \left( \begin{array}{c} 0 \\ 0 \\ 1 \\ 0 \end{array}\right)
+ \frac{\Delta _0}{2E} \left( \begin{array}{c} -\hat{k}_x+i\hat{k}_y \\ \hat{k}_z \\ 0 \\ 0 \end{array}\right).
\label{eq:psi2}
\eeq
Hence, it turns out that in superfluid states, $\underline{\mathcal{P}}^{\rm R}_{\pm}(E)$ projects a normal-state quasiparticle with a spin to the Nambu space. 
Equations (\ref{eq:psi1}) and (\ref{eq:psi2}) indicate that the pair potential $\Delta _0$ in the B-phase mixes the hole component $\chi(\hat{\bm k}) \equiv (-\hat{k}_x+i\hat{k}_y, \hat{k}_z)^{\rm T}$ that is the eigenvector of the chiral operator $\hat{\bm k}\cdot{\bm \sigma}$ with the eigenvalue $-1$. 
In Sec.~\ref{sec:exact}, we demonstrate that the projection operators $\underline{\mathcal{P}}^{\rm R}_{\pm}$ are useful for constructing the Bogoliubov spinors for surface bound states from the quasiclassical propagator. 

Following the ansatz in Ref.~\cite{eschrigPRB1999,eschrigPRB2000}, we parameterize the projection operators with complex spin matrices $a(\hat{\bm k},{\bm r};\omega_n)$ and $b(\hat{\bm k},{\bm r};\omega_n)$ as
\beq
\mathcal{P}_{+} = \left( 
\begin{array}{c} 1 \\ -b \end{array}
\right) \left( 1 - ab\right)^{-1}
\left( 1,  a  \right), 
\label{eq:P+} \\
\mathcal{P}_{-} = \left( 
\begin{array}{c} -a \\ 1 \end{array}
\right) \left( 1 - ba\right)^{-1}
\left( b, 1  \right).
\label{eq:P-}
\eeq
Using the identity $a(1+ba) \!=\! (1+ab)a$, one can confirm that the Ansatz for $\mathcal{P}_{\pm}$ in Eqs.~(\ref{eq:P+}) and (\ref{eq:P-}) satisfies the relations in Eq.~(\ref{eq:P}), implying that this parameterization automatically satisfies the normalization condition of $\underline{g}$. This parameterization considerably simplifies the transport equation (\ref{eq:eilen}) to the $2\times 2$ nonlinear differential equations. The equation of motion for the projectors is obtained from the quasiclassical transport equation (\ref{eq:eilen}). Substituting Eqs.~(\ref{eq:P+}) and (\ref{eq:P-}) into the equation of motion for the projectors, one can immediately extract the Riccati-type differential equation for the $2\!\times\! 2$ spin matrix $a \!\equiv\! a(\hat{\bm k},{\bm r}; \omega _n)$,
\beq
i{\bm v}_{\rm F}(\hat{\bm k}) \cdot {\bm \nabla} a + 2i\omega _n a + \Delta - a \Delta^{\dag}(-\hat{\bm k}) a + a \tilde{\nu}^{\prime} - \tilde{\nu} a = 0,
\label{eq:riccati}
\eeq
which is supplemented by initial conditions~\cite{eschrigPRB1999,eschrigPRB2000,nagatoJLTP1993}. The Riccati-type equation (\ref{eq:riccati}) is directly derived from the BdG equation (\ref{eq:bdg}) with the Andreev approximation by parametrizing the ratio of the quasiparticle wavefunction $u/v$ as the Riccati amplitude $a$~\cite{nagai,schopohl}. The equation for the Ricatti amplitudes $b$ is separated from that for $a$, which is obtained by using the symmetry 
\beq
a(\hat{\bm k},{\bm r};\omega _n) = b^{\ast}(-\hat{\bm k},{\bm r};\omega _n).
\label{eq:symmetryab}
\eeq
This relation implies that the quasiclassical propagators obey 
\beq
\underline{\tau}_x \underline{g}^{\ast}(\hat{\bm k},{\bm r};\omega _n)
\underline{\tau}_x = g(-\hat{\bm k},{\bm r};\omega _n),
\eeq
This represents the particle-hole symmetry given in Eq.~(\ref{eq:PHS}). For spatially uniform $\Delta \equiv \Delta(\hat{\bm k})$, the solution of the Riccati equation (\ref{eq:riccati}) is obtained as
\beq
a_{\rm bulk}(\hat{\bm k},\omega _n) =-\frac{\Delta (\hat{\bm k})}{i\omega_n + i\sqrt{\omega^2_n + \frac{1}{2}{\rm Tr}[\Delta(\hat{\bm k})\Delta^{\dag}(\hat{\bm k})]}}, 
\label{eq:abulk} \\
b_{\rm bulk}(\hat{\bm k},\omega _n) = \frac{\Delta^{\ast}(-\hat{\bm k})}{i\omega_n + i\sqrt{\omega^2_n + \frac{1}{2}{\rm Tr}[\Delta(\hat{\bm k})\Delta^{\dag}(\hat{\bm k})]}}.
\label{eq:bbulk}
\eeq 

Owing to the nonlinear term $a \Delta^{\dag} a$, Eq.~(\ref{eq:riccati}) becomes numerically stable along a trajectory in the direction of $\hat{\bm k}$. Likewise, the symmetry in Eq.~(\ref{eq:symmetryab}) indicates that the another Riccati amplitude $b(\hat{\bm k}) = a^{\ast}(-\hat{\bm k})$ is stable along trajectories in the direction of $-\hat{\bm k}$. The details on numerical calculation of Eq.~(\ref{eq:riccati}) in a restricted geometry are available in Refs.~\cite{mizushimaPRB2012,vorontsovPRB2003,tsutsumiPRB2011}.

The quasiclassical propagator is now reconstructed from the $2\!\times\! 2$ Riccati amplitudes $a$ and $b$ as
\beq
\underline{g}(\hat{\bm k},{\bm r}; \omega _n) = -i\pi \underline{N}
\left(
\begin{array}{cc}
1 + ab & 2a \\ -2b & -1 - ba
\end{array}
\right),
\label{eq:gfromab}
\eeq
where 
$\underline{N} = {\rm diag}[ (1 - ab)^{-1}, (1 - ba)^{-1}]$. Therefore, the Riccati amplitude $a$ is related to the off-diagonal propagator as 
\beq
a = - \frac{1}{i\pi}\left( 1 + \frac{i}{\pi}g \right)^{-1} f,
\eeq
which implies that $a$ is obtained by projecting the off-diagonal propagator to the particle-like subsector. In the same manner, the other Riccati amplitude $b$ is related to the hole-like projection of the off-diagonal propagator
\beq
b = \frac{1}{i\pi}\left(1 - \frac{i}{\pi}g\right)^{-1} \bar{f}.
\eeq

\subsection{Exact solutions for superfluid $^3$He-B}
\label{sec:exact}

Here, we analytically solve the Riccati-type transport equation for the superfluid $^3$He-B in the presence of a specular surface at $z=0$, where the B-phase is supposed to occupy the region of $z>0$. We also suppose the uniformity of the surface in the $x$-$y$ plane. Then, the specular surface imposes the following boundary condition on the quasiclassical propagator $\underline{g}(\hat{\bm k},{\bm r};\omega _n)=\underline{g}(\hat{\bm k},z;\omega _n)$, where 
the propagator along the trajectory $\hat{\bm k}$ matches that along the specularly scattered trajectory $\underline{\bm k}$
at the surface $z=0$:
\beq
\underline{g}(\hat{\bm k},z=0;\omega _n) = \underline{g}(\underline{\hat{\bm k}},z=0;\omega _n).
\eeq
The specularly scattered momentum $\underline{\hat{\bm k}}$ is defined as $\underline{\hat{\bm k}} \!=\! \hat{\bm k}-2\hat{\bm z}(\hat{\bm z}\cdot\hat{\bm k})$, where $\hat{\bm z}$ is a unit vector perpendicular to the surface. We also impose the following boundary condition on the propagator, 
\beq
\underline{g}(\hat{\bm k},z=\infty;\omega _n) = \underline{g}_{\rm bulk}(\hat{\bm k},\omega _n) ,
\eeq
where the propagator in the bulk B-phase, $\underline{g}_{\rm bulk}(\hat{\bm k},\omega _n)$, is given in Eq.~(\ref{eq:gbulk}). Then, the Riccati amplitudes $a$ and $b$ are matched at the surface as
\beq
a(\hat{\bm k},z=0;\omega _n) = a(\underline{\hat{\bm k}},z=0;\omega _n), \label{eq:bca1} \\
b(\hat{\bm k},z=0;\omega _n) = b(\underline{\hat{\bm k}},z=0;\omega _n). \label{eq:bcb1}
\eeq
The boundary condition at $z\rightarrow \infty$ is given with Eqs.~(\ref{eq:abulk}) and (\ref{eq:bbulk}) as 
\beq
a(\hat{\bm k},z=\infty;\omega _n) = a_{\rm bulk}(\hat{\bm k},\omega _n), \label{eq:bca2} \\
b(\hat{\bm k},z=\infty;\omega _n) = b_{\rm bulk}(\hat{\bm k},\omega _n). \label{eq:bcb2}
\eeq

Now, we employ the unitary transformation introduced in Eq.~(\ref{eq:unitaryB}), which reduces the B-phase pair potential to the chiral pairing form in Eq.~(\ref{eq:deltapm}). For simplicity, we neglect a magnetic field and Fermi liquid corrections, corresponding to Eq.~(\ref{eq:riccati}) with $\tilde{\nu} \!=\! \tilde{\nu}^{\prime} \!=\! 0$. Then, this unitary transformation considerably simplifies Eq.~(\ref{eq:riccati}) to 
\beq
v_{\rm F}\cos\theta _{\bm k} \frac{d}{dz} a_{\pm} = - 2\omega _n a_{\pm} +i \Delta _{\pm} 
+ i \Delta^{\ast}_{\pm} a^2_{\pm},
\label{eq:riccati2}
\eeq
where the $2\times 2$ Riccati amplitude $a$ is diagonalized by using the unitary transformation in Eq.~(\ref{eq:unitaryB}) to with two scalar functions $a_{\pm}$, 
\beq
S^{\dag}_{\phi _{\bm k}}U^{\dag}(\hat{\bm n},\varphi) a(\hat{\bm k},z;\omega _n) 
U^{\ast}(\hat{\bm n},\varphi)S^{\ast}_{\phi _{\bm k}} =
\left( \begin{array}{cc}a_+ & 0  \\ 0 & a_- \end{array}\right).
\eeq 
We here suppose the shperical Fermi surface, ${\bm v}_{\rm F}(\hat{\bm k}) \!=\! v_{\rm F}\hat{\bm k}$. Since Eq.~(\ref{eq:deltapm}) obeys $\Delta _- \!=\! - \Delta^{\ast}_+$, one finds $a_-(\hat{\bm k},z;\omega _n) \!=\! a^{\ast}_+(\hat{\bm k},z;\omega _n)$. 

The exact solution of the Riccati equation (\ref{eq:riccati2}) was first obtained in Ref.~\cite{schopohl} for the pair potential ansatz, $\Delta (z) = \tanh (z/\xi _0)$. This solution was developed in Ref.~\cite{tsutsumiJPSJ2012} to the more generic form of the pair potential, 
\beq
\Delta _{\pm}(\theta _{\bm k},z) = \Delta _0 \left[
\pm\cos\theta _{\bm k}\tanh\left( \frac{z}{\xi _0}\right) + i\sin \theta _{\bm k}
\right].
\label{eq:deltap}
\eeq
In particular, it is demonstrated in Ref.~\cite{tsutsumiJPSJ2012} that the ansatz in Eq.~(\ref{eq:deltap}) is a self-consistent solution in the weak coupling limit.

The solution of the Riccati amplitude is obtained as
\beq
a_{+}(z,\theta,\omega_n)=i\frac{\omega_n-\sqrt{\omega_n^2+\Delta _0^2}+\Delta _+(\theta _{\bm k},z)}
{\omega_n+\sqrt{\omega_n^2+\Delta_0^2}-\Delta^{\ast}_+(\theta _{\bm k},z)},
\label{RiccatiB}
\eeq
which satisfies the boundary conditions at $z=0$, Eqs.~(\ref{eq:bca1}) and (\ref{eq:bcb1}), and at $z=\infty$, Eqs.~(\ref{eq:bca2}) and (\ref{eq:bcb2}).
The self-consistent quasiclassical propagator is obtained by the unitary transformation of this solution as $\underline{g}=\underline{\mathcal{U}}(\hat{\bm n},\varphi)\mathcal{S}_{\phi _{\bm k}}\underline{\tilde{g}}\mathcal{S}^{\dag}_{\phi _{\bm k}}\underline{\mathcal{U}}^{\dag}(\hat{\bm n},\varphi)$, where $\underline{\tilde{g}}$ is obtained from Eq.~(\ref{RiccatiB}) with Eq.~(\ref{eq:gfromab}) and $b(\hat{\bm k},z;\omega _n)=a^{\ast}(-\hat{\bm k},z;\omega _n)$. 
Then, the diagonal component of the quasiclassical propagator for $^3$He-B in zero fields is 
\beq
g_0(\hat{\bm k},z;\omega _n) = - \frac{i\pi\omega _n}{\sqrt{\omega^2_n+\Delta^2_0}}\left[
1 + \frac{1}{2}\frac{\Delta^2_0\cos^2\theta _{\bm k}}{\omega^2_n+E^2_0(\hat{\bm k}_{\parallel})}
{\rm sech}^2\!\left( \frac{z}{\xi _0}\right)
\right], \\
\tilde{g}_{\parallel}(\hat{\bm k},z;\omega _n) = -\frac{\pi}{2\sqrt{\omega^2_n+\Delta^2_0}} \frac{\Delta^3_0\sin\theta _{\bm k}\cos^2\theta _{\bm k}}{\omega^2_n+E^2_0(\hat{\bm k}_{\parallel})}{\rm sech}^2\!\left( \frac{z}{\xi _0}\right), 
\label{eq:gvec}
\eeq
and $\tilde{g}_z(\hat{\bm k},z;\omega _n) = 0$. The spin part of the quasiclassical propagator, ${\bm g}$, is obtained from $\tilde{g}_{\parallel}$ and $\tilde{g}_z$ as 
\beq
g_{\mu}(\hat{\bm k},z;\omega _n) = R_{\mu \nu}(\hat{\bm n},\varphi) \tilde{g}_{\nu}(\hat{\bm k},z;\omega _n),
\eeq
where $\tilde{g}_{\nu}$ is defined as 
$(\tilde{g}_x,\tilde{g}_y,\tilde{g}_z) 
= (\tilde{g}_{\parallel}\sin\phi _{\bm k},-\tilde{g}_{\parallel}\cos\phi _{\bm k},\tilde{g}_z)$. The quasiparticle propagator derived here is deviated from that of bulk $^3$He-B in the surface region within $\xi _0$.

The retarded propagator $g^{\rm R}_0(E) \!=\! g^{\rm R}_0(\omega _n \rightarrow -i E+ 0_+)$ has poles at the dispersion of the surface Andreev bound states, $E_{0}(\hat{\bm k}_{\parallel})$, described in Eq.~(\ref{eq:E0a}). Using the exact solution on the quasiclassical propagator, we can calculate the local density of states for the bound state $|E|<\Delta_0$ as
\beq
\mathcal{N}(z,E)=-\frac{\mathcal{N}_{\rm F}}{\pi}\left\langle{\rm Im}g^{\rm R}_0(\hat{\bm k},z;E)\right\rangle _{\hat{\bm k}}
=\frac{\pi}{4}\mathcal{N}_{\rm F}\frac{|E|}{\Delta _0}\ {\rm sech}^2\!\left(\frac{z}{\xi _0}\right).
\label{LDOSMJB}
\eeq
The local density of states for the continuum state $|E|>\Delta _0$ is given as
\beq
\mathcal{N}(z,E) = \mathcal{N}_{\rm F}\left[\frac{|E|}{\xi (E)}
- \frac{1}{2}\!\left(\frac{|E|}{\xi (E)}-\frac{|E|}{\Delta_0}\tan^{-1}\!\frac{\Delta_0}{\xi (E)}\right)\!
\ {\rm sech}^2\!\left(\frac{z}{\xi_0}\right)\right], \nn \\
\label{LDOScontB}
\eeq
where $\xi(E)\equiv\sqrt{E^2-\Delta^2_0}$. The local density of states for the continuum state is also deformed from that in the bulk B-phase, $\mathcal{N}_{\rm bulk}(E) = \mathcal{N}_{\rm F}|E|/\sqrt{E^2-\Delta^2_0}$. This is because the density of states in the quasiclassical theory holds the sum rule $\frac{1}{E_{\rm c}}\int^{E_{\rm c}}_{-E_{\rm c}}\mathcal{N}(z,E)dE = \mathcal{N}_{\rm F}$ for $E_{\rm c}\gg \Delta _0$. The sum rule implies that the existence of the surface Majorana cone within $|E|<\Delta _0$ alters the density of states of the continuum states with $|E|>\Delta _0$ in the surface region. We will show below that the deviation of the surface density of states of the continuum states affects the spin current density. 

\begin{figure}
\begin{center}
\includegraphics[width=60mm]{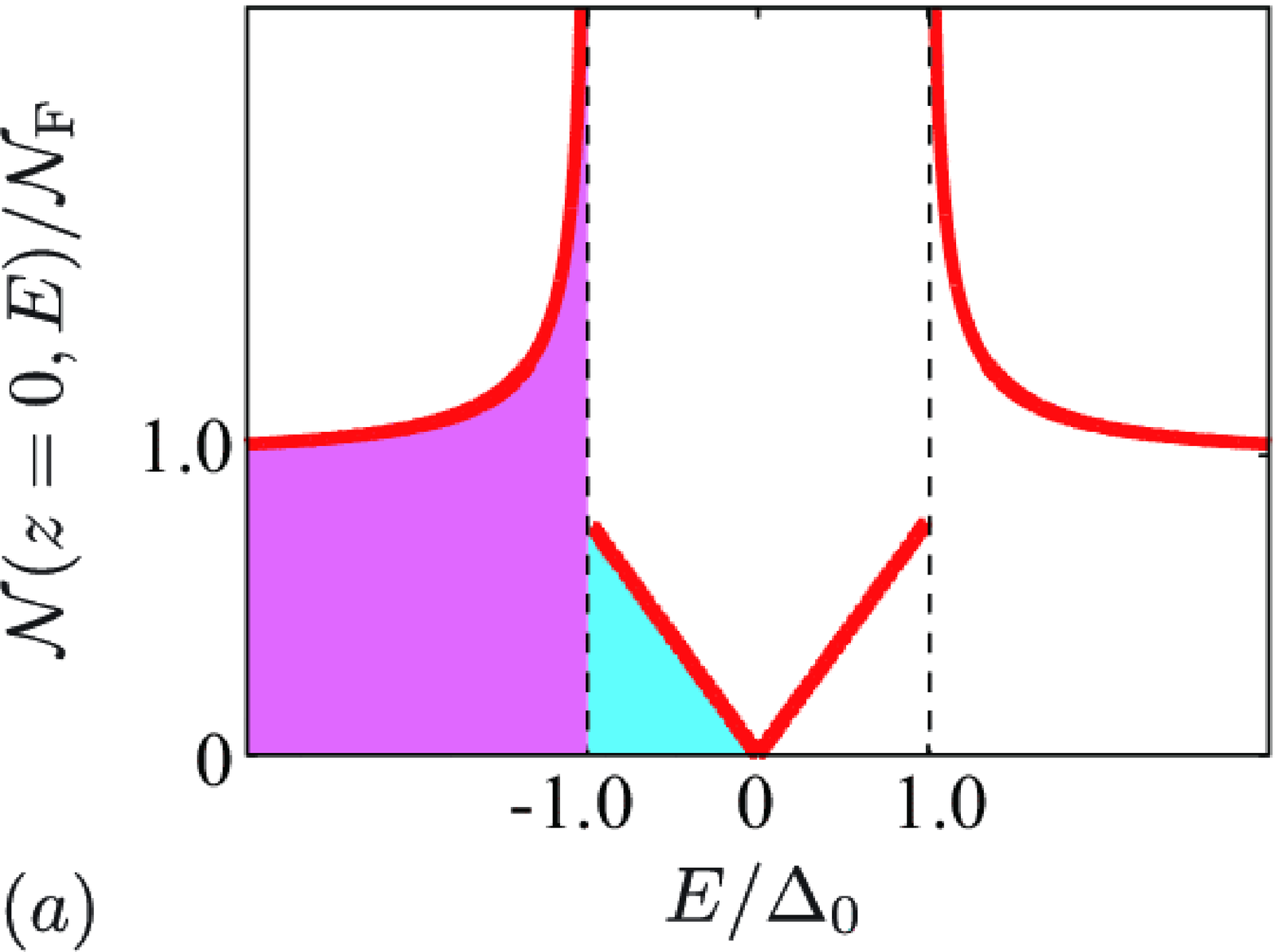}
\includegraphics[width=60mm]{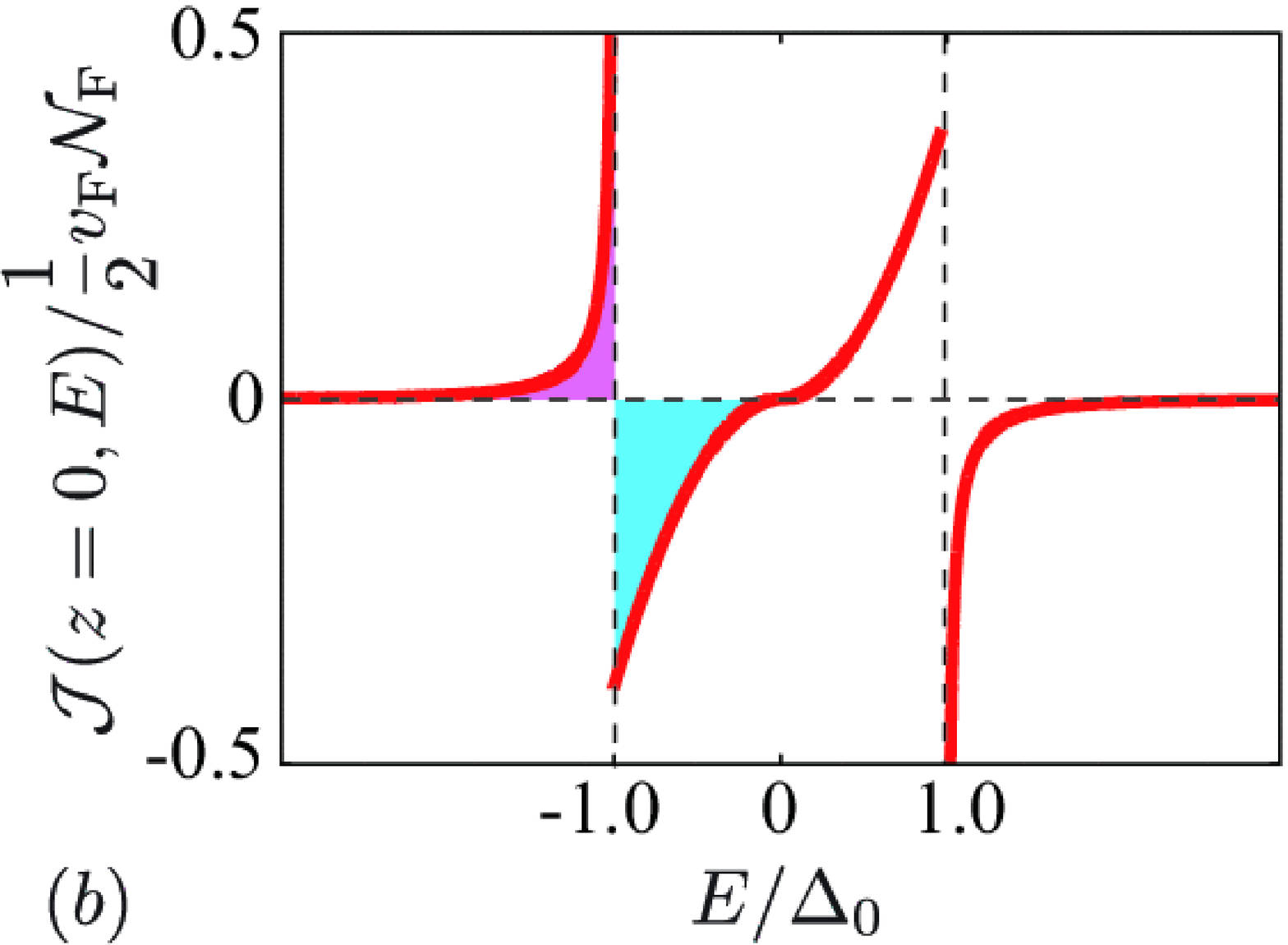}
\end{center}
\caption{
Local density of states $\mathcal{N}(z,E)$ (a) and spin current function (b) at the surface $z=0$ in $^3$He-B.
At a zero temperature, quasiparticles fill the colored states in (a).
The spin currents from the bound and continuum states are derived 
by integrating the blue and pink regions in (b), respectively.
Figures adapted from Ref.~\cite{tsutsumiJPSJ2012}.
}
\label{energyB}
\end{figure}

The local density of states at the surface $z=0$ is shown in Fig.~\ref{energyB}(a). Since the surface Majorana fermions have a two-dimensional relativistic dispersion, $E_0(k_x,k_y) \propto \sqrt{k^2_x+k^2_y}$, the surface density of states for the bound state yields a linear energy dependence with a slope $(\pi/4)\mathcal{N}_{\rm F}$. The linear dependence is also obtained by numerical calculation~\cite{mizushimaPRB2012,buchholtzPRB1981,tsutsumiPRB2011,nagatoJLTP1998}. The linear behavior of the low-energy density of states is responsible for the power-law behavior of the temperature-dependence of the specific heat in $^3$He-B confined in a slab geometry~\cite{mizushimaJLTP2011}, in contrary to the exponential behavior expected in the bulk B-phase. The large heat capacity in extremely low temperatures has recently been reported in Ref.~\cite{bunkov}, where the $10\%$ deviation from the bulk superfluid $^3$He heat capacity at $135 \mu {\rm K}$ has been observed. This deviation is in good agreement with the theoretical value of the heat capacity originating from the gapless surface states without fitting parameters in low temperatures. 

\subsection{Odd-frequency pairing and Majorana Ising spins}
\label{sec:oddising}

Here, we illustrate the odd-frequency pairing emergent in $^3$He-B and the identity between the odd-frequency pairing and surface density of states. We start with the exact solution of Eq.~(\ref{RiccatiB}) for the pair amplitudes of $^3$He-B at zero fields,
\beq
f_{\mu}(\hat{\bm k},z;\omega _n) = R_{\mu \nu}(\hat{\bm n},\varphi) \tilde{f}_{\nu}(\hat{\bm k},z;\omega _n),
\eeq
where $\tilde{f}_{\nu}$ is defined as 
$(\tilde{f}_x,\tilde{f}_y,\tilde{f}_z) = (\tilde{f}_{\parallel}\cos\phi _{\bm k},\tilde{f}_{\parallel}\sin\phi _{\bm k},\tilde{f}_z)$ and the each component is obtained with the exact solution of the Riccati equation as 
\beq
\tilde{f}_{\parallel}(\hat{\bm k},z;\omega _n) = \pi \frac{\Delta _0\sin\theta _{\bm k}}{\sqrt{\omega^2_n+\Delta^2_0}}
\left[
1 + \frac{1}{2} \frac{\Delta^2_0\cos^2\theta _{\bm k}}{\omega^2_n + E^2_0(\hat{\bm k}_{\parallel})}{\rm sech}^2\!\left( \frac{z}{\xi _0}\right)
\right], \label{eq:fpara} \\
\tilde{f}_z(\hat{\bm k},z;\omega _n) = \pi \frac{\Delta _0\cos\theta _{\bm k}}{\sqrt{\omega^2_n+\Delta^2_0}}
\left[ 
\tanh\!\left(\frac{z}{\xi}\right)
- \frac{1}{2}\frac{\omega _n \Delta _0\cos\theta _{\bm k}}{\omega^2_n+E^2_0(\hat{\bm k}_{\parallel})}
{\rm sech}^2\!\left( \frac{z}{\xi _0}\right)
\right]. \nn \\
\label{eq:fz}
\eeq

Using the analytic solution of the Cooper pair amplitude, we can reconstruct the ${\bm d}$-vector through the gap equation (\ref{eq:gap_qct}). We here suppose the simple $p$-wave interaction $V^{cd}_{ab}(\hat{\bm k}\cdot{\bm k}^{\prime}) = 3\Lambda \hat{\bm k}\cdot\hat{\bm k}^{\prime}$, where the coupling constant has the relation, $\Lambda^{-1} = \ln\frac{2\omega _{\rm c}}{\Delta _{0}}$. Then, the ${\bm d}$-vector at zero temperatures is calculated as~\cite{tsutsumiJPSJ2012}
\beq
d_x(\hat{\bm k},z) = \Delta _0 \hat{k}_x \left[ 
1 + \frac{\Lambda}{6}{\rm sech}^2\left( \frac{z}{\xi _0}\right)
\right], \\
d_y(\hat{\bm k},z) = \Delta _0 \hat{k}_y \left[ 
1 + \frac{\Lambda}{6}{\rm sech}^2\left( \frac{z}{\xi _0}\right)
\right], \\
d_z(\hat{\bm k},z) = \Delta _0 \hat{k}_z \tanh\left( \frac{z}{\xi _0}\right).
\eeq
This reduces to the pair potential ansatz in Eq.~(\ref{eq:deltap}). Therefore, the analytic solution of the propagator self-consistently satisfies the gap equation.

As shown in Eqs.~(\ref{eq:psi1}) and (\ref{eq:psi2}), in superfluid states, the projectors $\underline{\mathcal{P}}^{\rm R}_+(E)$ project a normal state with $\uparrow$ spin onto a quasiparticle with the energy $E$ in the Nambu space,
\beq
{\bm \varphi}_E = \underline{\mathcal{P}}^{\rm R}_+(E)
\left( \begin{array}{c} 1 \\ 0 \\ 0 \\ 0 \end{array}
\right) = \frac{1}{\pi}
\left( 
\begin{array}{c}
\pi + ig^{\rm R}_0 \\ 
e^{i\phi _{\bm k}}g^{\rm R}_{\parallel} \\
ie^{i\phi _{\bm k}}\bar{f}^{\rm R}_{\parallel} \\
-i\bar{f}^{\rm R}_z
\end{array}
\right).
\eeq
The particle-hole subspaces are coherently coupled through the anomalous part of the propagator, $f_{\mu}$. In addition, the chiral component is induced through $g_{\parallel}$ in the case of $^3$He-B. Wu and Sauls~\cite{wuPRB2013} demonstrated that the wavefunction of surface Majorana fermions in Eq.~(\ref{eq:varphi1}) can be extracted from this quasiclassical projectors. We can construct the Majorana Ising spins also from the exact solution on two branches $E_0(\hat{\bm k}_{\parallel})=\pm\Delta_0\sin\theta_{\bm k}$. By integrating over an infinitesimal bandwidth around the divergent bound state $[(-iE+0_+)^2+E_0^2]^{-1}\sim\pm\frac{i\pi}{2E_0}\delta(E\mp E_0)$,
\beq
{\bm \varphi}^{\pm}(\hat{\bm k}_{\parallel},z) = u(\theta_{\bm k},z)\mathcal{U}(\hat{\bm n},\varphi)
\left[e^{-i\phi_{\bm k}/2}{\bm \Phi}_+\mp e^{i\phi_{\bm k}/2}{\bm \Phi}_-\right],
\eeq
where the amplitude is
\beq
u(\theta_{\bm k},z)=\frac{\pi}{4}\Delta_0\cos\theta_{\bm k}{\rm sech}^2\left(\frac{z}{\xi_0}\right),
\eeq
and the spinors, ${\bm \Phi}_{\pm}$, are given by ${\bm \Phi}_+=(1,0,0,-i)^{\rm T}$ and ${\bm \Phi}_-=(0,i,1,0)^{\rm T}$. This is consistent with the wavefunction of the surface bound states in Eq.~(\ref{eq:varphi1}) obtained by the explicit calculation of the Andreev (BdG) equation.

The analytic solution for the off-diagonal propagator indicates that the odd-frequency pair amplitude is strongly constraint
\beq
\tilde{f}^{\rm OF}_{x} = \tilde{f}^{\rm OF}_y = 0, \hspace{3mm}
\tilde{f}^{\rm OF}_z \ne 0.
\eeq
This results in Eqs.~(\ref{eq:fEF}) and (\ref{eq:fOF}). In this solution with a specular surface, the odd-frequency pair amplitude at the surface is equivalent to the momentum resolved surface density of states $\mathcal{N}(\hat{\bm k},z;\omega _n)$ for the bound states $|E|<\Delta _0$,
\beq
\mathcal{N}(\hat{\bm k},z;E) = \frac{1}{\pi}\left| {\rm Re}{\bm f}^{{\rm OF}}(\hat{\bm k},z;E)\right|.
\label{eq:identity}
\eeq 
This relation reveals the identity between surface Majorana fermions and odd-frequency Cooper pair correlation. We would like to emphasize that the identity (\ref{eq:identity}) can be exactly held only for the exact zero energy $E=0$~\cite{higashitaniPRB2012}, while the identity for the {\it whole} spectrum of the surface Majorana cone is attributed to the pair potential ansatz in Eq.~(\ref{eq:deltap}) that is the self-consistent solution only in the weak-coupling limit. We also note that the identity between Majorana zero modes and odd-frequency Cooper pairs is clarified in a superconducting nanowire~\cite{asanoPRB2013} and chiral $p$-wave superconductor with a vortex in the quantum limit~\cite{dainoPRB2012}, which indicates the identity is held beyond the quasiclassical theory. 
Hence, the topologically protected surface states have multi-faceted properties, such as the Majorana fermion and odd-frequency pair. The physical consequence in the former is the Ising anisotropy of the surface spin susceptibility, while the latter is responsible for anomalous proximity effect.

\subsection{Surface spin current}

Using the analytic solution presented in the previous subsection, we examine the surface spin current in $^3$He-B. In the same manner with the argument on spin susceptibility in Sec.~\ref{sec:odd}, we first summarize the direct relation between the spin current and odd frequency pair $f^{\rm OF}$ in time-reversal invariant spin-triplet superfluids and superconductors. The spin current density $J^{\rm spin}_{\mu\nu}$ is defined with the quasiclassical propagator as
\beq
J^{\rm spin}_{\mu\nu}({\bm r}) = \frac{\mathcal{N}_{\rm F}}{2} {\rm Re} \left\langle 
v_{\nu}(\hat{\bm k})
g_{\mu}(\hat{\bm k},{\bm r};\omega _n)
\right\rangle _{\hat{\bm k},n}.
\label{eq:spincurrent}
\eeq
By using the symmetric properties of the quasiclassical propagator in Eqs.~(\ref{eq:sym1}), (\ref{eq:sym2}), and (\ref{eq:trs2}), the spin current for spin-triplet superfluids and superconductors is recast into~\cite{higashitaniJPSJ2014}
\beq
J^{\rm spin}_{\mu\nu}({\bm r}) = \frac{\mathcal{N}_{\rm F}}{4} {\rm Re} \left\langle 
v_{\nu}(\hat{\bm k})
\frac{i({\bm f}\times \bar{\bm f})_{\mu}}{g_0}
\right\rangle _{\hat{\bm k},n}.
\eeq
We here assume ${\bm v}(\hat{\bm k}) = - {\bm v}(-\hat{\bm k})$. For time-reversal invariant superfluids and superconductors, the propagator $g_0$ obeys the relation in Eq.~(\ref{eq:g0}). By using this, the spin current density at zero fields is expressed in terms of ETO pair amplitude $\tilde{\bm f}^{\rm EF}$ and OTE pair amplitude $\tilde{\bm f}^{\rm OF}$ as
\beq
\tilde{J}^{\rm spin}_{\mu\nu} ({\bm r})= R_{\mu\eta}(\hat{\bm n},\varphi)\tilde{J}^{\rm spin}_{\eta\nu}({\bm r})
\eeq
where $\tilde{J}^{\rm spin}_{\mu\nu}$ is the spin current density for $\hat{\bm n}=\hat{\bm z}$ and $\varphi = 0$. 
\beq
\tilde{J}^{\rm spin}_{\mu\nu}({\bm r}) = \frac{\mathcal{N}_{\rm F}}{2} {\rm Im}\left\langle 
v_{\nu}(\hat{\bm k})
\frac{(\tilde{\bm f}^{\rm EF}\times \tilde{\bm f}^{\rm OF})_{\mu}}{g_0}
\right\rangle _{\hat{\bm k},n}. 
\label{eq:spincurrent2}
\eeq
Hence, only $\tilde{J}^{\rm spin}_{xy}$ and $\tilde{J}^{\rm spin}_{yx}$ components remain nonzero,
\beq
\tilde{J}^{\rm spin}_{xy}(z,E) = - \tilde{J}^{\rm spin}_{yx}(z,E) \equiv \mathcal{J}(z,E),
\eeq
and otherwise $\tilde{J}_{\mu\nu}(z,E)=0$. Equation (\ref{eq:spincurrent2}) states that the surface bound states in time-reversal invariant superfluids and superconductors carry the surface spin current.

Substituting the exact solution shown in Eq.~(\ref{eq:gvec}) into Eq.~(\ref{eq:spincurrent}), we can calculate the spin current spectral function $\mathcal{J}(z,E)$
\beq
 \mathcal{J}(z,E)=\alpha\frac{E^2}{\Delta^2_0} 
 {\rm sech}^2\!\left(\frac{z}{\xi_0}\right),
 \label{eq:jsbound}
\eeq
for the bound state $|E|<\Delta_0$ and
\beq
\mathcal{J}(z,E)=-\frac{2}{3\pi}\alpha\left[\frac{\xi(E)}{\Delta _0}+\frac{2E^2}{\Delta _0\xi(E)}
-\frac{3E^2}{\Delta^2_0}\tan^{-1}\!\frac{\Delta _0}{\xi(E)}\right]\ {\rm sech}^2\!\left(\frac{z}{\xi _0}\right), \nn \\
\label{eq:jscont}
\eeq
for the continuum state $|E|>\Delta_0$, where $\alpha \equiv {\rm sgn}(E) \pi v_{\rm F}\mathcal{N}_{\rm F}/16$. 
This spin current function on the surface, $\mathcal{J}(z=0,E)$, is displayed in Fig.~\ref{energyB}(b),
where the contribution from surface bound states with $|E|\le \Delta _0$ is quadratic on $E$.
The continuum states with $|E|>\Delta _0$ also carry the finite spin current, $\mathcal{J}(z=0,E)$, whose sign is opposite to that from the surface bound states. The nonzero contribution of the continuum states originates in the fact that the surface density of states with $|E|>\Delta _0$ is deviated from that in the bulk $^3$He-B by the existence of low-lying bound states. The asymptotic behavior of the surface spin current function, $\mathcal{J}(z=0,E)$, at $|E/\Delta _0| \rightarrow \infty$ is estimated as
\beq
\mathcal{J}(z=0,E) \approx -\frac{v_{\rm F}\mathcal{N}_{\rm F}}{60}\left(\frac{\Delta _0}{E}\right)^3,
\eeq
implying that the contribution of higher-energy quasiparticles decreases with the same power law $\sim E^{-3}$ as the mass current in $^3$He-A~\cite{tsutsumiJPSJ2012}. This power law behavior indicates that higher-energy quasiparticles carrying the surface spin current are distributed neither in the narrow energy shell around the Fermi level nor up to the bottom of the Fermi sea. The spin (mass) current in $^3$He-B ($^3$He-A) is comprised of the wide range of energy eigenstates of the BdG Hamiltonian.

The total spin current flowing on the surface of $^3$He-B is defined by integrating the spectral function $J_{\rm spin}(T)\equiv \int^{\infty}_0 dz\int^{\infty}_{-\infty}dE\mathcal{J}(z,E)f(E,T)$. Substituting Eqs.~(\ref{eq:jsbound}) and (\ref{eq:jscont}) and integrating over $z$ and $E$, one obtains the total amount of the surface spin current at zero temperatures as 
\beq
J_{\rm spin}(0) = J_{\rm spin}^{\rm bound}+J_{\rm spin}^{\rm cont}
=-\frac{\kappa }{2\pi }\frac{n\hbar}{6}.
\label{Js}
\eeq
The quantum of circulation $\kappa=h/2m$ emerges from the ratio of coefficients between the spin and mass currents~\cite{tsutsumiJPSJ2012}, where $h$ is the Plank constant. The total spin current in Eq.~(\ref{Js}) is separated to the contributions carried by the surface Majorana cone and continuum states, $J^{\rm bound}_{\rm spin}$ and $J_{\rm spin}^{\rm cont}$, which can be estimated as   
\beq
J_{\rm spin}^{\rm bound} 
=-\frac{\pi }{8}\frac{\kappa}{2\pi }n\hbar,
\label{JsMJ} \\
J_{\rm spin}^{\rm cont}
=\frac{\pi }{8}\left(1-\frac{4}{3\pi }\right)\frac{\kappa }{2\pi }n\hbar.
\label{Jscont}
\eeq
This indicates ${\rm sgn}(J_{\rm spin}^{\rm bound}) = - {\rm sgn}(J_{\rm spin}^{\rm cont})$, which implies that the surface spin current carried by the surface Majorana cone with $|E|<\Delta _0$ flows in the opposite direction of that carried by continuum states with $|E| > \Delta _0$. 

\begin{figure}
\begin{center}
\includegraphics[width=60mm]{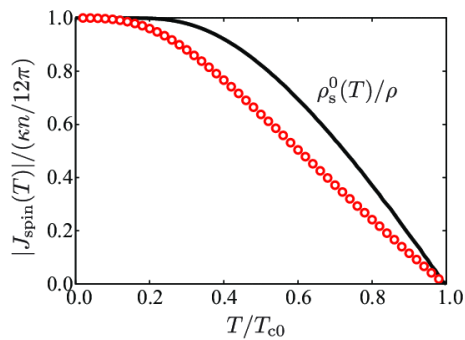}
\end{center}
\caption{
Temperature dependence of total spin current $J_{\rm spin}(T)$ (circles) and the superfluid density $\rho^0_{\rm s}(T)$ (solid line) in $^3$He-B. Figure adapted from Ref.~\cite{tsutsumiJPSJ2012}.
}
\label{temperatureB}
\end{figure}

Finally, we here mention the temperature dependence of the total spin current $J_{\rm spin}(T)$, which is calculated as
\beq
J_{\rm spin}(T) &=& \frac{2}{3}J_{\rm spin}(0)\frac{\pi T}{\Delta_0}
\sum_{n}\left[\frac{\sqrt{\omega_n^2+\Delta_0^2}}{\Delta_0}\right. \nn \\
&& \left.+2\frac{\omega_n^2}{\Delta_0\sqrt{\omega_n^2+\Delta_0^2}}
+3\frac{\omega_n^2}{\Delta_0^2}\ln\frac{\sqrt{\omega_n^2+\Delta_0^2}-\Delta_0}{|\omega_n|}\right].
\label{JsT}
\eeq
The temperature dependence of $J_{\rm spin}(T)$ is plotted in Fig.~\ref{temperatureB}
compared with the superfluid density $\rho_{\rm s}^0(T)$ (the solid curve).
It is seen that the temperature dependence of the net spin current is different from that of $\rho_{\rm s}^0$. This is because the low-temperature depletion of the total spin current is attributed to low-lying surface bound states.
There is no other low energy excitation. 
Expanding Eq.~(\ref{JsT}), one can derive the low-temperature behavior of the total spin current as~\cite{tsutsumiJPSJ2012,wuPRB2013}
\beq
J_{\rm spin}(T)=J_{\rm spin}(0)\left[1-C\left(\frac{\pi T}{\Delta _0}\right)^3+\mathcal{O}\left(\frac{\pi T}{\Delta _0}\right)^4\right],
\eeq
where the coefficient $C$ is fixed within $3/5\le C\le 1$ by the Euler-Maclaurin formula using up to the fourth Bernoulli number.
This $T^3$-power behavior of the depletion originates from the excitations of surface bound states that has a quadratic energy dependence in $\mathcal{J}(z,E)$ as shown in Fig.~\ref{energyB}(b).
The observation of the depletion could establish the existence of low-lying surface bound states.

Note that owing to the spontaneous breaking of the time-reversal symmetry, $^3$He-A is accompanied by the net mass current along the edge, which is carried by the gapless edge states. In the disk system with a radius much larger than the coherence length, angular momentum by the total surface mass current is $N\hbar/2$ at the zero temperature~\cite{stone} which corresponds to macroscopic intrinsic angular momentum by Cooper pairs~\cite{ishikawaPTP1977}, where $N$ is the number of $^3$He atoms in the system.
The total mass current decreases as $T^2$-power in low temperatures~\cite{tsutsumiJPSJ2012,saulsPRB2011}, which accidentally corresponds to the longitudinal superfluid density $\rho_{\rm sl}^0$~\cite{crossJLTP1975}.
The thermal depletion reflects the linear dispersion of the surface Majorana bound state that is flat to the nodal direction.


The surface current depletion of $T^3$-power in $^3$He-B and that of $T^2$-power in $^3$He-A are sensitive to thermal excitations in the surface Majorana bound state.
Detection of the spin current will be more difficult than that of the mass current because we have to separate spin states for the detection of the spin current.
Wu and Sauls~\cite{wuPRB2013} have pointed out that the surface Mojorana bound state in $^3$He-B can be observed by the thermal depletion of mass current.
Superfluid flow applied to $^3$He-B in a narrow channel induces mass current on surfaces of the channel owing to the time-reversal symmetry breaking.
The mass current decreases as $T^3$-power in low temperatures, which reflects the surface Majorana cone in $^3$He-B.



\section{Topological phase diagram in superfluid $^3$He}
\label{sec:numerical2}

In this section, we present the superfluid phase diagram of $^3$He in a restricted geometry. We will also illustrate the $H$- and $T$-dependence of spin susceptibility averaged over the slab, where the latter is associated with the NMR frequency shift and absorption. In order to discuss the thermodynamic stability and the phase diagram of superfluid $^3$He in a slab geometry, we estimate the thermodynamic functional within the quasiclassical approximation, 
\beq
\delta \Omega [\underline{g}]
= \frac{1}{2} \int^{1}_0 d\lambda {\rm Sp}^{\prime} \left\{ \left(\underline{\nu}+\underline{\Delta}\right)
\left( \underline{g}_{\lambda} - \frac{1}{2}\underline{g}  \right)
\right\} ,
\label{eq:omega}
\eeq
where we set $
{\rm Sp}^{\prime}\{ \cdots\} 
= {\mathcal N}_{\rm F} \int d{\bm r} \langle {\rm Tr}_4 \{ \cdots\} \rangle _{\hat{\bm k},\omega _n}$. 
The quasiclassical auxiliary function $\underline{g}_{\lambda}$ is obtained from the quasiclassical Eilenberger equation (\ref{eq:eilen}) with replacing $\underline{\nu}\!\rightarrow\! \lambda \underline{\nu}$ and $\underline{\Delta}\!\rightarrow\! \lambda \underline{\Delta}$ ($\lambda \!\in\![0,1]$), where the equation is solved once under a given self-energy but not self-consistently. The functional in Eq.~(\ref{eq:omega}) is obtained from the Luttinger-Ward thermodynamic functional (\ref{eq:LW}) associated with the Nambu-Gor'kov Green's function $\underline{G}$, whose detailed derivation is followed by the work in Ref.~\cite{vorontsovPRB2003}. Equation (\ref{eq:omega}) includes the influence of the condensation energy and quasiparticle excitations as well as the Fermi liquid corrections. 

\subsection{Thermodynamics in the absence of dipole-dipole interaction}

\begin{figure}[tb!]
\begin{center}
\includegraphics[width=75mm]{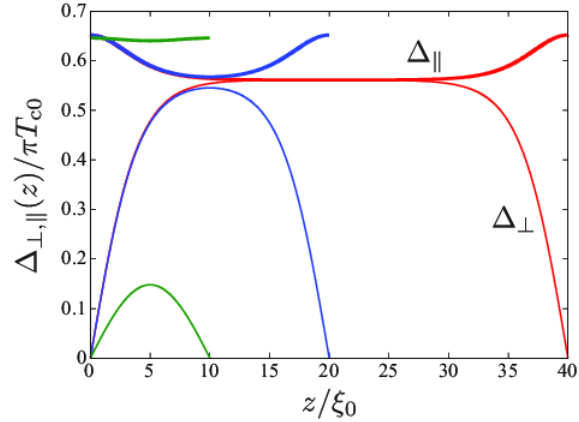}
\end{center}
\caption{Spatial profile of the pair potentials $\Delta _{\parallel}(z)$ (thick curves) and $\Delta _{\perp}(z)$ (thin curves) for $D/\xi _0=40$, $20$, and $10$ at zero fields.}
\label{fig:gapD}
\end{figure}

We fist summarize the pair breaking effect in a slab geometry and the superfluid phase diagram in the absence of a magnetic dipol-dipole interaction that is crucial for determining the topological phase transition at finite magnetic fields. As shown in previous sections, the specular surface is accompanied by the gapless surface bound states that behave as Majorana fermions. It was found by Buchholtz and Zwicknagl~\cite{buchholtzPRB1981} in 1981 that from the microscopic point of view, the surface states emerge on a specular surface of spin-triplet $p$-wave superconductors and superfluids. The existence of gapless bound states at the surface is fed back into the pair potential, which gives rise to a strong distortion of the surface pair potential. The existence and role of surface bound state were more explicitly discussed by Hara and Nagai~\cite{haraPTP1986}, who analyzed a $p$-wave polar state as an exactly solvable model. We display in Fig.~\ref{fig:gapD} the spatial profiles of the pair potentials in a slab geometry with different $D$'s. The pair potentials are the self-consistent solutions of the quasiclassical equation (\ref{eq:eilen}) and the gap equation (\ref{eq:gap_qct}) with Eq.~(\ref{eq:dvec_b2_initial}) at zero fields. Owing to the boundary condition at the specular surface in Eq.~(\ref{eq:bc}), $\Delta _{\parallel}(z)$ that is coupled to the momentum perpendicular to the surface must vanish at the surface, while the parallel component $\Delta _{\perp}(z)$ increases so as to gain the condensation energy. As $D$ decreases, the perpendicular component can not survive and the superfluid phase transition from the B to planar phase occurs.

In the basis of the quasiclassical theory, the quantitative phase diagram of the superfluid $^3$He in a restricted geometry have been clarified in Refs.~\cite{mizushimaPRB2012,haraJLTP1988,vorontsovPRB2003}. As shown in Fig.~\ref{fig:phase_slab}(b), there exists the critical thickness at which the B-phase undergoes the transition to the planar phase. At the weak coupling limit, the planar is energetically degenerate to the A-phase that is the time-reversal symmetry breaking phase with point nodes, while the strong coupling effect that is the spin-fluctuation feedback effect can stabilize the A-phase relative to the planar phase and the A-B phase boundary becomes the first order transition. 

The pair breaking effect and the enhancement of the surface density of states due to the surface bound state have been observed by several experiments, by using the NMR techniques~\cite{kawae,miyawaki,kawasaki,levitin,bennettJLTP2010,levitin2013,levitinPRL2013,ahonenJLTP1976,osheroff,ishikawaJLTP1989}, the motion of a vibrating wire resonator~\cite{castelijns}, transverse acoustic impedance~\cite{aokiPRL2005,saitoh,wada,murakawaPRL2009,murakawaJLTP2010,wasaiJLTP2010,murakawaJPSJ2011,okuda}, surface contributions of the heat capacity~\cite{choiPRL2006}, and anomalous attenuation of transverse sound~\cite{davisPRL2008}. Among them, Murakawa {\it et al.}~\cite{murakawaPRL2009,murakawaJPSJ2011} has observed the specularity dependence of the surface density of states by systematically controlling the surface specularity by coating the surface with $^4$He layers. More recently, the phase diagram was experimentally examined by using the well controlled confinement of nanofluidic samples~\cite{levitin2013}. Levitin {\it et al.}~\cite{levitin2013} confined a sample of the liquid $^3$He within a nanofluidic cavity of precisely defined geometry and utilized the SQUID-NMR technique that provides a fingerprint measurement for the order parameter distortion and phase diagram. Further details on detecting surface states and their Majorana nature will be discussed in Sec.~\ref{sec:detecting}.

\begin{figure}[tb!]
\begin{center}
\includegraphics[width=75mm]{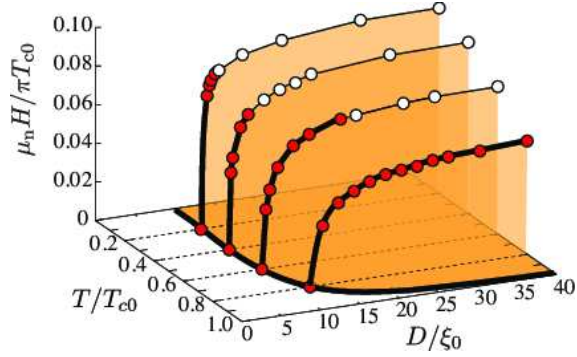}
\end{center}
\caption{Superfluid phase diagram in the space spanned by temperature $T$, perpendicular magnetic field $H$, and thickness $D$. The open (filled) circles and thin (thick) lines denote the first (second) order transition lines. The shaded area is occupied by the distorted BW state and the other is covered by the planar (or ABM) state. Figure adapted from Ref.~\cite{mizushimaPRB2012}.}
\label{fig:phase}
\end{figure}

The magnetic Zeeman field also induces the phase transition from the B-phase to A-phase (or planar phase). Figure~\ref{fig:phase} summarizes the phase diagram in a three-dimensional space spanned by the temperature $T$, perpendicular magnetic field $H$, and thickness $D$~\cite{mizushimaPRB2012}. In the region of the large thickness $D \!\gtrsim 11 \xi _0$ and low temperatures, the phase boundary $H _{\rm AB}$. is the first-order phase transition. As $D/ \xi_0$ increases, the first-order transition field $H _{\rm AB}$ slightly increases and reach saturation $\mu _{\rm n}H^{\ast}_{\rm AB}/\pi T_{\rm c0} \!=\! 0.095$ in the thermodynamic limit $D\!\gg\! \xi _0$. Using the parameters $T_{\rm c0} \!=\! 1{\rm mK}$ and the gyromagnetic ratio of $^3$He nuclei $\gamma \!=\! 2\mu _{\rm n}$, the critical field is estimated as $H_{\rm AB} \!\approx\! 0.35 {\rm T}$, which quantitatively agrees with Ref.~\cite{ashida} and the experimental data in Refs.~\cite{kyynarainen} and \cite{fisher}.

The bottom line in Fig.~\ref{fig:phase} describes the A-B phase transition in the absence of a magnetic field. The phase transition is found to be of the second order~\cite{haraJLTP1988,vorontsovPRB2003}. However, Vorontsov and Sauls predicted in Ref.~\cite{vorontsovPRL2007} that the vicinity of the second-order phase boundary around $D \!\sim\! 10\xi _0$ is covered by the new quantum crystalline phase, the so-called stripe phase. The stripe phase that spontaneously breaks the translational symmetry has not been observed in experiments. For simplicity, we below eliminate the possibility of the stripe phase from the phase diagram. The complete phase diagram under a magnetic field that takes account of the stripe phase remains as a future problem. 

We emphasize again that the phase diagram displayed in Fig.~\ref{fig:phase} does not takes account of the magnetic dipole-dipole interaction. Below, we will show that the dipole interaction changes the superfluid phase diagram in the region of low magnetic fields when the magnetic field is applied along the surface.

\subsection{Effect of dipole interaction: Ginzburg-Landau theory}
\label{sec:GL}

To examine the thermodynamic stability of the topological phase, we must take account of the magnetic dipole-dipole interaction that originates from the magnetic moment of $^3$He nuclei. The Hamiltonian for the nuclear-dipole interaction is given as 
\beq
\mathcal{H}_{\rm D} = \frac{\gamma^2}{2} \int d{\bm r}_1\int d{\bm r}_2 
Q_{\mu\nu}({\bm r}_{12}) S_{\mu}({\bm r}_1)S_{\nu}({\bm r}_2),
\label{eq:Hdip}
\eeq
where $\gamma = 2\mu _{\rm n}$ is the gyromagnetic ratio of the $^3$He nucleus and we have introduced the local spin operator $S_{\mu}({\bm r}) = \psi^{\dag}_a({\bm r})(\sigma _{\mu})_{ab}\psi _b ({\bm r})/2$. The dipole interaction potential $Q_{\mu\nu}({\bm r})$ is defined as 
$Q_{\mu\nu}({\bm r}) \!=\!  (\delta _{\mu\nu}- 3\hat{r}_{\mu}\hat{r}_{\nu})/r^3$.
Since the dipole interaction is much weaker than the pair interaction, it acts as a small perturbation on the order parameter. The residual interaction generates the spin-orbit coupling and reduces the full symmetry group in Eq.~(\ref{eq:full}) to $G_{\rm D} = {\rm SO}(3)_{{\bm L}+{\bm S}}\times {\rm U}(1)_{\phi} \times {\rm T}\times{\rm C}$. The spin-orbit interaction does not change the overall structure and remaining symmetry of the B-phase order parameter, while it imposes a constraint on the order parameter degenerate space. This implies that even in the bulk B-phase, the spin-orbit coupling imposes a strong constraint on the order parameters $\hat{\bm n}$ and $\varphi$. The order parameters must be determined by minimizing the magnetic dipole-dipole interaction. As shown in Eq.~(\ref{eq:sabs}), since the gapless nature of the surface bound states depends on $(\hat{\bm n},\varphi)$, the dipole interaction is crucial for quantitatively determining the gapless surface bound states and the stability of the topological phase.

To capture the essence of the interplay between the magnetic Zeeman field and dipol-dipole interaction, we here summarize the results in the basis of the Ginzburg-Landau theory. The analysis based on the Ginzburg-Landau theory initiated the theoretical studies for understanding the pair breaking effect and $\hat{\bm n}$-texture in restricted geometries~\cite{ambegaokar,brinkman,barton,smith,kjaldman,fujita,takagi,jacobsen,fetter,li,ullah,lin-liu,salomaa}. 

We start with the Ginzburg-Landau free energy functional $\mathcal{F}_{\rm GL}$~\cite{vollhardt},
\beq
\mathcal{F}_{\rm GL} = \int d{\bm r} \left( f_{\rm bulk} + f_{\rm grad} + f_{\rm mag} 
+ f_{\rm dip}\right),
\eeq
which holds the symmetry group $G$ in Eq.~(\ref{eq:full}). The bulk free energy and the gradient energy are given by
$f_{\rm bulk} \!=\!\alpha d^{\ast}_{\mu i}d_{\mu i}
+ \beta _1 d^{\ast}_{\mu i} d^{\ast}_{\mu i} d_{\nu j} d_{\nu j}
+ \beta _2 d^{\ast}_{\mu i} d_{\mu i} d^{\ast}_{\nu j} d_{\nu j}
+ \beta _3 d^{\ast}_{\mu i} d^{\ast}_{\nu i} d_{\mu j} d_{\nu j}
+ \beta _4 d^{\ast}_{\mu i} d_{\nu i} d^{\ast}_{\nu j} d_{\mu j}
+ \beta _5 d^{\ast}_{\mu i} d_{\nu i} d_{\nu j} d^{\ast}_{\mu j}$ and 
$f_{\rm grad} \!=\! K_1 \partial _i d^{\ast}_{\mu j} \partial _i d_{\mu j}
+ K_2 \partial _i d^{\ast}_{\mu j} \partial _j d_{\mu i}
+ K_3 \partial _i d^{\ast}_{\mu i} \partial _j d_{\mu j}$, respectively.
The phenomenological parameters satisfy the following relations in the weak coupling limit, 
$K_1 \!=\! K_2 \!=\! K_3 \!=\! K_0 \!=\! \mathcal{N}_F\xi^2_0/5$, 
$-2\beta _1 \!=\! \beta _2 \!=\! \beta _3 \!=\! \beta _4 \!=\! - \beta _5 \!=\! \frac{6}{5}\beta _0
= \frac{7 \zeta(3)N_{\rm F}}{120(\pi k _{\rm B}T_{\rm c0})^2}$,
and we have introduced $\alpha=-\frac{1}{3}\mathcal{N}_{\rm F}\left(1-\frac{T}{T_{\rm c0}}\right)$, and $\xi_0 = \frac{\hbar v_{\rm F}}{\pi k_{\rm B}T_{\rm c0}}\sqrt{\frac{7\zeta (3)}{48}}$. 
The magnetic field energy relevant to $^3$He in the equilibrium is the quadratic Zeeman energy that is given by 
\beq
f_{\rm mag} = g_{\rm m}H_{\mu}d^{\ast}_{\mu i} H_{\nu}d_{\nu i}.
\label{eq:FM}
\eeq
The factor $g_{\rm m}$ in Eq.~(\ref{eq:FM}) is given as 
\beq
g_{\rm m} = \frac{2}{3}\beta _0 \left( \frac{\mu _{\rm n}}{1+F^{\rm a}_0} \right)^2 
= \frac{7\zeta(3)\mathcal{N}_{\rm F}\gamma^2}{48[(1+F^{\rm a}_0)\pi k_{\rm B}T_{\rm c}]^2}.
\eeq 
We here neglect the higher-order correction to the weak-coupling theory that originates from the splitting of the Fermi surfaces and slightly shifts the pair interaction and $\mathcal{N}_{\rm F}$ for opposite spins~\cite{vollhardt,ambegaokar}.

We first ignore the dipole interaction, namely, $\mathcal{F}_{\rm dip}=0$. The spatial profile of the pair potential in the equilibrium is determined by minimizing the free energy functional as $\delta \mathcal{F}_{\rm GL}/\delta d^{\ast}_{\mu i} = 0$. The full numerical results of $\Delta _{\perp}(z)$ and $\Delta _{\parallel}(z)$ in a slab geometry are similar to those shown in Fig.~\ref{fig:gapD}. The perpendicular component $\Delta _{\perp}$ is suppressed by the pair breaking effect at the surfaces, while $\Delta _{\parallel}$ survives. Using the order parameter defined in Eqs.~(\ref{eq:dvec_slab}) and (\ref{eq:dvec_b2_initial}), one obtains the magnetic energy density in the Ginzburg-Landau regime as
\beq
f_{\rm mag} = g_{\rm m}H^2 \Delta^2_{\parallel} \left[
1 - \hat{\ell}^2_z(\hat{\bm n},\varphi) \left( 
1 - \eta^2(z)
\right)
\right].
\eeq
The function $\eta(z)\equiv\Delta _{\perp}(z)/\Delta _{\parallel}(z)$ denotes the ratio of the distorted pair potentials, where $\eta =1 $ is the isotropic BW state and $\eta = 0$ is the planar state. Since the pair breaking effect results in $\Delta _{\perp}(z) \le \Delta _{\parallel}(z)$ locally, one finds $0\le 1-\eta^2 \le 1$. This indicates that the magnetic field energy, $\mathcal{F}_{\rm mag}\equiv \int f_{\rm mag}(z)dz$, is minimized when $\hat{\ell}_z(\hat{\bm n},\varphi) = \pm 1$. Hence, the magnetic energy favors the non-topological phase without the hidden ${\bm Z}_2$ symmetry, in which surface Majorana fermions acquire a finite mass associated with the Zeeman energy.

Let us now consider the contribution of the dipole energy density $f_{\rm dip}$. The dipole energy within the Ginzburg-Landau regime is derived from the nuclear-dipole interaction Hamiltonian (\ref{eq:Hdip}) as $\langle\langle \mathcal{H}_{\rm D}\rangle\rangle$, where $\langle\langle \cdots \rangle\rangle$ is the thermal average defined in Eq.~(\ref{eq:G}). The dipole energy density is given as~\cite{leggettRMP,vollhardt,leggett1973,leggett1974}
\beq
f_{\rm dip} = \frac{1}{5}\lambda _{\rm D}\mathcal{N}_{\rm F} \left( 
d^{\ast}_{\mu\mu}d_{\nu\nu} +d^{\ast}_{\mu\nu}d_{\nu\mu} 
- \frac{2}{3}d^{\ast}_{\mu\nu}d_{\mu\nu}
\right).
\label{eq:fdip}
\eeq
Here, $\lambda _{\rm D}$ is a dimensionless dipole coupling parameter and approximately independent of pressure. The value is estimated as $\lambda _{\rm D} \approx 5 \times 10^{-7}$~\cite{vollhardt}.

We now determine the stable configuration of $(\hat{\bm n},\varphi)$ in a slab geometry without a magnetic Zeeman field. Substituting the pair potential in Eqs.~(\ref{eq:dvec_slab}) and (\ref{eq:dvec_b2_initial}) into Eq.~(\ref{eq:fdip}), the dipole energy density is recast into the following form:
\beq
f_{\rm dip}(z) &=& a(z) + \frac{1}{5}\lambda _{\rm D}\mathcal{N}_{\rm F}\Delta^2_{\parallel}(z)\nn \\
&& \times \left[
f_0(\varphi,z) + f_2(\varphi,z) \hat{n}^2_z + f_4(\varphi,z) \hat{n}^4_z
\right],
\eeq
where $a(z)$ is independent of both $\hat{\bm n}$ and $\varphi$. At zero field, the dipole energy depends on the spin-orbit angle $\varphi$ and $\hat{n}_z$. Here, we have introduced the coefficients: $f_0$ is the coefficient of the $\hat{n}_z$-independent term, 
\beq
f_0(\varphi,z) &=& 8 \left( \cos\varphi +\frac{1}{4}\right)^2 
- 8 \left(1-\eta (z)\right) \cos\varphi \left(\cos\varphi +\frac{1}{4}\right) \nn \\
&&+ 2 \left(1-\eta(z)\right)^2 \cos^2\varphi.
\eeq
The coefficients of the $\hat{n}^2_z$ and $\hat{n}^4_z$ terms are given as
\beq
f_2(\varphi,z) =- 2\left(1-\eta(z)\right) (1-\cos\varphi) \left[ 3+\left(2+\eta(z)\right)\cos\varphi\right], \\
f_4(\varphi,z) = 2\left(1-\eta(z)\right)^2 \left( 1-\cos\varphi\right)^2.
\eeq
The order parameters $(\hat{\bm n},\varphi)$ in the equilibrium are determined as a local minimum of the nuclear-dipole energy. For $^3$He confined in a slab geometry, the pair breaking effect at the surfaces distorts the isotropic pair potentials, leading to $\eta(z) < 1$. In this situation, one finds $f_2 < 0$ and $f_4 > 0$, which implies that the local minimum of $\mathcal{F}_{\rm dip}$ is located at a finite $\hat{n}_z$. Solving the set of equations, $\partial \mathcal{F}_{\rm dip}/\partial \varphi = 0$ and $\partial \mathcal{F}_{\rm dip}/\partial \hat{n}_z = 0$, one finds that the local minimum of $\mathcal{F}_{\rm dip}$ exists at 
\beq
\hat{\bm n} = (0,0,1),
\label{eq:n}
\eeq
and 
\beq
\varphi = \cos^{-1}\left( 
- \frac{1}{4}\frac{\langle \Delta _{\parallel}(z) \Delta _{\perp}(z)\rangle}{\langle \Delta^2_{\parallel}(z)\rangle}
\right),
\label{eq:varphi}
\eeq
where $\langle\cdots\rangle \equiv \frac{1}{D} \int^{D}_0 \cdots dz$ is the spatial average over the slab. For the bulk B-phase with $\Delta _{\parallel}=\Delta _{\perp}$, the angle $\varphi$ reduces to the so-called Leggett angle, $\varphi \approx -104^{\circ}$~\cite{leggett1973,leggett1974}. 

Hence, the dipolar field originating from $f_{\rm dip}$ tends to align $\hat{\bm n}$ to the $\hat{\bm z}$-axis. In the case of ${\bm H}\parallel\hat{\bm z}$ that the magnetic field is perpendicular to the surface, both the dipole field and magnetic Zeeman field favor the state with $\hat{\ell}_z=1$. This implies that an infinitesimal magnetic field, ${\bm H}\perp\hat{\bm z}$, destroys the topological phase and the surface Majorana fermions acquire an effective mass proportional to the Zeeman energy. 

\begin{figure}[tb!]
\begin{center}
\includegraphics[width=80mm]{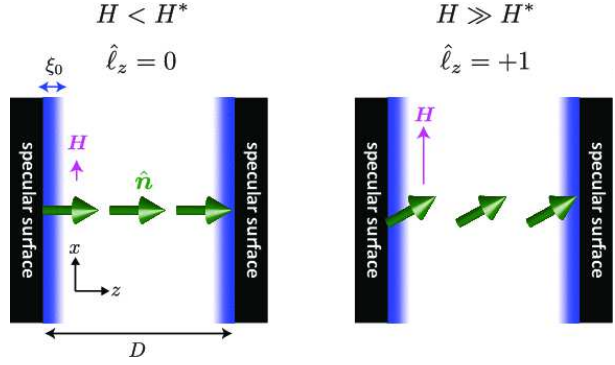}
\end{center}
\caption{Schematic picture of the orientation of the $\hat{\bm n}$-vector in the presence of a parallel magnetic field ${\bm H}$: $H< H^{\ast}$ (left) and $H\gg H^{\ast}$ (right), where $H^{\ast}$ is the critical field. In the case of $H \gg H^{\ast}$, the $\hat{\bm n}$-vector is oriented to the direction that maximize $\hat{\ell}_z$.}
\label{fig:nvec}
\end{figure}

The case of ${\bm H}\perp \hat{\bm z}$ that the magnetic field is parallel to the surface is remarkable. The $\hat{\bm n}$-vector is oriented to the $\hat{\bm z}$-axis when the dipolar field is much weaker than the Zeeman field. This configuration of Eqs.~(\ref{eq:n}) and (\ref{eq:varphi}) under ${\bm H}\perp \hat{\bm z}$ corresponds to the case of $\hat{\ell}_z=0$ in which the hidden ${\bm Z}_2$ symmetry is preserved. For a parallel field, therefore, the dipole interaction favors the topological phase protected by the hidden ${\bm Z}_2$ symmetry. For a magnetic field regime much stronger than the dipolar field, however, the magnetic field energy favors the situation of $\hat{\ell}_z = 1$ in which the ${\bm Z}_2$ symmetry is no longer held. Hence, as shown in Fig.~\ref{fig:nvec}, the analysis based on the Ginzburg-Landau theory indicates that there exists the critical magnetic field $H^{\ast}$ at which the topological phase transition occurs together with the spontaneous breaking of the hidden ${\bm Z}_2$ symmetry. 

In accordance with the Ginzburg-Landau analysis~\cite{vollhardt}, a characteristic field $H^{\ast}_{\rm GL}$ below which the $\hat{\bm n}$ is forced to be $\hat{\bm n}=\hat{\bm z}$ by the dipole interaction energy is given by 
\beq
H^{\ast}_{\rm GL} = \alpha \lambda^{1/2}_{\rm D}\frac{\pi k_{\rm B}T_{\rm c0}}{\gamma \hbar}(1+F^{\rm a}_0),
\eeq
where $\alpha =\sqrt{54/7\zeta(3)}$. The characteristic field $H^{\ast}_{\rm GL}$ is of order $25$G. This field is temperature-independent in the Ginzburg-Landau regime, but slightly depends on pressure.

\subsection{Topological phase diagram}

In the preceding subsection, we have illustrated that there exists a critical magnetic field $H^{\ast}$ in the basis of the Ginzburg-Landau theory. The theory does not take account of the information on low-lying quasiparticles that may be influential in the orientation of the $\hat{\bm n}$-vector. Beyond the Ginzburg-Landau theory, in this subsection, we utilize the quasiclassical Eilenberger theory that provides a tractable and quantitative scheme for understanding the interplay of the pair potential and quasiparticles.~\cite{eilenberger,serene}

To examine the topological phase diagram, we have to take account of the effects of the magnetic Zeeman field and nuclear-magnetic dipole interaction into the quasiclassical equation (\ref{eq:eilen}) and gap equation (\ref{eq:gap}). The effect of the magnetic dipole interaction in the superfluid $^3$He-B has been emphasized in spin dynamics~\cite{leggett1973,leggett1974,tewordtPLA1976,tewordtJLTP1979,schopohlJLTP1982,fishman1987}. Let us now start with the gap equation in terms of the quasiclassical propagators,
\beq
d_{\mu} (\hat{\bm k}, {\bm r}) 
= \frac{1}{2}(\sigma _{\mu} \sigma _y)^{\ast}_{ab}(\sigma _{\nu} \sigma _y)_{cd} 
 \left\langle V^{cd}_{ab}(\hat{\bm k}, \hat{\bm k}^{\prime})
f_{\nu}(\hat{\bm k}^{\prime},{\bm r};\omega _n)\right\rangle _{\hat{\bm k}^{\prime},n},
\label{eq:gapv1}
\eeq
where the repeated Roman indices imply the sum over spins $\uparrow$ and $\downarrow$. At the low pressure limit, the pair interaction of $^3$He atoms is described as 
\beq
V^{cd}_{ab}(\hat{\bm k}, \hat{\bm k}^{\prime})
= 3 |g| \hat{k}_{\mu}\hat{k}^{\prime}_{\mu} \delta _{ac} \delta_{bd} 
- Q_{\mu\nu}(\hat{\bm k},\hat{\bm k}^{\prime})(\sigma_{\mu})_{ac}(\sigma_{\nu})_{bd}.
\eeq
where the first term is the $p$-wave interaction with ${\rm SO}(3)_{\bm S}\!\times\!{\rm SO}(3)_{\bm L}\!\times\!{\rm U}(1)$ and the second term arises from the dipole-dipole interaction between $^3$He nuclei. The function $Q_{\mu\nu}(\hat{\bm k},\hat{\bm k}^{\prime})$ is obtained from the dipole interaction Hamiltonian (\ref{eq:Hdip}) as 
$
Q_{\mu\nu}({\bm k},{\bm k}^{\prime}) \!=\! g_{\rm D} R \int Q_{\mu\nu}({\bm r})e^{-i({\bm k}-{\bm k}^{\prime})\cdot{\bm r}}d{\bm r}
$ with ${\bm k} \!\approx\! \hat{\bm k}k_{\rm F}$. The factor $R$ includes the contributions of high energy quasiparticles. The dipole interaction, which reduces the ${\rm SO}(3)_{\bm S}\!\times\!{\rm SO}(3)_{\bm L}$ symmetry to ${\rm SO}(3)_{{\bm L}+{\bm S}}$, plays a crucial role on determining the critical field $H^{\ast}$ under a parallel magnetic field~\cite{mizushimaPRL2012}. 
The dipole potential $Q_{\mu\nu}({\bm k},{\bm k}^{\prime})$ can be expanded in terms of the partial wave series ($p$-, $f$-, and higher waves). However, since the pairing interaction between $^3$He atoms is dominated by the ${\rm SO}(3)_{\bm S}\!\times\!{\rm SO}(3)_{\bm L}\!\times\!{\rm U}(1)$ channel and the dipole interaction can be regarded as a small perturbation, we take account of only the $p$-wave contribution of $Q_{\mu\nu}(\hat{\bm k},\hat{\bm k}^{\prime})$. To this end, the gap equation (\ref{eq:gapv1}) reduces to~\cite{mizushima2014-2}
\beq
d_{\mu\nu}({\bm r}) 
&= & 3|g|\left\langle \hat{k}_{\nu}f_{\mu} \right\rangle _{\hat{\bm k},n}
-\tilde{g}_{\rm D}
\left( 1 + 3\delta _{\mu\nu} \right)
\left\langle \hat{k}_{\nu}{f}_{\mu} \right\rangle _{\hat{\bm k},n} \nn \\
&& - 3\tilde{g}_{\rm D}\left[
\left\langle \hat{k}_{\mu}{f}_{\nu} \right\rangle _{\hat{\bm k},n}
- \left\langle \hat{k}_{\nu}{f}_{\mu} \right\rangle _{\hat{\bm k},n}
\right].
\label{eq:gapv3}
\eeq
To determine the order parameters in the equilibrium, we first solve the set of the self-consistent equations (\ref{eq:eilen}) and (\ref{eq:gapv3}) with the normalization condition (\ref{eq:norm}) for a fixed $(\hat{\bm n},\varphi)$. Substituting the self-consistently calculated $\underline{g}$, $\underline{\nu}$, and $\underline{\Delta}$ into Eq.~(\ref{eq:omega}), we evaluate the thermodynamic potential for the given $(\hat{\bm n},\varphi)$. The order parameters $(\hat{\bm n},\varphi)$ in the equilibrium are determined by minimizing the thermodynamic potential. 

Before going to self-consistent calculation in a slab geometry, let us first solve the gap equation in the thermodynamic limit, where $d_{\mu\nu}({\bm r})$ is assumed to be spatially uniform and we ignore a magnetic field for simplicity. The quasiclassical propagator $\underline{g}$ at the limit is obtained in Eq.~(\ref{eq:gbulk}). Then, the gap equation (\ref{eq:gapv3}) is recast into 
\beq
d_{\mu\nu} \left( 1 - 3|g|J^{(2)}_{\nu}\right) 
= - \tilde{g}_{\rm D} \left(
3\delta _{\mu\nu}d_{\gamma\gamma}J^{(2)}_{\gamma} 
-2 d_{\mu\nu}J^{(2)}_{\nu}
+ 3 d_{\nu\mu}J^{(2)}_{\mu}
\right),
\label{eq:gap2}
\eeq
where 
$
J^{(n)}_{\mu}\equiv \pi \langle \hat{k}^n_{\mu}/[\omega^2_n+|{\bm d}(\hat{\bm k})|^2]^{1/2}\rangle _{\hat{\bm k},n}
$.
We regard the contribution of the dipole interaction as a small perturbation, which reduces the gap equation (\ref{eq:gap2}) to
\beq
d_{\mu\nu} \left( 1 - 3|g|J^{(2)}_{\nu}\right) 
= - \frac{\tilde{g}_{\rm D}}{3|g|} \left(
3\delta _{\mu\nu}d_{\gamma\gamma}
-2 d_{\mu\nu}
+ 3 d_{\nu\mu}
\right).
\label{eq:gap3}
\eeq
The higher order terms on $\tilde{g}_{\rm D}$ are neglected. Without loss of generality, the $\hat{\bm z}$-axis in the thermodynamic limit is set to be parallel to the $\hat{\bm n}$-vector. Substituting the order parameter of the distorted B-phase in Eq.~(\ref{eq:dvec_b2_initial}) into Eq.~(\ref{eq:gap3}), one obtains the set of three equations for the amplitudes $\Delta _{\parallel}$ and $\Delta _{\perp}$ and angle $\varphi$, 
\beq
\frac{\Delta _{\perp}}{\Delta _{\parallel}}\left( 1-3|g|J^{(2)}_z\right) 
= -\tilde{g}_{\rm D}J^{(2)}_z\left( 3\cos\!\varphi+2\frac{\Delta _{\perp}}{\Delta _{\parallel}}\right), 
\label{eq:gap4a} \\
1-\frac{3}{2}|g|(J^{(0)}-J^{(2)}_z)= 
\frac{\tilde{g}_{\rm D}}{2}(J^{(2)}_z-J^{(0)})\left( 7\cos\!\varphi + 3\frac{\Delta _{\perp}}{\Delta _{\parallel}}\right), 
\label{eq:gap4b} \\
1 - \frac{3}{2}|g|(J^{(0)}-J^{(2)}_z)
= -\frac{5}{2}\tilde{g}_{\rm D}(J^{(0)}-J^{(2)}_z).
\label{eq:gap4c}
\eeq
Equations (\ref{eq:gap4b}) and (\ref{eq:gap4c}) determine the relative angle $\varphi$ of the rotation matrix within the lowest order on $\tilde{g}_{\rm D}$ as
\beq
\varphi = \cos^{-1}\left(-\frac{1}{4}\frac{\Delta _{\perp}}{\Delta _{\parallel}}\right).
\eeq
This is consistent to the so-called Leggett angle that was obtained in Refs.~\cite{tewordtPLA1976,tewordtJLTP1979,schopohlJLTP1982,fishman1987} and from Eq.~(\ref{eq:varphi}). The orientation of the $\hat{\bm n}$-vector is determined by the competition between the dipole interaction, magnetic field, and pair breaking effect at the surface.

\begin{figure}[tb!]
\begin{center}
\includegraphics[width=75mm]{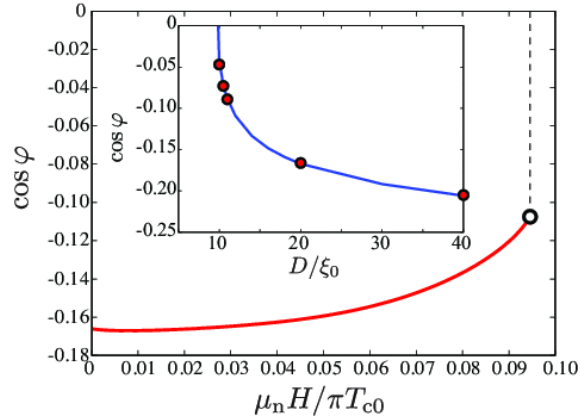}
\end{center}
\caption{Field-dependence of the stable Leggett angle $\varphi$ for ${\bm H}\parallel\hat{\bm z}$ at $T=0.2T_{\rm c0}$ and $D=20\xi _0$, where the $\hat{\bm n}$-vector is aligned to the $\hat{\bm z}$-axis. The first-order A-B phase transition occurs at $\mu _nH/\pi T_{\rm c0} \approx 0.09$ corresponding to $0.36$T. The inset of (a) shows the $D$-dependence of the Leggett angle $\varphi$ at $H=0$ and $T=0.2T_{\rm c0}$. The solid curve in the inset is obtained from Eq.~(\ref{eq:varphi}). Figures adapted from Ref.~\cite{mizushimaJPCS2012}.}
\label{fig:leggett}
\end{figure}

We now microscopically determine the angle $\varphi$ that minimizes the thermodynamic potential with the self-consistent solutions. First, we consider the case of a perpendicular magnetic field ${\bm H}\parallel\hat{\bm z}$ in a slab geometry. In this situation, the $\hat{\bm n}$-vector is always oriented to the surface normal direction, regardless of the value of $H$. The main panel of Fig.~\ref{fig:leggett} shows the field-dependence of $\varphi$ for ${\bm H}\parallel\hat{\bm z}$ and the inset is the $D$-dependence at zero fields, where we fix $T=0.2T_{\rm c0}$ and $D=20\xi _0$. As seen in the inset of Fig.~\ref{fig:leggett}, the $D$-dependence of $\varphi$ at zero fields is in good agreement with Eq.~(\ref{eq:varphi}) in the basis of the Ginzburg-Landau analysis. The angle $\varphi$ approaches zero at the critical thickness $D\approx 9.8\xi _0$ that the A-B phase transition occurs. As seen in the main panel of Fig.~\ref{fig:leggett}, the angle $\varphi$ is relatively insensitive to the increase of the applied field $H$.

\begin{figure}[t!]
\begin{center}
\includegraphics[width=85mm]{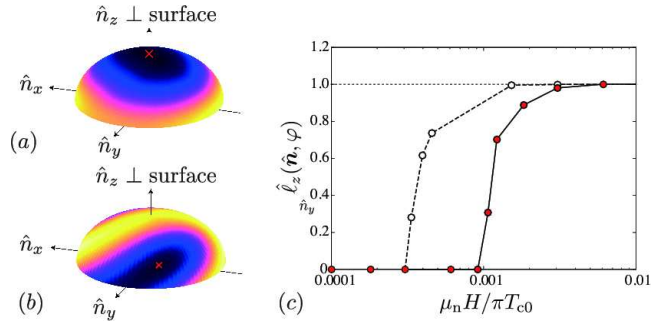}
\end{center}
\caption{Thermodynamic potential on the unit sphere of $\hat{\bm n}$, $\delta\Omega (\hat{\bm n})$, at $\mu _{\rm n}H/\pi T_{\rm c0}\!=\! 9.2\!\times\!10^{-4}$ (a) and $0.0061$ (b) where we fix $\varphi/\pi \!=\! - 0.5537$ that minimizes the dipole interaction at zero fields. We also set ${\bm H} \!\parallel\! \hat{\bm x}$ and $T/T_{\rm c0} \!=\! 0.2$. The bright (dark) color depicts the higher (lower) energy, and the crosses in (a) and (b) correspond to the minimum point of $\delta \Omega$. (c) Field dependence of $\hat{\ell}_z$ estimated with the stable $(\hat{\bm n},\varphi)$ for $\tilde{\Lambda}_{\rm D}/\Lambda^2 \!=\! 2 \!\times\! 10^{-4}$ (the solid line) and $2 \!\times\! 10^{-5}$ (the dashed line). Figures adapted from Ref.~\cite{mizushimaPRL2012}.}
\label{fig:n}
\end{figure}

Similarly with the Ginzburg-Landau regime, in the case of a parallel field ${\bm H}\parallel\hat{\bm x}$, there exists the critical field $H^{\ast}$ at which the $\hat{\bm n}$ changes the orientation from the surface normal. Figures~\ref{fig:n}(a) and \ref{fig:n}(b) depict the energy landscape $\delta \Omega$ on the unit sphere of $\hat{\bm n}$, where $\delta \Omega$ in Eq.~(\ref{eq:omega}) is estimated with the self-consistent solution of the quasiclassical propagator for a fixed $(\hat{\bm n},\varphi)$. As shown in Fig.~\ref{fig:leggett}, the angle $\varphi$ is insensitive to $H$ in the limit of the weak field. We therefore fix $\varphi$ to be the value that minimizes the thermodynamic potential at zero fields. The stable configuration of $\hat{\bm n}$ is determined as a consequence of the interplay between dipole interaction and Zeeman energy. In the weak field regime, as seen in Fig.~\ref{fig:n}(a), $\delta \Omega$ for a weak field has a minimum point at which the $\hat{\bm n}$-vector points to the $\hat{\bm z}$-axis. This corresponds to the case of $H<H^{\ast}$ with $\hat{\ell}_z (\hat{\bm n},\varphi)=0$ in Fig.~\ref{fig:nvec}. In contrast, in the relatively stronger field, as shown in Fig.~\ref{fig:n}(b), the $\hat{\bm n}$-vector tends to tilt from the surface normal direction. The resultant $\hat{\ell}_z$ becomes nonzero once $\hat{\bm n}$ tilts from the surface normal.

The field dependence of $\hat{\ell}_z$ estimated with the stable
configuration of $(\hat{\bm n},\varphi)$ is displayed in
Fig.~\ref{fig:n}(c). In the limit of the low field, $\hat{\ell}_z$ is
locked to be $\hat{\ell}_z \!=\! 0$, which ensures the existence of
surface Majorana fermions protected by the hidden ${\bm Z}_2$ symmetry. 
$\hat{\ell}_z$ stays zero up to the critical
value $\mu _{\rm n} H^{\ast}/\pi T_{\rm c0} \!\approx\! 0.001$, which is
consistent with the Ginzburg-Landau analysis in Sec.~\ref{sec:GL}. The critical field 
is estimated as $H^{\ast}\approx 20$-$30$G depending on pressure. 
For $H \!\ge\! H^{\ast}$, the topological order $\hat{\ell}_z$ becomes nonzero, which triggers off the spontaneous breaking of the hidden ${\bm Z}_2$ symmetry and simultaneously the topological phase transition, 
as discussed in Sec.~\ref{sec:hidden}. 
At the topological phase transition $H^{\ast}$, the surface Majorana fermion acquires the effective mass $\hat{\ell}_z \gamma H/2$ and the topological phase transition triggered by spontaneous symmetry breaking is not accompanied by closing the bulk gap.

\subsection{Surface spin susceptibility and odd-frequency pairing}

In the previous subsections, we have illustrated that in the superfluid $^3$He-B confined in a slab, there is the critical field $H^{\ast}$ at which the topological phase transition is concomitant with the spontaneous breaking of the hidden ${\bm Z}_2$ symmetry. The critical field is characterized by the topological order $\hat{\ell}_z$ and its amplitude parametrizes the effective mass gap that surface Majorana fermions acquire in the non-topological phase, as shown in Sec.~\ref{sec:hidden}. The nonzero value of $\hat{\ell}_z$ implies the tilting of the $\hat{\bm n}$-vector from the surface normal direction. In this subsections, we will examine a remarkable physical consequence of the topological order $\hat{\ell}_z$, that is the anomalous magnetic response of surface Majorana fermions and emergent odd-frequency even-parity pairing. 

As clarified in Secs.~\ref{sec:hidden} and \ref{sec:odd}, the Majorana fermion that is bound to the surface in the symmetry protected topological phase yields the multifaceted properties, the Ising magnetic response and the odd-frequency pairing. The former is a direct consequence of the symmetry protected topological phase with $\hat{\ell}_z=0$, which indicates that the surface Majorana fermion cannot be coupled to a parallel magnetic field and remains gapless unless $\hat{\ell}_z$ is nonzero. For a parallel field weaker than $H^{\ast}$, therefore, the surface Majorana fermion is not responsible for the enhancement of the surface spin susceptibility. As discussed in Sec.~\ref{sec:odd}, however, the odd-frequency pairing that is the another aspect of the surface states yields paramagnetic response, once the $\hat{\ell}_z$ becomes finite. Therefore, it is expected that the topological phase transition at $H=H^{\ast}$ is accompanied by the anomalous enhancement of the surface spin susceptibility. We here quantitatively examine the field dependence of the spin susceptibility in the basis of the fully self-consistent calculation of quasiclassical equations.

\begin{figure}[t!]
\begin{center}
\includegraphics[width=85mm]{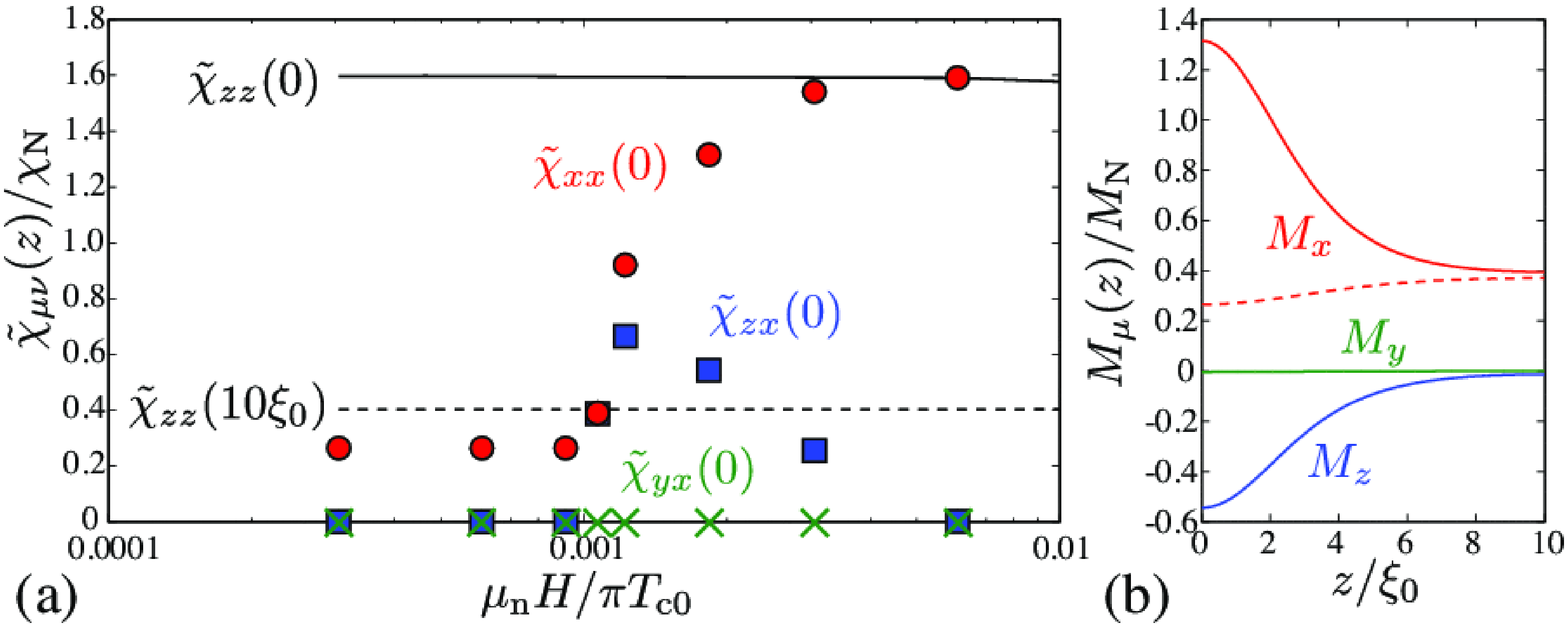}
\end{center}
\caption{(a) Field dependence of ${\chi}_{\mu \nu}(z)/\chi _{\rm N}$ at $T \!=\! 0.2T_{\rm c0}$. The solid (dashed) lines denote
${\chi}_{zz}(0)$ (${\chi}_{zz}(10\xi)$) for ${\bm H}\!\parallel\! \hat{\bm z}$ and the symbols correspond to ${\chi}_{\mu x}(0)$ for ${\bm H}\!\parallel\! \hat{\bm x}$. (b) $M_{\mu}(z)$ for ${\bm H} \!\parallel\! \hat{\bm x}$ at $\mu _{\rm n} H /\pi T_{\rm c0} \!=\! 9.2 \!\times\!10^{-4}$ (dashed line) and $0.0018$ (solid lines), where $M_{y,z}$ at $\mu _{\rm n} H /\pi T_{\rm c0} \!=\! 9.2 \!\times\!10^{-4}$ are zero. All data are taken with ${\Lambda}_{\rm D}/\Lambda^2 \!=\! 2 \!\times\! 10^{-4}$. Figures adapted from Ref.~\cite{mizushimaPRL2012}.}
\label{fig:mag}
\end{figure}

In Fig.~\ref{fig:mag}(a), we plot the field dependence of the local spin susceptibility on the surface, ${\chi}_{\mu x}(z \!=\! 0)$, where $\chi _{\mu\nu}$ is defined with the magnetization density $M_{\mu}(z)$ under a a magnetic field ${\bm H} \!\parallel\! \hat{\bm r}_{\nu}$ as ${\chi}_{\mu \nu}(z)/\chi _{\rm N} \!\equiv\! M_{\mu}(z)/M_{\rm N}$. The local magnetization density $M_{\mu}(z)$ is defined in Eq.~(\ref{eq:M}) with that in the normal state $M_{\rm N}$. It is seen from Fig.~\ref{fig:mag}(a) with the solid line that for ${\bm H} \!\parallel\! \hat{\bm z}$, the local spin susceptibility on the surface, ${\chi}_{zz}(0)$, is considerably enhanced, compared with ${\chi}_{zz}(z\!=\!10\xi )$ (the dashed line).

In contrast, when the parallel field (${\bm H} \!\parallel\! \hat{\bm x}$) is applied, the magnetization $M_{\mu}(z)$ on the surface is sensitive to the orientation of $\hat{\bm \ell}$. It is seen in Fig.~\ref{fig:mag}(b) with the dashed line that $M_x(z)$ at $ H \!=\! 9.2 \!\times\!10^{-4}\pi T_{\rm c0}/\mu _{\rm n}<H^{\ast}$ is strongly suppressed in the surface region. This implies that the surface Majorana fermion does not contribute to the magnetization $M_x(z)$ and is consistent with the property of the Majorana Ising spins discussed in Secs~\ref{sec:hidden} and \ref{sec:oddising}. 

In the relatively high field $H\!=\! 0.0018\pi T_{\rm c0}/\mu _{\rm n} > H^{\ast}$, however, $M_x(z)$ is enhanced around the surface, while the nonzero $M_z(z)$ that is the magnetic response perpendicular to ${\bm H}\!\parallel\! \hat{\bm x}$ emerges in the surface region. This emergence of $M_z(z)$ on the surface reflects the stable configuration of $(\hat{\bm n},\varphi)$, where $\hat{\ell}_z \!=\! R_{xz}(\hat{\bm n},\varphi)$ deviates from zero but is less than unity. As displayed in Figs.~\ref{fig:mag}(a) and \ref{fig:mag}(b), the magnetic field within the range of $0\!<\!\hat{\ell }_z \!<\! 1$ significantly induces $M_{z}(z)$ and ${\chi}_{zx}(z)$ on the surface, where the surface Majorana fermion opens a finite energy gap. In this situation, the hidden ${\bm Z}_2$ symmetry is no longer held and the winding number $w$ in Eq.~(\ref{eq:winding}) is ill-defined.

\begin{figure}[t!]
\begin{center}
\includegraphics[width=80mm]{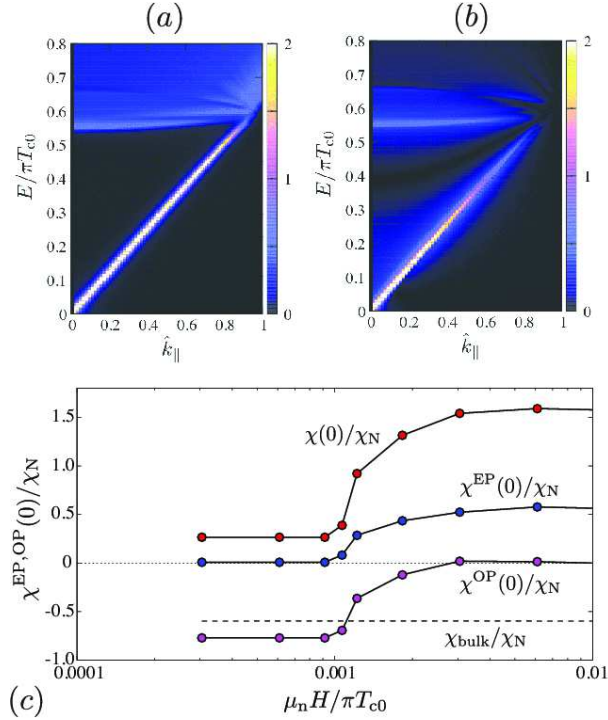}
\end{center}
\caption{(color online) Momentum resolved surface density of states $\mathcal{N}(\hat{\bm k},z=0;E)$ (a) and the OTE pair amplitude $|{\rm Re}f^{\rm OF}_z(\hat{\bm k},z=0;E)|$ (b). (c) Field-dependence of the surface spin susceptibilities, $\chi (0)$, $\chi^{\rm OP}(0)$, and $\chi^{\rm EP}(0)$ at $T=0.2T_{\rm c0}$. Figures adapted from Ref.~\cite{mizushima2014-2}.}
\label{fig:chi}
\end{figure}

Figures \ref{fig:chi}(a) and \ref{fig:chi}(b) show the momentum resolved surface density of states 
$\mathcal{N}(\hat{\bm k},z;E)$ defined in Eq.~(\ref{eq:dosk}) and the OTE pair amplitude $|{\rm Re}f^{\rm OF}_z(\hat{\bm k},z=0;E)|$, respectively. Here, we set $T=0.2T_{\rm c0}$ and $H = 0$, where $\hat{\bm \ell}_z = 0$ is favored. It is clearly seen that there exists the gapless surface bound state with the dispersion of the Majorana cone, $E({\bm k}_{\parallel}) = \Delta _0 {k}_{\parallel}/k_{\rm F}$. The momentum dependence of the OTE pairing traces that of the surface state $\mathcal{N}(\hat{\bm k},z;E)$, which indicates that the surface density of states is equivalent to the OTE pair amplitude, described in Eq.~(\ref{eq:equality}). We also confirm that $f^{\rm OF} _x = f^{\rm OF}_y = 0$, which is consistent to Eq.~(\ref{eq:fOF}). 

The field-dependence of the surface spin susceptibility at $T=0.2T_{\rm c0}$ is plotted in Fig.~\ref{fig:chi}(c). 
This numerically confirms the prediction obtained from the argument of the discrete symmetry in Sec.~\ref{sec:odd}. 
For time-reversal invariant superfluids, the surface spin susceptibility is generally divided into two contributions, $\chi _{\rm surf}= \chi _{\rm N}+\chi^{\rm EP} + \chi^{\rm OP}$, i.e., the contributions from even-parity and odd-parity pair amplitudes, $\chi^{\rm EP}$ and $\chi^{\rm OP}$. 

As discussed in Sec.~\ref{sec:odd}, the additional discrete symmetries preserved by the $^3$He-B impose strong constraint on the emergent Cooper pair amplitudes on the surface. As a consequence of the constraint, the surface spin susceptibility of the symmetry protected topological superfluid $^3$He-B is recast to $\chi _{\rm surf} = \chi _{\rm N} + \sqrt{1-\hat{\ell}^2_z}\chi^{\rm OP} + \hat{\ell}_z\chi^{\rm EP}$ as shown in Eq.~(\ref{eq:chi_final}) and the contribution of the OTE pairing is parameterized by the topological order $\hat{\ell}_z$. In the symmetry protected topological phase with $\hat{\ell}_z=0$, therefore, only the ETO pair amplitudes can contribute to $\chi _{\rm surf}$ and the OTE pairing is not coupled to the applied magnetic field ${\bm H}\parallel\hat{\bm x}$. The emergent ETO pairing at the surface is constrained by the discrete symmetry associated with the $\pi$-rotation in the spin space as ${\bm f}^{\rm EF}=(f^{\rm EF}_x,f^{\rm EF}_y,0)$. This implies that the ${\bm d}$-vector has the component parallel to the applied magnetic field, ${\bm d}\cdot{\bm H}\neq 0$. Therefore, the emergent ETO pairing suppresses the surface spin susceptibility as well as that in the bulk, $\chi _{\rm surf} =\chi _{\rm N} + \chi^{\rm OP} < \chi_{\rm N}$.

For the non-topological phase with $\hat{\ell}_z\neq 0$ corresponding to $H>H_{\rm c}$, however, only the OTE pairing $f^{\rm OP}_{\mu}$ is responsible for the surface spin susceptibility, while the ETO pairing is not coupled to ${\bm H}$. In accordance with the Ginzburg-Landau analysis in Sec.~\ref{sec:odd}, the OTE pairing yields paramagnetic response to the applied field, resulting in $\chi^{\rm EP}>0$. In contrast, for ${\bm H}\perp\hat{\bm z}$, $\hat{\ell}_z=1$ corresponds to ${\bm f}^{\rm EF}\cdot{\bm H}=0$, implying ${\bm d}\perp{\bm H}=0$. Hence, ETO pairing does not contribute to the total spin susceptibility and $\chi \approx \chi _{\rm N} + \chi^{\rm EP} > \chi _{\rm N}$. As shown in Fig.~\ref{fig:chi}(c), the OTE pair amplitudes maintain the paramagnetic response even in low temperatures beyond the Ginzburg-Landau regime. Hence, the surface spin susceptibility anomalously enhances at the critical field $H_{\rm c}$ and the drastic change of the magnetic response is attributed to the change of the Cooper pair amplitudes that couple to the applied field. Owing to the paramagnetic response of the OTE pair amplitudes, the resultant spin susceptibility at the surface exceeds $\chi _{\rm N}$.

\begin{figure}[t!]
\begin{center}
\includegraphics[width=75mm]{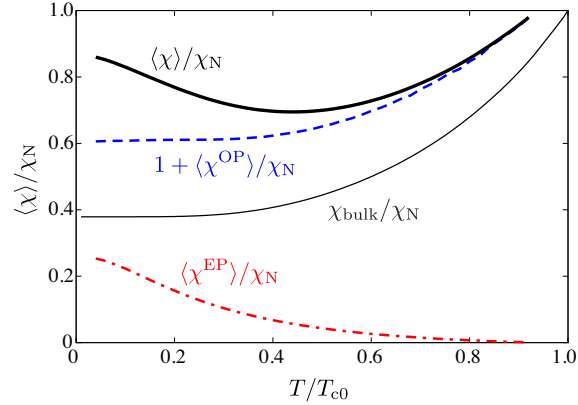}
\end{center}
\caption{(color online) Temperature-dependence of the spatially averaged spin susceptibilities, $\langle\chi \rangle$, $\langle\chi^{\rm OP}\rangle$, and $\langle\chi^{\rm EP}\rangle$ at $\mu _{\rm n} H = 0.009 \pi T_{\rm c0}$ and $D=20\xi _0$. We also plot the spin susceptibility of the bulk $^3$He-B in Eq.~(\ref{eq:chibulkB}), $\chi _{\rm bulk}$. Figure adapted from Ref.~\cite{mizushima2014-2}.}
\label{fig:chiT}
\end{figure}

We plot in Fig.~\ref{fig:chiT} the temperature dependence of the spatially averaged spin susceptibility, 
$\langle \chi \rangle \equiv \frac{1}{D}\int^{D}_0 \chi (z) dz$, 
at $\mu _{\rm n}H = 0.009\pi T_{\rm c0}$ corresponding to the non-topological phase. For comparison, we plot the spin susceptibility in the bulk B-phase given with the Fermi liquid parameter $F^{\rm a}_0$ by
\beq
\chi _{\rm bulk}= \frac{(1+F^{\rm a}_0)[2+Y(T)]}{3+F^{\rm a}_0[2+Y(T)]}\chi _{\rm N},
\label{eq:chibulkB}
\eeq
where $Y(T)$ is the Yosida function~\cite{vollhardt}. The nonlinear effect of the Zeeman magnetic field on the spin susceptibility was investigated by Fishman and Sauls~\cite{fishman1986} for the bulk $^3$He-B and in Ref.~\cite{mizushimaPRB2012} for a restricted geometry.

It is seen in Fig.~\ref{fig:chiT} that the $T$-dependence of $\langle \chi \rangle $ in a slab exhibits the non-monotonic behavior, where there exists a critical temperature below which $\langle \chi \rangle$ increases as $T$ decreases. We now identify that the increase of $\langle \chi\rangle$ in the low temperature regime reflects the coupling of the applied field to the OTE pairing that yields paramagnetic response. In high temperature regime, the continuum states with $E>\Delta$ dominate the spin susceptibility, whose temperature dependence is characterized by the Yosida function. As $T$ decreases, however, the OTE pairing gradually grows, while the contributions from the continuum states exponentially decreases. Hence, the increase of the averaged spin susceptibility $\langle \chi \rangle$ in the low $T$ region of Fig.~\ref{fig:chiT} indicates the enhancement of local magnetization density at the surface, while the behavior in the high $T$ regime is dominated by the magnetization density in the central region of the system. This non-monotopic behavior of $\langle\chi\rangle$ may be observable only in the non-topological phase. Since the OTE pairing is not responsible for the susceptibility in the symmetry protected topological phase within $H<H^{\ast}$, the $T$-dependence follows that of $\chi _{\rm N}+\langle \chi^{\rm OP}\rangle$ in Fig.~\ref{fig:chiT}. 

According to the sum rule, the static spin susceptibility $\langle\chi \rangle$ is obtained by integrating the absorptive part of the dynamical spin susceptibility over all the frequency~\cite{leggett1973,leggett1974}. Hence, the temperature- and field-dependences of $\langle \chi \rangle$ are detectable through NMR experiments~\cite{ahonenJLTP1976}. The field and temperature dependences of $\langle \chi \rangle$ may unveil the surface state of the symmetry protected topological superfluid $^3$He-B. 

\section{Detecting Majorana fermions}
\label{sec:detecting}

In this paper, it has been demonstrated that the nontrivial topological superfluidity of $^3$He-B confined in a slab geometry is protected by the time-reversal symmetry, when a magnetic Zeeman field is absent. The surface is accompanied by the helical Majorana fermion with the gapless energy $E_{\rm surf}({\bm k}) = \frac{\Delta _0}{k_{\rm F}} |{\bm k}_{\parallel}|$. The remarkable consequence of helical Majorana fermions is that the low-energy surface density of states is not coupled to the local density fluctuation, and yields the Ising anisotropic magnetic response. In addition, the surface states exhibits the odd-frequency Cooper pair amplitudes that are responsible for the anomalously large magnetic response when the Majorana fermion acquires an effective mass generated by the spontaneous breaking of the hidden ${\bm Z}_2$ symmetry. Here, we summarize the effort that has been made for seeking Majorana fermions in superfluid $^3$He.

\subsection{Heat Capacity}

\begin{figure}[tb!]
\begin{center}
\includegraphics[width=65mm]{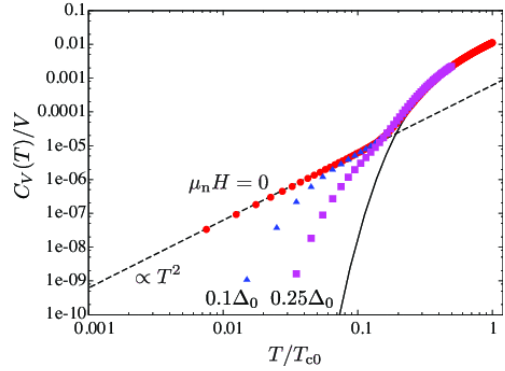}
\caption{Specific heat $C(T)$ for $k_{\rm F}\xi\!=\! 20$ with and without a magnetic field parallel to $z$-axis, where the both axes are plotted with the logarithmic scale. The solid line depicts the BCS-like exponential behavior in the bulk region. Figure adapted from Ref.~\cite{mizushimaJLTP2011}.}
\label{fig:heat}
\end{center}
\end{figure}

In the previous sections, we have clarified that the surface bound states with and without a magnetic field is describable with the $2+1$-dimensional Majorana equation and its dispersion is isotropic and linear on $|{\bm k}_{\parallel}|$ (see Fig.~\ref{fig:heat}(a)). The two-dimensional relativistic dispersion is responsible for the linear behavior of the low-energy density of states at the specular surface, 
\beq
\mathcal{N}(z=z_{\rm surf},E) \propto |E|.
\eeq
The local density of states, $\mathcal{N}(z,E) \!=\! \alpha |E|$, at the surface gives rise to a power-law behavior of the specific heat, $C(T) \propto T^2$, for low temperatures $T\!\ll\! T_c$. This is distinctive from the bulk contribution leading to the BCS-like exponential behavior of $C(T)$. The temperature dependence of the heat capacity of $^3$He-B confined in a slab geometry is plotted in Fig.~\ref{fig:heat}, where the low-temperature $C(T)$ is deviated from the exponential behavior and follows the power-law behavior $T^2$. A magnetic field sufficiently larger than the critical field $H^{\ast}$ (i.e., $\hat{\ell}_z=1$) makes a finite energy gap in the Majorana cone. As shown in Fig.~\ref{fig:heat}, the resultant $C(T)$ is deviated from the power-law behavior, which reflects the existence of multiple gap scale $\Delta _0$ and $\gamma H/2$. Note that since the low-energy density of states is sensitive to the condition of the surface and the diffusive surface considerably increases the amount of density of states at the zero energy~\cite{vorontsovPRB2003,nagai,nagatoJLTP1998,zhangPLA1988}. 

The first precise measurement of the surface specific heat was performed in the experimental group in Northwestern University~\cite{choiPRL2006}, which reported the clear deviation of $C(T)$ from that of the bulk $^3$He-B in the vicinity of the transition temperature $T_{\rm c0}$. The measurement of $C(T)$ in lower temperatures down to $T=135~\mu{\rm K}$ was performed in Ref.~\cite{bunkov} by reanalyzing the experimental data in Ref.~\cite{bunkov2}. They observed a $10~\%$ deviation from the heat capacity of the bulk superfluid $^3$He-B at $T=135~\mu{\rm K}$. The deviation is attributed to the contribution from the surface Majorana cone. 
 
\subsection{Transverse acoustics}

The transverse zero sound is a manifestation that $^3$He is a strongly correlated Fermi liquid in low temperatures. The transverse sound in the collisionless regime was originally predicted by Landau in 1957, which is the propagation of a transverse oscillation of the Fermi surface~\cite{landau}. The restoring force is provided by strong quasiparticle interactions through the current density fluctuation. Although the transverse wave has been experimentally examined in the normal $^3$He that is well describable as a Fermi liquid, the observation still remains controversial~\cite{ketterson,ketterson2,flowers}. 

The dispersion and attenuation of transverse sound are determined by the conservation law of momentum, 
\beq
\omega \delta j_{\mu}({\bm q},\omega) - \frac{1}{m}\delta \Pi _{\mu\nu}({\bm q},\omega)q_{\nu} = 0.
\label{eq:momconv}
\eeq
where $\delta {\bm j}$ is the fluctuation of the current density and $\delta \Pi _{\mu\nu}$ is the momentum tensor. This is supplemented by the quasiclassical transport equation that generalizes the Landau's kinematic equation in superfluids. In accordance with the Fermi liquid theory, there exist two different sound waves: Longitudinal and transverse sound waves. The longitudinal sound propagates in the quantum liquid through the local density fluctuation. Since the wavelength is much shorter than the superfluid coherence length and no low-lying fermionic excitations exist in the bulk $^3$He-B, the attenuation of longitudinal sound has been established as a high resolution spectroscopy for low-lying bosonic collective modes in the bulk $^3$He-B. Owing to the Majorana nature of surface states, however, the Majorana fermion might not be coupled to the longitudinal sound wave. 

The other sound wave, the transverse sound, propagates the transverse current fluctuation and its coupling to the surface state is not forbidden by the Majorana nature. In particular, in the superfluid phase, the transverse current propagates as sound wave owing to the coupling to low-lying bosonic collective modes~\cite{moores,sauls1999}. The transverse current is obtained by projecting the current fluctuation to the transverse circular polarization vector, $\hat{\bm e}^{(\pm)}=\frac{1}{\sqrt{2}}(\hat{\bm x}\pm i {\bm y})$ as $\delta j_{\pm} \equiv \delta{\bm j}\cdot\hat{\bm e}^{(\pm)\ast}$, where $\hat{\bm x}$ and $\hat{\bm y}$ are unit vectors normal to the propagation direction $\hat{\bm q}\equiv {\bm q}/|{\bm q}|$. The dispersion of transverse current propagation is obtained from Eq.~(\ref{eq:momconv}) as~\cite{moores,sauls1999}
\beq
\frac{\omega}{qv_{\rm F}}
= \frac{2}{5}
\left( 1 + \frac{F^{\rm s}_1}{3}\right)
\left\{ \frac{1}{\sqrt{2}}
\frac{\delta {g}^{(2,\pm)}_0(q,\omega)}{\delta {g}^{(1,\pm)}_0(q,\omega)}
\right\}
\eeq
The restoring force $\delta {g}^{(2,\pm)}_0/\delta {g}^{(1,\pm)}_0$ is determined by the momentum tensor coupled to the transverse current fluctuation $\delta j_{\pm}\propto \delta {g}^{(1,\pm)}_0$. The momentum tensor $\delta g^{(2,\pm 1)}_0$ is obtained by solving the quasiclassical transport equation, which is given for the bulk $^3$He-B as
\beq
\delta g^{(2,\pm 1)} _0
&=& \frac{F^{\rm s}_1}{3+F^{\rm s}_1} \left( \frac{qv_{\rm F}}{\omega}\right)
\xi _1 (q,\omega) \left\{ 
\frac{1}{\sqrt{2}}\delta g^{(1,\pm 1)}_0 (q,\omega)
\right\} \nn \\
&& + \left( \frac{\omega}{2\Delta}\right)\Lambda _1 (q,\omega)
\mathcal{D}^-_{2\pm 1}(q,\omega).
\eeq
The restoring force is contributed from the order parameter fluctuation $\mathcal{D}^-_{2\pm 1}$ in addition to quasiparticle interactions with $F^{\rm s}_1$. The collective modes $\mathcal{D}^-_{2\pm 1}$ coupled to the current fluctuation are categorized to $(J,M_J)^K = (2,\pm 1)^-$ modes in terms of the total angular momentum of the Cooper pair, $J$, the projection along the $\hat{\bm q}$-axis, $M_J = - J ,\cdots, +J$, and the parity under particle-hole conversion, $K=\pm $. The modes are called the squashing mode, whose dispersion $\Omega _{2-}(q)$ at zero fields is given in the bulk $^3$He-B as
$[ \Omega^-_{2 \pm 1}(q)]^2 = \frac{12}{5}\Delta^2_0 + \frac{2}{5}(v_{\rm F}q)^2$.
%
To this end, the dispersion of transverse sound wave is 
\beq
\left(\frac{\omega}{qv_{\rm F}}\right)^2
= \frac{F^{\rm s}_1}{15}\rho _{\rm n}(\omega)
+ \frac{2F^{\rm s}_1}{75}\rho _{\rm s}(\omega)
\frac{\omega^2}{(\omega + i\Gamma)^2 - [\Omega^-_{2\pm 1}(q)]^2}. 
\label{eq:dispTSW}
\eeq
The response function $\rho _{\rm s}(\omega)$ is the generalized Tsuneto function, and $\rho _{\rm n}(\omega)$ and $\rho _{\rm s}(\omega)$ reduce to noncondensate and condensate densities in the limit of $\omega \rightarrow 0$, respectively. The restoring forces for transverse sound in $^3$He-B are therefore contributed from the bosonic collective modes in addition to the quasiparticle interactions.

The dispersion of transverse sound is independent of its circulation when the time-reversal symmetry is preserved. The magnetic Zeeman field, however, gives rise to the splittings of $J=2^-$ collective modes as $\Omega _{2-}(q,H) = \Omega _{2-}(q,0)  + M_J g_{2-}\omega _{\rm L}$, where $g_{2-}$ is the $g$-factor for $J=2^-$ modes and $\omega _{\rm L}$ is the effective Larmor frequency~\cite{tewordtJLTP1979,saulsPRL1982}. The dispersion of the left (right) circularly polarized wave is obtained from Eq.~(\ref{eq:dispTSW}) by replacing $\Omega^-_{2\pm 1}(q)$ to $\Omega^-_{2\pm 1}(q,H)$. The field-induced mode splittings are responsible for the circular birefringence of transverse waves, where the phase velocity $\mathcal{C}_+ \equiv {\rm Re}[\omega _+/q]$ for the left circularly polarized wave is deviated by the magnetic field from that for the right circularly polarized wave, $\mathcal{C}_+\neq\mathcal{C}_-$. The circular birefringence gives rise to the acoustic analogue of the magneto-optic Faraday effect in which the direction of a linearly polarized wave rotates along the direction of propagation with the period proportional to the inverse of the Zeeman splitting~\cite{moores,sauls1999}. 

The acoustic Faraday effect was first observed in Ref.~\cite{lee1999}. The Faraday effect in transverse sound has been established as a high-resolution spectroscopy for low-lying bosonic excitation spectra in the bulk superfluid $^3$He-B~\cite{davis1,davis2,davis3,davis4,collett}, because the quasiparticle states are empty up to the threshold of the pair breaking. In particular, the phase velocity of transverse sound waves observed in experiments is in quantitatively good agreement with that calculated from Eq.~(\ref{eq:dispTSW}), while the unexpected behavior of the attenuation was observed in the frequency range $1.6\lesssim \omega/\Delta _0 \lesssim 2.0$ in Ref.~\cite{davis3}. The attenuation becomes saturated to the temperature independent value at low temperatures that is anomalously larger than that expected from theoretical calculation in Eq.~(\ref{eq:dispTSW}). Since the fraction of thermally excited quasiparticles exponentially decreases in low temperatures, the damping mechanism of transverse sound through the coupling to background quasiparticles can be ruled out. Hence, the experimental observation in Ref.~\cite{davis3} suggests that the anomalous attenuation might be attributed to the coupling of transverse sound waves with surface Andreev bound states, namely, the Majorana cone.


Using the AC-cut quartz transducers immersed in liquid $^3$He, Aoki {\it et al.}~\cite{aokiPRL2005} measured the complex transverse acoustic impedance of the superfluid $^3$He-B. The impedance is defined as the ratio of the shear stress $\Pi _{xz}$ to the wall velocity $u_x$, $Z=Z^{\prime}+iZ^{\prime\prime}\equiv \Pi _{xz}/u_x$, where the surface is set to be in the $x$-$y$ plane. In a pure $^3$He system, the wall that corresponds to the surface of the transducer is fully diffusive. In the diffusive limit, it has been predicted that surface bound states form a nearly flat band within the low-energy region of $|E|\le\Delta^{\ast}$~\cite{nagatoJLTP1998,vorontsovPRB2003,nagai}. The surface density of states has a very sharp edge at $E=\Delta^{\ast}$ that is smaller than the bulk energy gap $\Delta _0$. The temperature dependence of the complex impedance observed two different energy scales $\Delta^{\ast}$ and $\Delta _0$~\cite{aokiPRL2005,saitoh,wada,murakawaPRL2009,murakawaJLTP2010,wasaiJLTP2010,murakawaJPSJ2011,okuda} as a kink and peak in $Z^{\prime}(T)$ and $Z^{\prime\prime}(T)$.

The remarkable point is that the surface specularity is controllable by coating the wall of the transducer with thin layers of $^4$He. The roughness on the surface can be reduced by $^4$He atoms that are selectively absorbed onto the wall. The surface acoustic impedance measurement under well controlled surface conditions has revealed the spectroscopic details of the surface bound states, such as the dispersion and the surface-condition dependence of the surface states. With increasing the specularity, they observed the surface condition dependence of surface bound states that is the reduction of the zero energy density of states and the behavior of $\Delta^{\ast}-\Delta _0 \rightarrow 0$~\cite{murakawaPRL2009,murakawaJPSJ2011,okuda}. All the measurements are well explained by the quasiclassical Keldysh theory with random $S$-matrix model for surface roughness~\cite{nagatoJLTP2007,nagai}.

\subsection{Quantized thermal Hall conductivity}
\label{sec:thermal}

In Sec.~\ref{sec:topoMajo}, we have illustrated that there exist $2+1$ massless or massive Majorana fermions in the surface of $^3$He confined in a slab geometry and the low-energy surface states are described by the effective action $\mathcal{S}_{\rm surf}$ in Eq.~(\ref{eq:action}). The effective mass is associated with the spontaneous breaking of the hidden ${\bm Z}_2$ symmetry. The nontrivial property of the massive Majorana fermions is characterized by the topological invariant $N_{2}$ introduced in Eq.~(\ref{eq:topologicalN}).

Here let us remind a similar situation in topological insulators: Three-dimensional topological insulators are accompanied by two-dimensional Dirac fermions. When magnetic impurities are sprinkled in the surface region, the Dirac fermion acquires an effective mass. Similarly with $^3$He-B confined in a slab under a magnetic field, the massive Dirac fermion yields the nontrivial topological property characterized by $N_{2}={\rm sgn}(M)/2$, where $M$ is the effective mass of the Dirac fermion. This is responsible for the half-quantum Hall effect that the Hall conductivity $\sigma _{\rm H}$ is quantized in units of $e^2/h$ as $\sigma _{\rm H} = N_{2}e^2/h$~\cite{qiRMP2011}. This is a manifestation of topological nontrivial nature of surface Dirac fermions emergent in three-dimensional topological insulators. 

For superconductors and superfluids, however, the ${\rm U}(1)$ gauge symmetry is spontaneously broken. This implies that the Hall conductivity is not quantized even if $N_{2}$ is nonzero. Instead, in the case of unconventional superconductors, it has been pointed out that the spin Hall conductivity is quantized when the spin-rotation symmetry is preserved~\cite{read,senthil}. However, this cannot be applied to the case of the superfluid $^3$He-B, because the surface Majorana fermion spontaneously breaks the ${\rm SU}(2)$ symmetry in the spin space. Even though the ${\rm U}(1)$ gauge symmetry and spin rotation symmetry are absent, massive two-dimensional Majorana fermions that are bound to the surface of $^3$He-B carry the Hall component of the nondissipative thermal transport. This is the consequence of the energy conservation. Hence, the massive Majorana fermion carries the quantization of the thermal Hall conductivity~\cite{read,wangPRB2011,ryuPRB2012,nomuraPRL2012,shiozakiPRL2013},
\beq
\kappa _{\rm H} = N_{2} \frac{\pi^2 k^2_{\rm B}}{6h}T.
\eeq
Note that Shiozaki and Fujimoto~\cite{shiozakiPRL2013} clarified the relation between bulk winding number of three-dimensional topological superconductors and the thermal response. 

This quantized transport quantity manifests the nontrivial topological property of massive Majorana fermions emergent in $^3$He-B. However, we also notice that in contrast to superconductors, Nambu-Goldstone modes remain gapless in the case of superfluids. The order parameter fluctuation modes contribute to the thermal conductivity through the vertex corrections~\cite{graf}, which may deviate $\kappa _{\rm H}$ from the quantized value. We also notice that the thermal response of the $^3$He-A in a thin film that is a time-reversal breaking topological superfluid is characterized by the Chern number~\cite{sumiyoshi}.





\subsection{Spin dynamics in a restricted geometry}

Probing the spin dynamics has been a fingerprint to determine the structure of the Cooper pair states of superfluid $^3$He. The theoretical study on the spin dynamics was initiated by Leggett who proposed the coupled equations of motions for the spin and order parameters of the superfluid, ${\bm S}$ and ${\bm d}$, respectively~\cite{leggett1974,leggett-takagi1,leggett-takagi2}. The theory has succeeded in explaining the NMR properties of the superfluid $^3$He in both the linear and nonlinear regimes~\cite{leggettRMP,wheatleyRMP}. The keys to understand them are the nuclear dipole-dipole interaction and the spontaneous breaking of relative spin-orbit rotation symmetry. The contributions of all the spins that have no correlation on their directions average to zero and the resulting magnetic field generated by the dipole interaction vanishes in the lowest order of the perturbation theory with respect to the nuclear dipole constant $g_{\rm D}\equiv \mu^2_{\rm n}/a^3$, where $a$ is the mean interatomic distance. As a result, the NMR frequency in the normal $^3$He is the Larmor frequency $\omega _{\rm L}=\gamma H$, where $\gamma$ is the gyromagnetic ratio of $^3$He nuclei. In the superfluid phases of $^3$He, however, the spin rotation symmetry is spontaneously broken. The symmetry breaking generates the nuclear dipolar field, which is responsible for a large shift of the NMR frequency~\cite{leggett1973}. Indeed, the longitudinal frequency shift in the $^3$He-B is given by $\Omega _{\rm B}\equiv\omega - \omega _{\rm L}  = (3\pi\gamma^2\Delta^2_{\rm B}(T)/2g^2\chi _{\rm B})^{1/2}$ which is distinguishable from that of the $^3$He-A, $\Omega _{\rm A}$, as $\Omega^2_{\rm B}/\Omega^2_{\rm A}=5/2$~\cite{leggett1974}.

It has also been clarified that the Leggett equation well describes the nontrivial nonlinear phenomenon that is the ringing of magnetization after a sudden change of an applied field, $H\rightarrow H+\delta H$. The linear and nonlinear ringing phenomena without damping have been theoretically studied by Maki and Tsuneto for the A phase~\cite{makiPTP1974}, Brinkman~\cite{brinkmanPL1974} and Maki and Hu~\cite{makiJLTP1975v1,makiJLTP1975v2}, independently, for the B phase. For the bulk $^3$He-B where $\hat{\bm n}=\hat{\bm z}$ and $\varphi = \cos^{-1}(-1/4)$ in the equilibrium at zero fields, the change of the magnetization after a sudden application of $\delta H$ along the $\hat{\bm z}$-axis generates a torque on the spin axis, while the $\hat{\bm n}$-vector is fixed to be parallel to the field ${\bm H}\parallel\hat{\bm z}$, i.e.,  $\partial _t \hat{\bm n}(t) = 0$. The angle $\varphi(t)$ oscillates with the longitudinal resonance frequency $\Omega _{\rm B}$, when $\delta H \ll \Omega _{\rm B}/\gamma$~\cite{makiJLTP1975v1}. In the opposite limit where $\delta H \gg \Omega _{\rm B}/\gamma $, the ringing frequency approaches $\delta \omega =\gamma \delta H$. Webb {\it et al.}~\cite{webbPL1974,webbPRL1974} first observed the linear and nonlinear ringing phenomena in both $^3$He-A and B confined in a cylindrical container with a magnetic field parallel to the wall. The ringing mode is generated by using the technique of a sudden change of the field. The experimental results are in qualitatively good agreement with the theoretical predictions. 

A remarkable observation was made by Webb {\it et al.}~\cite{webbJLTP1977}, who experimentally determined the critical field above which the $\hat{\bm n}$-vector is tilted from the surface normal. They extracted the temperature dependence of $H^{\ast}$ in the vicinity of $T_{\rm c}$ through the systematic studies of the properties of the ``wall-pinned'' ringing mode in $^3$He-B. Here, the liquid $^3$He is confined in a single slab cavity (a long rectangular cavity with $1.0 {\rm mm}\times 10.0 {\rm mm} \times 23 {\rm mm}$) with the thickness $1.0$ mm, and both $H$ and $\delta H$ are parallel to the walls. Hence, the critical field observed by Webb {\it et al.}~\cite{webbJLTP1977} is nothing but $H^{\ast}$ at which the topological phase transition occurs together with the spontaneous symmetry breaking (see Figs.~\ref{fig:phase_topo} and \ref{fig:nvec}). 

The wall-pinned ringing mode in the $^3$He-B was first predicted by Brinkman~\cite{brinkmanPL1974} and Maki and Hu~\cite{makiJLTP1975v1,makiJLTP1975v2}, independently. The damping effects were taken into account by Leggett~\cite{leggettPRL1975} and Maki and Ebisawa~\cite{makiPRB1976}. 
Let us now consider the experimental situation in Ref.~\cite{webbJLTP1977}, where a static magnetic filed is applied along the wall, ${\bm H}\parallel\hat{\bm x}$. In the low field regime, $H \ll H^{\ast}$, the $\hat{\bm n}$-vector and $\varphi$ are forced by the dipole interaction energy to be $\hat{\bm n}\parallel\hat{\bm z}$ and $\varphi _{\rm L}=\varphi = \cos^{-1}(-1/4)$, while the spin is parallel to the applied field ${\bm S}\parallel\hat{\bm x}$. This configuration corresponds to the symmetry protected topological phase with $\hat{\ell}_z=0$ (see Fig.~\ref{fig:nvec}). The wall-pinned ringing mode is generated by a sudden removal of the static field. In the limit of $H,\delta H \ll \Omega _{\rm B}/\gamma$, since the gain of the kinetic energy provided by the field change, $E_{\rm kin}=\frac{1}{2}\chi (\delta H)^2$, is much smaller than the dipole interaction energy (\ref{eq:fdip}), the spin dynamics is constrained to be on the local minima of the dipole interaction energy so that $\varphi (t) = \varphi _{\rm L}$ is fixed for all time. The Leggett equation with the constraint has a solution with $\partial _t (\hat{\bm n}\cdot{\bm S})=0$ and $\partial _t {\bm S}\propto \hat{\bm n}$~\cite{brinkmanPL1974}. The wall-pinned mode corresponds to the mutual rotation of $\hat{\bm n}$ and ${\bm S}$ where the total magnetization is conserved. In the weak field limit, the magnetization harmonically oscillates with the ringing frequency $\omega _r = \sqrt{2/5} (\gamma \delta H)$. 

In Ref.~\cite{webbJLTP1977}, the wall-pinned mode is no longer observed for $H>H^{\ast}$, which implies that $\hat{\bm n}$ is tilted by the magnetic field energy from the surface normal and the non-topological phase without the ${\bm Z}_2$ symmetry (i.e., $\hat{\ell}_z = +1$) is realized. The critical field observed in Ref.~\cite{webbJLTP1977} is around 10 G in the vicinity of $T_{\rm c}$, which is the same order as $H^{\ast}\sim 20$-$30$ G obtained from the microscopic calculation in Sec.~\ref{sec:numerical2}. However, we would like to mention that the experiments were done in the narrow temperature range $1-T/T_{\rm c} \lesssim 0.015$ and the surface is not coated by the $^4$He layer, i.e., the diffusive surface. The observation of $H^{\ast}$ in the whole temperature region and the effect of the surface condition remain as unresolved problems. 

For $H\gg H^{\ast}$, the $\hat{\bm n}$-vector texture that satisfies $\hat{\ell}_z = \pm 1$ was observed by transverse NMR measurements in a parallel plate geometry~\cite{ahonenJLTP1976,ishikawaJLTP1989}. In both the experiments, the magnetic field is applied along the plates. NMR techniques have been developed to reveal the phase diagram in a restricted geometry~\cite{kawae,miyawaki,kawasaki,levitin,bennettJLTP2010,levitin2013,levitinPRL2013}. Most recently, Levitin {\it et al.}~\cite{levitin2013,levitinPRL2013} has succeeded in uncovering the $\hat{\bm n}$-textures and the confinement-induced order parameter distortion in a thin slab well-controlled surface condition. In these experiments, a magnetic field is perpendicular to the surface, where the ${\bm Z}_2$ symmetry that is the combination of the time-reversal and spin $\pi$-rotation is explicitly broken. Using a sensitive SQUID NMR spectrometer, Levitin {\it et al.}~\cite{levitin2013,levitinPRL2013} observed in the B-phase positively and negatively shifted NMR signals. The former is attributed to the configuration of $\hat{\ell}_z = +1$ that minimizes both the magnetic and dipole energies, while the negative shift is explained by the configuration of $\hat{\ell}_z = -1$ where $\hat{\bm n} \perp \hat{\bm H}\parallel\hat{\bm z}$ and $\varphi = \pi$. The latter configuration does not minimize the nuclear dipole energy, but may be stable in a magnetic field much higher than the dipolar field ($H \gg H_{\rm D}\sim 50$G).

\subsection{Electron spin relaxation}

As discussed in Sec.~\ref{sec:majorana}, the helical Majorana fermion is insensitive to the local density fluctuation. This implies that the Majorana fermion does not alter the mobility of electron bubbles injected into the free surface of $^3$He~\cite{ikegami}. The Ising-like magnetic response of the surface bound states is the direct consequence of nontrivial topological property of the superfluid $^3$He-B. Chung and Zhang~\cite{chungPRL2009} proposed the most direct way to detect the Majorana Ising spin through an electron spin relaxation experiment. The experimental setup may be realized by injecting electron bubbles below the free surface of the liquid $^3$He. The spin of the injected electron bubbles interact with the spins of $^3$He nuclei through the magnetic dipole-dipole interaction. Therefore, the relaxation of the spins of injected electron bubbles reflects the anomalous spin anisotropy of low-energy surface bound states, when the electrons are injected in the vicinity of the surface. Owing to the Majorana Ising spin, the spin relaxation time $T_1$ is determined by the orientation of the applied Zeeman field as 
\beq
T^{-1}_1 \propto \sin^2\theta _{\bm H},
\label{eq:T1}
\eeq 
in low temperatures~\cite{chungPRL2009,mizushimaJLTP2011}, where $\theta _{\bm H}$ is the angle of the applied field tilted from the surface normal (the geometry is depicted in Fig.~\ref{fig:phase_slab}). Hence, this offers a spectroscopy for spin susceptibility of topologically protected Majorana fermons that are bound to the surface. 

The local relaxation time $T_1({\bm r},\theta _{\bm H})$ is obtained from the spin-spin correlation function $\chi _{\mu\nu}({\bm r}_1,{\bm r}_2;\omega _n) \!\equiv\! \langle\langle S_{\mu}({\bm r}_1)S_{\nu}({\bm r}_2) \rangle\rangle _{\omega _n} $ with the local spin operator $S_{\mu}({\bm r})$. The explicit expression for $T^{-1}_1$ is given as~\cite{chungPRL2009,mizushimaJLTP2011}
\beq
\frac{1}{T_1({\bm r},\theta _{\bm H})} \!=\! T\lim _{\omega \rightarrow 0}
\frac{{\rm Im}\chi _{\theta _{\bm H}}({\bm r},{\bm r};\omega)}{\omega},
\eeq
where $\chi _{\theta _{\bm H}}$ is defined as $\chi _{\theta _{\bm H}} \equiv \chi _{xx}\cos^2\theta _{\bm H} + \chi _{yy}+ \chi _{zz}\sin^2\theta _{\bm H}$. This formalism takes account of all quasiparticle states including the continuum states. For simplicity, we here assume a contact interaction between injected electrons and $^3$He atoms. We also consider a weak filed regime in which the $\hat{\bm n}$-vector is aligned along the $\hat{\bm z}$-axis that is normal to the surface. In accordance with the Majorana Ising nature in Eqs.~(\ref{eq:MIS}) and (\ref{eq:chiM}), then, only the $\chi _{zz}$ component remains nonzero, resulting in Eq.~(\ref{eq:T1}). 

\begin{figure}[tb!]
\begin{center}
\includegraphics[width=120mm]{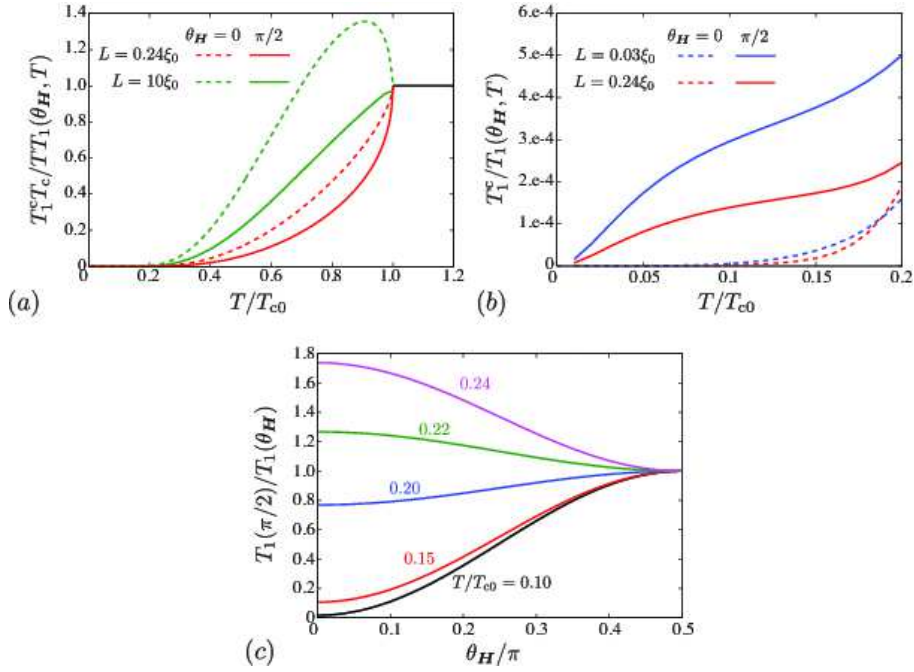}
\end{center}
\caption{Relaxation time $T_1(\theta _{\bm H})$ of injected electrons through the interaction with $^3$He atoms: (a) $T$-dependence of $1/TT_1(\theta _{\bm H})$, (b) $1/T_1(\theta _{\bm H})$, and (c) $\theta _{\bm H}$-dependence at $L \!=\! 0.24\xi$. Here, $L$ denotes the depth of the injected electrons from the surface. In (a), $T_1(\theta _{\bm H})$ is scaled by the relaxation time at $T\!=\! T_{c0}$, $T^{c}_1$. The parameters are set to be $k_{\rm F} \xi _0 \!=\! 20$ and $D=20\xi _0$. Figures adapted from Ref.~\cite{mizushimaJLTP2011}.}
\label{fig:t1}
\end{figure}

We present in Fig.~\ref{fig:t1} the numerical results on $T^{-1}_1(\theta _{\bm H})$ obtained from the full numerical calculation of the BdG equation with a specular boundary condition. Figure~\ref{fig:t1}(a) shows the temperature-dependence of $1/TT_1(\theta _{\bm H})$ for $L=0.24 \xi _0$ and $10 \xi_0$, where $L$ is the depth of the injected electrons from the surface. The coherence peak is observed only in the case of $\theta _{\bm H}=0$ and $L=10\xi_0$. This corresponds to the situation that the electrons are injected in the central region of the slab and the applied field perpendicular to the surface opens a finite energy gap in the Majorana cone. The coherence peak that characterizes the full gap nature of the low-energy quasiparticle states is however fragile against the orientation of the magnetic field. Indeed, the coherence peak disappears in the case of $\theta _{\bm H}=\pi/2$ where helical Majorana fermions remain gapless.  It is seen from Fig.~\ref{fig:t1}(b) that $T^{-1}_1$ is strongly suppressed in the low temperature region, when the DC field is applied along the surface normal direction ($\theta _{\bm H}=0$). As ${\bm H}$ is tilted from the surface normal direction, the relaxation time $T_1$ decreases, which indicates that the surface Majorana fermions contribute to the spin relaxation of electron bubbles.

This anisotropy is well explained by the concept of the Majorana Ising spin that leaves only $\chi _{zz}$ finite in $\chi _{\theta _{\bm H}}$. The $\theta _{\bm H}$-dependence is displayed in Fig.~\ref{fig:t1}(c), where $T^{-1}_1(\theta _{\bm H}) \!\propto\! \sin^2\theta _{\bm H}$ is observed in low temperature regime $T\!\lesssim\! 0.2T_c$. The contribution of the continuum states with $|E| \!>\! \Delta _0$ is negligibly small in this temperature regime. As temperature increases, however, the anisotropic behavior of $T_1$ is spoilt by the thermal excitation of the continuum states. It should be mentioned that the coherence peak which appears in $T^{-1}_1(\theta _{\bm H})$ of the bulk $^3$He-B around $T\!=\! T_c$ disappears in $T^{-1}_1(\theta _{\bm H})$ in the surface the surface~\cite{mizushimaJLTP2011}.

The coupling of a magnetic impurity to Majorana fermions bound at the edge of two-dimensional topological superconductors was also studied by Shindou {\it et al.}~\cite{shindouPRB2010}. They clarified that when a magnetic field is perpendicular to the orientation of the  Majorana Ising spin, i.e., $\theta _{\bm H}=\pi/2$, the systems can be mapped onto an Ohmic dissipative two-level system that shows the anisotropic Kondo effect. The impurity spin has a anisotropic and singular magnetic response due to the quantum dissipation from the background Majorana Ising spins.





\section{Concluding remarks}
\label{sec:summary}

The superfluid $^3$He-B confined in a slab geometry offers a prototypical system to study the interplay between topology and symmetry. In this paper, we have emphasized that the superfluid possesses unique topological phenomena associated with the intertwining of the topological phase transition with the spontaneous symmetry breaking. The topological superfluid $^3$He-B is accompanied by helical Majorana fermions that are bound the surface. We have unveiled the multifaceted properties of the gapless surface states as the symmetry-protected Majorana fermion in Secs.~\ref{sec:topo} and \ref{sec:field}, the Andreev bound state in Sec.~\ref{sec:andreev}, and odd-frequency pair amplitudes in Secs.~\ref{sec:numerical} and \ref{sec:exact2}. Based on symmetry consideration and microscopic calculation, we offer the complete topological phase diagram of the superfluid $^3$He-B confined in a slab geometry under a parallel magnetic field. It is demonstrated that there is the critical field at which the topological phase transition takes place together with the spontaneous symmetry breaking. The helical Majorana fermions and odd-frequency Cooper pair amplitudes emergent in the surface give rise to the anomalous magnetic response and anomalous quantum criticality at the critical field.

Lastly, we would like to mention the issues of which we do not take account here: The effect of surface boundary condition~\cite{zhangPRB1987,thuneberg1992,buchholtz1979,buchholtz1986,buchholtz1991,buchholtz1993,nagatoJLTP1996,nagatoJLTP1998,nagatoJLTP2007} and the possibility of the stripe phase in the vicinity of the A-B phase boundary~\cite{vorontsovPRL2007}. In the absence of a magnetic field, the surface density of states in the low energy region is considerably enhanced by the diffusive surface~\cite{vorontsovPRB2003,nagatoJLTP1998}. The low-energy density of states filled in by the skew scattering of the quasiparticle at the rough surface might drastically change the temperature- and field-dependences of the spin susceptibility. Note that the specularity of the surface of $^3$He can be experimentally controlled by coating it with $^4$He layers~\cite{murakawaPRL2009,murakawaJPSJ2011}. Furthermore, it has been predicted that the vicinity of the A-B phase transition ($D\!\sim\! 10\xi _0$) is occupied by the stripe phase~\cite{vorontsovPRL2007}, when the magnetic field is absent. However, the robustness of the stripe phase against a Zeeman field and surface roughness is not trivial, which remains as a future problem. 

Although in this paper we focus on the topological superfluidity of the superfluid $^3$He-B, other anisotropic superfluid states are competitive to the isotropic BW state. The ABM and planar states can be stabilized in a thin film that the thick is comparable with the superfluid coherence length~\cite{mizushimaPRB2012,buchholtzPRB1981,haraPTP1986,haraJLTP1988,vorontsovPRB2003,vorontsovPRL2007,nagai,tsutsumiJPSJ2010,tsutsumiPRB2011}, while the one-dimensional polar phase may be stabilized in a narrow cylinder~\cite{fetter,aoyama}. The topology of the ABM state that spontaneously breaks time-reversal symmetry is characterized by the mirror Chern number~\cite{uenoPRL2013,SatoPhysicaE2014}. The zero-energy flat band appears in the edge, vortex, and domain wall~\cite{tsutsumiJPSJ2010,silaevPRB2012,tsutsumiPRL2008,kopninPRB1991,volovikJETP1994,misirpashaev}, which is protected by topologically stable Fermi points in momentum space~\cite{volovikJETP2011,heikkilaJETP2011}. Furthermore, the integer and half-quantized vortices are accompanied by chiral Majorana fermions protected by the mirror reflection symmetry that yield non-Abelian statistics~\cite{ivanovPRL2001,SatoPhysicaE2014}. The planar state preserving time-reversal symmetry has the additional discrete symmetry that is a combination of $\pi$ spin rotation and $\pi/2$ phase rotation. The discrete symmetry modifies the topological property from that of the BW state~\cite{makhlin}. 

The superfluid $^3$He confined in a slab geometry can be a treasure house of topological superfluidity and exotic quasiparticles. In Sec.~\ref{sec:detecting}, we have discussed several ways to detect the manifestation of the topological superfluidity and exotic quasiparticles, especially in the $^3$He-B. This includes the longitudinal and transverse sound waves, spin wave, and thermal transport. It has been well recognized in the bulk $^3$He that such nonequilibrium Fermi liquid behavior in the collisionless regime is intertwining with quasiparticles and low-lying bosonic modes that are the Nambu-Goldstone and Higgs modes. Hence, understanding such intertwining effect in quantum nonequilibrium phenomena is indispensable for detecting and manipulating topological quantum phenomena in the superfluid $^3$He, such as helical/chiral Majorana fermions and spin/mass current.

\ack


We acknowledge fruitful collaboration with M. Ichioka and T. Kawakami. We also gratefully thank W. P. Halperin, S. Higashitani, O. Ishikawa, K. Nagai, M. Nitta, R. Nomura, Y. Okuda, J. A. Sauls, J. Saunders, and Y. Tanaka for various fruitful discussions and comments. This work was supported by JSPS (Nos.~25800199, 25287085, and 26400360) and ``Topological Quantum Phenomena'' (No.~22103005 and 25103716) KAKENHI on innovation areas from MEXT.




\end{document}